\documentclass[rmp,aps,11pt,a4paper]{revtex4}

\usepackage{graphics}
\usepackage{latexsym}  
\usepackage{epsfig}

\renewcommand{\a}{\alpha}
\renewcommand{\b}{\beta}

\renewcommand{\l}{\lambda}

\newcommand{\beq}{\begin{equation}} \newcommand{\eeq}{\end{equation}}
\newcommand{\bea}{\begin{eqnarray}} \newcommand{\eea}{\end{eqnarray}}
\newcommand{\be}{\begin{enumerate}} \newcommand{\ee}{\end{enumerate}}
\newcommand{\bi}{\begin{itemize}} \newcommand{\ei}{\end{itemize}}
\newcommand{\ba}{\begin{array}} \newcommand{\ea}{\end{array}}
\newcommand{\bc}{\begin{center}} \newcommand{\ec}{\end{center}}

\newcommand{\Tr}{\mathop{\rm Tr}}           
\newcommand{\tr}{\mathop{\rm tr}}           
\newcommand{\half}{\textstyle {1\over2} \displaystyle}    
\newcommand{\third}{\textstyle {1\over3} \displaystyle}   
\newcommand{\quarter}{\textstyle {1\over4} \displaystyle} 
\newcommand{\sixth}{\textstyle {1\over6} \displaystyle}   
\newcommand{\eigth}{\textstyle {1\over8} \displaystyle}   
\newcommand{\twoth}{\textstyle {2\over3} \displaystyle}   
\newcommand{\thrqu}{\textstyle {3\over4} \displaystyle}   

\def\lsim{\mathrel{\rlap{\lower4pt\hbox{\hskip1pt$\sim$}}
    \raise1pt\hbox{$<$}}}           
\def\gsim{\mathrel{\rlap{\lower4pt\hbox{\hskip1pt$\sim$}}
    \raise1pt\hbox{$>$}}}           

\newcommand{\Dslash}{\not{\hbox{\kern-2pt $\partial$}}} 
\newcommand{\dslash}{\not{\hbox{\kern-4pt $D$}}} 

\def\inbar{\,\vrule height1.5ex width.4pt depth0pt}
\def\IR{\relax{\rm I\kern-.18em R}}
\def\IC{\relax\hbox{$\inbar\kern-.3em{\rm C}$}}
\def\Tr{\textrm{Tr}~}

\begin{document}
\title{Discrete and Continuum Quantum Gravity}

\author{Herbert W. Hamber}
\email{Herbert.Hamber@aei.mpg.de}
\affiliation{Max Planck Institute for Gravitational Physics (Albert Einstein Institute), 
\\ D-14476 Potsdam, Germany }

\begin{abstract} 
I review discrete and continuum approaches to quantized gravity,
based on the covariant Feynman path integral approach.
\end{abstract}                                                                 

\maketitle


\pagebreak[4]

\tableofcontents

\pagebreak[4]

\section{INTRODUCTION}
\label{sec:intro}

In this review article I will attempt to cover key aspects and open issues related to a consistent lattice regularized formulation of quantum gravity. Such a formulation can be viewed as an important and perhaps essential step towards a quantitative, controlled investigation of the physical content of the theory. The main emphasis of the present review will therefore rest on discrete and continuum space-time formulations of quantum gravity based on the covariant Feynman path integral approach, and their mutual interrelation. 

The first part of the review will introduce the basic elements of a covariant formulation of continuum quantum gravity, with special emphasis on those issues which bear some immediate relevance for the remainder of the work. These  include a discussion of the nature of the spin-two field, its wave equation and possible gauge choices, the Feynman propagator, the coupling of a spin two field to matter, and the implementation of a consistent local gauge invariance to all orders, ultimately leading to the Einstein action. Additional terms in the gravitational action, such as the cosmological constant and higher derivative contributions, are naturally introduced at this stage, and play some role later in the context of a full quantum theory.

A section on the perturbative (weak field) expansion introduces the main aspects of the background field method as applied to gravity, including the choice of field parametrization and gauge fixing terms. Later the results on the structure of one- and two-loop divergences in pure gravity are discussed, leading up to the conclusion of perturbative non-renormalizability for the Einstein theory in four dimensions. The relevant one-loop and two-loop counterterms will be recalled. One important aspect that needs to be emphasized is that these perturbative methods generally rely on a weak field expansion for the metric fluctuations, and are therefore not well suited for investigating the (potentially physically relevant) regime where metric fluctuations can be large.

Later the Feynman path integral for gravitation is introduced, in analogy with the closely related case of Yang-Mills theories. This, of course, brings up the thorny issue of the gravitational functional measure, expressing Feynman's sum over geometries, as well as important aspects related to the convergence of the path integral and derived quantum averages, and the origin of the conformal instability affecting the Euclidean case. An important point that needs to be emphasized is the strongly constrained nature of the theory, which depends, in the absence of matter, and as in pure Yang-Mills theories, on a single dimensionless parameter $G \lambda$, besides a required short distance cutoff.

Since quantum gravity is not perturbatively renormalizable, the following question arises naturally: what other theories are not perturbatively
renormalizable, and what can be done with them?
The following parts will therefore summarize the methods of the $2+\epsilon$ expansion for gravity, an expansion in the deviation of the space-time dimensions from two, where the gravitational coupling is dimensionless and the theory appears therefore power-counting renormalizable. 
As initial motivation, but also for illustrative and pedagogical purposes, the non-linear sigma model will be introduced first.
The latter represents a reasonably well understood perturbatively non-renormalizable theory above two dimensions, characterized by a rich two-phase structure, and whose scaling properties in the vicinity of the fixed point can nevertheless be accurately computed (via the $2+\epsilon$ expansion, as well as by other methods, including most notably the lattice) 
in three dimensions, and whose universal predictions are known to compare favorably with experiments. 
In the gravity context, to be discussed next, the main results of the perturbative expansion are the existence of a nontrivial ultraviolet fixed point close to the origin above two dimensions (a phase transition in statistical field theory language), and the predictions of universal scaling exponents 
in the vicinity of this fixed point. 

The next sections deal with the natural lattice discretization for quantum gravity based on Regge's simplicial formulation, with a primary focus on the physically relevant four-dimensional case. 
The starting point there is a description of a discrete manifold in terms
of edge lengths and incidence matrices, then moving on to a description of curvature in terms of deficit angles, thereby offering a re-formulation of continuum gravity in terms of the discrete Regge action and ensuing lattice field equations. 
The direct and clear correspondence between lattice quantities (edges, dihedral angles, volumes, deficit angles, etc.) and continuum operators (metric, affine connection, volume element, curvature tensor etc.) will be emphasized all along.  The latter will be useful in defining, as an example, discrete formulations of curvature squared terms which arise in higher derivative gravity theories, or more generally as radiatively induced corrections.
An important element in this lattice-to-continuum correspondence will be the development of the lattice weak field expansion, allowing in this context again a clear and precise identification between lattice and continuum degrees of freedom, as well as their gauge invariance properties, as illustrated in the weak field limit by the arbitrariness in the assignments of edge lengths used to cover a given physical geometry. 
The lattice analogues of gravitons arise naturally, and their 
transverse-traceless nature (in a suitable gauge) can easily be made manifest.

When coupling matter fields to lattice gravity one needs to introduce new fields localized on vertices, as well as appropriate dual volumes which enter the definition of the kinetic terms for those fields. 
The discrete re-parametrization invariance properties of the discrete matter action will be described next. In the fermion case, it is necessary (as in the continuum) to introduce vierbein fields within each simplex, and then use an appropriate spin rotation matrix to relate spinors between neighboring simplices. In general the formulation of fractional spin fields on a simplicial lattice could have some use in the lattice discretization of supergravity theories.
At this point it will also be useful to compare, and contrast, Regge's simplicial formulation to other discrete approaches to quantum gravity, such as the hypercubic (vierbien-connection) lattice formulation, and fixed-edge-length approaches such as dynamical triangulations.

The next sections deals with the interesting problem of what gravitational observables should look like, i.e. which expectation values of operators (or ratios thereof) have meaning and physical interpretation in the context of a manifestly covariant formulation, specifically in a situation where metric fluctuations are not necessarily bounded. 
Such averages naturally include expectation values of the (integrated) scalar curvature and other related quantities (involving for example curvature squared terms), as well as correlations of operators at fixed geodesic distance, sometimes referred to as bi-local operators.
Another set of physical averages refer to the geometric nature of space-time itself, such as the fractal dimension. 
One more set of physical observables correspond to the gravitational analog of the Wilson loop (providing information about the parallel transport of vectors, and therefore on the effective curvature, around large near-planar loops), and the correlation between particle world-lines (providing information about the static gravitational potential).
It is reasonable to expect that these quantities will play an important role in the physical characterization of the two phases of gravity, as seen both in the $2+\epsilon$ and in the lattice formulation in four dimensions.

There are reasons to believe that ultimately the investigation of a strongly coupled regime of quantum gravity, where metric fluctuations cannot be assumed to be small, requires the use of numerical methods applied to the lattice theory. 
A discrete formulation combined with numerical tools can therefore be viewed
as an essential step towards a quantitative and controlled investigation of the physical content of the theory:
in the same way that a discretization of a complicated ordinary differential equation can be viewed as a mean to determine the properties of its solution with arbitrary accuracy.
These methods are outlined next, together with a summary of the main lattice results, suggesting the existence of two phases (depending on the value of the bare gravitational coupling) and in agreement with the qualitative
predictions of the $2+\epsilon$ expansion. 
Specifically one finds a weak coupling degenerate polymer-like phase, and a strong coupling smooth phase with bounded curvatures in four dimensions. 
The somewhat technical aspect of the determination of universal critical
exponents and non-trivial scaling dimensions, based on finite size methods, 
is outline, together with a detailed (but by now standard) discussion of
how the lattice continuum limit has to be approached in the vicinity of 
a non-trivial ultraviolet fixed point.

The determination of non-trivial scaling dimensions in the vicinity of the fixed point opens the door to a discussion of the renormalization group properties of fundamental couplings, i.e. their scale dependence, as well as the emergence of physical renormalization group invariant quantities, such as the gravitational correlation length and the closely related gravitational condensate.
Such topics will be discussed next, with an eye towards perhaps more physical applications. 
These include a discussion on the physical nature of the expected quantum corrections to the gravitational coupling, based, in part on an analogy to qed and qcd, on the effects of a virtual graviton cloud (as already suggested in
the $2+\epsilon$ expansion context), and of how the two phases of lattice gravity relate to the two opposite scenarios of gravitational screening (for weak coupling, and therefore unphysical due to the branched polymer nature of this phase) versus anti-screening (for strong coupling, and therefore physical).

A final section touches on the general problem of formulating running gravitational couplings in a context that does not assume weak gravitational fields and close to flat space at short distance. 
The discussion includes a brief presentation on the topic of covariant running of $G$ based on the formalism of non-local field equations, with the scale dependence of $G$ expressed through the use of a suitable covariant d'Alembertian. Simple applications to standard metrics (static isotropic and homogeneous isotropic) are briefly summarized and their potential physical consequences and interpretation elaborated.

The review will end with a general outlook on future prospects for lattice studies of quantum gravity, some open questions and work that can be done to help elucidate the relationship between discrete and continuum models, such as extending the range of problems addressed by the lattice, and providing new impetus for further developments in covariant continuum quantum gravity. 

Notation: Throughout this work, unless stated otherwise, the same notation
is used as in (Weinberg, 1973), with the sign of the Riemann tensor reversed. The signature is therefore $-,+,+,+$. In the Euclidean case $t=-i\tau$ 
the flat metric is of course the Kronecker $\delta_{\mu\nu}$, with
the same conventions as before for Riemann.

\section{CONTINUUM FORMULATION }
\label{sec:continuum}

\subsection{General Aspects}
\label{sec:formulation}

The Lagrangian for the massless spin-two field can be constructed in
close analogy to what one does in the case of electromagnetism.
In gravity the electromagnetic interaction $ e \, j \cdot A$ is
replaced by a term 
\beq
\half \, \kappa \, h_{\mu\nu} (x) \, T^{\mu\nu} (x)
\label{eq:t-source}
\eeq
where $\kappa$ is a constant to be determined later,
$T^{\mu\nu} $ is the conserved
energy-momentum tensor
\beq
\partial_\mu \, T^{\mu\nu}(x) \, = \, 0
\eeq
associated with the sources,
and $ h_{\mu\nu} (x) $ describes the gravitational field.
It will be shown later that $\kappa$ is related
to Newton's constant $G$ by $\kappa = \sqrt{16 \pi G}$.

\subsubsection{Massless Spin Two Field}
\label{sec:spin2}

As far as the pure gravity part of the action is
concerned, one has in principle four
independent quadratic terms one can construct out
of the first derivatives of $h_{\mu\nu}$, namely
\bea
\partial_\sigma h_{\mu\nu} \, \partial^\sigma h^{\mu\nu} &&
\;\; , \;\;\;\; 
\partial^\nu h_{\mu\nu} \, \partial_\sigma h^{\mu\sigma}
\;\; ,
\nonumber \\
\partial^\nu h_{\mu\nu} \, \partial^\mu h_{\sigma}^{\;\;\sigma} &&
\;\; , \;\;\;\; 
\partial^\mu h_{\nu}^{\;\;\nu} \, \partial_\mu h_{\sigma}^{\;\;\sigma}
\;\; .
\eea
The term 
$\partial_\sigma h_{\mu\nu} \, \partial^\nu h^{\mu\sigma}$
need not be considered separately, as it can be shown
to be equivalent to the second term in the above list,
after integration by parts.
After combining these four terms into an action
\bea
\int dx & [ & a \, \partial_\sigma h_{\mu\nu} \, \partial^\sigma h^{\mu\nu}
+ b \, \partial^\nu h_{\mu\nu} \, \partial_\sigma h^{\mu\sigma}
\nonumber \\
&& + \, c \, \partial^\nu h_{\mu\nu} \, \partial^\mu h_{\sigma}^{\;\;\sigma}
+ d \, \partial^\mu h_{\nu}^{\;\;\nu} \, \partial_\mu h_{\sigma}^{\;\;\sigma}
\nonumber \\
&& + \half \, \kappa \, h_{\mu\nu} \, T^{\mu\nu} \, ] 
\eea
and performing the required variation with respect to 
$h_{\alpha\beta}$, one obtains for the field equations
\bea
& 2 \, a &  
\partial_\sigma \partial^\sigma h_{\alpha\beta} 
\nonumber \\
+ & b & \! ( \,
\partial_\beta \partial^\sigma h_{\alpha\sigma} 
+ \partial_\alpha \partial^\sigma h_{\beta\sigma} \, ) 
\nonumber \\
+ & c & \! ( \,
\partial_\alpha \partial_\beta h_{\sigma}^{\;\;\sigma}
+ \eta_{\alpha\beta} \,
\partial_\mu \partial_\nu h^{\mu\nu} \, ) 
\nonumber \\
+ & 2 \, d &  
\eta_{\alpha\beta} \, \partial_\mu \partial^\mu h_{\sigma}^{\;\;\sigma}
\nonumber \\
&& = \, \half \, \kappa \; T_{\alpha\beta}
\label{eq:h-field}
\eea
with $\eta_{\alpha\beta}= {\rm diag} (-1,1,1,1)$.
Consistency requires that the four-divergence of
the above expression give zero on both sides,
$\partial^\beta (\dots )=0$.
After collecting terms of the same type, one is led to the three
conditions
\bea
( 2 a + b ) \; \partial_\sigma \partial^\sigma \partial^\beta \,
\, h_{\alpha\beta} \, = \, 0
\nonumber \\
( b + c ) \; \partial_\alpha \partial^\beta \partial^\sigma \,
h_{\beta\sigma} \, = \, 0
\nonumber \\
( c + 2 d ) \; \partial_\alpha \partial_\beta \partial^\beta \,
h_{\sigma}^{\;\;\sigma}  \, = \, 0
\eea
with unique solution (up to an overall constant, which can be
reabsorbed into $\kappa$)
$a=-\quarter$, $b=\half$, $c=-\half$ and $d=\quarter$.
As a result, the quadratic part of the Lagrangian for the
pure gravitational field is given by 
\bea
{\cal L}_{sym} & = &
- \, \quarter \, \partial_\sigma h_{\mu\nu} \, \partial^\sigma h^{\mu\nu}
+ \, \half \, \partial^\nu h_{\mu\nu} \, \partial_\sigma h^{\mu\sigma}
\nonumber \\
&& - \, \half \,
\partial^\nu h_{\mu\nu} \, \partial^\mu h_{\sigma}^{\;\;\sigma}
+ \, \quarter \, 
\partial^\mu h_{\nu}^{\;\;\nu} \, \partial_\mu h_{\sigma}^{\;\;\sigma}
\label{eq:h-action}
\eea

\subsubsection{Wave Equation}
\label{sec:wave}

One notices that the field equations of Eq.~(\ref{eq:h-field}) take on a 
particularly simple form if one introduces
trace reversed variables $\bar h_{\mu\nu}(x) $,
\beq
\bar h_{\mu\nu} = h_{\mu\nu} - \half \, \eta_{\mu\nu} \, h_\sigma^{\;\;\sigma}
\eeq
and
\beq
\bar T_{\mu\nu} = T_{\mu\nu} - \half \, \eta_{\mu\nu} \, T_\sigma^{\;\;\sigma}
\label{eq:t-bar}
\eeq
In the following it will be convenient to write the trace as 
$h = h_{\;\;\sigma}^\sigma$ so that $\bar h^\sigma_{\;\;\sigma} = - h$, 
and define the d'Alembertian as
$\Box = \partial_\mu \partial^\mu = \nabla^2 - \partial_t^2 $.
Then the field equations become simply 
\beq
\Box \, h_{\mu\nu} 
- 2 \, \partial_\nu \, \partial^\sigma \,
\bar h_{\mu\sigma} \, = \, - \kappa \, \bar T_{\mu\nu}
\label{eq:h-field1}
\eeq
One important aspect of the field equations is that
they can be shown to be invariant under a local gauge
transformation of the type
\beq
h_{\mu\nu}'  \, = \, h_{\mu\nu} 
+ \partial_\mu \, \epsilon_\nu
+ \partial_\nu \, \epsilon_\mu
\label{eq:h-gauge}
\eeq
involving an arbitrary gauge parameter $\epsilon_\mu (x)$.
This invariance is therefore analogous to the local
gauge invariance in QED,
$A_{\mu}' =  A_{\mu} + \partial_\mu \epsilon $.
Furthermore, it suggests choosing a suitable gauge (analogous to
the familiar Lorentz gauge $\partial^\mu A_\mu =0$)
in order to simplify the field equations, for example
\beq
\partial^\sigma \, \bar h_{\mu\sigma} = 0
\label{eq:h-gauge-fix}
\eeq
which is usually referred to as the harmonic gauge
condition.
Then the field equations in this gauge become simply
\beq
\Box \, h_{\mu\nu} \, = \, - \kappa \, \bar T_{\mu\nu}
\label{eq:h-field2}
\eeq
These can then be easily solved in momentum space
($\Box \rightarrow -k^2$) to give
\beq
h_{\mu\nu} \, = \, \kappa \, { 1 \over k^2 } \, \bar T_{\mu\nu}
\label{eq:h-source}
\eeq
or, in terms of the original $T_{\mu\nu}$,
\beq
h_{\mu\nu} \, = \, \kappa \, { 1 \over k^2 } \, 
( \, T_{\mu\nu} - \half \, \eta_{\mu\nu} T_{\sigma}^{\;\;\sigma} \, )
\eeq
It should be clear that this gauge is particularly convenient for
practical calculations, since then graviton propagation
is given simply by a factor of $1/k^2$;
later on gauge choices will be introduced where this is
no longer the case.

Next one can compute the amplitude for the interaction of 
two gravitational sources characterized by energy-momentum
tensors $T$ and $T'$.
From Eqs.~(\ref{eq:t-source}) and (\ref{eq:h-source}) one has
\beq
\half \, \kappa \, T_{\mu\nu}' \,  h^{\mu\nu} 
= \half \, \kappa^2 \, 
T_{\mu\nu}' \,  { 1 \over k^2 } \, \bar T^{\mu\nu}
\eeq
which can be compared to the electromagnetism
result $ j_{\mu}' { 1 \over k^2 } \, j^{\mu}$.

To fix the value of the parameter $\kappa$ it is easiest to
look at the static case, for which the only
non-vanishing component of $T_{\mu\nu}$ is $T_{00}$.
Then
\beq
\half \, \kappa^2 \, 
T_{00}' \;  { 1 \over k^2 } \; \bar T^{00}
\, = \, \half \, \kappa^2 \, 
T_{00}' \;  { 1 \over k^2 } \; 
( T_{00} - \half \, \eta_{00} T_{0}^{\;\;0} )
\label{eq:h-int}
\eeq
For two bodies of mass $M$ and $M'$ the static
instantaneous amplitude (by inverse Fourier
transform, thus replacing 
${4 \pi \over {\bf k}^2 } \rightarrow { 1 \over r}$)
then becomes
\beq
- \, \half \, \kappa^2 \, \half \, M' { 1 \over 4 \pi r } \, M'
\eeq
which, by comparison to the expected Newtonian
potential energy $-GMM'/r$, gives the desired
identification $\kappa = \sqrt{16 \pi G}$.

The pure gravity part of the action in Eq.~(\ref{eq:h-action})
only propagates transverse traceless modes (shear waves).
These correspond quantum mechanically to a particle of zero mass and spin two, 
with two helicity states $h=\pm 2$,
as shown for example in (Weinberg, 1973) by looking at the nature
of plane wave solutions $h_{\mu\nu}(x) = e_{\mu\nu} \, e^{i k \cdot x}$
to the wave equation in the harmonic gauge.
Helicity 0 and $\pm 1$ appear initially, but can be made to vanish by
a suitable choice of coordinates.

\begin{figure}[h]
\epsfig{file=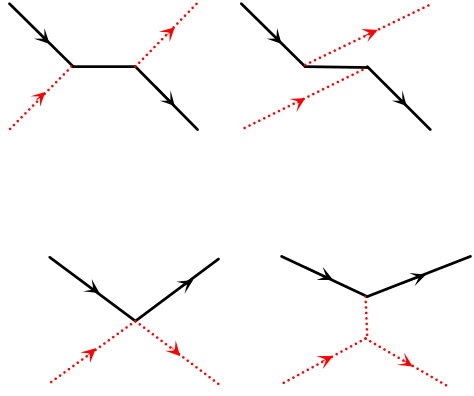,width=8cm}
\caption{Lowest order diagrams illustrating the gravitational analog
to Compton scattering. 
Continuous lines indicate a matter particle, short dashed lines a 
graviton.
Consistency of the theory requires that the two bottom diagrams
be added to the two on the top.}
\label{fig:compton}
\end{figure}

One would expect the gravitational field $h_{\mu\nu}$ to carry energy
and momentum, which would be described by a 
tensor $\tau_{\mu\nu} (h) $.
As in the case of electromagnetism, where one has
\beq
T_{\alpha\beta}^{(em)} = F_{\alpha\gamma} F_{\beta}^{\;\;\gamma}
- \quarter \, \eta_{\alpha\beta} 
F_{\gamma\delta} F^{\gamma\delta} \;\; ,
\eeq
one would also expect such a tensor to be quadratic
in the gravitational field $h_{\mu\nu}$.
A suitable candidate for the energy-momentum tensor
of the gravitational field is
\beq
\tau_{\mu\nu} = {1 \over 8 \pi G } \,
\left ( - \quarter \, h_{\mu\nu} \,
\partial^\lambda \partial_\lambda h^{\sigma}_{\;\;\sigma} 
\, + \, \dots \right )
\eeq
where the dots indicate 37 possible additional terms, involving 
schematically, either terms of the type $h \, \partial^2 h$,
or of the type $(\partial h)^2$.
Such a $\tau_{\mu\nu}$ term would have to be added on the r.h.s.
of the field equations in Eq.~(\ref{eq:h-field1}), and would 
therefore act as an additional source for the gravitational field
(see Fig.~\ref{fig:compton}).
But the resulting field equations would then no
longer invariant under Eq.~(\ref{eq:h-gauge}),
and one would have to change therefore the gauge
transformation law by suitable terms of order $h^2$,
so as to ensure that the new field equations
would still satisfy a local gauge invariance.
In other words, all these complications arise because
the gravitational field carries energy and momentum, and
therefore gravitates.




Ultimately, a complete and satisfactory answer to these recursive
attempts at constructing a consistent, locally gauge 
invariant, theory of the $h_{\mu\nu}$ field is found
in Einstein's non-linear General Relativity theory, as shown in
(Feynman, 1962; Boulware and Deser, 1969).
The full theory is derived from the Einstein-Hilbert action 
\beq
I_E \; = \; { 1 \over 16 \pi G } \int d x \, \sqrt{g (x)} \, R(x)
\label{eq:eh-action}
\eeq
which generalized Eq.~(\ref{eq:h-action}) beyond the
weak field limit.
Here $\sqrt{g}$ is the square root of the determinant of the metric
field $g_{\mu\nu} (x)$, with $g = - \det g_{\mu\nu}$,
and $R$ the scalar curvature.
The latter is related to the Ricci tensor $R_{\mu\nu}$ 
and the Riemann tensor $R_{\mu\nu\lambda\sigma}$ by
\bea
R_{ \mu \nu } & = & g^{ \lambda \sigma }  R_{ \lambda \mu \sigma \nu }
\nonumber \\
R & = & g^{ \mu \nu }  g^{ \lambda \sigma } 
R_{ \mu \lambda \nu \sigma }
\eea
where $g^{\mu\nu}$ is the matrix inverse of $g_{\mu\nu}$,
\beq
g^{\mu\lambda} \, g_{\lambda\nu} = \delta^{\mu}_{\;\;\nu}
\eeq
In terms of the affine connection $\Gamma^\lambda_{\mu\nu}$,
the Riemann tensor $ R_{ \; \mu \nu \sigma }^\lambda (x) $ is given
by
\beq
R_{ \; \mu \nu \sigma }^\lambda \; = \;
\partial_\nu \Gamma_{ \mu \sigma }^\lambda 
- \partial_\sigma \Gamma_{ \mu \nu }^\lambda
+ \Gamma_{ \mu \sigma }^\eta  \Gamma _{ \nu \eta }^\lambda
- \Gamma_{ \mu \nu }^\eta  \Gamma_{ \sigma \eta }^\lambda
\eeq
and therefore
\beq
R_{\mu \nu} \; = \;
\partial_\sigma \Gamma_{ \mu \nu }^\sigma -
\partial_\nu \Gamma_{ \mu \sigma }^\sigma 
+ \Gamma_{ \sigma \lambda }^\lambda \Gamma_{\mu\nu}^\sigma
- \Gamma_{ \sigma \nu }^\lambda \Gamma _{ \mu \lambda }^\sigma \;\; ,
\eeq
with the affine connection $\Gamma_{\mu\nu}^\lambda (x) $
in turn constructed from components of the metric field $g_{\mu\nu}$  
\beq
\Gamma_{ \mu \nu }^\lambda = \half \; g^{ \lambda \sigma }
\Bigl ( \partial_\mu g_{ \nu \sigma } 
+ \partial_\nu g_{ \mu \sigma } 
- \partial_\sigma g_{ \mu \nu } \Bigr )
\eeq
The following algebraic symmetry properties of the Riemann tensor
will be of use later 
\beq
R_{ \mu \nu \lambda \sigma } = - R_{ \nu \mu \lambda \sigma }
= -  R_{ \mu \nu \sigma \lambda } = R_{ \nu \mu \sigma \lambda }
\eeq
\beq 
R_{ \mu \nu \lambda \sigma } = R_{ \lambda \sigma \mu \nu }
\eeq
\beq
R_{ \mu \nu \lambda \sigma } + R_{ \mu \lambda \sigma \nu } +
R_{ \mu \sigma \nu \lambda } = 0
\eeq
In addition, the components of the Riemann tensor satisfy the 
differential Bianchi identities
\beq
\nabla_\alpha R_{\mu\nu\beta\gamma} +
\nabla_\beta  R_{\mu\nu\gamma\alpha} +
\nabla_\gamma R_{\mu\nu\alpha\beta} = 0
\eeq
with $\nabla_\mu$ the covariant derivative.
It is known that these, in their contracted form, ensure
the consistency of the field equations.
From the expansion of the Einstein-Hilbert
gravitational action in powers of the deviation of
the metric from the flat metric $\eta_{\mu\nu}$,
using 
\bea
R_{\mu\nu} & = & \half (  \partial^2 h_{\mu\nu}
- \partial_\alpha \partial_\mu h^{\alpha}_{\;\;\nu} 
- \partial_\alpha \partial_\nu h^{\alpha}_{\;\;\nu}
+ \partial_\mu \partial_\nu h^{\alpha}_{\;\;\alpha} ) + O(h^2)
\nonumber \\
R & = & \partial^2 h^\mu_{\;\;\mu} 
- \partial_{\alpha} \partial_\mu h^{\alpha\mu} + O(h^2)
\label{eq:h-ricci}
\eea
one has for the action contribution
\bea
\sqrt{g} \, R & = & - \quarter \, 
\partial_\sigma h_{\mu\nu} \, \partial^\sigma h^{\mu\nu}
+ \half \, \partial^\nu h_{\mu\nu} \, \partial_\sigma h^{\mu\sigma}
\nonumber \\
&& - \half \,
\partial^\nu h_{\mu\nu} \, \partial^\mu h_{\sigma}^{\;\;\sigma}
+ \quarter \, 
\partial^\mu h_{\nu}^{\;\;\nu} \, \partial_\mu h_{\sigma}^{\;\;\sigma}
+ O(h^3)
\label{eq:h-eh}
\eea
again up to total derivatives.
This last expression is in fact the same as Eq.~(\ref{eq:h-action}).
The correct relationship between the original graviton field
$h_{\mu\nu}$
and the metric field $g_{\mu\nu}$ is 
\beq
g_{\mu\nu} (x) = \eta_{\mu\nu} + \kappa \, h_{\mu\nu} (x)
\eeq
If, as is often customary,
one rescales $ h_{\mu\nu}$ in such a way that the $\kappa$ factor
does not appear on the r.h.s., then both the $g$
and $h$ fields are dimensionless.

The weak field invariance properties of the gravitational action of
Eq.~(\ref{eq:h-gauge}) are replaced in the general theory by
general coordinate transformations $x^{\mu} \rightarrow {x'}^{\mu}$,
under which the metric transforms as a covariant second rank tensor 
\beq
g_{\mu\nu}' (x') = 
{\partial x^\rho   \over \partial {x'}^\mu } \,
{\partial x^\sigma \over \partial {x'}^\nu } \;
g_{\rho\sigma} (x)
\label{eq:metric-law}
\eeq
which leaves the infinitesimal proper time interval
$d \tau$ with
\beq
d \tau^2 = - g_{\mu\nu} \, d x^\mu d x^\nu
\label{eq:proper-time}
\eeq
invariant.
In their infinitesimal form, coordinate transformations
are written as
\beq
x'^{\mu} = x^{\mu} + \epsilon^\mu (x)
\label{eq:x-law}
\eeq
under which the metric at the same point $x$ 
then transforms as 
\beq
\delta g_{ \mu \nu } (x) = -
g_{ \lambda \nu } (x) \, \partial_\mu \epsilon^\lambda (x) 
- g_{\lambda \mu } (x) \, \partial_\nu \epsilon^\lambda (x)
- \epsilon^\lambda  (x) \, \partial_\lambda g_{ \mu \nu } (x) 
\eeq
and which is usually referred to as the Lie derivative of $g$.
The latter generalizes the weak field gauge invariance property
of Eq.~(\ref{eq:h-gauge}) to all orders in $h_{\mu\nu}$.

For infinitesimal coordinate transformations, one can gain some additional
physical insight by decomposing the derivative of the small coordinate change
$\epsilon_\mu$ in Eq.~(\ref{eq:x-law}) as
\beq
{ \partial \epsilon_\mu \over \partial x^\nu } = 
s_{\mu\nu} + a_{\mu\nu} + t_{\mu\nu}
\eeq
with
\bea
s_{\mu\nu} & = & 
{\textstyle {1 \over d} \displaystyle}  
\, \eta_{\mu\nu} \, \partial \cdot \epsilon
\nonumber \\
a_{\mu\nu} & = & 
\half \, ( \partial_\mu \epsilon_\nu - \partial_\nu \epsilon_\mu ) 
\nonumber \\
t_{\mu\nu} & = & \half \, ( \partial_\mu \, \epsilon_\nu + \partial_\nu \, \epsilon_\mu 
- {\textstyle {2 \over d} \displaystyle}
\, \eta_{\mu\nu} \, \partial \cdot \epsilon )
\eea
Then $s_{\mu\nu}(x)$ can be thought of describing local scale transformations,
$a_{\mu\nu}(x)$ is written in terms of an antisymmeric tensor and
therefore describes local rotations, while $t_{\mu\nu}(x)$ contains a traceless
symmetric tensor and describes local shears.

Since both the scalar curvature $R(x)$ and the volume element 
$ d x  \sqrt{ g (x)}$ are separately invariant under the
general coordinate transformations of Eqs.~(\ref{eq:metric-law})
and (\ref{eq:x-law}), both of the following action contributions are
acceptable
\bea
&& \int  d x  \, \sqrt{ g(x) } 
\nonumber \\
&& \int  d x  \, \sqrt{ g(x) } \; R(x)
\eea
the first being known as the cosmological constant
contribution (as it represents the total space-time
volume). 
In the weak field limit, the first, cosmological constant term
involves 
\beq
\sqrt g = 1 + \half \, h_\mu^{\;\;\mu}
+ \eigth \, h_\mu^{\;\;\mu} h_\nu^{\;\;\nu}
- \quarter \, h_{\mu\nu} h^{\mu\nu} + O(h^3)
\label{eq:h-cosm}
\eeq
which is easily obtained from the matrix formula
\beq
\sqrt{ \det g} = \exp ( \half \tr \ln g )
= \exp [ \half \tr \ln ( \eta + h  ) ]
\eeq
after expanding out the exponential in powers of
$h_{\mu\nu}$.
We have also reverted here to the more traditional
way of performing the weak field expansion (i.e. without factors
of $\kappa$),
\bea
g_{\mu\nu} & = & \eta_{\mu\nu} + h_{\mu\nu}
\nonumber \\
g^{\mu\nu} & = & \eta^{\mu\nu} 
- h_{\mu\nu} + h_\mu^{\;\;\alpha} \, h_{\alpha\nu} + \dots
\label{eq:metric-wfe}
\eea
with $\eta_{\mu\nu}$ the flat metric.
The reason why such a $\sqrt{g}$ cosmological consant
term was not originally
included in the construction of the Lagrangian
of  Eq.~(\ref{eq:h-action}) is that it does not contain
derivatives of the $h_{\mu\nu}$ field.
It is in a sense analogous to a mass term, without giving
rise to any breaking of local gauge invariance.

In the general theory, the energy-momentum tensor
for matter $T_{\mu\nu}$ is most suitably defined
in terms of the variation of the matter action $I_{\rm matter}$,
\beq
\delta I_{\rm matter} = 
\half \, \int d x \, \sqrt{g} \; \delta g_{\mu\nu} \, T^{\mu\nu}
\label{eq:en-mom-matter}
\eeq 
and is conserved if the matter action is a scalar,
\beq
\nabla_{\mu} \, T^{\mu\nu} = 0
\eeq
Variation of the gravitational Einstein-Hilbert action
of Eq.~(\ref{eq:eh-action}), with the matter part added,
then leads to the field equations
\beq
R_{\mu\nu} - \half \, g_{\mu\nu} R + \lambda g_{\mu\nu} = 8 \pi G \, T_{\mu\nu}
\eeq
Here we have also added a cosmological constant term, with 
a scaled cosmological constant $\lambda= \lambda_0 / 16 \pi G$,
which follows from adding to the gravitational
action a term $ \lambda_0 \int \sqrt{g}$
\footnote{
The present experimental value for Newton's constant is 
$ \hbar G / c^3 = (1.61624(12)\times 10 ^{-33} cm)^2 $.
Recent observational evidence [reviewed in (Damour,2007)] suggests a
non-vanishing positive
cosmological constant $\lambda$, corresponding to a vacuum density
$\rho_{vac} \approx (2.3 \times 10^{-3} eV )^4 $ 
with $\rho_{vac}$ related to $\lambda$ by $\lambda = 8 \pi G \rho_{vac} / c^4 $.
As can be seen from the field equations, $\lambda$ has the same
dimensions as a curvature.
One has from observation $ \lambda \sim  1/(10^{28} cm )^2 $,
so this new curvature length scale is comparable
to the size of the visible universe $\sim 4.4 \times 10^{28} cm$.
}.

One can exploit the freedom under general coordinate
transformations $x'^\mu = f(x^\mu)$ to impose a suitable
coordinate condition, such as
\beq
\Gamma^{\lambda} \equiv g^{\mu\nu} \, \Gamma^\lambda_{\mu\nu} =0
\eeq
which is seen to be equivalent to the following gauge
condition on the metric 
\beq
\partial_{\mu} ( \sqrt{g} \, g^{\mu\nu} ) = 0
\label{eq:harm-gauge}
\eeq
and therefore equivalent, in the weak field limit,
to the harmonic gauge condition introduced previously
in Eq.~(\ref{eq:h-gauge-fix}).

\subsubsection{Feynman Rules}
\label{sec:fey}

The Feynman rules represent the standard way to do perturbative
calculations in quantum gravity.
To this end one first expands again the action out in powers of the
field $h_{\mu\nu}$ and separates out the quadratic part, which gives
the graviton propagator, from the rest of the Lagrangian which gives
the $O(h^3), O(h^4) \dots$ vertices.
To define the graviton propagator one also requires the addition
of a gauge fixing term and the associated Faddeev-Popov ghost
contribution (Feynman, 1962; Faddeev and Popov, 1968).
Since the diagrammatic calculations are performed using
dimensional regularization, one first needs to define the theory
in $d$ dimensions;
at the end of the calculations
one will be interested in the limit $d \rightarrow 4$.

So first one expands around the $d$-{\it dimensional} 
flat Minkowski space-time metric, with signature given by
$\eta_{\mu\nu}={\rm diag}(-1,1,1,1, \dots)$.
The Einstein-Hilbert action in $d$ dimensions is given by
a generalization of Eq.~(\ref{eq:eh-action})
\beq
I_{\rm E} = {1\over 16\pi G} \int d^d x \,
\sqrt{g(x)} \, R(x)\, \;\; ,
\label{eq:eh-action-d}
\eeq
with again $g(x)=-{\rm det}(g_{\mu\nu})$ and $R$ the scalar curvature;
in the following it will be assumed, at least initially, that the bare 
cosmological constant $\lambda_0$ is zero.
The simplest form of matter coupled in an invariant way to gravity is a set of
spinless scalar particles of mass $m$, with action 
\beq
I_{\rm m} \; = \; \half \int d^d x \;
\sqrt{g(x)} \,
\left [ \, - g^{\mu \nu} (x) \, 
\partial_{\mu} \phi (x) \, \partial_{\nu} \phi (x) \; - \;
m^2 \, \phi^2 (x) \, \right ] \;\; .
\label{eq:scalar-action}
\eeq
In Feynman diagram perturbation theory the metric $g_{\mu\nu} (x) $
is expanded around the flat metric $\eta_{\mu\nu}$,
by writing again
\beq
g_{\mu\nu}(x) \; = \; \eta_{\mu\nu} + \sqrt{16 \pi G } \, h_{\mu\nu}(x)
\;\; .
\eeq
The quadratic part of the Lagrangian [see Eq.~(\ref{eq:h-action})] is then 
\beq
{\cal L} = 
- \quarter \, \partial_\mu h_{\alpha\beta} \, \partial^\mu h^{\alpha\beta}
+ \eigth \, ( \partial_\mu h^\alpha_{\;\;\alpha} )^2
+ \half \, C_\mu^2 
+ \half \, \kappa \, h_{\mu\nu} \, T^{\mu\nu}
+ {\cal L}_{\rm gf} + \dots
\label{eq:h-quadr-tv}
\eeq
where the dots indicate terms that are either total derivatives,
or higher order in $h$.
A suitable gauge fixing term $C_\mu$ is given by
\beq
C_\mu \equiv \partial_\alpha h^\alpha_{\;\;\mu} 
- \half \, \partial_\mu h^\alpha_{\;\;\alpha}
\label{eq:gauge-fix}
\eeq
Without such a term the quadratic part of the
gravitational Lagrangian of Eq.~(\ref{eq:h-action})
would contain a zero mode 
$h_{\mu\nu} \sim \partial_\mu \epsilon_\nu + \partial_\nu \epsilon_\mu $,
due to the gauge invariance of Eq.~(\ref{eq:h-gauge}), which
would make the graviton propagator ill defined.

The gauge fixing contribution ${\cal L}_{\rm gf}$ itself will be written
as the sum of two terms,
\beq
{\cal L}_{\rm gf} = - \half \, C_\mu^2 + {\cal L}_{\rm ghost}
\eeq
with the first term engineered so as to conveniently cancel the
$+ \half C_\mu^2$ in Eq.~(\ref{eq:h-quadr-tv}) and thus give a
well defined graviton propagator.
Note incidentally that this gauge is {\it not} the harmonic gauge
condition of Eq.~(\ref{eq:h-gauge-fix}), and is usually
referred to instead as the DeDonder gauge.
The second term is determined as usual from the variation of
the gauge condition under an infinitesimal gauge transformation
of the type in Eq.~(\ref{eq:h-gauge})
\beq
\delta C_\mu = \partial^2 \epsilon_\mu + O(\epsilon^2 )
\eeq
which leads to the lowest order ghost Lagrangian
\beq
{\cal L}_{\rm ghost} = 
- \, \partial_\mu \bar \eta_\alpha \, \partial^\mu \eta^\alpha
+ O (h^2)
\label{eq:ghost}
\eeq
where $\eta_\alpha$ is the spin-one anticommuting ghost field,
with propagator
\beq
D^{(\eta)}_{\mu\nu} (k) = { \eta_{\mu\nu} \over k^2 }
\eeq
In this gauge the graviton propagator is finally
determined from the surviving quadratic
part of the pure gravity Lagrangian, which is 
\beq
{\cal L}_0 = 
- \quarter \, \partial_\mu h_{\alpha\beta} \, \partial^\mu h^{\alpha\beta}
+ \eigth \, ( \partial_\mu h^\alpha_{\;\;\alpha} )^2
\label{eq:h-quadr-gf-tv}
\eeq
The latter can be conveniently re-written in terms of a matrix $V$
\beq
{\cal L}_0 = - \half \, \partial_\lambda h_{\alpha\beta} \, 
V^{\alpha\beta\mu\nu} \, \partial^{\lambda} h_{\mu\nu} 
\label{eq:h-quadr-gf-tv1}
\eeq
with 
\beq
V_{\alpha\beta\mu\nu}
= \half \, \eta_{\alpha\mu} \eta_{\beta\nu} 
- \quarter \, \eta_{\alpha\beta} \eta_{\mu\nu}
\eeq
The matrix $V$ can easily be inverted, 
for example by re-labelling rows and columns via the correspondence
\beq
11 \rightarrow 1, \;\; 22 \rightarrow 2, \;\; 33 \rightarrow 3, \, \dots \,
12 \rightarrow 5, \;\; 13 \rightarrow 6, \;\; 14 \rightarrow 7 \;\; \dots
\eeq
and the graviton Feynman propagator in $d$ dimensions is 
then found to be of the form 
\beq
D_{\mu\nu\alpha\beta}(k) =
{ \eta_{\mu\alpha} \eta_{\nu\beta} +
\eta_{\mu\beta} \eta_{\nu\alpha} -
{ 2 \over d-2 } \eta_{\mu\nu} \, \eta_{\alpha\beta} 
\over k^2 }
\label{eq:grav-prop}
\eeq
with a suitable $i \epsilon$ prescription to correctly integrate
around poles in the complex $k$ space.
Equivalently the whole procedure could have been performed
from the start with an Euclidean metric 
$\eta_{\mu\nu} \rightarrow \delta_{\mu\nu}$
and a complex time coordinate $t=-i\tau$ with hardly any changes
of substance.
The simple pole in the graviton propagator at $d=2$ serves as a reminder
of the fact that, due to the Gauss-Bonnet identity, the
gravitational Einstein-Hilbert action of Eq.~(\ref{eq:eh-action-d})
becomes a topological invariant in two dimensions.

Higher order correction in $h$ to the Lagrangian for pure gravity
then determine to order $h^3$ the three-graviton vertex, to order
$h^4$ the four-graviton vertex, and so on.
Because of the $\sqrt{g}$ and $g^{\mu\nu}$ terms in the action, there are
an infinite number of vertices in $h$.

Had one included a cosmological constant term as in Eq.~(\ref{eq:h-cosm}),
which can also be expressed in terms of the matrix $V$ as
\beq
\sqrt{g} \; = \; 1 + \half \, h_{\mu\mu} -
\half h_{\alpha\beta} V^{\alpha\beta\mu\nu} h_{\mu\nu} + O(h^3) ,
\eeq
then the the expression in Eq.~(\ref{eq:h-quadr-gf-tv1}) would have read 
\beq
{\cal L}_0 =  \lambda_0 ( 1 + \kappa \, \half h^\alpha_{\;\;\alpha} )
+ \half \, h_{\alpha\beta} \, 
V^{\alpha\beta\mu\nu} \, ( \partial^2 + \lambda_0 \kappa^2 ) h_{\mu\nu} 
\label{eq:h-quadr-gf-tv2}
\eeq
with $\kappa^2 = 16 \pi G$.
Then the graviton propagator would have been remained the same,
except for the replacement $k^2 \rightarrow k^2 - \lambda_0 \kappa^2 $.
In this gauge it would correspond to the exchange of 
a particle of mass $\mu^2 = - \lambda_0 \kappa^2$.
The term linear in $h$ can be interpreted as a uniform constant
source for the gravitational field.
But one needs to be quite careful, since 
for non-vanishing cosmological constant 
flat space $g_{\mu\nu} \sim \eta_{\mu\nu}$ is no
longer a solution of the vacuum field equations
and the problem becomes a bit more subtle: one needs to expand
around the correct vacuum solutions in the presence of
a $\lambda$-term, which are no longer constant.

Another point needs to be made here.
One peculiar aspect of perturbative gravity is that there is no unique way
of doing the weak field expansions, and one can have therefore
different sets of Feynman rules, even apart from the choice of
gauge condition, depending on how one chooses to do the expansion
for the metric.

For example, the structure of the scalar field action
of Eq.~(\ref{eq:scalar-action}) suggests to define instead
the small fluctuation graviton field $h_{\mu\nu}(x)$ via
\beq
\tilde g^{\mu\nu}(x) \equiv 
g^{\mu\nu}(x)
\sqrt{g(x)} = \eta^{\mu\nu} + K \; h^{\mu\nu}(x) \;\; .
\label{eq:g-tilde}
\eeq
with $K^2 = 32 \pi G$ (Faddeev and Popov, 1973; Capper et al, 1973).
Here it is $h^{\mu\nu}(x) $ that should be referred to as "the
graviton field".
The change of variables from the $g_{\mu\nu}$'s to the 
$g^{\mu\nu}(x)\sqrt{g(x)}$'s involves a Jacobian, which can be taken
to be one in dimensional regularization.
There is one obvious advantage of this expansion over the previous one,
namely that it leads
to considerably simpler Feynman rules, both for the graviton vertices
and for the scalar-graviton vertices, which can be advantageous
when computing one-loop scattering amplitudes of scalar
particles (Hamber and Liu, 1985).
Even the original gravitational action has a simpler form in terms
of the variables of Eq.~(\ref{eq:g-tilde}) as shown originally 
in (Goldberg, 1958).

\begin{figure}[h]
\epsfig{file=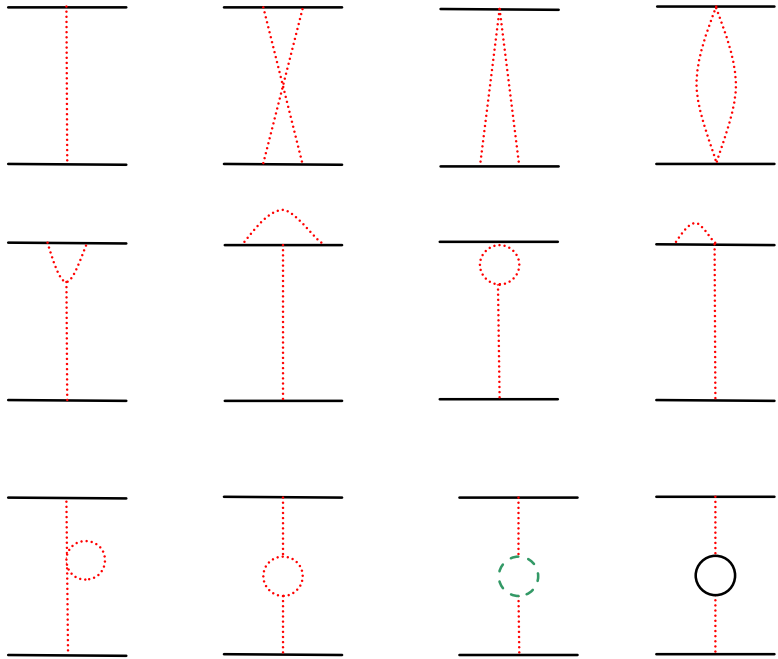,width=12cm}
\caption{Lowest order diagrams illustrating modifications
to the classical gravitational potential due to graviton
exchange.
Continuous lines denote a spinless heavy matter particle, short dashed lines a 
graviton and the long dashed line the ghost loop.
The last diagram shows the scalar matter loop contribution.}
\label{fig:potdiags}
\end{figure}

Again, when performing Feynman diagram perturbation theory
a gauge fixing term needs to be added in order to define
the propagator, for example of the form 
\beq
{ 1 \over K^2 }
\left ( \partial_\mu \sqrt{g} \, g^{\mu\nu} \right )^2 \;\; ,
\eeq
In this new framework the bare graviton propagator is given simply by
\beq
D_{\mu\nu\alpha\beta}(k) =
{ \eta_{\mu\alpha} \eta_{\nu\beta} +
\eta_{\mu\beta} \eta_{\nu\alpha}
- \eta_{\mu\nu} \eta_{\alpha\beta} 
\over 2 \, k^2 }
\label{eq:grav-prop-1}
\eeq
which should be compared to Eq.~(\ref{eq:grav-prop}) (the extra factor
of one half here is just due to the convention in the choice of $K$).
One notices that now there are no factors of $1/(d-2)$ for the graviton
propagator in $d$ dimensions.
But such factors appear instead in the expression
for the Feynman rules for the graviton vertices, and such $(d-2)^{-1}$
pole terms appear therefore regardless of the choice of expansion field.
For the three-graviton and two ghost-graviton vertex the
relevant expressions are quite complicated.
The three-graviton vertex is given by
\bea
&& U(q_1,q_2,q_3)_{\a_1 \b_1, \a_2 \b_2, \a_3 \b_3} = \nonumber \\
&& - { K \over 2 } \Bigl [
q^2_{(\a_1} q^3_{\b_1)} 
\left (
2 \eta_{\a_2(\a_3} \eta_{\b_3)\b_2} - {\textstyle {2 \over d-2} \displaystyle}
\eta_{\a_2\b_2} \eta_{\a_3\b_3}
\right ) \nonumber \\
&& \;\;\;\;\;\; + 
q^1_{(\a_2} q^3_{\b_2)} 
\left (
2 \eta_{\a_1(\a_3} \eta_{\b_3)\b_1} - {\textstyle {2 \over d-2} \displaystyle}
\eta_{\a_1\b_1} \eta_{\a_3\b_3}
\right ) \nonumber \\
&& \;\;\;\;\;\; +
q^1_{(\a_3} q^2_{\b_3)} 
\left (
2 \eta_{\a_1(\a_2} \eta_{\b_2)\b_1} - {\textstyle {2 \over d-2} \displaystyle}
\eta_{\a_1\b_1} \eta_{\a_2\b_2}
\right )  \nonumber \\
&& \;\;\;\;\;\; +
2 q^3_{(\a_2} \eta_{\b_2)(\a_1} \eta_{\b_1)(\a_3}  q^2_{\b_3)} +
2 q^1_{(\a_3} \eta_{\b_3)(\a_2} \eta_{\b_2)(\a_1}  q^3_{\b_1)} +
2 q^2_{(\a_1} \eta_{\b_1)(\a_3} \eta_{\b_3)(\a_2}  q^1_{\b_2)}  \nonumber \\
&& \;\;\;\;\;\; +
q^2 \cdot q^3  \left ( 
{\textstyle {2 \over d-2} \displaystyle} \eta_{\a_1(\a_2} \eta_{\b_2)\b_1} \eta_{\a_3\b_3} +
{\textstyle {2 \over d-2} \displaystyle} \eta_{\a_1(\a_3} \eta_{\b_3)\b_1} \eta_{\a_2\b_2} -
2 \eta_{\a_1(\a_2} \eta_{\b_2)(\a_3} \eta_{\b_3)\b_1} \right ) \nonumber \\
&& \;\;\;\;\;\; +
q^1 \cdot q^3  \left ( 
{\textstyle {2 \over d-2} \displaystyle} \eta_{\a_2(\a_1} \eta_{\b_1)\b_2} \eta_{\a_3\b_3} +
{\textstyle {2 \over d-2} \displaystyle} \eta_{\a_2(\a_3} \eta_{\b_3)\b_2} \eta_{\a_1\b_1} -
2 \eta_{\a_2(\a_1} \eta_{\b_1)(\a_3} \eta_{\b_3)\b_2} \right ) \nonumber \\
&& \;\;\;\;\;\; +
q^1 \cdot q^2  \left ( 
{\textstyle {2 \over d-2} \displaystyle} \eta_{\a_3(\a_1} \eta_{\b_1)\b_3} \eta_{\a_2\b_2} +
{\textstyle {2 \over d-2} \displaystyle} \eta_{\a_3(\a_2} \eta_{\b_2)\b_3} \eta_{\a_1\b_1} -
2 \eta_{\a_3(\a_1} \eta_{\b_1)(\a_2} \eta_{\b_2)\b_3} \right )
\Bigr ] \;\; . \nonumber \\
\eea
The ghost-graviton vertex is given by
\beq
V(k_1,k_2,k_3)_{\a \b, \l \mu} = K \left [
- \eta_{\l(\a} k_{1 \b)} k_{2\mu } 
+ \eta_{\l \mu} k_{2(\a)} k_{3\b)} 
\right ] \;\; ,
\eeq
and the two scalar-one graviton vertex is given by
\beq
{ K \over 2 } \left ( p_{1 \mu} p_{2 \nu} + p_{1 \nu} p_{2 \mu} -
{ 2 \over d-2} \; m^2 \; \eta_{\mu\nu} \right ) \;\; ,
\eeq
where the $p_1,p_2$ denote the four-momenta of the incoming and outgoing
scalar field, respectively.
Finally the two scalar-two graviton vertex is given by
\beq
{ K^2 m^2 \over 2(d-2) } \left ( \eta_{\mu \lambda} \eta_{\nu \sigma}
+ \eta_{\mu \sigma} \eta_{\nu \lambda} 
- { 2 \over d-2 } \; \eta_{\mu \nu} \eta_{\lambda \sigma} \right )
\;\; , 
\eeq
where one pair of indices $(\mu,\nu)$ is associated with one graviton line,
and the other pair $(\lambda,\sigma)$ is associated with the other graviton
line.
These rules follow readily from the expansion of the gravitational action to
order $G^{3/2}$ ($K^3$), and of the scalar field action to order
$G$ ($K^2$), as shown in detail in (Capper et al, 1973).
Note that the poles in $1/(d-2)$ have disappeared from the propagator,
but have moved to the vertex functions.
As mentioned before, they reflect the kinematic singularities that arise
in the theory as $d \rightarrow 2$ due to the Gauss-Bonnet identity.
As an illustration, Fig.~\ref{fig:potdiags} shows the lowest order
diagrams contributing to the static potential between two massive spinless
sources (Hamber and Liu, 1995).




\subsubsection{One-Loop Divergences}
\label{sec:one-loop}

Once the propagators and vertices have been defined, one can
then proceed as in QED and Yang-Mills theories and evaluate the
quantum mechanical one loop corrections. 
In a renormalizable theory with a dimensionless coupling,
such as QED and Yang-Mills theories, one has that the
radiative corrections lead to charge, mass and field re-definitions.
In particular, for the pure $SU(N)$ gauge action one finds
\beq
I_{YM} = - { 1 \over 4 g^2 \, N } \, \int d x \, 
\tr F_{\mu\nu}^2 
\; \rightarrow \;
- { 1 \over 4 g_R^2 \, N } \, \int d x \, 
\tr F_{R\;\; \mu\nu}^2 
\eeq
so that the form of the action is preserved by the renormalization
procedure: no new interaction terms such as $ (D_\mu F^{\mu\nu})^2 $ need to
be introduced in order to re-absorb the divergences.

In gravity the coupling is dimensionful, $G \sim \mu^{2-d}$, and 
one expects trouble already on purely dimensional grounds,
with divergent one loop corrections proportional to
$G \Lambda^{d-2} $ where $\Lambda$ is an ultraviolet cutoff
\footnote{
Indeed it was noticed very early on in the development of renormalization
theory that perturbatively non-renormalizible theories would involve couplings
with negative mass dimensions, and for which cross sections would
grow rapidly with energy (Sakata, Umezawa and Kamefuchi, 1952).
It had originally been suggested by Heisenberg (Heisenberg, 1938) that
the relevant mass scale appearing in such interactions with dimensionful 
coupling constants should be used to set an upper energy limit on the
physical applicability of such theories.
}. 
Equivalently, one expects to lowest order bad ultraviolet behavior
for the running Newton's constant at large momenta,
\beq
{ G(k^2) \over G } \sim 1 + {\rm const.} \; G \, k^{d-2} + \, O(G^2)
\eeq
These considerations also suggest that perhaps ordinary Einstein gravity
is perturbatively renormalizable in the traditional sense in
two dimensions, an issue to which we will return later in Sect.~\ref{sec:graveps}.

A more general argument goes as follows. 
The gravitational action contains the scalar curvature $R$ which
involves two derivatives of the metric.
Thus the graviton propagator in momentum space will go like $1/k^2$,
and the vertex functions like $k^2$.
In $d$ dimensions each loop integral with involve a momentum
integration $d^d k$, so that the superficial degree of divergence ${\cal D}$
of a Feynman diagram with $V$ vertices, $I$ internal lines and $L$ loops
will be given by
\beq
{\cal D} = d \, L + 2 V - 2 I
\eeq
The topological relation involving $V$, $I$ and $L$ 
\beq
L \, = \, 1 + I - V
\eeq
is true for any diagram, and yields
\beq
{\cal D} = 2 + (d-2) \, L
\eeq
which is independent of the number of external lines.
One concludes therefore that for $d>2$ the degree of divergence increases with
increasing loop order $L$.

The most convenient tool to determine the structure of the
divergent one-loop corrections to Einstein gravity is the
background field method (DeWitt, 1967; 't Hooft and Veltman, 1974) combined
with dimensional regularization, wherein ultraviolet divergences
appear as poles in $\epsilon=d-4$
\footnote{The second reference uses a complex time (Euclidean) $x_0=i c t$
notation that differs from the one used here.}.
In non-Abelian gauge theories the background field method greatly
simplifies the calculation of renormalization factors, while
at the same time maintaining explicit gauge invariance.

The essence of the method is easy to describe: one replaces
the original field appearing in the classical action by $A+Q$,
where $A$ is a classical background field and $Q$ the quantum fluctuation.
A suitable gauge condition is chosen (the background gauge),
such that manifest gauge invariance is preserved for the
background $A$ field.
After expanding out the action to quadratic order in the $Q$ field,
the functional integration over $Q$ is performed, leading
to an effective action for the background $A$ field.
From the structure of this effective action the renormalization
of the couplings, as well as possible additional counterterms,
can then be read off. 
In the case of gravity it is in fact sufficient to look at the
structure of those terms appearing in the effective action
which are quadratic in the background field $A$.
A very readable introduction to the background field method as applied
to gauge theories can be found in (Abbot, 1981).

Unfortunately perturbative calculations in gravity are rather cumbersome
due to the large number of indices and contractions, so the
rest of this section is only intended more as a general outline,
with the scope of hopefully providing some of the flavor
of the original calculations.
The first step consists in the replacement
\beq
g_{\mu\nu} \; \rightarrow \; \bar g_{\mu\nu} = g_{\mu\nu} + h_{\mu\nu}
\label{eq:h-back}
\eeq
where now $g_{\mu\nu} (x) $ is the classical background field and 
$h_{\mu\nu}$ the quantum field, to be integrated over.
To determine the structure of one loop divergences
it will often be sufficient to consider at the very end just the case of
a flat background metric, $g_{\mu\nu} = \eta_{\mu\nu}$, or a small
deviation from it.

After a somewhat tedious calculation one finds for the bare action
\beq
{\cal L} = \sqrt{g} \, [ \, c_0 + c_1 R \, ]
\eeq
expanded out to quadratic order in $h$ 
\bea
{\cal L} = && \sqrt{g} \, [ c_0 \{ 1 
+ \half h^{\alpha}_{\;\;\alpha}  
- \quarter h^{\alpha}_{\;\;\beta} h^\beta_{\;\;\alpha}
+ \eigth h^{\alpha}_{\;\;\alpha} h^\beta_{\;\;\beta} \}
\nonumber \\
& + & c_1 \{ R - \half h^{\alpha}_{\;\;\alpha} R 
+ h^{\alpha}_{\;\;\beta} R^\beta_{\;\;\alpha}
- \eigth R \, h^{\alpha}_{\;\;\alpha} h^\beta_{\;\;\beta}
+ \quarter R \, h^{\alpha}_{\;\;\beta} h^\beta_{\;\;\alpha}
\nonumber \\
& - & h^{\nu}_{\;\;\beta} h^\beta_{\;\;\alpha} R^\alpha_{\;\;\nu}
+ \half h^{\alpha}_{\;\;\alpha} h^\nu_{\;\;\beta} R^\beta_{\;\;\nu}
- \quarter \nabla_\nu h^{\alpha}_{\;\;\beta} \nabla^\nu
h^{\beta}_{\;\;\alpha}
\nonumber \\
& + & \nabla_\nu h^{\alpha}_{\;\;\alpha} \nabla^\nu
h^{\beta}_{\;\;\beta}
- \half \nabla_\beta h^{\alpha}_{\;\;\alpha} \nabla^\mu
h^{\beta}_{\;\;\mu}
+ \half \nabla^\alpha h^{\nu}_{\;\;\beta}
\nabla_\nu h^{\beta}_{\;\;\alpha} \} ]
\nonumber \\
\label{eq:h-background}
\eea
up to total derivatives.
Here $\nabla_\mu$ denotes a covariant derivative with
respect to the metric $g_{\mu\nu}$.
For $g_{\mu\nu}=\eta_{\mu\nu}$ the above expression coincides with
the weak field Lagrangian contained in 
Eqs.~(\ref{eq:h-action}) and (\ref{eq:h-quadr-tv}),
with a cosmological constant term added, as given in 
Eq.~(\ref{eq:h-cosm}).

To this expression one needs to add the gauge fixing and ghost contributions,
as was done in Eq.~(\ref{eq:h-quadr-tv}).
The background gauge fixing term used is 
\beq
- \half C_\mu^2 = - \half \sqrt{g} \,
( \nabla_\alpha h^\alpha_{\;\;\mu} - \half \nabla_\mu h^\alpha_{\;\;\alpha} )
( \nabla_\beta h^{\beta\mu} - \half \nabla^\mu h^\beta_{\;\;\beta} )
\label{eq:gauge-fix-back}
\eeq
with a corresponding ghost Lagrangian 
\beq
{\cal L}_{\rm ghost} = \sqrt{g} \, \bar \eta_\mu ( 
\partial_\alpha \partial^\alpha \eta^\mu 
- R^\mu_{\;\;\alpha} \eta^\alpha )
\label{eq:ghost-back}
\eeq
The integration over the $h_{\mu\nu}$ field can then be performed
with the aid of the standard Gaussian integral formula
\beq
\ln \int [d h_{\mu\nu} ] \exp \{ - \half h \cdot M(g) \cdot h - N (g) \cdot h \}
= \half N (g) \cdot M^{-1}(g) \cdot N(g) - \half \tr \ln M(g) + {\rm const.}
\label{eq:gauss-int}
\eeq
leading to an effective action for the $g_{\mu\nu}$ field.
In practice one is only interested in the divergent part, which
can be shown to be local.
Specific details of the functional measure over metrics $[d g_{\mu\nu}]$
are not deemed to be essential at this stage, as in perturbation theory one
is only doing Gaussian integrals, with $h_{\mu\nu}$ ranging
from $- \infty$ to $+\infty$. 
In particular when using dimensional regularization one uses
the formal rule
\beq
\int d^d k = (2 \pi )^d \delta^{(d)} (0) = 0
\label{eq:delta0}
\eeq
which leads to some technical simplifications but
obscures the role of the measure.

In the flat background field case case $g_{\mu\nu}=\eta_{\mu\nu}$,
the functional integration over the $h_{\mu\nu}$ fields would have been
particularly simple, since then one would be using
\beq
h_{\mu\nu}(x) \, h_{\alpha\beta} (x') \;\; \rightarrow \;\;
< \! h_{\mu\nu}(x) \, h_{\alpha\beta} (x') \! > \; = \; G_{\mu\nu\alpha\beta} 
(x,x') \;\; ,
\eeq
with the graviton propagator $G(k)$ given in Eq.~(\ref{eq:grav-prop}).
In practice, one can use the expected generally covariant structure of the 
one-loop divergent part 
\beq
\Delta {\cal L}_g \propto \sqrt{g} 
\left ( \alpha \, R^2 + \beta \, R_{\mu\nu} R^{\mu\nu} \right ) \;\; ,
\eeq
with $\alpha$ and $\beta$ some real parameters,
as well as its weak field form, obtained from 
\bea
R^2 & = & \partial^2 h^\mu_{\;\;\mu} \partial^2 h^\alpha_{\;\;\alpha}
- 2 \partial^2 h^\mu_{\;\;\mu} \partial_{\alpha} \partial_\beta h^{\alpha\beta}
+ \partial_{\alpha} \partial_\beta h^{\alpha\beta}
\partial_{\mu} \partial_\nu h^{\mu\nu}
\nonumber \\
R_{\alpha\beta} R^{\alpha\beta}
& = & \quarter ( 
\partial^2 h^\mu_{\;\;\mu} \partial^2 h^\alpha_{\;\;\alpha}
+ \partial^2 h_{\mu\alpha} \partial^2 h^{\mu\alpha}
- 2 \partial^2 h^\mu_{\;\;\mu}
\partial_{\alpha} \partial_{\beta} h^{\alpha\beta}
\nonumber \\
&& - 2 \partial_{\alpha} \partial_\nu h^\nu_{\;\;\mu}
\partial_{\alpha} \partial_\beta h^{\mu\beta}
+ 2 \partial_{\mu} \partial_\nu h^{\mu\nu}
\partial_{\alpha} \partial_\beta h^{\alpha\beta}
\nonumber \\
\label{eq:h-r2}
\eea
[compare with Eq.~(\ref{eq:h-ricci})], combined  with some
suitable special choices for the background metric, such as 
$g_{\mu\nu} (x) = \eta_{\mu\nu} f(x)$, to further simplify the calculation.
This eventually determines the required one-loop counterterm 
for pure gravity to be
\beq
\Delta {\cal L}_g = { \sqrt{g} \over 8 \pi^2 (d-4) }
\left ( {1 \over 120 } R^2 + { 7 \over 20 } R_{\mu\nu} R^{\mu\nu} \right )
\label{eq:one-loop-div}
\eeq
For the simpler case of classical gravity coupled invariantly to a
single real quantum scalar field one finds
\beq
\Delta {\cal L}_g = { \sqrt{g} \over 8 \pi^2 (d-4) }
{1 \over 120 } \left ( \half \, R^2 + R_{\mu\nu} R^{\mu\nu} \right )
\eeq
The complete set of one-loop divergences, computed using the
alternate method of the heat
kernel expansion and zeta function regularization
\footnote{
The zeta-function regularization (Hawking, 1977) involves studying
the behavior of the function
$ \zeta (s) = \sum_{n=0}^\infty ( \lambda_n )^{-s} $,
where the $\lambda_n$'s are the eigenvalues of the second
order differential operator $M$ in question.
The series will converge for $s>2$, and can be used for an
analytic continuation to $s=0$, which then leads to the formal result
$ \log ( \det M ) = \log \prod_{n=0}^\infty \lambda_n = - \zeta ' (0) $.
}
close to four dimensions, can be found in the comprehensive
review (Hawking, 1977) and further references therein.
In any case one is led to conclude that pure quantum gravity
in four dimensions is not perturbatively renormalizable:
the one-loop divergent part contains local operators which were not
present in the original Lagrangian. 
It would seem therefore that these operators would have to be
added to the bare ${\cal L}$, so
that a consistent perturbative renormalization program can be developed
in four dimensions.

There are two interesting, and interrelated, aspects of the result
of Eq.~(\ref{eq:one-loop-div}).
The first one is that for pure gravity the divergent part vanishes
when one imposes the tree-level equations of motion $R_{\mu\nu}=0$:
the one-loop divergence vanishes on-shell.
The second interesting aspect is that the specific structure of the
one-loop divergence is such that its effect can actually
be re-absorbed into a field redefinition,
\bea
g_{\mu\nu} & \rightarrow & g_{\mu\nu} + \delta g_{\mu\nu} 
\nonumber \\
&& \;\;\;\; \delta g_{\mu\nu} \, \propto \, 
{ 7 \over 20 } R_{\mu\nu} - { 11 \over 60 } R \, g_{\mu\nu}
\eea
which renders the one-loop amplitudes finite for pure gravity.
Unfortunately this hoped-for mechanism does not seem to work to two loops,
and no additional miraculous cancellations seem to occur there.
At two loops one expects on general grounds terms of the type
$\nabla^4 R$, $R \nabla^2 R$ and $R^3$. 
It can be shown that the first class of terms reduce to total derivatives,
and that the second class of terms can also be made to vanish on shell
by using the Bianchi identity.
Out of the last set of terms, the $R^3$ ones, one can show ('t Hooft, 2002)
that there are potentially 20 distinct contributions, of which 19 
vanish on shell (i.e. by using the tree level field equations $R_{\mu\nu}=0$).
An explicit calculation then shows that a new non-removable on-shell
$R^3$-type divergence arises in pure gravity at two loops
(Goroff and Sagnotti, 1985; van de Ven, 1992) from the only possible
surviving non-vanishing counterterm, namely
\beq
\Delta {\cal L}^{(2)} \; = \; {\sqrt{g} \over (16 \pi^2)^2 (4-d) } \,
{ 209 \over 2880} \,
R_{\mu\nu}^{\;\;\;\;\rho\sigma} \, R_{\rho\sigma}^{\;\;\;\;\kappa\lambda} 
\, R_{\kappa\lambda}^{\;\;\;\;\mu\nu} \;\; .
\eeq
To summarize, radiative corrections to pure Einstein gravity without
a cosmological constant term induce one-loop $R^2$-type divergences
of the form
\beq
\Gamma_{div}^{(1)} \; = \; {1 \over d-4} \, { \hbar \over 16 \pi^2 } 
\int d^4 x \sqrt{g} \; \left ( 
{ 7 \over 20 } \, R_{\mu\nu} \, R^{\mu\nu} \, + \, { 1 \over 120 } \, R^2 
\right ) \;\; ,
\eeq
and a two-loop non-removable on-shell $R^3$-type divergence of the type
\beq
\Gamma_{div}^{(2)} \; = \; {1 \over d-4} \, { 209 \over 2880} \,
{ \hbar^2 \, G \over (16 \pi^2)^2 } 
\int d^4 x \sqrt{g} \; 
R_{\mu\nu}^{\;\;\;\;\rho\sigma} \, R_{\rho\sigma}^{\;\;\;\;\kappa\lambda} 
\, R_{\kappa\lambda}^{\;\;\;\;\mu\nu} \;\; .
\eeq
which present an almost insurmountable obstacle to the traditional
perturbative renormalization procedure in four dimensions.
One can therefore attempt to summarize the situation so far as follows:

\begin{enumerate}

\item[$\circ$]
In principle perturbation theory in $G$ in provides a clear, covariant 
framework in which radiative corrections to gravity can be computed in a 
systematic loop expansion.
The effects of a possibly non-trivial gravitational measure do not show up
at any order in the weak field expansion, and radiative
corrections affecting the renormalization of the cosmological constant,
proportional to $\delta^d (0)$, are set to zero in dimensional regularization.

\item[$\circ$]
At the same time at every order in the loop expansion new invariant
terms involving higher derivatives of the metric are generated,
whose effects cannot be simply re-absorbed into a re-definition of
the original couplings.
As expected on the basis of power-counting arguments, the theory
is not perturbatively renormalizable in the traditional sense in four dimensions
(although it seems to fail this test by a small measure in lowest
order perturbation theory).

\item[$\circ$]
The standard approach based on a perturbative expansion of the pure Einstein
theory in four dimensions is therefore not convergent 
(it is in fact badly divergent), and represents therefore a temporary dead end.
  
\end{enumerate}

\subsubsection{Higher Derivative Terms}
\label{sec:hdqg}

In the previous section it was shown that quantum corrections to the
Einstein theory 
generate in perturbation theory $R^2$-type terms in four dimensions.
It seems therefore that, for the consistency of the perturbative renormalization group approach in four dimension, these terms would have to be included from the start, at the level of the bare microscopic action.
Thus the main motivation for studying gravity with higher derivative terms
is that it might cure the problem of ordinary quantum
gravity, namely its perturbative non-renormalizabilty in four dimensions.
This is indeed the case, in fact one can prove that higher derivative
gravity (to be defined below) is perturbatively renormalizable
to all orders in four dimensions. 

At the same time new issues arise, which will be detailed below.
The first set of problems has to do with the fact that, quite generally
higher derivative theories with terms of the type
$\phi \, \partial^4 \phi$ suffer from potential unitarity problems,
which can lead to physically unacceptable negative probabilities.
But since these are genuinely dynamical issues, it will be difficult
to answer them satisfactorily in perturbation theory.
In non-Abelian gauge theories one can use higher derivative terms,
instead of the more traditional dimensional continuation, to regulate ultraviolet divergences (Slavnov, 1973), and higher derivative terms
have been used successfully for some time in lattice regulated field
theories (Symanzik, 1983).
In these approaches the coefficient of the higher derivative terms
is taken to zero at the end.
The second set of issues is connected with the fact that the theory
is asymptotically free in the higher derivative couplings,
implying an infared growth which renders the perturbative estimates
unreliable at low energies, in the regime of perhaps greatest physical interest. 
Note that higher derivative terms arise in string theory as well
(F\"orger, Ovrut, Theisen and Waldram, 1996).

Let us first discuss the general formulation.
In four dimensions possible terms quadratic in the curvature are
\bea
&& \int  d^4 x  \, \sqrt g  \; R^2
\nonumber \\
&& \int  d^4 x  \, \sqrt g  \; R_{ \mu \nu }  R^{ \mu \nu }
\nonumber \\
&& \int  d^4 x  \, \sqrt g  \; R_{ \mu \nu \lambda \sigma }  R^{ \mu \nu 
\lambda \sigma }
\nonumber \\
&& \int  d^4 x  \, \sqrt g  \; C_{ \mu \nu \lambda \sigma }  C^{ \mu \nu 
\lambda \sigma }
\nonumber \\
&& \int  d^4 x  \, \sqrt g \;
\epsilon^{ \mu \nu \kappa \lambda }  \epsilon^{ \rho \sigma \omega \tau }
\; R_{ \mu \nu \rho \sigma }  R^{ \kappa \lambda \omega \tau }
= 128 \pi^2  \, \chi
\nonumber \\
&& \int  d^4 x  \, \sqrt g  \; 
\epsilon^{ \rho \sigma \kappa \lambda } \;
R_{ \mu \nu \rho \sigma } R_{ \;\;\;\;  \kappa \lambda }^{ \mu \nu }
= 96 \pi^2 \, \tau
\label{eq:curv2}
\eea
where $\chi$ is the Euler characteristic and $\tau$ the Hirzebruch signature.
It will be shown below that these quantities are not all independent.
The Weyl conformal tensor is defined in $d$ dimensions as
\bea
C_{ \mu \nu \lambda \sigma } & = &
R_{ \mu \nu \lambda \sigma } - 
{\textstyle { 2 \over d-2 } \displaystyle}
( g_{\mu [ \lambda } \, R_{ \sigma ] \nu } -
g_{\nu [ \lambda } \, R_{ \sigma ] \mu } )
\nonumber \\
&& + {\textstyle { 2 \over (d-1)(d-2) } \displaystyle}  
\, R  \, g_{ \mu [ \lambda }  g_{ \sigma ] \nu }
\eea
where square brackets denote antisymmetrization.
It is called conformal because it can be shown to be invariant
under conformal transformations of the metric, 
$g_{\mu\nu} (x)\rightarrow \Omega^2 (x) \, g_{\mu\nu} (x) $.
In four dimensions one has
\beq
C_{ \mu \nu \lambda \sigma } =
R_{ \mu \nu \lambda \sigma } -
R_{ \lambda [ \mu }  g_{ \nu ] \sigma } -
R_{ \sigma  [ \mu }  g_{ \nu ] \lambda } +
\third  R  \, g_{ \lambda [ \mu }  g_{ \nu ] \sigma }
\eeq
The Weyl tensor can be regarded as the traceless part of the
Riemann curvature tensor,
\beq
g^{ \lambda \sigma }  C_{ \lambda \mu \sigma \nu }
= g^{ \mu \nu }  g^{ \lambda \sigma } 
C_{ \mu \lambda \nu \sigma } = 0 \;\; .
\eeq
and on-shell the Riemann tensor in fact coincides with the Weyl tensor.
From the definition of the Weyl tensor one infers
in four dimensions the following curvature-squared identity
\beq
R_{ \mu \nu \lambda \sigma }  R^{ \mu \nu \lambda \sigma } =
C_{ \mu \nu \lambda \sigma } C^{ \mu \nu \lambda \sigma }
+ 2 \, R_{ \mu \nu }  R^{ \mu \nu } - \third  R^2
\eeq
Some of these results are specific to four dimensions.
For example, in three dimensions the Weyl tensor
vanishes identically and one has
\beq
R _ { \mu \nu \lambda \sigma }  R ^ { \mu \nu \lambda \sigma } -
4  R _ { \mu \nu }  R ^ { \mu \nu } - 3  R ^ 2  \; = \; 0
\;\;\;\;\;\;
C _ { \mu \nu \lambda \sigma }  C ^ { \mu \nu \lambda \sigma } = 0 ,
\eeq
In four dimensions the expression for the Euler characteristic
can be written equivalently as
\beq
\chi = {1 \over 32 \pi^2 }  \int d^4 x  \sqrt g  \;
\Bigl [ R_{ \mu \nu \lambda \sigma }  R^{ \mu \nu \lambda \sigma } - 4  
R_{ \mu \nu }  R^{ \mu \nu } + R^2 \Bigr ]
\eeq
The last result is the four-dimensional analog of the two-dimensional
Gauss-Bonnet formula
\beq
\chi = {1 \over 2 \pi} \int d^2 x  \, \sqrt g  \; R 
\eeq
where $\chi=2(g-1)$ and $g$ is the genus of the surface (the number of
handles).
For a manifold of fixed topology one can therefore use in four dimensions
\beq
R_{ \mu \nu \lambda \sigma } R^{ \mu \nu \lambda \sigma } = 4  \,
R_{ \mu \nu }  R^{ \mu \nu } - R^2 + {\rm const.}
\label{eq:rie2-4d}
\eeq
and
\beq
C _ { \mu \nu \lambda \sigma }  C ^ { \mu \nu \lambda \sigma } 
= 2 \, ( R_{ \mu \nu } R^{ \mu \nu } - \third R^2 ) + {\rm const.}
\label{eq:weyl-2}
\eeq
Thus only {\it two} curvature squared terms for the gravitational
action are independent in four dimensions (Lanczos, 1938), which can
be chosen, for example, to be $R^2$ and $R_{\mu\nu}^2$.
Consequently the most general curvature squared 
action in four dimensions can be written as
\beq
I = \int  d^4 x  \, \sqrt g \; \Bigl [ \lambda_0 + k \, R +
a \, R_{ \mu \nu }  R^{ \mu \nu } 
- \third \,( b+a ) R^2 \Bigr ]
\label{eq:hdqg}
\eeq
with $k=1/16 \pi G$, and up to boundary terms. 
The case $b=0$ corresponds, by virtue of Eq.~(\ref{eq:weyl-2}),
to the conformally invariant, pure Weyl-squared case.
If $b<0$ then around flat space one encounters a tachyon at tree
level.
It will also be of some interest later that in the Euclidean case 
(signature $++++$) the full gravitational
action of Eq.~(\ref{eq:hdqg}) is positive for $a>0$, $b<0$
and $\lambda_0 > - 3/4 b (16 \pi G)^2 $.

Curvature squared actions for classical gravity were originally considered
in (Weyl, 1922) and (Pauli, 1956).
In the sixties it was argued that the higher derivative action of Eq.~(\ref{eq:hdqg}) should be power counting renormalizable
(Utiyama and DeWitt, 1961).
Later it was proven to be renormalizable to all orders in
perturbation theory (Stelle, 1977).
Some special cases of higher derivative theories have been shown
to be classically equivalent to scalar-tensor theories (Whitt 1984).

One way to investigate physical properties of higher derivative theories
is again via the weak field expansion.
In analyzing the particle content 
it is useful to introduce a set of spin projection
operators (Arnowitt, Deser and Misner, 1958; van Nievenhuizen, 1973), quite
analogous to what is used in describing transverse-traceless (TT)
modes in classical gravity (Misner, Thorne and Wheeler, 1973).
These projection operators then 
show explicitly the unique decomposition of
the continuum gravitational action for linearized gravity into spin two
(transverse-traceless) and spin zero (conformal mode) parts.
The spin-two projection operator $P^{(2)}$ is defined in $k$-space as
\bea
P^{(2)}_{\mu\nu\alpha\beta} = &&  {1 \over 3 k^2 }
\left ( k_\mu k_\nu \eta_{\alpha\beta} +
k_\alpha k_\beta \eta_{\mu\nu} \right ) 
\nonumber \\
& - & {1 \over 2 k^2 } \left ( k_\mu k_\alpha \eta_{\nu\beta}
+ k_\mu k_\beta \eta_{\nu\alpha}
+ k_\nu k_\alpha \eta_{\mu\beta}
+ k_\nu k_\beta \eta_{\mu\alpha} \right )  
\nonumber \\
& + & {2 \over 3 k^4 } k_\mu k_\nu k_\alpha k_\beta
+ {1 \over 2} \left ( \eta_{\mu\alpha} \eta_{\nu\beta}
+ \eta_{\mu\beta} \eta_{\nu\alpha} \right )
- {1 \over 3 } \eta_{\mu\nu} \eta_{\alpha\beta}  \;\; ,
\label{eq:s2}
\eea
the spin-one projection operator $P^{(1)}$ as
\bea
P^{(1)}_{\mu\nu\alpha\beta} = && {1 \over 2 k^2 } \left (
k_\mu k_\alpha \eta_{\nu\beta} +
k_\mu k_\beta  \eta_{\nu\alpha} +
k_\nu k_\alpha \eta_{\mu\beta} +
k_\nu k_\beta  \eta_{\mu\alpha} \right )
\nonumber \\
& - & {1 \over k^4 } k_\mu k_\nu k_\alpha k_\beta
\label{eq:s1}
\eea
and the spin-zero projection operator $P^{(0)}$ as
\bea
P^{(0)}_{\mu\nu\alpha\beta} = & - & {1 \over 3 k^2 } 
\left ( k_\mu k_\nu \eta_{\alpha\beta} +
k_\alpha k_\beta \eta_{\mu\nu} \right )
\nonumber \\
& + & {1 \over 3 } \eta_{\mu\nu} \eta_{\alpha\beta} 
+ {1 \over 3 k^4 } k_\mu k_\nu k_\alpha k_\beta  \;\; .
\label{eq:s0}
\eea
It is easy to check that the sum of the three spin projection operators
adds up to unity
\beq
P^{(2)}_{\mu\nu\alpha\beta} \; + \; 
P^{(1)}_{\mu\nu\alpha\beta} \; + \; 
P^{(0)}_{\mu\nu\alpha\beta} \; = \; 
{1 \over 2} \left ( \eta_{\mu\alpha} \eta_{\nu\beta}
\; + \; \eta_{\mu\beta} \eta_{\nu\alpha} \right )  \;\; .
\eeq
These projection operators then
allow a decomposition of the gravitational field $h_{\mu\nu}$
into three independent modes.
The spin two or transverse-traceless part
\beq
h^{TT}_{\mu\nu} \; = \;  
P^\alpha_{\;\;\mu} P^\beta_{\;\;\nu} h_{\alpha\beta}  
- \third P_{\mu\nu} P^{\alpha\beta} h_{\beta\alpha}
\eeq
the spin one or longitudinal part
\beq
h^{L}_{\mu\nu} \; = \; h_{\mu\nu} 
- P^\alpha_{\;\;\mu} P^\beta_{\;\;\nu} h_{\alpha\beta}
\eeq
and the spin zero or trace part
\beq
h^{T}_{\mu\nu} \; = \; \third P_{\mu\nu} P^{\alpha\beta} h_{\alpha\beta}
\eeq
are such that their sum gives the original field $h$
\beq
h \; = \; h^{TT} + h^{L} + h^{T}  \;\; ,
\eeq
with the quantity $ P_{\mu\nu} $ defined as
\beq
P_{\mu\nu} \, = \, \eta_{\mu\nu} - { 1 \over \partial^2 } \; 
{ \partial_\mu \partial_\nu }
\label{eq:proj}
\eeq
or, equivalently, in $k$-space
$P_{\mu\nu} = \eta_{\mu\nu} - k_\mu k_\nu / k^2 $.

One can learn a number of useful aspects of the theory by looking
at the linearized form of the equations of motion.
As before, the linearized form of the action is obtained by setting
$g_{\mu\nu}=\eta_{\mu\nu}+h_{\mu\nu}$ and expanding in $h$.
Besides the expressions given in Eq.~(\ref{eq:h-ricci}), one needs
\bea
\sqrt{g} \; R^2 & = & (  \partial^2 h^\lambda_{\;\;\lambda}
- \partial_\lambda \partial_\kappa h^{\lambda\kappa} )^2 + O(h^3)
\nonumber \\
\sqrt{g} \; R_{\lambda\mu\nu\kappa} R^{\lambda\mu\nu\kappa}
& = & \quarter \,
(  \partial_\mu \partial_\kappa h_{\nu\lambda} 
+ \partial_\lambda \partial_\nu h_{\mu\kappa}
- \partial_\lambda \partial_\kappa h_{\mu\nu} 
- \partial_\mu \partial_\nu h_{\kappa\lambda} )^2 + O(h^3) \; ,
\eea
from which one can then obtain, for example from Eq.~(\ref{eq:rie2-4d}),
an expression for $\sqrt{g} \, (R_{\mu\nu})^2$,
\bea
\sqrt{g} \; R_{\alpha\beta} R^{\alpha\beta}
& = & \quarter ( 
\partial^2 h^\mu_{\;\;\mu} \partial^2 h^\alpha_{\;\;\alpha}
+ \partial^2 h_{\mu\alpha} \partial^2 h^{\mu\alpha}
- 2 \partial^2 h^\mu_{\;\;\mu}
\partial_{\alpha} \partial_{\beta} h^{\alpha\beta}
\nonumber \\
&& - 2 \partial_{\alpha} \partial_\nu h^\nu_{\;\;\mu}
\partial_{\alpha} \partial_\beta h^{\mu\beta}
+ 2 \partial_{\mu} \partial_\nu h^{\mu\nu}
\partial_{\alpha} \partial_\beta h^{\alpha\beta} ) + O(h^3)
\nonumber \\
\eea
Using the three spin projection operators defined previously, the action for linearized gravity without a cosmological
constant term, Eq.~(\ref{eq:h-action}), can then be re-expressed as
\beq
I_{\rm lin} \, = \, \quarter \, k \int d x \;
h^{\mu\nu} \, [ P^{(2)} - 2 P^{(0)} ]_{\mu\nu\alpha\beta} \; 
\partial^2 \, h^{\alpha\beta}
\label{eq:wfe2}
\eeq
Only the $P^{(2)}$ and $P^{(0)}$ projection operators for the spin-two
and spin-zero modes, respectively, appear in the action for
the linearized gravitational field;
the spin-one gauge mode does not enter the linearized action.
Note also that the spin-zero mode enters with the wrong sign (in the linearized action it appears as a ghost contribution), but to this order 
it can be removed by a suitable choice of gauge in which the trace
mode is made to vanish, as can be seen, for example, from
Eq.~(\ref{eq:h-field2}).

It is often stated that higher derivative theories suffer from
unitarity problems.
This is seen as follows.
When the higher derivative terms are included, the corresponding linearized
expression for the gravitational action becomes
\bea
I_{\rm lin} = \half \int  d x  & \{ & 
h^{\mu\nu} \, [ \half \, k + \half \, a \,
(- \partial^2 ) ] ( - \partial^2 ) \, 
P^{(2)}_{\mu\nu\rho\sigma} \, h^{\rho\sigma}
\nonumber \\
& + & h^{\mu\nu} \, [ - k - 2 \, b \,
(- \partial^2 ) ] (- \partial^2 ) \, 
P^{(0)}_{\mu\nu\rho\sigma} \, h^{\rho\sigma} \,
\}
\label{eq:hdqg-lin}
\eea
Then the potential problems with unitarity and ghosts at ultrahigh energies, 
say comparable to the Planck mass $ q \sim 1/G$, can be seen by examining the graviton propagator (Salam and Strathdee, 1978).
In momentum space the free graviton propagator for higher derivative gravity and $\lambda_0 = 0$ can be written as
\beq
k \, < h_{ \mu \nu } (q) \, h_{ \rho \sigma } (-q) > \; = 
{ 2  P_{ \mu \nu \rho \sigma }^{(2)} \over  q^2 + { a \over k} q^4 } +
{  P_{ \mu \nu \rho \sigma }^  {(0)} \over  - q^2 - { 2 b \over k} q^4 }
+ {\rm gauge \; terms}
\eeq
The first two terms on the r.h.s. can be decomposed as
\beq
2  P_{ \mu \nu \rho \sigma }^{(2)} \,
\left [ { 1 \over  q^2 } - { 1 \over q^2 + { k \over a }  } \right ] +
P_{ \mu \nu \rho \sigma }^{(0)} \,
\left [ - { 1 \over  q^2 } + { 1 \over q^2 + { k \over 2 b } } \right ] 
\label{eq:hdqg-prop}
\eeq
One can see that, on the one hand, the higher derivative terms improve the
ultraviolet
behavior of the theory,  since the propagator now falls of as $1/q^4$
for large $q^2$.
At the same time, the theory appears to contain a spin-two ghost of mass 
$ m_2 = \mu / \sqrt a $ and a spin-zero particle of mass 
$ m_0 = \mu / \sqrt { 2 b } $.
Here we have set $\mu = 1/\sqrt(16 \pi G) $, which is of the order of the Planck mass ($1/\sqrt{G/ \hbar c } = 1.2209 \times 10^{19} GeV/c^2$).
For $b<0$ one finds a tachyon pole, which seems, for the time
being, to justify the original choice of $b>0$ in Eq.~(\ref{eq:hdqg}).

Higher derivative gravity theories also lead to modifications to
the standard Newtonian potential, even though such deviations only
become visible at very
short distances, comparable to the Planck length 
$l_P = \sqrt{ \hbar G / c^3 } = 1.61624 \times 10^{-33} cm$.
In some special cases they can be shown to be classically equivalent to
scalar-tensor theories without higher derivative terms (Whitt, 1984).
The presence of massive states in the tree level graviton
propagator indicates short distance deviations from the static Newtonian
potential of the form
\beq
h_{ 0 0 } \, \sim \, { 1 \over r } -
{ 4 \over 3 } \, { e^{ - m_2  r } \over r } +
{ 1 \over 3 } \, { e^{ - m_0  r } \over r } 
\label{eq:hdqg-pot}
\eeq
Moreover in the extreme case corresponding to the absence of the
Einstein term ($k=0$) the potential is linear in $r$;
but in this limit the theory is strongly infrared divergent, and it
is not at all clear whether weak coupling perturbation theory
is of any relevance.

In the quantum theory perturbation theory is usually performed
around flat space, which requires $\lambda_0 = 0$, or around
some fixed classical background.
One sets again 
$g_{\mu\nu} \rightarrow \bar g_{\mu \nu} = g_{\mu \nu} + h_{\mu \nu}$ and expands the higher derivative action in powers of $h_{\mu \nu}$.
If $\lambda_0$ is nonzero, one has to expand around a solution of the classical
equations of motion for higher derivative gravity with a $\lambda$-term
(Barth and Christensen, 1983), and the solution will no longer be constant
over space-time.
The above expansion is consistent with the assumption that the two higher
derivative couplings $a$ and $b$ are large, since in
such a limit one is close to flat space.
One-loop radiative corrections then show that the theory is
asymptotically free in the higher derivative couplings $a$ and $b$
(Julve and Tonin, 1978; Fradkin and Tseytlin, 1981; 
Avramidy and Barvinsky, 1985).

The calculation of one-loop quantum fluctuation effects proceeds
in a way that is similar to the pure Einstein gravity case.
One first decomposes the metric field as a classical background part
$g_{\mu\nu}(x)$ and a quantum fluctuation part $h_{\mu\nu}(x)$
as in Eq.~(\ref{eq:h-back}), and then expands the classical action
to quadratic order in $h_{\mu\nu}$, with gauge fixing and ghost
contributions added, similar to those in Eqs.~(\ref{eq:gauge-fix-back}) and
(\ref{eq:ghost-back}), respectively.
The first order variation of the action of Eq.~(\ref{eq:hdqg})
gives the field equations
for higher derivative gravity in the absence of sources,
\bea
{ \partial I \over \partial g^{\mu\nu} } & = &
{ 1 \over \kappa^2 } \sqrt{g} \, ( R^{\mu\nu} - \half g^{\mu\nu} R ) +
\half \lambda_0 \sqrt{g} \, g^{\mu\nu} 
\nonumber \\
&& + a \, \sqrt{g} \; [ \, \twoth (1+\omega) \, R \, 
( R^{\mu\nu} - \quarter \, g^{\mu\nu} R )
\nonumber \\
&& + \half \, g^{\mu\nu} R_{\alpha\beta} R^{\alpha\beta}
- 2 R^{\mu\alpha\nu\beta} R_{\alpha\beta} 
+ \third ( 1 - 2 \omega ) \, \nabla^\mu \nabla^\nu R 
\nonumber \\
&& - \Box R^{\mu\nu} + \sixth ( 1 + 4 \omega ) \, g^{\mu\nu} \, \Box R \, ] = 0
\eea
where we have set for the ratio of the two higher derivative
couplings $\omega = b/a $.

The second order variation is done similarly.
It then allows the Gaussian integral over the quantum fields to be performed
using the formula of Eq.~(\ref{eq:gauss-int}).
One then finds that the one-loop effective action,
which depends on $g_{\mu\nu}$ only, can be expressed as
\beq
\Gamma = \half \tr \ln F_{mn} - 
\tr \ln Q_{\alpha\beta} - \half \tr \ln c^{\alpha\beta}
\label{eq:hdqg-gamma}
\eeq
with the quantities $F_{nm}$ and $Q_{\alpha\beta}$ defined by
\bea
F_{nm} & = & { \delta^2 I \over \delta g^m \, \delta g^n } +
{ \delta \chi_\alpha \over \delta g^m } \,
c^{\alpha\beta} \, { \delta \chi_\beta \over \delta g^n }
\nonumber \\
Q_{\alpha\beta} & = & { \delta \chi_\alpha \over \delta g^m } \, \nabla_\beta^m
\eea
A shorthand notation is used here, where spacetime and internal indices
are grouped together so that $g^m = g_{\mu\nu} (x)$.
$\chi_\alpha$ are a set of gauge conditions, $c^{\alpha\beta}$ is a nonsingular functional matrix fixing the gauge, and the $\nabla^i_\alpha$ are the local generators of the group of general coordinate transformations, 
$\partial^i_\alpha f^\alpha = 2 g_{\alpha ( \mu } \nabla_{\nu ) } f^\alpha (x)$.

Ultimately one is only interested in the divergent part of the effective
one-loop action.
The method of extracting the divergent part out of the determinant
(or trace ) expression in Eq.~(\ref{eq:hdqg-gamma}) is similar to
what is done, for example, in QED to evaluate the contribution of the fermion
vacuum polarization loop to the effective action.
There, after integrating out the fermions, one obtains a functional
determinant of the massless Dirac operator $ \dslash (A) $ in
an external $A_\mu$ field,
\beq
\tr \ln \dslash (A) - \tr \ln \Dslash \, = \,
{ c \over \epsilon } \, \int d^4 x \,
A^\mu \, ( \eta_{\mu\nu} \, \partial^2 - \partial_\mu \partial_\nu ) \, A^\nu
+ \dots
\eeq
with $c$ a calculable numerical constant. 
The trace needs to be regulated, and one way of doing it is via the integral
representation
\beq
\half \tr \ln \dslash^{\, 2} (A) = 
- \half \int_\eta^\infty { d t \over t } \,
\tr \exp \, [ - t \, \dslash^2 (A) ]
\eeq
with $\eta$ a cutoff that is sent to zero at the end of the
calculation.
For gauge theories a detailed discussion can be found 
for example in (Rothe, 1985), and references therein.
In the gravity case further discussions and more results can be found in
(De Witt 1965; 't Hooft and Veltman 1974; Gilkey 1975; Christensen and
Duff 1979) and references therein.

In the end, by a calculation similar to the one done in the pure 
Einstein gravity case, one finds that the one-loop contribution to the 
effective action contains for $d \rightarrow 4$ a divergent term of the form
\beq
{\Delta \cal L} = { \sqrt{g}  \over 16 \pi^2 (4-d) } \, \left \{ \,
\beta_2 \, ( R_{\mu\nu}^2 - \third R^2 ) 
+ \beta_3 \, R^2 + \beta_4 \, R + \beta_5 \, \right \}
\eeq
with the coefficients for the divergent parts given by
\bea
&& \beta_2 = {133 \over 10}
\nonumber \\
&& \beta_3 = { 10 \over 9 } \, \omega^2 + { 5 \over 3} \, \omega + { 5 \over 36}
\nonumber \\
&& \beta_4 = { 1 \over a \kappa^2 } \,
\left ( {10 \over 3} \, \omega - { 13 \over 6 } - { 1 \over 4 \omega } \right )
\nonumber \\
&& \beta_5 = { 1 \over a^2 \kappa^4 } \,
\left ( { 5 \over 2 } + {1 \over 8 \omega^2 } \right ) +
{ \tilde \lambda \over a \kappa^4 } \, 
\left ( {56 \over 3 } + { 2 \over 3 \omega } \right )
\eea
Here $\omega \equiv b/a$ and $\tilde \lambda$ 
is the dimensionless combination of the cosmological and Newton's constant 
$\tilde \lambda \equiv \half \lambda_0 \kappa^4 $ with $\kappa^2 = 16 \pi G$.
A divergence proportional to the topological invariant $\chi$ with coefficient
$\beta_1$ has not been included, as it only adds a field-independent 
constant to the action for a manifold of fixed topology.
Also $\delta^{(4)}(0)$-type divergences possibly originating from 
a non-trivial functional measure over the $g_{\mu\nu}$'s have been set to zero.

The structure of the ultraviolet divergences (which for an explicit momentum 
cutoff 
$\Lambda$ would have appeared as $1 / \epsilon \leftrightarrow \ln \Lambda $)
allow one to read off immediately the renormalization group 
$\beta$-functions for the various couplings.
To this order, the renormalization group equations for
the two higher derivative couplings $a$ and $b$ and the 
dimensionless ratio of cosmological and Newton's
constant $\tilde \lambda $ are 
\bea
&& { \partial a  \over  \partial t } = \beta_2 + \dots
\nonumber \\
&& { \partial \omega \over \partial t } = 
- { 1 \over a } \, ( 3 \beta_3 + \omega \beta_2 ) + \dots
\nonumber \\ 
&& { \partial \tilde \lambda \over \partial t } = 
\half \kappa^4 \, \beta_5 + 2 \tilde \lambda \kappa^2 \, \beta_4 + \dots
\label{eq:hdqg-rg}
\eea
with the dots indicating higher loop corrections.
Here $t$ is the logarithm of the relevant energy scale, 
$t = (4 \pi)^{-2} \ln ( \mu / \mu_0 )$, with $\mu$ a
momentum scale $q^2 \approx \mu^2$, and $\mu_0$ some fixed reference scale.
It is argued furthermore by the quoted authors that only the quantities
$\beta_2$, $\beta_3$ and the combination
$\kappa^4 \beta_5 + 4 \tilde \lambda \kappa^2 \beta_4$ are gauge independent,
the latter combination appearing in the renormalization group
equation for $\tilde \lambda (t)$
(this is a point to which we shall return later, as it follows quite
generally from the properties of the gravitational action, and
therefore from the gravitational functional integral, under a field
rescaling, see Sect.~\ref{sec:graveps}).

The perturbative scale dependence of the couplings $a(\mu)$, $b(\mu)$ and
$\tilde \lambda (\mu)$ follows from integrating the
three differential equations in Eq.~(\ref{eq:hdqg-rg}).
The first renormalization group equation is easily integrated, and
shows the existence of an ultraviolet fixed point at $a ^{-1} = 0$;
the one-loop result for the running coupling $a$ is simply given by
$a(t) = a(0) + \beta_2 \, t  $, or 
\beq
a^{-1} ( \mu ) \; \mathrel{\mathop\sim_{ \mu \rightarrow \infty }} \;
{ 16 \pi^2 \over \beta_2  \ln ( \mu / \mu_0 ) }
\eeq
with $\mu_0$ a reference scale.
It suggests that the effective higher derivative coupling
$a (\mu) $ increases at short distances, but decreases
in the infrared regime $\mu \rightarrow 0$.
But one should keep in mind
that the one loop results are reliable at best only at very short
distances, or large energy scales, $t \rightarrow \infty$.
At the same time these results seem physically reasonable,
as one would expect curvature squared terms to play less of
a role at larger distances, as in the classical theory.

The scale dependence of the other couplings is a bit more complicated.
The equation for $\omega (t)$ exhibits two fixed points at
$\omega_{uv} \approx -0.0229$ and $\omega_{ir} \approx -5.4671$; in
either case this would correspond to a higher derivative action with
a positive $R^2$ term.
It would also give rise to rapid short distance
oscillations in the static
potential, as can be seen for example from Eq.~(\ref{eq:hdqg-pot}) 
and the definition of $ m_0 = \mu / \sqrt { 2 b } $.
The equation for $\tilde \lambda (t)$ gives a solution to
one-loop order $\tilde \lambda (t) \sim {\rm const.} \, t^q $ with
$q \approx 0.91$, suggesting that the effective gravitational constant,
in units of the cosmologial constant, decreases at large distances.
The experimental value for Newton's constant
$ \hbar G /c^3 = (1.61624 \times 10 ^{-33} cm)^2 $
and for the scaled cosmological constant
$ G \lambda_0 \sim  1/(10^{28} cm )^2 $ is such
that the observed dimensionless ratio between the two is very
small, $ G^2 \lambda_0 \sim 10^{-120}$.
In the present model is seems entirely unclear how such a small ratio
could arise from perturbation theory alone.

At short distances the dimensionless coupling 
$\tilde \lambda \sim \lambda_0 G^2 $ seems to increases rapidly, 
thus partially invalidating the conclusions of a weak field expansion
around flat space, which are based generally on the assumption of small
$G$ and $\lambda_0$.
At the same time, the fact that the higher derivative coupling $a$
grows more rapidly
in the ultraviolet than the coupling $\tilde \lambda$ can
be used retroactively at least as a partial justification for
the flat space expansion,
in which the cosmological and Einstein terms are treated perturbatively.
Ultimately the resolution of such delicate and complex
issues would presumably require
the development of the perturbative expansion not around flat
space, but more appropriately around the de Sitter
metric, for which $R= 2 \lambda_0 / \kappa^2 $.
Even then one would have to confront such genuinely
non-perturbative issues, such as what happens to the spin-zero 
ghost mass, whether the ghost poles gets shifted away from the real axis
by quantum effects, and what the true ground state of the theory looks
like in the long distance, strong fluctuation regime not accessible by perturbation theory.

What is also a bit surprising is that higher derivative gravity, to
one-loop order, does not exhibit a nontrivial ultraviolet point
in $G$, even though such a fixed point is clearly present in the $2+\epsilon$
expansion (to be discussed later) at the one- and two-loop order,
as well as in the lattice regularized theory in four dimensions
(also to be discussed later).
But this could just reflect a limitation of the one-loop calculation; to
properly estimate the uncertainties of the perturbative
results in higher derivative
gravity and their potential physical implications a two-loop calculation
is needed, which hopefully will be performed in the near future.

To summarize, higher derivative gravity theories based on $R^2$-type terms
are perturbatively renormalizable, but exhibit some short-distance oddities in
the tree-level spectrum, associated with either ghosts or tachyons.
Their perturbative (weak field) treatment suggest that the higher
derivative couplings are only relevant at short distances, comparable
to the Planck length, but the general evolution of the couplings 
away from a regime where perturbation theory is reliable remains
an open question, which perhaps will never be answered satisfactorily
in perturbation theory, if non-Abelian gauge theories, which are also
asymptotically free, are taken as a guide.

\subsubsection{Supergravity}
\label{sec:super}

An alternative approach to the vexing problem of ultraviolet divergences
in perturbative quantum gravity (and for that matter, in any field theory) is to
build in some additional degree of symmetry, such that loop effects 
acquire reduced divergence properties, or even become finite.
One such approach, based on the invariance under local supersymmetry transformation, adds to the
Einstein gravity Lagrangian a spin-$3/2$ gravitino field, whose
purpose is to exactly cancel the loop divergences in the Einstein
contributions.
The enhanced symmetry is built in to ensure that such a cancellation
does not just occur at one loop order, but propagates to every
order of the loop expansion.
The intent of this section is more to provide the general flavor of such
an approach, and illustrate supergravity theories by a few
specific examples of suitable actions.
The reader is then referred to the vast literature on the subject
for further examples, as well as contemporary leading candidate theories.

In the simplest scenario, one adds to gravity a spin-${3\over2}$ fermion
field with suitable symmetry properties.
A generally covariant action describing the interaction of
vierbein fields $e^a_\mu (x)$ 
(with the metric field given by $g_{\mu\nu} = e^a_\mu e_{a \nu}$) 
and Rarita-Schwinger spin-${3\over2}$ fields $\psi_\mu (x)$, subject to the
Majorana constraints $\psi_\rho = C \bar \psi_\rho^T $,
was originally given in (Ferrara, Freedman and van Nieuwenhuizen, 1976).
In the second order formulation it contains three contributions
\beq
I \, = \, \int d^4 x \; ( {\cal L}_2 + {\cal L}_{3/2} + {\cal L}_4 )
\label{eq:sugra}
\eeq
with the usual Einstein term
\beq
{\cal L}_2 = { 1 \over 4 \kappa^2 } \, \sqrt{g} \, R \; ,
\eeq
the gravitino contribution
\beq
{\cal L}_{3/2} = - \half \, \epsilon^{\lambda\rho\mu\nu} \,
\bar \psi_\lambda \gamma_5 \gamma_\mu D_\nu \psi_\rho  \; ,
\eeq
and a quartic fermion self-interaction
\bea
{\cal L}_4 = - & { 1 \over 32 \kappa^2 \, \sqrt{g} } &
\, (
\epsilon^{\tau\alpha\beta\nu} \epsilon_\tau^{\;\;\gamma\delta\mu} + 
\epsilon^{\tau\alpha\mu\nu} \epsilon_\tau^{\;\;\gamma\delta\beta} -
\epsilon^{\tau\beta\mu\nu} \epsilon_\tau^{\;\;\gamma\delta\alpha} )
\nonumber \\
& \times & 
( \bar \psi_\alpha \gamma_\mu \psi_\beta ) 
( \bar \psi_\gamma \gamma_\nu \psi_\delta ) 
\eea
The covariant derivative defined as
\beq
D_\nu \psi_\rho = \partial_\nu \psi_\rho 
- \Gamma^\sigma_{\nu\rho} \, \psi_\sigma
+ \half \, \omega_{\nu a b} \, \sigma^{ab} \, \psi_\rho
\eeq
involves the standard affine connection $\Gamma^\sigma_{\nu\rho}$,
as well as the vierbein connection
\bea
\omega_{\nu \, a b} = & \half & [ e_a^{\;\;\mu} ( \partial_\nu \, e_{b \mu}
- \partial_\mu \, e_{b \nu} ) +
e_a^{\;\;\rho} e_b^{\;\;\sigma} 
( \partial_\sigma \, e_{c \rho} ) \, e^c_{\;\;\nu} ]
\nonumber \\
& - & ( a \leftrightarrow b ) 
\eea
with Dirac spin matrices
\beq
\sigma_{ab} = \quarter \, [ \gamma_a , \gamma_b ] \; .
\eeq
One can show that the combined Lagrangian is invariant, 
up to terms of order $(\psi)^5$, 
under the simultaneous transformations
\bea
&& \delta e^a_{\;\;\mu} (x) = i \kappa \, 
\bar \epsilon (x) \, \gamma^a \, \psi_\mu (x)
\nonumber \\
&& \delta g_{\mu\nu} = i \kappa \, \bar \epsilon (x) \, [ 
\gamma_\mu \, \psi_\nu (x) + \gamma_\nu \, \psi_\mu ]
\nonumber \\
&& \delta \psi_\mu (x) = \kappa^{-1} D_\mu \, \epsilon (x)
+ \quarter \, i \kappa \,
( 2 \, \bar \psi_\mu \gamma_a \psi_b + \bar \psi_a \gamma_\mu \psi_b ) 
\, \sigma^{ab} \, \epsilon (x)
\eea
where $\epsilon(x)$ in an arbitrary Majorana spinor.

The action of Eq.~(\ref{eq:sugra}) can be written equivalently 
in first order form (Deser and Zumino, 1976) as
\beq
I \, = \, \int d^4 x \, \left ( \quarter \kappa^{-2} \, e \, R 
- \half \, \epsilon^{\lambda\rho\mu\nu} \,
\bar \psi_\lambda \, \gamma_5 \, \gamma_\mu \, D_\nu \, \psi_\rho \right )
\label{eq:sugra1}
\eeq
with $e_{a\mu}$ the vierbein with $e_{a\mu} e^a_{\nu} = g_{\mu\nu}$, and
\beq
e = \det e_{a \mu} \; , \;\;\;\;\;\; R = e_a^{\;\;\mu} e_b^{\;\;\nu}
R_{\mu\nu}^{\;\;\;\;ab} \;\; .
\eeq
The covariant derivative $D_\mu$ on $\psi_\nu$ is defined in terms of
its spin-$\half$ part only
\beq
D_\mu = \partial_\mu - \half \, \omega_{\mu \,  a b} \, \sigma^{ab} \; .
\eeq
and is related to the curvature tensor via the commutator identity
\beq
[ D_\mu , D_\nu ] = - \half \, R_{\mu\nu \, ab} \, \sigma^{ab}
\eeq
The first order action in Eq.~(\ref{eq:sugra1}) is invariant under
\bea
&& \delta e^a_{\;\;\mu} (x) = i \kappa \, 
\bar \epsilon (x) \, \gamma^a \, \psi_\mu (x)
\nonumber \\
&& \delta \psi_\mu (x) = \kappa^{-1} \, D_\mu \, \epsilon (x)
\nonumber \\
&& \delta \omega_\mu^{\;\;ab}= B_\mu^{\;\;ab}
- \half \, e_\mu^{\;\;b} \, B_c^{\;\;ab}
+ \half \, e_\mu^{\;\;a} \, B_c^{\;\;bc}
\eea
with the quantity $B$ defined as
\beq
B_a^{\lambda\mu} = i \, \epsilon^{\lambda\mu\nu\rho}
\; \bar \epsilon \,
\gamma_5 \, \gamma_a \, D_\nu \, \psi_\rho
\eeq
and $\epsilon(x)$ in an arbitary local Majorana spinor.
In the first order formulation the vierbeins $e_{a\mu} (x)$, the 
connections $\omega_\mu^{\;\;ab}(x) $ and the Majorana
vector-spinors $\psi_\mu (x)$ are supposed to be varied
independently. 

In (D'Eath, 1984; 1994) the ${\cal N}=1$ supergravity action
is written as  
\beq
I \, = \, \int d^4 x \, ( {\cal L}_2 + {\cal L}_{3/2} )
\label{eq:sugra2}
\eeq
with
\beq
{\cal L}_2 = { 1 \over 8 \kappa^2 } \,
\epsilon^{\mu\nu\rho\sigma} \,
\epsilon_{abcd} \, e^a_{\;\;\mu} \, e^b_{\;\;\nu} \,
R^{cd}_{\;\;\;\;\rho\sigma}
\eeq
\beq
{\cal L}_{3/2} = - \half \, \epsilon^{\mu\nu\rho\sigma} (
\bar \psi_\mu \, e^a_{\;\;\nu} \, \bar \sigma_a \, D_\rho \, \psi_\sigma -
D_\rho \, \bar \psi_\mu \, e^a_{\;\;\nu} \, \bar \sigma_a \, \psi_\sigma )
\eeq
in terms of the Weyl spinor gravitino fields $\psi_{A\mu}$
and $\bar \psi_{A'\mu}$ and the vierbein field $e^a_{\;\;\mu}$.
The $\epsilon$'s are Levi-Civita symbols with curved (up)
and flat (down) indices respectively.
The quantities $\bar \sigma_a $ represent a curved space generalization
of the Pauli matrices discussed in (Carroll et al, 1994).

The original motivation for the supergravity action
of Eqs.~(\ref{eq:sugra}) or (\ref{eq:sugra1}) was 
that, just like ordinary source-free gravity is ultraviolet
finite on-shell because of the identity relating 
the invariant $(R_{\mu\nu\rho\,\sigma})^2$ to  
$(R_{\mu\nu})^2$ and $R^2$, identities among invariants
constructed out of $\psi_\mu$ and the strong constraints of
supersymmetry would ensure one-loop, and higher, renormalizabilty
of supergravity.
There are reasons to believe that the triviality results found originally
in globally supersymmetric theories (Nicolai, 1984) will not carry over
into theories with local supersymetry.

It was shown originally in (Grisaru, van Nieuwenhuizen and Vermaseren, 1976) and (Grisaru, 1976) 
that the original supergravity theory is finite to at least two loops.
But most likely it fails to be finite at three loops 
(Deser, Kay and Stelle, 1977).
As a consequence, more complex theories were devised to avoid the
three-loop catastrophe.
A new formulation, ${\cal N}=4$ extended supergravity based on an $SO(4)$
symmetry, was suggested in 
(Das 1977; Cremmer and Scherk, 1977; Nicolai and Townsend, 1981).
This theory now contains vector, spinor and scalar particle in addition to
the gravitino and the graviton.
Specifically, the theory contains a vierbein field $e_{a\mu}$,
four spin-${3 \over 2}$ Majorana fields $\psi_\mu^i$, four spin-${1\over2}$
Majorana fields $\xi^i$, six vector fields $A^{ij}_\mu$, a scalar field
$A$ and a pseudoscalar field $B$, all massless, for a grand total
of 53 independent terms in the Lagrangian.
Subsequently ${\cal N}=8$ supergravity was proposed, based on the even
larger group $SO(8)$ (Cremmer and Julia, 1978).
The enlarged theory now contains one graviton, 8 gravitinos, 
28 vector fields, 56 Majorana spin-$\half$ fields and 70 scalar
fields, all massless.
In general, $SO({\cal N})$ supergravity contains ${\cal N}$ gravitinos,
$\half {\cal N}({\cal N}-1)$ gauge fields, as well as
several spin-$\half$ Majorana fermions and complex scalars.
The $SO({\cal N})$ symmetry here is one which rotates, for example, the
${\cal N}$ gravitinos into each other.
In (Christensen, Duff, Gibbons and Ro\u cek, 1980) it was shown that in
general such theories are finite at one loop order for ${\cal N}>4$.
For ${\cal N} >8 $ these theories become less viable since
one then has more than one graviton, which leads to paradoxes, 
as well as particles with spin $j>2$. 

It is beyond our scope here to go any more deeply in the issue of
the origin of such intriguing ultraviolet cancellations.
But, as perhaps the simplest and most elementary motivation, one
can use the Nielsen-Hughes formula (Nielsen, 1980; Hughes, 1981)
for the one-loop $\beta$-function contribution from a particle of spin $s$
\beq
\beta_0 \, = \, - (-1)^{2 s} \, [ (2 s)^2 - \third ]
\eeq
to verify, by virtue of the particle multiplicities given above,
that for example for ${\cal N}=4$ the lowest order divergences cancel
\beq
\beta_0 \, = \, {\textstyle{
- { 47 \over 3} \cdot 1 
+ { 26 \over 3} \cdot 4
- { 11 \over 3} \cdot 6
+ { 2 \over 3} \cdot 4
+ { 1 \over 3} \cdot 1 
}\displaystyle} \, = \, 0
\eeq
For ${\cal N}=8$ one has a similar complete cancellation
\beq
\beta_0 \, = \, {\textstyle{
- { 47 \over 3} \cdot 1
+ { 26 \over 3} \cdot 8
- { 11 \over 3} \cdot 28
+ { 2 \over 3} \cdot 56
+ { 1 \over 3} \cdot 35 
} \displaystyle} \, = \, 0
\eeq
Still, the issue of perturbative ultraviolet finiteness of these theories
remains largely an open question, in part due to the daunting complexity
of higher loop calculations, even though one believes
that the high level of symmetry should ensure the cancellation
of ultraviolet divergences to a very high order (at least up to seven loops).
Recently it was suggested, based on the correspondence between
${\cal N}=8$ supergravity and ${\cal N}=4$ super Yang-Mills theory
and the cancellations which arise at one and higher loops,
that supergravity theories might be finite to all orders 
in the loop expansion (Bern, Dixon and Roiban, 2006).

One undoubtedly very attractive feature of supergravity theories is that
they lead naturally
to a small, or even vanishing, renormalized cosmological constant $\lambda_0$. 
Due to the high level of symmetry, quartic and quadratic
divergences in this quantity are expected to cancel exactly between bosonic
and fermionic
contributions, leaving a finite or even zero result.
The hope is that some of these desirable features will survive
supersymmetry breaking, a mechanism eventually required in order
to remove, or shift to a high mass, the so far unobserved
supersymmetric partners of the standard model particles.

\subsection{Feynman Path Integral Formulation}
\label{sec:path}

So far the discussion of quantum gravity has focused almost entirely on perturbative aspects, where the gravitational coupling $G$ is assumed
to be weak, and the weak field expansion based on
$\bar g_{\mu\nu}= g_{\mu\nu}+ h_{\mu\nu}$ can be performed with some
degree of reliability.
At every order in the loop expansion the problem then reduces to the
systematic evaluation of an increasingly complex sequence of Gaussian
integrals over the (small) quantum fluctuation $h_{\mu\nu}$.

But there are reasons to expect that non-perturbative effects play an
important role in quantum gravity.
Then an improved formulation
of the quantum theory is required, which does not rely exclusively
on the framework of a perturbative expansion.
Indeed already classically a black hole solution can hardly be
considered a small perturbation of flat space.
Furthermore, the fluctuating metric field $g_{\mu\nu}$ is dimensionless
and carries therefore no natural scale.
For the simpler cases of a scalar field and non-Abelian gauge theories 
a consistent
non-perturbative formulation based on the Feynman path integral 
has been known for some time and is well developed.
Combined with the lattice approach, it provides an effective and powerful
tool for systematically investigating non-trivial
strong coupling behavior, such as confinement and chiral symmetry breaking.
These phenomena are known to be generally inaccessible in weak coupling 
perturbation theory.
In addition, the Feynman path integral approach
provides a manifestly
covariant formulation of the quantum theory, without the need for an
artificial $3+1$ split required by the more traditional canonical approach,
and the ambiguities that may follow from it.
In fact, as will be seen later, in its non-perturbative lattice formulation no gauge fixing is required.

\begin{figure}[h]
\epsfig{file=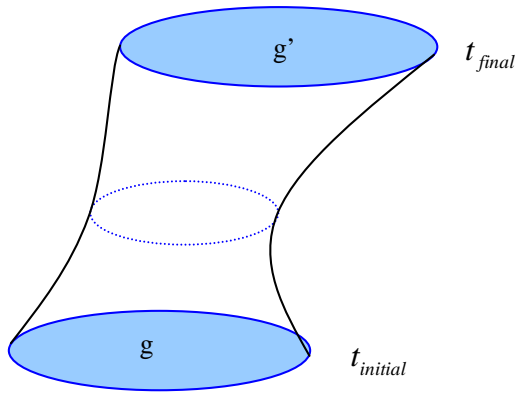,width=8cm}
\caption{Quantum mechanical amplitude of transitioning from an initial
three-geometry described by $g$ at time $t_{initial}$ to a final three-geometry
described by $g'$ at a later time $t_{final}$.
The full amplitude is a sum over all intervening metrics connecting the two bounding three-surfaces, weighted by $\exp (i I / \hbar)$ where
$I$ is a suitably defined gravitational action.}
\label{fig:amplitude}
\end{figure}

In a nutshell, the Feynman path integral formulation for pure quantum
gravitation can be expressed in the functional integral formula
\beq
Z = \int_ {\rm geometries }  e^{ \, { i \over \hbar} I_{\rm geometry} } \;\; ,
\label{eq:fey-path}
\eeq
(for an illustration see Fig.~\ref{fig:amplitude}),
just like the Feynman path integral for a non-relativistic quantum mechanical 
particle (Feynman, 1951; Feynman and Hibbs, 1962) expresses
quantum-mechanical amplitudes in terms of sums over paths
\beq
A  ( i \rightarrow f ) = \int_ {\rm paths }  
e^{ { \, i \over \hbar} I_{\rm path} } \;\; .
\eeq
What is the precise meaning of the expression in Eq.~(\ref{eq:fey-path})?
The remainder of this section will be devoted to discussing attempts at
a proper definition of the gravitational path integral of 
Eq.~(\ref{eq:fey-path}).
A modern rigorous discussion of path integrals in quantum mechanics and
(Euclidean) quantum field theory can be found, for example, in
(Albeverio and Hoegh-Krohn, 1976), (Glimm and Jaffe, 1981), and
(Zinn-Justin, 2002).

\subsubsection{Sum over Paths}
\label{sec:pathnon}

Already for a non-relativistic particle the path integral needs to be
defined quite carefully by discretizing the time
and introducing a short distance cutoff.
The standard procedure starts from the quantum-mechanical 
transition amplitude
\beq
A ( q_i , t_i \rightarrow q_f , t_f ) \; = \;
< q_f | \, e^{ - {i \over \hbar} H (t_f - t_i) } \, | q_i >
\eeq
and subdivides the time interval into $n+1$ segments of size $\epsilon$
with $t_f = (n+1) \epsilon + t_i $.
Using completeness of the coordinate basis $| q_j >$ at all
intermediate times, one obtains the textbook
result, here for a non-relativistic particle described 
by a Hamiltonian $H(p,q)=p^2/(2m) + V(q)$,
\bea
A ( q_i , t_i \rightarrow q_f , t_f ) & = &
\lim_{n \rightarrow \infty} \int_{-\infty}^{\infty} \prod_{j=1}^n
{ d q_j \over \sqrt{2 \pi i \hbar \epsilon / m } } 
\nonumber \\
& & \;\;\;\; \times 
\exp \left \{ 
{ i \over \hbar} \, \sum_{j=1}^{n+1} \epsilon \left [
\half \, m \, \left ( { q_j - q_{j-1} \over \epsilon } \right )^2 
- V \left ( { q_j + q_{j-1} \over 2 } \right ) 
\right ] \right \}
\label{eq:path-nonrel}
\eea
The expression in the exponent is recognized as a discretized
form of the classical action.
The above quantum-mechanical amplitude $A$ is usually written in shorthand as
\beq
A ( q_i , t_i \rightarrow q_f , t_f ) =
\int_{q_i(t_i)}^{q_f(t_f)} [ d q ] \,
\exp \left \{ 
{ i \over \hbar} \int_{t_i}^{t_f} dt \, L (q,\dot q)
\right \}
\label{eq:path-nonrel1}
\eeq
with $L= \half \, m \dot q^2 - V(q)$ the Lagrangian for the particle.
Thus what appears in the exponent is the classical action
\beq
I = \int_{t_i}^{t_f} dt \, L (q,\dot q)
\eeq
associated with a given trajectory $q(t)$, connecting
the initial coordinate $q_i(t_i)$ with the final one $q_f(t_f)$.
Then $[dq]$ is the functional measure over paths $q(t)$, as
spelled out explicitly in the precise lattice definition of
Eq.~(\ref{eq:path-nonrel}).
One advantage associated with having the classical action 
appear in the quantum mechanical amplitude is that all
the symmetries of the theory are manifest in the Lagrangian form.
The symmetries of the Lagrangian then have direct implications
for the study of quantum mechanical amplitudes.
A stationary phase approximation to the path integral, valid in the
limit $\hbar \rightarrow 0$, leads to the least action principle
of classical mechanics
\beq
\delta \, I = 0
\eeq
In the above derivation it is not necessary to use a uniform lattice
spacing $\epsilon$; one could have used as well a non-uniform
spacing $\epsilon_i = t_i - t_{i-1}$ but the result would have been
the same in the limit $ n \rightarrow \infty $ (in analogy with
the definition of the Riemann sum for ordinary integrals).
Since quantum mechanical paths have a zig-zag nature and
are nowhere differentiable,
the mathematically correct definition should be taken from the
finite sum in Eq.~(\ref{eq:path-nonrel}).
In fact it can be shown that differentiable paths have zero measure
in the Feynman path integral: already for the non-relativistic
particle most of the contributions
to the path integral come from paths that are far from smooth
on all scales (Feynman and Hibbs, 1963), the so-called Wiener
paths in turn related to Brownian motion.
In particular, the derivative $\dot q(t)$ is not always
defined, and the correct definition for the path integral
is the one given in Eq.~(\ref{eq:path-nonrel}).
A very complete and contemporary reference to the many
applications of path integrals
to non-relativistic quantum systems and statistical physics 
can be found in two recent books (Zinn-Justin, 2003; Kleinert, 2006).

As a next step, one can generalized the Feynman path integral construction to
$N$ particles with coordinates $q_i (t)$ ($i=1,N$), and finally to the 
limiting case of continuous fields $\phi(x)$.
If the field theory is defined from the start on a lattice, then the 
quantum fields are defined on suitable lattice points as $\phi_i$.

\subsubsection{Eulidean Rotation}
\label{sec:eucl}

In the case of quantum fields, one is generally interested in
the vacuum-to-vacuum amplitude, which requires
$t_i \rightarrow - \infty $ and $t_f \rightarrow + \infty $.
Then the functional integral with sources is of the form
\beq
Z [J] = \int [d \phi ] \exp \left 
\{ i \int d^4 x [ {\cal L}(x) + J(x) \phi (x) ] \right \}
\label{eq:z-mink}
\eeq
where $[d \phi] = \prod_x d \phi(x) $, and ${\cal L}$ 
the usual Lagrangian density for the scalar field,
\beq
{\cal L} = - \half \, [(\partial_\mu \phi )^2 - \mu^2 \, \phi^2 
- i \epsilon \, \phi^2 ]  - V(\phi)
\eeq
However even with an underlying lattice discretization, the integral
in Eq.~(\ref{eq:z-mink})
is in general ill-defined without a damping factor, 
due to the $i$ in the exponent (Zinn-Justin, 2003).

Advances in axiomatic field theory 
(Osterwalder and Schrader, 1973; Glimm and Jaffe 1974; Glimm and Jaffe, 1981) indicate that
if one is able to construct a well defined field theory in Euclidean
space $x=(\bf x, \tau)$ obeying certain axioms, then there
is a corresponding field theory in Minkowski space $({\bf x}, t)$ 
with
\beq
t = - \, i \, \tau
\eeq
defined as an analytic continuation of the Euclidean theory,
such that it obeys the Wightmann axioms (Streater and Wightman, 2000).
The latter is known as the {\it Euclidicity Postulate},
which states that the Minkowski
Green's functions are obtained by analytic continuation 
of the Green's function derived from the Euclidean functional.
One of the earliest discussion of the connection between Euclidean
and Minkowski filed theory can be found in (Symanzik, 1969).
In cases where the Minkowski theory appears pathological, the
situation generally does not improve by rotating
to Euclidean space.
Conversely, if the Euclidean theory is pathological, the problems
are generally not removed by considering the Lorentzian case.
From a constructive field theory point of view it seems difficult
for example to make sense, for either signature, out of one of
the simplest cases: a scalar
field theory where the kinetic term has the wrong sign 
(Gallavotti, 1985).

Then the Euclidean functional integral with sources is defined as
\beq
Z_E [J] = \int [d \phi ] \exp \left 
\{ - \int d^4 x [ {\cal L}_E (x) + J(x) \phi (x) ] \right \}
\label{eq:z-eucl}
\eeq
with $\int {\cal L}_E$ the Euclidean action, and
\beq
{\cal L}_E = \half (\partial_\mu \phi )^2 + \half \, \mu^2 \phi^2 + V(\phi)
\eeq
with now
$ (\partial_\mu \phi)^2 = (\nabla \phi)^2 + (\partial \phi / \partial \tau)^2 $.
If the potential $V(\phi)$ is bounded from below, then the
integral in Eq.~(\ref{eq:z-eucl}) is expected to be convergent.
In addition, the Euclidicity Postulate determines the correct
boundary conditions to be imposed on the propagator 
(the Feynman $i \epsilon$ prescription).
Euclidean field theory has a close and deep connection with statistical field 
theory and critical phenomena, whose foundations are surveyed for example 
in the monographs of (Parisi, 1982) and (Cardy, 1996).

Turning to the case of gravity, it should be clear that to all orders
in the weak field expansion there is really no difference of
substance between the Lorentzian (or pseudo-Riemannian) and the
Euclidean (or Riemannian) formulation.
Indeed most, if not all, of the perturbative calculations in the
preceding sections could have been carried out with the 
Riemannian weak field expansion about flat Euclidean space
\beq
g_{\mu\nu} = \delta_{\mu\nu} + h_{\mu\nu}
\eeq
with signature $++++$, or about some suitable classical Riemannian
background manifold, without any change of substance in the results.
The structure of the divergences would have been identical,
and the renormalization group properties of the coupling
the same (up to the trivial replacement of say the Minkowski momentum
$q^2$ by its Euclidean expression $q^2 = q_0^2 + {\bf q}^2 $ etc.).
Starting from the Euclidean result, the analytic continuation of
results such as Eq.~(\ref{eq:hdqg-rg}) to the pseudo-Riemannian
case would have been trivial.

\subsubsection{Gravitational Functional Measure}
\label{sec:measure}

It is still true in function space that one needs a metric
before one can define a volume element.
Therefore, following De Witt (De Witt 1962), one 
needs first to define an invariant norm for metric deformations  
\beq
\Vert \delta g \Vert^2 \, = \, 
\int d^d x \, \delta g_{\mu \nu}(x) \,
G^{\mu \nu, \alpha \beta} \bigl ( g(x) \bigr ) \,
\delta g_{\alpha \beta}(x) \;\; ,
\label{eq:dw-def}
\eeq
with the inverse of the super-metric $G$ given by the ultra-local
expression
\beq
G^{\mu \nu, \alpha \beta} \bigl ( g(x) \bigr ) \, = \, 
\half \, \sqrt{g(x)} \, \left [ \,
g^{\mu \alpha}(x) g^{\nu \beta}(x) +
g^{\mu \beta}(x) g^{\nu \alpha}(x) + \lambda \,
g^{\mu \nu}(x) g^{\alpha \beta}(x) \, \right ]
\label{eq:dw-super}
\eeq
with $\lambda$ an arbitrary real parameter.
The De Witt supermetric then defines a suitable volume element $\sqrt{G}$
in function space, such that the functional measure over
the $g_{\mu\nu}$'s taken on the form
\beq
\int [d \, g_{\mu\nu} ] \, \equiv \, \int \, \prod_x \, 
\Bigl [ \, \det G(g(x)) \, \Bigr ]^{1/2} \,
\prod_{\mu \geq \nu} d g_{\mu \nu} (x) \;\; .
\label{eq:dw-det0}
\eeq
The assumed locality of the super-metric $G^{\mu \nu, \alpha \beta} (g(x)) $ 
implies that its determinant is a local function of $x$ as well.
By a scaling argument given below
one finds that, up to an inessential multiplicative
constant, the determinant of the supermetric is given by
\beq
\det G (g(x)) \propto (1 + \half \, d \, \lambda ) \,
\bigl [ \, g(x) \, \bigr ]^{ (d-4)(d+1) /4 }
\;\; .
\label{eq:dw-det}
\eeq
which shows that one needs to impose the condition
$\lambda \neq - 2 / d $ in order to avoid the vanishing of $\det G$.
Thus the local measure for the Feynman path integral for pure gravity
is given by
\beq
\int \, \prod_x \, \bigl [ g(x) \bigr ]^{ (d-4)(d+1)/8 } \,
\prod_{\mu \ge \nu} \, d g_{\mu \nu} (x)
\label{eq:dw-dewitt}
\eeq
In four dimensions this becomes simply
\beq
\int [d \, g_{\mu\nu} ] \,  = \, 
\int \, \prod_x \, \prod_{\mu \ge \nu} \, d g_{\mu \nu} (x)
\label{eq:dw-dewitt-4d}
\eeq
However it is not obvious that the above construction is unique.
One could have defined, instead of Eq.~(\ref{eq:dw-super}),
$G$ to be almost the same, but without the $\sqrt{g}$ factor in front,
\beq
G^{\mu \nu, \alpha \beta} \bigl ( g(x) \bigr ) \, = \, 
\half \, \left [ \,
g^{\mu \alpha}(x) g^{\nu \beta}(x) +
g^{\mu \beta}(x) g^{\nu \alpha}(x) + \lambda \,
g^{\mu \nu}(x) g^{\alpha \beta}(x) \, \right ]
\label{eq:dw-super-mis}
\eeq
Then one would have obtained
\beq
\det G (g(x)) \propto (1 + \half \, d \, \lambda ) \,
\bigl [ \, g(x) \, \bigr ]^{ - (d+1) }
\;\; ,
\label{eq:dw-det-mis}
\eeq
and the local measure for the path integral for gravity
would have been given now by
\beq
\int \, \prod_x \, \bigl [ g(x) \bigr ]^{ -(d+1)/2 } \,
\prod_{\mu \ge \nu} \, d g_{\mu \nu} (x) \;\; .
\label{eq:dw-misn}
\eeq
In four dimensions this becomes
\beq
\int [d \, g_{\mu\nu} ] \,  = \, 
\int \, \prod_x \, \bigl [ g(x) \bigr ]^{ -5/2 } \,
\prod_{\mu \ge \nu} \, d g_{\mu \nu} (x)
\label{eq:dw-misn-4d}
\eeq
which was originally suggested in (Misner, 1957).

One can find in the original reference an argument suggesting that the
last measure is unique, provided the product $\prod_x$ is interpreted
over 'physical' points, and invariance is imposed at one and the same
'physical' point.
Furthermore since there are $d(d+1)/2$ independent components of the metric in $d$ dimensions, the Misner measure is seen to be invariant under a
re-scaling $g_{\mu\nu} \rightarrow \Omega^2 g_{\mu\nu} $ of the metric
for any $d$, but as a result is also found to be singular at small $g$.
Indeed the De Witt measure of Eq.~(\ref{eq:dw-dewitt}) and
the Misner scale invariant measure of Eqs.~(\ref{eq:dw-misn})
and (\ref{eq:dw-misn-4d}) 
could be just as well regarded as two special cases of a slightly
more general supermetric
$G$ with prefactor $\sqrt{g}^{(1-\omega)}$, with $\omega=0$
and $\omega=1$ corresponding to the original De Witt and Misner measures,
respectively.

The power in Eqs.~(\ref{eq:dw-det}) and (\ref{eq:dw-dewitt}) 
can be found for example as follows.
In the Misner case, Eq.~(\ref{eq:dw-misn}),
the scale invariance of the functional measure follows directly from the
original form of the supermetric $G(g)$ in Eq.~(\ref{eq:dw-super-mis}),
and the fact that the metric $g_{\mu\nu}$ has $\half d(d+1)$ 
independent components in $d$ dimensions.
In the DeWitt case one rescales the matrix $G(g)$ by a factor
$\sqrt{g}$.
Since $G(g)$ is a $ \half d(d+1) \times \half d(d+1)$ matrix, its
determinant is modified by an overall factor of $g^{d(d+1)/4}$.
So the required power in the functional measure 
is $ - \half (d+1) + \eigth d(d+1)= \eigth (d-4)(d+1)$, in agreement with
Eq.~(\ref{eq:dw-dewitt}).

Furthermore, one can show that if one introduces an $n$-component
scalar field $\phi(x)$ in the functional integral, it leads to
further changes in the gravitational measure.
First, in complete analogy to the gravitational case, one has for the scalar field deformation 
\beq
\Vert \delta \phi \Vert^2 \, = \, 
\int d^d x \, \sqrt{g(x)} \, \bigl ( \delta \phi (x) \bigr )^2  \;\; ,
\label{eq:dw-scalar}
\eeq
and therefore for the functional measure over $\phi$ one has the expression
\beq
\int [ d \phi ] \, = \,
\int \prod_x \, 
\bigl [ \sqrt{g(x)} \bigr ]^{n/2} \, \prod_x \, d \phi(x) \;\; .
\eeq
The first factor clearly represents an additional contribution
to the gravitational measure.
One can indeed verify that one just followed the correct procedure,
by evaluating for example the scalar functional integral in the
large mass limit,
\beq
\int \prod_x \, \bigl [ \sqrt{g(x)} \bigr ]^{n / 2} \, \prod_x \, d \phi(x) \,
\exp \left ( - \half m^2 \int \sqrt{g} \, \phi^2 \right ) \, = \,
\left ( \frac{ 2 \pi}{m^2} \right )^{n V /2} \, = \, {\rm const.}
\eeq
so that, as expected, for a large scalar mass $m$ the field $\phi$
completely decouples,
leaving the dynamics of pure gravity unaffected.

These arguments would lead one to suspect that the volume factor 
$g^{\sigma /2}$, when included 
in a slightly more general gravitational functional measure of the form
\beq
\int [d \, g_{\mu\nu} ] \,  = \, 
\prod_x \, \left [ g(x) \right ]^{\sigma / 2} \, 
\prod_{ \mu \ge \nu } \, d g_{ \mu \nu } (x) \;\; ,
\label{eq:gen-meas}
\eeq
perhaps does not play much of a role after all, at least
as far as physical properties are concerned.
Furthermore, in $d$ dimensions the $\sqrt{g}$ volume factors are
entirely absent ($\sigma=0$) if one chooses $\omega = 1- 4/d$, which
would certainly seem the simplest choice from a practical point of view.

When considering a Hamiltonian approach to quantum gravity, one finds
a rather different form for the functional measure (Leutwyler, 1964), which now
includes non-covariant terms.
This is not entirely surprising, as the introduction of a Hamiltonian requires
the definition of a time variable and therefore a preferred direction, and a specific choice of gauge.
The full invariance properties of the original action are no longer manifest
in this approach, which is further reflected in the use of a rigid lattice
to properly define and regulate the Hamiltonian path integral, allowing
subsequent formal manipulations to have a well defined meaning.
In the covariant approach one can regard formally the measure contribution as effectively a modification of the Lagrangian, leading to an $L_{eff}$.
The additional terms, if treated consistently will result in a modification of the Hamiltonian, which therefore in general will not be of the form one would have naively guessed from the canonical rules (Abers, 2004).
One can see therefore that the possible original measure ambiguity found in the covariant approach is still present in the canonical formulation.
One new aspect of the Hamiltonian approach is though that conservation of probability, which implies the unitarity of the scattering matrix, can further
restrict the form of the measure, if such a requirement is pushed down all the way to the cutoff scale (in a simplicial lattice context, the latter would be equivalent to the requirement of Osterwalder-Schrader reflection positivity at the cutoff scale). 
Whether such a requirement is physical and meaningful in a geometry that is strongly fluctuating at short distances, and for which a notion of time and
orthogonal space-like hypersurfaces is not necessarily well defined, remains an open question, and perhaps mainly an academic one. 
When an ultraviolet cutoff is introduced (without which the theory would not
be well defined), one is after all concerned in the end only with distance scales which are much larger than this short distance cutoff. 

Along these lines, the following argument supporting the possible irrelevance
of the measure parameter $\sigma$ can be given (Faddeev and Popov, 1973;
Fradkin and Vilkovisky, 1973).
Namely, one can show that the gravitational functional measure
of Eq.~(\ref{eq:gen-meas}) is invariant under infinitesimal general
coordinate transformations, irrespective of the value of $\sigma$.
Under an infinitesimal change of coordinates 
${x'}^{\mu} = x^{\mu} + \epsilon^{\mu} (x)$
one has
\beq
\prod_x \;
\left [ g(x) \right ]^{\sigma / 2}
\, \prod_{ \mu \ge \nu } \, d g_{ \mu \nu } (x)
\; \mathrel{\mathop\rightarrow} \;
\prod_x \,
\left ( \det { \partial {x'}^{\beta} \over \partial {x}^{\alpha} }
\right )^{\gamma} \;
\left [ g(x) \right ]^{\sigma / 2}
\; \prod_{ \mu \ge \nu } \, d g_{ \mu \nu } (x) \;\; .
\eeq
with $\gamma$ a power that depends on $\sigma$ and the dimension.
But for an infinitesimal coordinate transformations the additional factor
is equal to one,
\beq
\prod_x \;
\left ( \det { \partial {x'}^{\beta} \over \partial {x}^{\alpha} }
\right )^{\gamma} \,
\, = \,
\prod_x \,
\bigl [ \det ( \delta_{\alpha}^{\;\; \beta} + \partial_{\alpha}
\epsilon^{\beta} ) \bigr ]^{\gamma} \;
\; = \; 
\exp \left \{ \, \gamma \, \delta^d (0) \int d^d x \; \partial_{\alpha}
\epsilon^{\alpha} \right \} \, = \, 1 \;\; .
\eeq
and we have used
\beq
a^d \sum_x \; \rightarrow \; \int d^d x
\eeq
with lattice spacing $a=\pi/\Lambda$ and momentum cutoff
$\Lambda$ [see Eq.~(\ref{eq:delta0})].
So in some respects it appears that $\sigma$ can be compared to
a gauge parameter.

In conclusion, there is no clear a priori way of deciding between the
various choices
for $\sigma$, and the evidence so far suggests that it may very well turn
out to be an irrelevant parameter.
The only constraint seems that the regularized gravitational
path integral should be well defined,
which would seem to rule out singular measures, which need
additional regularizations at small volumes.
It is noteworthy though that the $g^{\sigma/2}$ volume term in the measure is 
completely local and contains no derivatives.
Thus in perturbation theory it cannot affect the propagation
properties of gravitons, and only contributes ultralocal
$\delta^d(0)$ terms to the effective action, as can be seen from
\beq
\prod_x \, \bigl [ g(x) \bigr ]^{\sigma/2} 
= \exp \left \{ \half \, \sigma \, \delta^d (0) 
\int d^d x \, \ln g(x) \right \} \;\; .
\eeq
with
\beq
\ln g(x) = \half \, h_\mu^{\;\;\mu} - 
\quarter \, h_{\mu\nu} h^{\mu\nu} + O(h^3)
\eeq
which follows from the general formal expansion formula for an 
operator ${\bf M} \equiv 1 + {\bf K }$
\beq
\tr \ln ( 1 + {\bf K} ) = \sum_{n=1}^\infty { (-1)^{n+1} \over n }
\, \tr {\bf K}^n \;\; .
\label{eq:tracelog}
\eeq
which is valid provided the traces of all powers of ${\bf K}$ exist.
On a spacetime lattice one can interpret the delta
function as an ultraviolet cutoff term, $\delta^d (0) \approx \Lambda^d$.
Then the first term shifts the vacuum solution and the second one modifies the
bare cosmological constant.
To some extent these type of contributions can be regarded as similar to
the effects arising from a renormalization of the cosmological
constant, ultimately affecting only the distribution of local volumes.
So far numerical studies of the lattice models to be discussed later
show no evidence of any sensitivity of the critical exponents
to the measure parameter $\sigma$.

Later in this review (Sect.~\ref{sec:lattmeas}) we will again return to the issue of the
functional measure for gravity in possibly the only context where it can be
posed, and to some extent answered, satisfactorily:
in a lattice regularized version of quantum gravity, going back
to the spirit of the original definition of Eq.~(\ref{eq:path-nonrel}).

In conclusion, the Euclidean Feynman path integral for pure
Einstein gravity with a cosmological constant term is given by 
\beq
Z_{cont} \; = \; \int [ d \, g_{\mu\nu} ] \; \exp \Bigl \{
- \lambda_0 \, \int d x \, \sqrt g \, + \, 
{ 1 \over 16 \pi G } \int d x \sqrt g \, R \Bigr \} \;\; .
\label{eq:zcont}
\eeq
It involves a functional integration over all metrics, with measure 
given by a suitably regularized form of 
\beq
\int [ d \, g_{\mu\nu} ] \; = \; \int \prod_x \; 
\left ( g(x) \right )^{ \sigma / 2 } \;
\prod_{\mu \ge \nu} \, d g_{\mu \nu} (x)
\label{eq:gen-meas1}
\eeq
as in Eqs.~(\ref{eq:dw-dewitt}), (\ref{eq:dw-misn}) and
(\ref{eq:gen-meas}).
For geometries with boundaries, further terms should be added
to the action, representing the effects of those boundaries.
Then the path integral will depend in general on some
specified initial and final three-geometry
(Hartle and Hawking 1977).

\subsubsection{Conformal Instability}
\label{sec:conformal}

Euclidean quantum gravity suffers potentially from a disastrous problem
associated with the conformal instability: the presence of kinetic
contributions to the linearized action entering with the wrong sign.

As was discussed previously in Sec.~\ref{sec:hdqg}, the action for linearized gravity
without a cosmological constant term, Eq.~(\ref{eq:h-action}), 
can be conveniently written using the three spin projection operators
$ P^{(0)} , P^{(1)}, P^{(2)} $  as 
\beq
I_{\rm lin} \, = \, {k \over 4 } \int d x \;
h^{\mu\nu} \, [ P^{(2)} - 2 P^{(0)} ]_{\mu\nu\alpha\beta} \; 
\partial^2 \, h^{\alpha\beta}
\eeq
so that the spin-zero mode enters with the wrong sign, or what is normally
referred to as a ghost contribution.
Actually to this order it can be removed by a suitable choice of gauge,
in which the trace mode is made to vanish, as can be seen, for example,
in Eq.~(\ref{eq:h-field2}).
Still, if one were to integrate in the functional integral over the spin-zero
mode, one would have to distort the integration contour to complex values,
so as to render the functional integral convergent.

The problem is not removed by introducing higher derivative terms, as
can be seen from the action for the linearized theory of Eq.~(\ref{eq:hdqg-lin}),
\bea
I_{\rm lin} = \half \int  d x  & \{ & 
h^{\mu\nu} \, [ \half \, k + \half \, a \,
(- \partial^2 ) ] ( - \partial^2 ) \, 
P^{(2)}_{\mu\nu\rho\sigma} \, h^{\rho\sigma}
\nonumber \\
& + & h^{\mu\nu} \, [ - k - 2 \, b \,
(- \partial^2 ) ] (- \partial^2 ) \, 
P^{(0)}_{\mu\nu\rho\sigma} \, h^{\rho\sigma} \,
\}
\eea
as the instability reappears for small momenta, where the higher derivative
terms can be ignored (see for example Eq.~(\ref{eq:hdqg-prop})).
There is a slight improvement, as the instability is cured for large momenta,
but it is not for small ones.
If the perturbative quantum calculations can be used as a guide,
then at the fixed points one has $b<0$, corresponding to a tachyon pole
in the spin-zero sector, which would  indicate 
further perturbative instabilities.
Of course in perturbation theory there never is a real problem, 
with or without higher derivatives, as one can just
define Gaussian integrals by a suitable analytic continuation.

But the instability seen in the weak field limit is not an artifact
of the weak field expansion.
If one attempts to write down a path integral for pure gravity of the form
\beq
Z = \int [ d \, g_{\mu\nu} ]  \;  e^{ - I_E }
\eeq
with an Euclidean action
\beq
I_E = \lambda_0 \, \int d x \, \sqrt g \, - \, 
{ 1 \over 16 \pi G } \int d x \sqrt g \, R 
\eeq
one realizes that it too appears ill defined due to the fact that the scalar
curvature can become arbitrarily positive, or negative.
In turn this can be seen as a direct consequence of the fact that while
gravitational radiation has positive energy, gravitational potential
energy is negative because gravity is attractive.
To see more clearly that the gravitational action can be made arbitrarily
negative consider the conformal transformation 
$ \tilde g_{\mu \nu} = \Omega ^2 g_{\mu \nu} $ where $\Omega$ 
is some positive function.
Then the Einstein action transforms into
\beq
I_E ( \tilde g ) = - { 1 \over 16 \pi G }  \int  d ^ 4 x  \sqrt g \;
( \Omega^2 R + 6 \; g^{\mu \nu} \partial_\mu \Omega \, \partial_\nu \Omega ) \;\; .
\eeq
which can be made arbitrarily negative by choosing a rapidly varying conformal
factor $\Omega$.
Indeed in the simplest case of a metric $g_{\mu\nu} = \Omega ^2 \eta_{\mu \nu}$
one has 
\beq
\sqrt{g} \, ( R -2 \lambda ) \, = \,  6 \, g^{\mu \nu} \partial_\mu \Omega \, \partial_\nu \Omega  \, - \, 2 \lambda \Omega^4
\eeq
which looks like a $\lambda \phi^4$ theory but with the wrong sign for the
kinetic term.
The problem is referred to as the conformal instability of the classical
Euclidean gravitational action (Hawking 1977).
The gravitational action is unbounded from below,
and the functional integral is possibly divergent, depending
on the detailed nature of the gravitational measure contribution
$[dg_{\mu\nu}]$,
more specifically its behavior in the regime of strong fields and
rapidly varying conformal factors.

A possible solution to the unboundedness problem has been described by
Hawking, who suggests performing the integration over all metrics by first
integrating over conformal factors by distorting the integration contour
in the complex plane to avoid the unboundedness problem, followed by
an integration over conformal equivalence classes of metrics
(Gibbons and Hawking, 1977; Hawking, 1978; Gibbons, Hawking and Perry, 1978;
Gibbons and Perry, 1978).
Explicit examples have been given where manifestly convergent Euclidean functional integrals have been formulated in terms of physical (transverse-traceless) degrees
of freedom, where the weighting can be shown to arise from a manifestly
positive action (Schleich, 1985; Schleich 1987).
A similar convergent procedure seem obtainable for some so-called minisuperspace 
models, where the full functional integration over the fluctuating metric
is replace by a finite dimensional integral over a set of parameters
characterizing the reduced subspace of the metric in question, see for
example (Barvinsky, 2007).
But it is unclear how this procedure can be applied outside perturbation
theory, where it not obvious how such a split for the metric should be
performed.

An alternate possibility is that the unboundedness of the classical Euclidean
gravitational action (which in the general case is certainly physical, and
cannot therefore be simply removed by a suitable choice of gauge) is not
necessarily an obstacle to defining the quantum theory.
The quantum mechanical attractive Coulomb well problem has, for zero orbital
angular momentum or in the one-dimensional case, a similar type of instability,
since the action there is also unbounded from below.
The way the quantum mechanical treatment ultimately evades the problem
is that the particle has a vanishingly small probability amplitude to
fall into the infinitely deep well.
In other words, the effect of quantum mechanical fluctuations in
the paths (their zig-zag motion) is just as important as the fact 
that the action is unbounded.
Not unexpectedly, the Feynman path integral solution of the Coulomb
problem requires again first the introduction of a lattice, and
then a very careful treatment of the behavior close to the singularity
(Kleinert, 2006).
For this particular problem one is of course aided by the fact that
the exact solution is known from the Schr\"odinger theory.

In quantum gravity the question regarding the conformal instability 
can then be rephrased in a similar way: Will the quantum fluctuations 
in the metric be strong enough so that physical excitations will
not fall into the conformal well?
Phrased differently, what is the role of a non-trivial 
gravitational measure, giving rise to a density of states $n(E)$ 
\beq
Z \, \propto \, \int_0^\infty \, dE \; n(E) \; e^{-E} \; ,
\eeq
regarding the issue of ultimate convergence (or divergence) of the
Euclidean path integral.
Of course to answer such questions satisfactorily one needs a
formulation which is not restricted to small fluctuations
and to the weak field limit.
Ultimately in the lattice theory the answer is yes, for
sufficiently strong coupling $G$ (Hamber and Williams, 1984; Berg 1985).

\subsection{Gravity in $2+\epsilon$ Dimensions}
\label{sec:epsilon}

In the previous sections it was shown that pure Einstein gravity is not perturbatively renormalizable in the traditional sense in four dimensions.
To one-loop order higher derivative terms are generated, which, when included
in the bare action, lead to potential unitarity problems, whose proper
treatment most likely lies outside the perturbative regime.
The natural question then arises: Are there any other
field theories where the standard perturbative treatment fails, 
yet for which one can find alternative methods and from them
develop consistent predictions?
The answer seems unequivocally yes.
Outside of gravity, there are two notable examples of field theories, the non-linear sigma model and the self-coupled fermion model, which are
not perturbatively renormalizable for $d>2$, and yet lead to
consistent and in some instances testable predictions above $d=2$.

The key ingredient to all of these results is, as originally recognized by Wilson, the existence of a non-trivial ultraviolet
fixed point, a phase transition in the statistical field theory context,
with non-trivial universal scaling dimensions (Wilson, 1971; Wilson and Fisher 1972; Wilson, 1973; Gross, 1976).
Furthermore, three quite different theoretical approaches are available
for comparing predictions: the $2+\epsilon$ expansion, the large-$N$
limit, and the lattice approach.
Within the lattice approach, several additional techniques are available:
the strong coupling expansion, the weak coupling expansion and the
numerically exact evaluation of the path integral.
Finally, the results for the non-linear sigma model in the
scaling regime around the non-trivial ultraviolet fixed point can be compared
to high accuracy satellite experiments on three-dimensional systems, 
and the results agree in some cases to several decimals.

The next three sections will therefore discuss these models from 
the perspective of those results which will have some relevance 
later for the gravity case. 
Of particular interest are predictions for universal
corrections to free field behavior, for the scale
dependence of couplings, and the role of the non-perturbative
correlation length which arises in the strong coupling regime.

Later sections will then discuss the $2+\epsilon$ expansion
for gravity, and what can be learned from it by comparing
it to the analogous expansion in the non-linear sigma model.
The similarity between the two models is such that they both
exhibit a non-trivial ultraviolet fixed point, a two-phase
structure, non-trivial exponents and scale-dependent couplings.

\subsubsection{Perturbatively Non-renormalizable Theories: The Sigma Model}

\label{sec:sigma}

The $O(N)$-symmetric non-linear $\sigma$-model provides an instructive and
rich example of a theory which, above two dimensions, is not 
perturbatively renormalizable in the traditional sense, and
yet can be studied in a controlled way in the context of Wilson's 
$2+\epsilon$ expansion.
Such framework provides a consistent way to calculate nontrivial
scaling properties
of the theory in those dimensions where it is not perturbatively
renormalizable (e.g. $d=3$ and $d=4$), which can then be compared
to non-perturbative
results based on the lattice theory, as well as to experiments,
since in $d=3$ the model describes either a ferromagnet or superfluid helium
in the vicinity of its critical point.
In addition, the model can be solved exactly in the large $N$ limit for 
any $d$, without any reliance on the $2+\epsilon$ expansion.
Remarkably, in all three approaches it exhibits a non-trivial
ultraviolet fixed point at some coupling $g_c$ (a phase transition
in statistical mechanics language), separating a weak coupling massless
ordered phase from a massive strong coupling phase.

The non-linear $\sigma$-model is described by an $N$-component field $\phi_a$
satisfying a unit constraint $\phi^2(x)=1$,
with functional integral given by
\bea
Z [J] \, & = & \, \int [ \, d \phi \, ] \, 
\prod_x \, \delta \left [ \phi (x) \cdot \phi (x) - 1 \right ] \,
\nonumber \\
&& \times \,
\exp \left ( - \, { \Lambda^{d-2} \over g } \, S(\phi) 
\, + \int d^d x \; J(x) \cdot \phi(x) \,  \right )
\nonumber \\
\label{eq:sigma-cont}
\eea
The action is taken to be $O(N)$-invariant
\beq
S(\phi) \, = \, \half \,
\int d^d x \; \partial_\mu \phi (x) \cdot \partial_\mu \phi (x)
\eeq
$\Lambda$ here is the ultraviolet cutoff and $g$ the bare
dimensionless coupling at the cutoff scale $\Lambda$;
in a statistical field theory context $g$ plays the role of a temperature.

In perturbation theory one can eliminate one $\phi$ field by 
introducing a convenient parametrization for the unit sphere,
$\phi (x) = \{ \sigma(x),{\bf \pi}(x) \}$ where $\pi_a$ is an
$N-1$-component field, and then solving locally for $\sigma(x)$
\beq
\sigma (x) \, = \, [ \, 1 - {\bf \pi}^2 (x) \, ]^{1/2}
\eeq
In the framework of perturbation theory in $g$ 
the constraint $ |{\bf \pi} (x)| <1$
is not important as one is restricting the fluctuations
to be small.
Nevertheless the ${\bf \pi}$ integrations will be extended from
$- \infty$ to $+\infty$, which reduces the development
of the perturbative expansion to a sequence of Gaussian integrals.
Values of ${\bf \pi} (x) \sim 1 $ give exponentially small
contributions of order $\exp (- {\rm const.} / g ) $ which are
therefore negligible to any finite order in perturbation theory.

In term of the ${\bf \pi}$ field the original action $S$ becomes
\beq
S({\bf \pi}) \, = \, \half \,
\int d^d x \, \left [ \,
( \partial_\mu {\bf \pi} )^2 
+ { ( {\bf \pi} \cdot \partial_\mu {\bf \pi} )^2 \over 1 - {\bf \pi}^2 }
\, \right ] 
\eeq
The change of variables from $\phi(x)$ to ${\bf \pi}(x) $ also gives
rise to a Jacobian
\beq
\prod_x \left [ \, 1 - {\bf \pi}^2 \, \right ]^{-1/2} \, \sim \;
\exp \left [ \, - \half \, \delta^d (0) \int d^d x \,
\ln ( 1 - {\bf \pi}^2 ) \, \right ]
\label{eq:jacobian}
\eeq
which is necessary for the cancellation of spurious tadpole
divergences.
The combined functional integral for the unconstrained
${\bf \pi}$ field is then given by
\beq
Z [ {\bf J} ] = \int [ \, d {\bf \pi} \, ] \, 
\exp \left ( - \, { \Lambda^{d-2} \over g } \, S_{0}( {\bf \pi} ) 
\, + \int d^d x \; {\bf J(x)} \cdot {\bf \pi} (x) \right )
\label{eq:sigma-cont1}
\eeq
with 
\bea
S_{0}( {\bf \pi} ) & = & 
\, \half \int d^d x \, \left [ 
( \partial_\mu {\bf \pi} )^2 
+ { ( {\bf \pi} \cdot \partial_\mu {\bf \pi} )^2 \over 1 - {\bf \pi}^2 }
\right ] 
\nonumber \\
&& + \, \half \, \delta^d (0) \int d^d x \; \ln ( 1 - {\bf \pi}^2 ) 
\eea
In perturbation theory the above action is expanded out in powers
of ${\bf \pi}$.
The propagator for the ${\bf \pi}$ field can be read off
from the quadratic part of the action,
\beq
\Delta_{ab} (k^2) \, = \, { \delta_{ab} \over k^2 }
\eeq
In the weak coupling limit the ${\bf \pi}$ fields correspond
to the Goldstone modes of the spontaneously broken $O(N)$
symmetry, the latter being broken spontaneously by
a non-vanishing vacuum expectation value $\langle {\bf \pi} \rangle \neq 0$.

Since the ${\bf \pi}$ field has mass dimension $\half (d-2)$, and the
interactions $\partial^2 {\bf \pi}^{2n}$ consequently has dimension
$n (d-2) +2 $, one finds that the theory is renormalizable in
$d=2$ and perturbatively non-renormalizable above $d=2$.
Furthermore, in spite of the theory being non-polynomial, it
can still be renormalized via the introduction of only two
renormalization constant, the coupling renormalization being
given by a constant $Z_g$ and the wavefunction renormalization by a
constant ${\bf \pi}$ by $Z$.
Potential infrared problems due to massless propagators
are handled by introducing an
external $h$-field term for the original composite $\sigma$ field,
which then acts as a mass term for the ${\bf\pi}$ field,
\bea
h \! \int d^d x \, \sigma (x)  & = &
h \! \int d^d x \, [ \, 1 - {\bf \pi}^2 (x) \, ]^{1/2}
\nonumber \\
& = & \int d^d x \, [ \, h - \half h \, {\bf \pi}^2 (x) \, + \dots \, ]
\eea
A proof can be found (David, 1982) that all $O(N)$ invariant Green's
are infrared finite in the limit $h \rightarrow 0$.

One can write down the same field theory on a lattice, where it
corresponds to the $O(N)$-symmetric classical Heisenberg
model at a finite temperature $T \sim g$.
The simplest procedure is to introduce a hypercubic lattice
of spacing $a$, with sites labeled by integers 
${\bf n}=(n_1 \dots n_d)$,
which introduces an ultraviolet cutoff $\Lambda \sim \pi/a$.
On the lattice field derivatives are replaced by finite differences
\beq
\partial_\mu \phi (x) \; \rightarrow \; 
\Delta_\mu \phi ( {\bf n} ) = 
{ \phi ( {\bf n} + {\bf \mu} ) - \phi ( {\bf n} ) \over a }
\eeq
and the discretized path integral then reads
\bea
Z[ {\bf J} ] & = & \, \prod_{\bf n} d \phi ( {\bf n} )
\, \delta [ \phi^2 ( {\bf n} ) -1 ] \,
\nonumber \\
& \times &  \exp \left [ - {a^{2-d} \over 2 g} \sum_{\bf n, \mu}
\left ( \Delta_\mu \phi ( {\bf n} ) \right )^2 +
\sum_{\bf n} {\bf J(x)} \cdot \phi (x) \right ]
\nonumber \\
\label{eq:sigma-lattice}
\eea
The above expression is recognized as the partition
function for a ferromagnetic $O(N)$-symmetric spin
system at finite temperature. 
Besides ferromagnets, it can be used to describe systems which
are related to it by universality, such as superconductors and
superfluid helium transitions, whose critical behavior is described by a 
complex phase, and which are therefore directly connected to the
plane rotator $N=2$, or $U(1)$, model.

In addition the lattice model of Eq.~(\ref{eq:sigma-lattice})
provides an explicit regularization for
the continuum theory, which makes expressions like the
one in Eq.~(\ref{eq:jacobian}) acquire a well defined meaning.
It is in fact the only regularization which allows a discussion
of the role of the measure in perturbation theory (Zinn-Justin, 2002).
At the same time it provides an ultraviolet regularization for
perturbation theory, and allows for various non-perturbative
calculations, such as power series expansions in three
dimensions and explicit numerical integrations of the path integral
via Monte Carlo methods.

In two dimensions one can compute the renormalization of the coupling
$g$ from the action of Eq.~(\ref{eq:sigma-cont1}) and one finds
after a short calculation (Polyakov 1975) for small $g$
\beq
{ 1 \over g (\mu) } \; = \;  { 1 \over g} \, 
+ \, { N - 2 \over 8 \pi } \, \ln { \mu^2 \over \Lambda^2 } 
\, + \, \dots
\eeq
where $\mu$ is an arbitrary momentum scale.
Physically one can view the origin of the factor of $N-2$ in the fact that
there are $N-2$ directions in which the spin can experience
rapid small fluctuations perpendicular to its average
slow motion on the unit sphere, and that only these fluctuations
contribute to leading order (Kogut 1979).

In two dimensions the quantum correction (the second term on the r.h.s.)
increases the value of the effective coupling at low momenta (large
distances), unless $N=2$ in which case the correction vanishes.
In fact the quantum correction can be shown to vanish to all orders
in this case; the vanishing of the $\beta$-function in two dimensions
for the $O(2)$ model is true only in perturbation theory,
for sufficiently strong coupling a phase transition appears,
driven by the unbinding of vortex pairs (Kosterlitz and Thouless, 1973).
For $N>2$ as $g(\mu)$ flows toward increasingly strong coupling it
eventually leaves the regime where perturbation theory can
be considered reliable.
But for bare $g \approx 0$ the quantum correction is negligible
and the theory is scale invariant around the origin: 
the only fixed point of the renormalization group, 
at least in lowest order perturbation theory, is at $g=0$.
For fixed cutoff $\Lambda$, the theory is weakly coupled at
short distances but strongly coupled at large distances.
The results in two dimensions for $N>2$ are qualitatively very similar to
asymptotic freedom in four-dimensional $SU(N)$ Yang-Mills theories.

\begin{figure}[h]
\epsfig{file=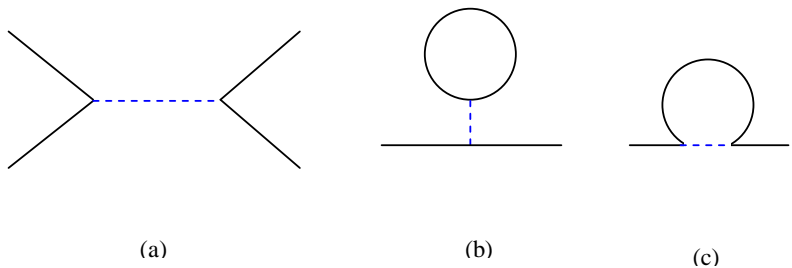,width=12cm}
\caption{One-loop diagrams giving rise to coupling and field renormalizations
in the non-linear $\sigma$-model.
Group theory indices $a$ flow along the thick lines,
dashed lines should be contracted to a point.}
\label{fig:sigma-one}
\end{figure}

Above two dimensions, $d-2=\epsilon >0$ and one can redo the same
type of perturbative calculation to determine the coupling renormalization.
The relevant diagrams are shown in Fig.~\ref{fig:sigma-one}.
One finds for the effective coupling $g_e$, i.e. the
coupling which includes the leading radiative correction
(using dimensional regularization, which is more convenient than
an explicit ultraviolet cutoff $\Lambda$ for performing actual 
perturbative calculations),
\beq
{ 1 \over g_e } \; = \;  { \Lambda^\epsilon \over g }
\, \left [ 1 \, - \, { 1 \over \epsilon } \, { N - 2 \over 2 \pi } \, g  
\, + \, O(g^2) \, \right ]
\eeq
The requirement that the dimensionful effective coupling $g_e$ be 
defined independently of the scale $\Lambda$ is expressed as
$ \Lambda { d \over d \Lambda} g_e =0$, and 
gives for the Callan-Symanzik $\beta$-function 
(Callan, 1970; Symanzik, 1970) for $g$
\beq
\Lambda \, { \partial \, g \over \partial \, \Lambda } \; = \; 
\beta (g) \; = \; 
\epsilon \, g - { N - 2 \over 2 \pi } \, g^2 
\, + \, O \left ( g^3, \epsilon g^2 \right ) 
\label{eq:beta-nonlin}
\eeq
The above $\beta$-function determines the scale dependence
(at least in perturbation theory) of $g$ for an arbitrary scale,
which from now on will be denoted as $\mu$.
Then the differential equation  
$ \mu{ \partial g \over \partial \mu } = \beta (g (\mu) )$ 
uniquely determines how $g (\mu) $ flows as a function of
momentum scale $\mu$.
The scale dependence of $g(\mu)$ is such that if the initial
$g$ is less than the ultraviolet fixed point value $g_c$, with
\beq
g_c \, = \, { 2 \pi \epsilon \over N - 2 } \, + \, \dots
\label{eq:gc}
\eeq
then the coupling will flow towards the Gaussian fixed
point at $g=0$.
The new phase that appears when $\epsilon >0$ and
corresponds to a low temperature, spontaneously
broken phase with finite order parameter.
On the other hand if $ g > g_c$ then the coupling
$g(\mu)$ flows towards
increasingly strong coupling, and eventually
out of reach of perturbation theory.
In two dimensions the $\beta$-function has no zero
and only the strong coupling phase is
present.
 
The one-loop running of $g$ as a function of a sliding momentum scale $\mu=k$
and $\epsilon>0$ can be obtained by integrating 
Eq.~(\ref{eq:beta-nonlin}).
One finds
\beq
g(k^2) \; = \; { g_c \over 1 \, \pm \, a_0 \, (m^2 / k^2 )^{(d-2)/2} } 
\label{eq:grun-nonlin} 
\eeq
with $a_0$ a positive constant and $m$ a mass scale;
the combination $ a_0 m^{d-2}$ is just the integration
constant for the differential equation, which we prefer
to split here in a momentum scale and a dimensionless coefficient
for reasons that will become clear later.
The choice of $+$ or $-$ sign is determined from whether one is
to the left (+), or to right (-) of $g_c$, in which case
$g(k^2)$ decreases or, respectively, increases as one flows away
from the ultraviolet fixed point.
The renormalization group invariant mass scale $\sim m$
arises here as an arbitrary
integration constant of the renormalization group equations,
and cannot be determined from perturbative arguments alone.
It should also be clear that
multiplying both sides of Eq.~(\ref{eq:grun-nonlin}) by
the ultraviolet cutoff factor $\Lambda^{2-d}$ to get
back the original dimensionful coupling multiplying
the action $S(\phi)$ in Eq.~(\ref{eq:sigma-cont}) does not
change any of the conclusions.

Note that the result of Eq.~(\ref{eq:grun-nonlin}) is quite different
from the naive expectation based on straight perturbation
theory in $d>2$ dimensions (where the theory is not perturbatively
renormalizable)
\beq
{ g(k^2) \over g } \sim 1 + {\rm const.} \; g \, k^{d-2} + O(g^2)
\eeq
which gives a much worse ultraviolet behavior.
The existence of a non-trivial ultraviolet fixed point alters
the naive picture and drastically improves the ultraviolet behavior.

For large $g$ one can easily see, for example from the structure
of the lattice action in Eq.~(\ref{eq:sigma-lattice}),
that correlation functions must decay exponentially at
large separations.
In the strong coupling limit, spins separated by a 
distance $|x|$ will fluctuate
in an uncorrelated fashion, unless they are connected
by a minimal number of link contributions from the 
action.
One expects therefore for the lattice
connected correlation function of two $\phi$ fields,
separated by a lattice distance $n a$,
\beq
< \! \pi ( n a ) \, \pi (0) \! >_c 
\; \mathrel{\mathop\sim_{ n \rightarrow \infty }} \;
\left ( { 1 \over g } \right )^{n}
\eeq
which can be re-written in continuum language
\beq
< \! \pi (x) \, \pi (0) \! >_c 
\; \mathrel{\mathop\sim_{ |x| \rightarrow \infty }} \;
\exp (- |x| / \xi)
\label{eq:two-point}
\eeq
with $m= 1/\xi = \Lambda \, [ \ln g + O(1/g)]$
and $\Lambda = 1/a$.
From the requirement that the correlation length $\xi$ be
a physical quantity independent of scale, 
and consequently a renormalization group invariant,
\beq
\Lambda \, { d \over d \, \Lambda } \, 
m ( \Lambda, g (\Lambda ) ) \, = \, 0 \;\; ,
\label{eq:m-scale}
\eeq
one obtains in the strong coupling limit
\beq
\beta (g ) \, = \, - g \ln g \, + \, O(1/g) \;\; .
\eeq

If in Eq.~(\ref{eq:grun-nonlin}) one sets the momentum
scale $k$ equal to the cutoff scale $\Lambda$ and solves
for $m$ in the strong coupling phase one obtains
\beq
{ g_c \over g (\Lambda ) } = 1 - a_0 
\left ( { m^2 \over \Lambda^2 } \right )^{(d-2)/2}
\eeq
and therefore for $m$ in terms of the
bare coupling $ g \equiv g(\Lambda) $
\beq
m(g) = \Lambda \, \left ( { g_c \over a_0 } \right )^{2/(d-2)} 
\left ( { 1 \over g_c } - { 1 \over g } \right )^{1/(d-2)}
\label{eq:m-coeff}
\eeq
This last result shows the following important fact:
if $m$ is identified with the
inverse of the correlation length $\xi$ (which can be extracted
non-perturbatively,
for example from the exponential decay of correlation
functions, and is therefore generally unambiguous), then the
calculable constant relating $m$ to $g$ in Eq.~(\ref{eq:m-coeff})
uniquely determines the coefficient $a_0$
in Eq.~(\ref{eq:grun-nonlin}).
For example, in the large $N$ limit the value for $a_0$ will
be given later in Eq.~(\ref{eq:m-largen1}).

In general one can write down the complete renormalization group
equations for the cutoff-dependent $n$-point functions 
$\Gamma^{(n)} (p_i,g,h,\Lambda)$ (Brezin and Zinn-Justin, 1976;
Zinn-Justin, 2002).
For this purpose one needs to define the renormalized
truncated $n$-point function $\Gamma_r^{(n)}$,
\beq
\Gamma_r^{(n)} (p_i,g_r,h_r,\mu) = 
Z^{n/2} (\Lambda / \mu ,g) \, \Gamma^{(n)} (p_i,g,h,\Lambda) 
\eeq
where $\mu$ is a renormalization scale, and the constants
$g_r$, $h_r$ and $Z$ are defined by
\bea
g & = &(\Lambda / \mu)^{d-2} Z_g \, g_r  \;\;\;\;\; {\bf \pi} (x) = Z^{1/2} {\bf \pi}_r (x)
\nonumber \\
h & = & Z_h h_r \;\;\;\;\;\;\;\;\;\;\;\;\;\; Z_h = Z_g / \sqrt{Z}
\eea
The requirement that the renormalized $n$-point function
$\Gamma_r^{(n)}$ be independent of the cutoff $\Lambda$
then implies
\beq
\left [ \Lambda { \partial \over \partial \Lambda } +
\beta (g) { \partial \over \partial g }
- { n \over 2 } \zeta (g) 
+ \rho (g) \, h \, { \partial \over \partial h}
\right ]
\Gamma^{(n)} (p_i,g,h,\Lambda) =0
\eeq
with the renormalization group
functions $\beta (g)$, $\zeta (g)$ and $\rho(g)$
defined as
\bea
\Lambda { \partial \over \partial \Lambda } \vert_{\rm ren. fixed} \; g
& = & \beta (g)
\nonumber \\
\Lambda { \partial \over \partial \Lambda } \vert_{\rm ren. fixed} \, (- \ln Z)
& = & \zeta (g)
\nonumber \\
2 - d + \half \zeta (g) + { \beta (g) \over g } & = & \rho (g) \;\; .
\label{eq:rg-npoint}
\eea
Here the derivatives of the bare coupling $g$, of the ${\bf \pi}$-field
wave function
renormalization constant $Z$ and of the external field $h$
with respect to the cutoff $\Lambda$ are evaluated
at fixed renormalized (or effective) coupling, at the
renormalization scale $\mu$.

To determine the renormalization group functions
$\beta (g)$, $\zeta (g)$ and $\rho(g)$ one can in
fact follow a related but
equivalent procedure, in which, instead of requiring the renormalized
$n$-point functions $\Gamma^{(n)}_r$ to be independent of the
cutoff $\Lambda$ at fixed renormalization scale $\mu$ as in
Eq.~(\ref{eq:rg-npoint}), one imposes that the {\it bare}
$n$-point functions $\Gamma^{(n)}$ be independent of the
renormalization scale  $\mu$ at {\it fixed} cutoff $\Lambda$.
One can show (Brezin, Le Guillou and Zinn-Justin, 1978)
that the resulting renormalization group functions are
identical to the previous ones, and that one can obtain
the scale dependence of the couplings (i.e. $\beta(g)$)
either way.
Physically the latter way of thinking is perhaps more suited
to a situation where one is dealing with
a finite cutoff theory, where the ultraviolet cutoff 
$\Lambda$ is fixed
and one wants to investigate the scale (momentum)
dependence of the couplings, for example $g(k^2)$.

For our purposes it will sufficient to look, in the zero-field
case $h=0$, at the $\beta$-function of Eq.~(\ref{eq:beta-nonlin})
which incorporates, as should already be clear from the
result of Eq.~(\ref{eq:rg-npoint}), a tremendous amount
of information about the model.
Herein lies the power of the renormalization group: the knowledge 
of a handful of functions [$\beta(g), \zeta(g)$] is sufficient
to completely determine the momentum dependence of
all $n$-point functions $\Gamma^{(n)} (p_i,g,h,\Lambda)$.

One can integrate the $\beta$-function equation 
in Eq.~(\ref{eq:beta-nonlin}) to obtain the renormalization
group invariant quantity
\beq
\xi^{-1} (g) = m(g) = {\rm const.} \; \Lambda \, 
\exp \left ( - \int^g { dg' \over \beta (g') } \right )
\label{eq:xi-beta}
\eeq  
which is identified with the correlation length appearing,
for example, in Eq.~(\ref{eq:two-point}).
The multiplicative constant in front of the expression on the right
hand side arises as an integration constant, and
cannot be determined from perturbation theory in $g$. 
Conversely, it is easy to verify that $\xi $ is indeed
a renormalization group invariant, 
$\Lambda { d \over d \Lambda } \, \xi ( \Lambda , g (\Lambda) ) = 0 $,
as stated previously in Eq.~(\ref{eq:m-scale}). 

In the vicinity of the fixed point at $g_c$ one can do the
integral in Eq.~(\ref{eq:xi-beta}), using Eq.~(\ref{eq:gc})
and the resulting linearized expression for the
$\beta$-function in the vicinity of the non-trivial
ultraviolet fixed point,
\beq
\beta (g) 
\; \mathrel{\mathop\sim_{ g \rightarrow g_c }} \;
\beta ' (g_c) \, (g - g_c) \, + \, \dots 
\label{eq:beta-lin}
\eeq
and one finds
\beq
\xi^{-1} (g) = m(g) \propto \,
\Lambda \,  | \, g - g_c \, |^{\nu }
\label{eq:m-sigma}
\eeq
with a correlation length exponent 
$\nu = - 1 / \beta'(g_c) \sim 1 / (d-2) + \dots$.
Thus the correlation length $\xi(g)$
diverges as one approaches the fixed point at $g_c$. 

In general the existence of a non-trivial ultraviolet fixed point
implies that the large momentum behavior above two dimensions
is not given by naive perturbation theory; it is
given instead by the critical behavior of the 
renormalized theory.
In the weak coupling, small $g$ phase the scale $m$ can be regarded
as a crossover scale between the free field behavior at large
distance scales and the critical behavior which sets in at
large momenta. 

In the non-linear $\sigma$-model another quantity 
of physical interest is the function
$M_0 (g)$, 
\beq
M_0 (g) = \exp \left [ 
- \half \int_0^g dg' \, { \zeta (g') \over \beta (g') }
\right ]
\eeq
which is proportional to the order parameter (the magnetization)
of the non-linear $\sigma$-model.
To one-loop order one finds 
$\zeta (g) = {1 \over 2 \pi} (N-1) g + \dots$
and therefore
\beq
M_0 (g) = {\rm const.} \, ( g_c - g )^\beta
\eeq
with $\beta = \half \nu ( d-2+\eta)$
and $\eta=\zeta(g_c)-\epsilon$.
To leading order in the $\epsilon$ expansion one
has for the anomalous dimension of the 
${\bf \pi}$ field $\eta= \epsilon / (N-2) + O (\epsilon^2)$.
In gauge theories, including gravity, there is no local
order parameter, so this quantity has no obvious
generalization there.  

In general the $\epsilon$-expansion is only expected to
be asymptotic.
This is already seen from the expansion for $\nu$ 
which has recently been computed to four loops 
(Hikami and Brezin 1978, Bernreuther and Wegner 1986,
Kleinert 2000)
\bea
\nu^{-1} = & \epsilon & + \, { \epsilon^2 \over N-2 }
+ { \epsilon^3 \over 2 (N-2) }
\nonumber \\
& - &[ 30 -14 N + N^2 + (54 - 18 N ) \zeta (3) ]
\nonumber \\
& \times & { \epsilon^4 \over 4 (N-2)^3 }
+ \dots
\label{eq:nu-eps}
\eea
which needs to be summed by Borel-Pad\'e methods to obtain
reliable results in three dimensions.
For example, for $N=3$ one finds in three dimensions 
$\nu \approx 0.799$, which can be compared to the 
$4-\epsilon$ result for the $\lambda \phi^4$ theory
to five loops $\nu \simeq 0.705$,
to the seven-loop perturbative expansion 
for the $\lambda \phi^4$ theory directly in $3d$
which gives $\nu \simeq 0.707$, with
the high temperature series result $\nu \simeq 0.717$ and the
Monte Carlo estimates $\nu \simeq 0.718$, as
compiled for example in a recent comprehensive
review (Guida and Zinn-Justin, 1997). 

There exist standard methods to deal with asymptotic
series such as the one in Eq.~(\ref{eq:nu-eps}).
To this purpose one considers a general series
\beq
f(g) = \sum_{n=0}^\infty \, f_n \, g^n
\eeq
and defines its Borel transform as
\beq
F(b) = \sum_{n=0}^\infty \, {f_n \over n! } \, b^n
\eeq
One can attempt to sum the series for $F (b)$
using Pad\'e methods and conformal transformations. 
The original function $f(g)$ is then recovered by
performing an integral over the Borel transform
variable $b$
\beq
f(g) = {\textstyle {1 \over g} \displaystyle}
\,
\int_0^\infty \, db \, e^{-b/g} \, F ( b )
\eeq
where the familiar formula
\beq
\int_0^\infty \, d z \, z^n \, e^{ - z / g } = n! \, g^{n+1}
\eeq
has been used.
Bounds on the coefficients $f_n$ suggest that
in most cases $F(z)$ is analytic in a circle
of radius $a$ around the origin,
and that the integral will converge for $|z|$
small enough, within a sector $| \arg z| < \alpha /2$ 
with typically $\alpha \ge \pi$
(Le Guillou and Zinn-Justin, 1990).

The first singularity along the positive real axis
is generally referred to as an infrared renormalon,
and is expected to be, in the $2d$
non-linear $\sigma$-model, at $b=1/2 \beta_0$
where $\beta_0= (N-2)/ 2 \pi $,
and gives rise to
non-analytic corrections of order $\exp (- 2 \pi / (N-2) g )$
(David, 1982).
Such non-analytic contributions presumably account for the
fact (Cardy and Hamber, 1980) that the $N=2$ model has a
vanishing $\beta$-function to all orders in $d=2$,
and yet has non-trivial finite exponents in $d=3$,
in spite of the result of Eq.~(\ref{eq:nu-eps}).
Indeed the $2+\epsilon$ expansion is not particularly
useful for the special case of the $N=2$ $\sigma$-model.
Then the action is simply given by
\beq
S(\theta) = { \Lambda^{d-2} \over 2 g } \, 
\int d^d x \, \left [ \, \partial_\mu \theta (x) \, \right ]^2
\eeq 
with $\phi_1 (x) = \sin \theta (x) $ and 
$\phi_2 (x) = \cos \theta (x)$, describing the
fluctuations of a planar spin in $d$ dimensions.
The $\beta$-function of 
Eq.~(\ref{eq:beta-nonlin}) then vanishes identically in $d=2$,
and the corrections to $\nu$ diverge for $d>2$, as in
Eq.~(\ref{eq:nu-eps})).
Yet this appears to be more a pathology of the perturbative
expansion in $\epsilon$, since after all the lattice model of 
Eq.~(\ref{eq:sigma-lattice}) is still well defined, and so
is the $4-\epsilon$ expansion for the continuum
linear $O(N)$ $\sigma$-model.
Thus, in spite of the model being again not perturbatively
renormalizable in $d=3$, one can still develop, for these
models, the full machinery of the renormalization group
and compute the relevant critical exponents.

Perhaps more importantly, a recent space shuttle experiment
(Lipa et al 2003) has succeeded in measuring the specific heat
exponent $\alpha=2-3\nu$ of superfluid Helium (which is supposed
to share the same universality class as the $N=2$ 
non-linear $\sigma$-model, with the complex phase of
the superfluid condensate acting as the order parameter)
to very high accuracy 
\beq
\alpha = - 0.0127 (3)
\eeq 
Previous theoretical predictions for the $N=2$ model
include the most recent
four-loop $4-\epsilon$ continuum result $\alpha = - 0.01126 (10)$
(Kleinert, 2000),
a recent lattice Monte Carlo estimate $\alpha = -0.0146(8)$ (Campostrini
et al, 2001), and the lattice variational renormalization group
prediction $\alpha=-0.0125 (39)$ (Hamber, 1981).

Perhaps the message one gains from this rather lengthy 
discussion of the non-linear $\sigma$-model in $d>2$ is that:

\begin{enumerate}

\item[ $\circ$ ]
The model provides a specific example of a theory which is not
perturbatively renormalizable in the traditional sense, and
for which the naive perturbative expansion in fixed dimension
leads to uncontrollable divergences and inconsistent results;

\item[ $\circ$ ]
Yet the model can be constructed perturbatively
in terms of a double expansion in $g$ and $\epsilon=d-2$.
This new perturbative expansion, combined with the renormalization
group, in the end provides explicit and detailed information
about universal scaling properties of the theory in the
vicinity of the non-trivial ultraviolet point at $g_c$;

\item[ $\circ$ ]
The continuum field theory predictions obtained
this way generally agree, for distances much larger than
the cutoff scale, with lattice results, and,
perhaps more importantly, with high precision experiments on systems 
belonging to the same universality class of the $O(N)$
model.
  
\end{enumerate}

\subsubsection{Non-linear Sigma Model in the Large-N Limit}
\label{sec:largen}

A rather fortunate circumstance is represented by the fact that
in the large $N$ limit the non-linear $\sigma$-model can be solved exactly.
This allows an independent verification of the correctness of the 
general ideas developed in the previous section, as well as a
direct comparison of explicit results for universal
quantities.
The starting point is the functional integral of Eq.~(\ref{eq:sigma-cont}),
\beq
Z = \int [ \, d \phi (x) \, ] \, 
\prod_x \, \delta \left [ \phi^2 (x) - 1 \right ] \,
\exp \left ( - \, S(\phi)  \right )
\eeq
with
\beq
S(\phi) \, = \, { 1 \over 2 T } \,
\int d^d x \; \partial_\mu \phi (x) \cdot \partial_\mu \phi (x)
\eeq
The constraint on the $\phi$ field can be implemented via
an auxiliary Lagrange multiplier field $\alpha (x)$.
One writes
\beq
Z = \int [ \, d \phi (x) ] \, [ d \alpha (x) ] \,
\exp \left ( - \, S(\phi,\alpha)  \right )
\eeq
with
\beq
S(\phi,\alpha) \, = \, { 1 \over 2 T } \, \int d^d x \, 
\left [
( \partial_\mu \phi (x) )^2 \, + \, \alpha (x) (\phi^2 (x) -1 ) 
\right ]
\label{eq:sigma-cont2}
\eeq
Since the action is now quadratic in $\phi(x)$ one can integrate
over $N-1$ $\phi$-fields (denoted previously
by ${\bf \pi}$).
The resulting determinant is then re-exponentiated, and one
is left with a functional integral over the remaining
first field $\phi_1 (x) \equiv \sigma (x)$,
as well as the Lagrange multiplier field $\alpha(x)$,
\beq
Z = \int [ \, d \sigma (x) \, d \alpha (x) ] \,
\exp \left ( - \, S_N (\phi,\alpha)  \right )
\label{eq:z-largen}
\eeq
with now
\bea
S_N (\phi,\alpha) & = & { 1 \over 2 T } \, \int d^d x \, 
\left [
( \partial_\mu \sigma )^2 + \alpha (\sigma^2 -1 ) 
\right ]
\nonumber \\
&& + \half (N-1) \tr \ln [ - \partial^2 + \alpha ]
\label{eq:s-largen}
\eea
In the large $N$ limit one can neglect, to leading order,
fluctuations in the $\alpha$ and $\sigma$ fields.
For a constant $\alpha$ field, $< \! \alpha (x) \! > = m^2 $,
the last (trace) term can be written in momentum space as
\beq
\half (N-1) \int^\Lambda { d^d k \over (2 \pi)^d } 
\ln ( k^2 + m^2 )
\eeq
which makes the evaluation of the trace straightforward.
As should be clear from Eq.~(\ref{eq:sigma-cont2}),
the parameter $m$ can be interpreted as the mass of the
$\phi$ field.
The functional integral in Eq.~(\ref{eq:z-largen})
can then be evaluated by the saddle point method.
It is easy to see from Eq.~(\ref{eq:s-largen}) 
that the saddle point conditions are
\bea
\sigma^2 & = & 1 - (N-1) \Omega_d (m) \, T
\nonumber \\
m^2 \sigma & = & 0
\label{eq:saddle}
\eea
with the function $\Omega_d (m)$ given by the integral
\beq
\Omega_d = \int^\Lambda { d^d k \over (2 \pi)^d } \, { 1 \over k^2 + m^2 }
\eeq
The latter can be evaluated in terms of a hypergeometric function,
\beq
\Omega_d = { 1 \over 2^{d-1} \pi^{d/2} \Gamma (d/2) } 
\, { \Lambda^d \over m^2 d } \; \,
{}_2 F_1 \left [
1, { d \over 2 }; \, 1 + { d \over 2 } ; \, - { \Lambda^2 \over m^2 }
\right ]
\eeq
but here one only really needs it in the large
cutoff limit, $ m \ll \Lambda$, in which case one finds the more
tractable expression
\beq
\Omega_d (m) - \Omega_d (0) = m^2 [ c_1 m^{d-4} + c_2 \Lambda^{d-4} 
+ O ( m^2 \Lambda^{d-6} ) ]
\eeq
with $c_1$ and $c_2$ some $d$-dependent coefficients.

From Eq.~(\ref{eq:saddle}) one notices that at weak coupling
and for $d>2$ a non-vanishing $\sigma$-field expectation
value implies that $m$, the mass of the ${\bf \pi}$ field, 
is zero.
If one sets $(N-1) \Omega_d (0) = 1 / T_c  $, one can
then write the first expression in Eq.~(\ref{eq:saddle}) as
\beq
\sigma (T) \, = \, \pm \, [ 1 - T / T_c ]^{1/2}
\label{eq:magnetiz}
\eeq
which shows that $T_c$ is the critical coupling at
which the order parameter $\sigma$ vanishes. 

Above $T_c$ the order parameter $\sigma$ vanishes, and $m (T) $ is obtained,
from Eq.~(\ref{eq:saddle}), by the solution of the nonlinear
gap equation
\beq
{ 1 \over T } = (N-1) 
\int^\Lambda { d^d k \over (2 \pi)^d } \, { 1 \over k^2 + m^2 }
\eeq
Using the definition of the critical coupling $T_c$, one can
now write, for $2<d<4$, for the common mass
of the $\sigma$ and ${\bf \pi}$ fields
\beq
m (T) \, \mathrel{\mathop\sim_{ m \ll \Lambda }} \,
\left ( { 1 \over T_c } - { 1 \over T } \right )^{1/(d-2)}
\label{eq:m-largen}
\eeq
which gives for the correlation length exponent
the non-gaussian value $\nu=1/(d-2)$,
with the gaussian value $\nu=1/2$ being recovered as
expected at $d=4$ (Wilson and Fisher 1972).
Note that in the large $N$ limit the constant of proportionality
in Eq.~(\ref{eq:m-largen}) is completely determined by the explicit
expression for $\Omega_d (m)$.

Perhaps one of the most striking aspects of the non-linear sigma model
above two dimensions is that all particles are massless in perturbation theory,
yet they all become massive in the strong coupling phase $T>T_c$, with
masses proportional to the non-perturbative scale $m$.

Again one can perform a renormalization group analysis
as was done in the previous section in the context of the
$2+\epsilon$ expansion.
To this end one defines dimensionless coupling
constants $g= \Lambda^{d-2} T$ and $g_c = \Lambda^{d-2} T_c$
as was done in Eq.~(\ref{eq:sigma-cont}).
Then the non-perturbative result of Eq.~(\ref{eq:m-largen}) becomes
\beq
m (g) \; \simeq \;  
c_d \cdot \Lambda \left ( { 1 \over g_c } - { 1 \over g } \right )^{1/(d-2)}
\label{eq:m-largen1}
\eeq
with the numerical coefficient given by
$c_d =[ \frac{1}{2} (d-2) \pi \vert \csc \left(\frac{d \pi }{2}\right) \vert 
]^{ { 1 \over d-2} }$.
One welcome feature of this large-$N$ result is the
fact that it provides an explicit value for the 
coefficient in Eq.~(\ref{eq:m-coeff}), namely
\beq
c_d = \, \left ( { g_c \over a_0 } \right )^{1/(d-2)}  \;\; ,
\eeq
and thereby for the numerical factor $a_0$ appearing in 
Eqs.~(\ref{eq:m-coeff}) and (\ref{eq:grun-nonlin}).

Again the physical, dimensionful mass $m$ in Eqs.~(\ref{eq:m-largen})
or (\ref{eq:m-largen1})
is required to be scale- and cutoff-independent as in
Eq.~(\ref{eq:m-scale})
\beq
\Lambda \, { d \over d \Lambda } \, m (\Lambda , g (\Lambda)) = 0
\label{eq:mass-indep}
\eeq
or, more explicitly, using the expression for
$m$ in Eq.~(\ref{eq:m-largen1}),
\beq
\left [
\Lambda \, { \partial \over \partial \Lambda } 
\, + \, \beta (g) \, { \partial \over \partial g }
\right ] \, \Lambda \, 
( {1 \over g_c } - {1 \over g } )^{1/(d-2)} = 0
\eeq
which implies for the $O(N)$ $\beta$-function in the large $N$
limit the simple result
\beq
\beta (g) = (d-2) \, g \, ( 1 - g / g_c )
\label{eq:beta-largen}
\eeq
The latter is valid again in the vicinity of the fixed point at $g_c$,
due to the assumption, used in Eq.~(\ref{eq:m-largen}), of $m \ll \Lambda$.
Note that it vanishes in $d=2$, and for $g=0$, in agreement with
the $2+\epsilon$ result of Eq.~(\ref{eq:beta-nonlin}).
Furthermore Eq.~(\ref{eq:beta-largen}) gives the momentum dependence of the
coupling at fixed cutoff.
After integration, one finds for the momentum ($\mu$) dependence of the coupling
at fixed cutoff $\Lambda$
\beq
{ g(\mu) \over g_c } = { 1 \over 1 - c \,( \mu_0 / \mu )^{d-2} } \approx 
1 + c \, ( \mu_0 / \mu )^{d-2} + \dots
\eeq
with $c \mu_0^{d-2}$ the integration constant.
The sign of $c$ then depends on whether one is on the right ($c>0$) or
on the left ($c<0$) of the ultraviolet fixed point at $g_c$.

\begin{figure}[h]
\epsfig{file=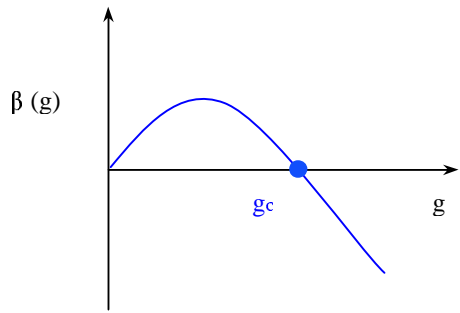,width=8cm}
\caption{The $\beta$-function for the non-linear $\sigma$-model
in the large-$N$ limit for $d>2$.}
\label{fig:beta-nonlin}
\end{figure}

One notices therefore again that the general shape of $\beta(g)$ is of
the type shown in Fig.~\ref{fig:beta-nonlin},
with $g_c$ a stable non-trivial UV fixed point, and $g=0$ and $g=\infty$
two stable (trivial) IR fixed points. 
Once more, at the critical point $g_c$ the $\beta$-function
vanishes and the theory becomes scale invariant.
Furthermore one can check that again $\nu = - 1/\beta'(g_c)$
where $\nu$ is the exponent in Eq.~(\ref{eq:m-largen}).
As before one can can re-write the physical mass $m$ for $2<d<4$ as
\beq
\xi^{-1} (g) = m (g)
\propto \Lambda \exp \left ( - \int^g { dg' \over \beta (g') } \right )
\eeq
as was done previously in Eq.~(\ref{eq:xi-beta}). 

Another general lesson one learns is that that Eq.~(\ref{eq:mass-indep}),
\beq
\left [
\Lambda { \partial \over \partial \Lambda } 
\, + \, \beta (g) \, { \partial \over \partial g }
\right ] \, m( \Lambda , g (\Lambda) )  = 0
\eeq
can be used to provide a non-perturbative definition
for the $\beta$-function $\beta(g)$.
If one sets $ m = \Lambda F(g)$, with $F(g)$ 
a dimensionless function of $g$, then one has the simple result
\beq
\beta (g) = - { F(g) \over F'(g) }
\eeq
Thus the knowlegde of the dependence of the mass gap $m$
on the bare coupling $g$ fixes the shape of the $\beta$
function, at least in the vicinity of the fixed point.
It should be clear then that the definition of the $\beta$-function per se,
and therefore the scale dependence of $g(\mu)$ which
follows from it [as determined from the solution of the
differential equation
$\mu { \partial g \over \partial \mu } = \beta (g(\mu))$]
is {\it not} necessarily tied to perturbation theory.

When $N$ is large but finite, one can develop a systematic $1/N$
expansion in order to evaluate the corrections to the picture
presented above (Zinn-Justin, 2002). 
Corrections to the exponents are known up to order $1/N^2$, but
the expressions are rather complicated for arbitrary $d$
and will not be reproduced here.
In general it appears that the $1/N$ expansion is only asymptotic,
and somewhat slowly convergent for useful values of $N$
in three dimensions.

\subsubsection{Self-coupled Fermion Model}
\label{sec:largenf}

Finally it would seem worthwhile to mention another example of
a theory which naively is not perturbatively renormalizable in $d>2$,
and yet whose critical properties can be worked out both in the
$2+\epsilon$ expansion, and in the large $N$ limit. 
It is described by an $U(N)$-invariant action
containing a set of $N$ massless self-coupled Dirac fermions 
(Wilson, 1973; Gross and Neveu, 1974)
\beq
S( \psi, \bar \psi) = - \int d^d x [
\bar \psi \cdot \Dslash \, \psi 
+ \half \Lambda^{d-2} \, u \, ( \bar \psi \cdot \psi )^2 ] \;\; .
\eeq
In even dimensions the discrete chiral symmetry 
$\psi \rightarrow \gamma_5 \psi$, 
$\bar \psi \rightarrow - \bar \psi \gamma_5 $
prevents the appearance of a fermion mass term.
Interest in the model resides in the fact that it exhibits 
a mechanism for dynamical mass generation and 
chiral symmetry breaking.

In two dimensions the fermion self-coupling constant
is dimensionless, and after 
setting $d=2+\epsilon$ one is again ready to develop the
full machinery of the 
perturbative expansion in $u$ and $\epsilon$, as was done
for the non-linear $\sigma$-model, since
the model is again believed
to be multiplicatively renormalizable in the framework
of the $2+\epsilon$ expansion.
For the $\beta$-function one finds to three loops
\beq
\beta (u) = \epsilon u 
- { \bar N -2 \over 2 \pi } u^2
+ { \bar N -2 \over 4 \pi^2 } u^3
+ { ( \bar N -2 ) ( \bar N - 7 ) \over 32 \pi^3 } u^4 + \dots
\label{eq:beta-gn}
\eeq
with the parameter $\bar N = N \tr {\bf 1} $, where the
last quantity is the identity matrix in the $\gamma$-matrix
algebra. 
In two dimensions $\bar N = 2 N$ and 
the model is asymptotically
free; for $\bar N=2$ the interaction is proportional
to the Thirring one and the $\beta$ function vanishes
identically.

As for the case of the non-linear $\sigma$-model,
the solution of the renormalization group
equations involves an invariant scale, which
can be obtained (up to a constant which cannot
be determined from perturbation theory alone) 
by integrating Eq.~(\ref{eq:beta-gn})
\beq
\xi^{-1} (u) = m (u) = {\rm const. } \, \Lambda
\exp \left [ - \int^u { d u' \over \beta (u') } \right ]
\eeq
In two dimensions this scale is, to lowest order
in $u$, proportional to
\beq
m(u) \, 
\mathrel{\mathop\sim_{ u \rightarrow 0 }} \,
\Lambda \, \exp \left [ 
- { 2 \pi \over ( \bar N - 2 ) u } 
\right ]
\eeq
and thus non-analytic in the bare coupling $u$.
Above two dimensions a non-trivial ultraviolet
fixed point appears at
\beq
u_c = { 2 \pi \over \bar N - 2 } \, \epsilon
+ { 2 \pi \over ( \bar N - 2 )^2 } \, \epsilon^2
+ { ( \bar N + 1 ) \pi \over 2 ( \bar N - 2 )^3 } \, \epsilon^3
+ \dots
\eeq
In the weak coupling phase $u<u_c$ the fermions stay
massless and chiral symmetry is unbroken, 
whereas in the strong coupling phase
$u>u_c$ (which is the only phase present in $d=2$)
chiral symmetry is broken, a fermion condensate
arises and a non-perturbative fermion mass is generated.
In the vicinity of the ultraviolet fixed point
one has for the mass gap 
\beq
m (u) \, \mathrel{\mathop\sim_{ u \rightarrow u_c }} \,
\Lambda \, \left ( u - u_c \right )^\nu
\label{eq:m-gn}
\eeq
up to a constant of proportionality,
with the exponent $\nu$ given by
\beq
\nu^{-1} \equiv - \beta' (u_c) = \epsilon - 
{ \epsilon^2 \over \bar N - 2  } 
- { ( \bar N - 3 ) \pi \over 2 ( \bar N - 2 )^2 } \, \epsilon^3
+ \dots
\eeq
The rest of the analysis proceeds in a way that, at
least formally, is virtually identical to
the non-linear $\sigma$-model case.
It need not be repeated here, as one can just take
over the relevant formulas for the renormalization
group behavior of $n$-point functions, for the running
of the couplings etc.

The existence of a non-trivial ultraviolet fixed point
implies that the large momentum behavior above two dimensions
is not given by naive perturbation theory; it is
given instead by the critical behavior of the 
renormalized theory, in accordance with 
Eq.~(\ref{eq:beta-gn}).
In the weak coupling, small $u$ phase the scale $m$ can be regarded
as a crossover scale between the free field behavior at large
distance scales and the critical behavior which sets in at
large momenta.

Finally, the same model can be solved exactly in the large $N$ limit.
There too one can show that the model is characterized
by two phases, a weak coupling phase where the fermions are
massless and a strong coupling phase in which a
chiral symmetry is spontaneously broken.

\subsubsection{The Gravitational Case}
\label{sec:graveps}

In two dimensions the gravitational coupling becomes dimensionless, 
$G\sim \Lambda^{2-d}$,
and the theory appears perturbatively renormalizable.
In spite of the fact that the gravitational action reduces to a topological
invariant in two dimensions, it would seem meaningful to try to construct, 
in analogy to what was suggested originally for scalar field theories 
(Wilson, 1973), the theory perturbatively as a double series in 
$\epsilon=d-2$ and $G$.

One first notices though that in pure Einstein gravity, with Lagrangian density 
\beq
{\cal L} = - { 1 \over 16 \pi G_0} \, \sqrt{g} \, R \;\; ,
\eeq
the bare coupling $G_0$ can be completely reabsorbed by a field redefinition
\beq
g_{\mu\nu} = \omega \, g_{\mu\nu}'
\label{eq:metric-scale}
\eeq
with $\omega$ is a constant, and thus
the renormalization properties of $G_0$ have no physical meaning
for this theory.
This simply follows from the fact that $\sqrt{g} R$ is homogeneous in $g_{\mu\nu}$, which is quite different from the Yang-Mills case. 
The situation changes though when one introduces a second dimensionful
quantity to compare to.
In the pure gravity case this contribution is naturally
supplied by the cosmological constant term proportional to $\lambda_0$,
\beq
{\cal L} = - { 1 \over 16 \pi G_0} \, \sqrt{g} \, R \, + \, \lambda_0 \sqrt{g}
\eeq
Under a rescaling of the metric as in Eq.~(\ref{eq:metric-scale}) one
has
\beq
{\cal L} = - { 1 \over 16 \pi G_0} \, \omega^{d/2-1} \, \sqrt{g'} \, R' 
\, + \, \lambda_0 \, \omega^{d/2} \, \sqrt{g'}
\label{eq:rescale}
\eeq
which is interpreted as a rescaling of the two bare couplings
\beq
G_0 \rightarrow \omega^{-d/2+1} G_0 \; , \;\;\;\; 
\lambda_0 \rightarrow \lambda_0 \, \omega^{d/2}
\eeq
leaving the dimensionless combination $G_0^d \lambda_0^{d-2}$ unchanged.
Therefore only the latter combination has physical meaning in pure gravity.
In particular, one can always choose the scale $\omega = \lambda_0^{-2/d}$
so as to adjust the volume term to have a unit coefficient.
More importantly, it is physically
meaningless to discuss separately the renormalization properties
of $G_0$ and $\lambda_0$, as they are individually gauge-dependent
in the sense just illustrated.
These arguments should clarify why in the following it will be sufficient
at the end to just focus on the renormalization properties of 
one coupling, such as Newton's constant $G_0$.

In general it is possible at least in principle to define quantum gravity
in any $d>2$.
There are $d(d+1)/2$ independent components of the
metric in $d$ dimensions, and the same number of algebraically independent
components of the Ricci tensor appearing in the field equations. 
The contracted Bianchi identities reduce the
count by $d$, and so does general coordinate invariance,
leaving $d(d-3)/2$ physical gravitational
degrees of freedom in $d$ dimensions.
At the same time, four space-time dimensions is known to be the lowest dimension
for which Ricci flatness does not imply the vanishing of the gravitational
field, $R_{\mu\nu\lambda\sigma}=0$, and therefore the first dimension
to allow for gravitational waves and their quantum counterparts, gravitons.

In a general dimension the position space tree-level graviton propagator
of the linearized theory, given in $k$-space in Eq.~(\ref{eq:grav-prop}),
can be obtained by Fourier transform and is proportional to 
\beq
\int d^d k \, { 1 \over k^2 } \, e^{i \, k \cdot x } \; = \;
{ \Gamma \left ( {d - 2 \over 2} \right ) \over 4 \, \pi^{d/2} \,
( x^2 )^ { d /2 - 1} } \; .
\eeq
The static gravitational potential is then proportional to the spatial
Fourier transform
\beq
V(r) \, \propto \, 
\int d^{d-1} {\bf k} \, { e^{i {\bf k} \cdot {\bf x} } \over {\bf k}^2 }
\, \sim \, { 1 \over r^{d-3} }
\eeq
and can be shown to vanish in $d=3$.
To show this one needs to compute the analog of Eq.~(\ref{eq:h-int})
in $d$ dimensions, which is
\beq
- \, {\kappa^2 \over 2 } \, \int d^d x \,
\left [ T_{\mu\nu} \, \Box^{-1} \, T^{\mu\nu} \, - \, 
(d-2)^{-1} \, T_{\mu}^{\;\;\mu} \, \Box^{-1} \, T_{\nu}^{\;\;\nu} \right ]
\nonumber \\
\; \rightarrow \; 
- \, {d-3 \over d-2 } \, {\kappa^2 \over 2 } \, \int d^{d-1} x \,
T^{00} \, {\cal G} \, T^{00} ,
\eeq
where the Green's function $\cal{G}$ is the static limit of $1 / \Box$,
and $\kappa^2 = 16 \pi G$.
The above result then shows that there are no Newtonian forces in 
$d$=2+1 dimensions
(Deser, Jackiw and 't Hooft, 1984; Deser and Jackiw, 1984).
The only fluctuations left in $3d$ are associated with the scalar curvature
(Deser, Jackiw and Templeton, 1982).

The $2+\epsilon$ expansion for pure gravity then proceeds as follows.
First the gravitational part of the action
\beq
{\cal L} = - { \mu^\epsilon \over 16 \pi G} \, \sqrt{g} \, R \;\; ,
\label{eq:l-pure}
\eeq
with $G$ dimensionless and $\mu$ an arbitrary momentum scale, 
is expanded by setting
\beq
g_{\mu\nu} \, \rightarrow \, \bar g_{\mu\nu} = g_{\mu\nu} \, + \, h_{\mu\nu}
\eeq
where $g_{\mu\nu}$ is the classical background field and $h_{\mu\nu}$
the small quantum fluctuation.
The quantity ${\cal L}$ in Eq.~(\ref{eq:l-pure}) is naturally identified with
the bare Lagrangian, and the scale $\mu$ with a microscopic ultraviolet
cutoff $\Lambda$, the inverse lattice spacing in a lattice formulation.
Since the resulting perturbative expansion is generally reduced to
the evaluation of Gaussian integrals, the original constraint (in the Euclidean
theory)
\beq
\det g_{\mu\nu} (x) \, > 0 
\label{eq:vol-cont}
\eeq
is no longer enforced (the same is not true in the lattice
regulated theory, where it plays an important role, see the discussion
following Eq.~(\ref{eq:tieq-1d})).

A gauge fixing term needs to be added, in the form of a background
harmonic gauge condition,
\beq
{\cal L}_{gf} = \half \alpha 
\sqrt{ g} \,  g_{\nu\rho}
\left ( 
\nabla_\mu h^{\mu\nu} - \half \beta g^{\mu\nu} \partial_\mu h 
\right )
\left ( 
\nabla_\lambda h^{\lambda\rho} - \half \beta g^{\lambda\rho} \partial_\lambda h 
\right )
\label{eq:l-gauge}
\eeq
with $ h^{\mu\nu} = g^{\mu\alpha} g^{\nu\beta} h_{\alpha\beta} $,
$h = g^{\mu\nu} h_{\mu\nu}$ and $ \nabla_\mu $ the 
covariant derivative with respect to the background metric
$g_{\mu\nu}$.
The gauge fixing term also gives rise to a Faddeev-Popov ghost
contribution ${\cal L}_{ghost}$ containing the ghost field
$\psi_\mu$, so that the total
Lagrangian becomes ${\cal L}+{\cal L}_{gf}+{\cal L}_{ghost}$.

In a flat background, $ g^{\mu\nu} = \delta^{\mu\nu}$,
one obtains from the quadratic part of the Lagrangian of
Eqs.~(\ref{eq:l-pure}) and (\ref{eq:l-gauge}) 
the following expression for the graviton propagator
\bea
&& < \! h_{\mu\nu} (k) h_{\alpha\beta} (-k) \! > \, =
\nonumber \\
&& {1 \over k^2 } ( \delta_{\mu\alpha} \delta_{\nu\beta} +
\delta_{\mu\beta} \delta_{\nu\alpha} )
- { 2 \over d-2 } { 1 \over k^2 } \, \delta_{\mu\nu} \delta_{\alpha\beta}
\nonumber \\
&&- \left ( 1 \! - \! {1 \over \alpha } \right ) \! {1 \over k^4 }
( \delta_{\mu\alpha} k_\nu k_\beta + 
\delta_{\nu\alpha} k_\mu k_\beta  +
\delta_{\mu\beta} k_\nu k_\alpha  +
\delta_{\nu\beta} k_\mu k_\alpha )
\nonumber \\
&&+ { 1 \over d-2 } { 4 (\beta-1) \over \beta - 2 } { 1 \over k^4 }
( \delta_{\mu\nu} k_\alpha k_\beta + 
\delta_{\alpha\beta} k_\mu k_\nu )
\nonumber \\
&&+ { 4 (1-\beta) \over (\beta-2)^2 } \left [ 
2 - { 3 - \beta \over \alpha } - { 2 (1-\beta) \over d-2 } \right ]
{1 \over k^6 } k_\nu k_\nu k_\alpha k_\beta 
\nonumber \\
\label{eq:prop-2d}
\eea
Normally it would be convenient to choose a gauge $\alpha$=$\beta=1$,
in which case only the first two terms for the graviton
propagator survive.
But here it might be advantageous
to leave the two gauge parameters unspecified,
so that one can later show explicitly the gauge
independence of the final result. 
In particular the gauge parameter $\beta$ is related
to the gauge freedom associated with the possibility,
described above, of rescaling the metric $g_{\mu\nu}$. 
Note also the presence of kinematical poles
in $\epsilon=d-2$ in the second, fourth and fifth term
for the graviton propagator.

\begin{figure}[h]
\epsfig{file=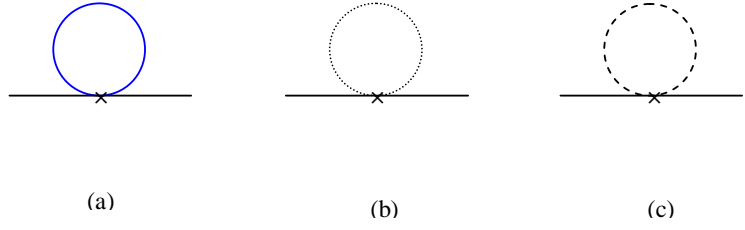,width=12cm}
\caption{One-loop diagrams giving rise to coupling renormalizations
in gravity. From left to right, graviton loop, ghost loop and scalar matter
loop.}
\label{fig:gravity-one}
\end{figure}

To illustrate explicitly the mechanism of coupling renormalization,
the cosmological term will be discussed first, since
the procedure is a bit simpler.
The cosmological term $\sqrt{g}$ is first expanded  
by setting $g_{\mu\nu} = g_{\mu\nu} + h_{\mu\nu}$
with a flat background $ g_{\mu\nu} = \delta_{\mu\nu}$.
One has
\beq
\sqrt{g} = 1 + \half h - \quarter h_{\mu\nu} h^{\mu\nu} + \eigth h^2 + O(h^3)
\eeq
with $h= h_{\;\;\mu}^\mu$.
Terms linear in the fluctuation $h_{\mu\nu}$ are dropped,
since in a properly chosen background such terms are expected to be absent.
The one-loop correction to the $1$ term in the above expression
is then given by the tadpole diagrams for the two quadratic
terms,
\beq
- \quarter h_{\mu\nu} h^{\mu\nu} + \eigth h^2 \, \rightarrow \, 
- \quarter \, < \! h_{\mu\nu} h^{\mu\nu} \! > + \eigth \, < \! h^2 \! >
\eeq
These are easily evaluated using the graviton propagator of
Eq.~(\ref{eq:prop-2d}).
For the one loop divergences 
(see Fig.~\ref{fig:gravity-one}) associated with the $\sqrt{g}$ term
one then obtains
\beq
\lambda_0 \rightarrow \lambda_0 \left [
1 - \left ( 
{ a_1 \over \epsilon } + { a_2 \over \epsilon^2 } \right ) G \right ]
\label{eq:a-co}
\eeq
with coefficients
\bea
a_1 & = & - { 8 \over \alpha } + 8 { (\beta-1)^2 \over (\beta-2)^2 } 
+ 4 {( \beta -1 ) (\beta -3) \over \alpha (\beta-2)^2 }
\nonumber \\
a_2 & = & 8 { (\beta-1)^2 \over (\beta-2)^2 }
\label{eq:a-coeff}
\eea
One notices that the kinematic singularities in the graviton
propagator, proportional to $1/(d-2)$, can combine with
the one loop ultraviolet divergent part of momentum integrals,
as in
\beq
{1 \over \epsilon} \, \int { d^d k \over (2 \pi)^d } { 1 \over k^2 }
\, \sim \, {1 \over \epsilon^2 }
\eeq
to give terms of order $1/\epsilon^2$ in Eq.~(\ref{eq:a-co}).
Generally it is better to separate the ultraviolet divergence from
the infrared one, by using for example the following regulated integral
\beq
\int { d^d k \over (2 \pi)^d } \, { 1 \over (k^2 + \mu^2)^a }
\, = \, {1 \over (4 \pi)^{d/2} } \,
{ \Gamma (a-d/2) \over \Gamma (a) } \, ( \mu^2 )^{d/2-a}
\eeq
for $a=1$ and $\mu \rightarrow 0$.

One can then follow the same procedure for the $\sqrt{g}\, R$ term.
First one needs to expand the Einstein term to quadratic order
in the quantum field $h_{\mu\nu}$
\bea
&& \sqrt{\bar g} \, \bar R = \sqrt{ g} \, R
\nonumber \\
& + &
\sqrt{ g} \, \bigl \{
\quarter \nabla_\rho h^\mu_{\;\;\nu} \nabla^\rho h^\nu_{\;\;\mu}
- \half \nabla_\nu h^\nu_{\;\;\mu} \nabla_\rho h^{\rho\mu}
+ \half R^\sigma_{\;\;\rho\mu\nu} h^{\rho}_{\;\;\sigma} h^{\mu\nu}
\bigr \} + \dots
\nonumber \\
\eea
where $\nabla_\mu $ denotes the covariant derivative with respect
to the background metric $ g_{\mu\nu}$.
The complete expansion was given previously in Eq.~(\ref{eq:h-background}).
The same expansion then needs to be done for the gauge fixing term
of Eq.~(\ref{eq:l-gauge}) as well, and furthermore it is again
convenient to choose as a background field 
the flat metric $ g_{\mu\nu} = \delta_{\mu\nu}$.
For the one loop divergences associated with the $\sqrt{g}R $ term
one then finds
\beq
{ \mu^\epsilon \over 16 \pi G}
\rightarrow
{ \mu^\epsilon \over 16 \pi G} \left ( 1 - { b \over \epsilon } \, G \right )
\eeq
with the coefficient $b$ given by (Gastmans et al 1977, Christensen et al 1978)
\beq
b = { 2 \over 3 } \cdot 19 + { 4 (\beta-1)^2 \over (\beta-2)^2 }
\label{eq:b-coeff}
\eeq
Thus the one-loop radiative corrections modify the total Lagrangian to
\beq
{\cal L} \rightarrow
- { \mu^\epsilon \over 16 \pi G} \left ( 1 - { b \over \epsilon } G \right )
\sqrt{g} R 
+ \lambda_0 \left [ 1 - \left (  
{ a_1 \over \epsilon } + { a_2 \over \epsilon^2 } \right ) G \right ] \! \sqrt{g}
\eeq
Next one can make use of the freedom to rescale the metric,
by setting
\beq
\left [ 1 - \left (  
{ a_1 \over \epsilon } + { a_2 \over \epsilon^2 } \right ) G \right ] \sqrt{g}
= \sqrt{g'}
\label{eq:scaled-cosm}
\eeq
which restores the original unit coefficient for the cosmological constant term.
The rescaling is achieved by the following field redefinition
\beq
g_{\mu\nu} = 
\left [ 1 - \left (  
{ a_1 \over \epsilon } + { a_2 \over \epsilon^2 } \right ) G \right ]^{-2/d} 
\, g'_{\mu\nu}
\eeq
Hence the cosmological term is brought back into the standard form
$\lambda_0 \sqrt{g'}$, and one obtains for the
complete Lagrangian to first order in $G$
\beq
{\cal L} \rightarrow
- { \mu^\epsilon \over 16 \pi G} 
\left [ 1 - { 1 \over \epsilon } ( b - \half a_2 ) G \right ]
\sqrt{g'} R' 
+ \lambda_0 \sqrt{g'}
\eeq
where only terms singular in $\epsilon$ have been retained.
From this last result one can finally read off the renormalization of 
Newton's constant
\beq
{ 1 \over G} \rightarrow { 1 \over G }
\left [ 1 - { 1 \over \epsilon } ( b - \half a_2 ) \, G \right ]
\label{eq:g-ren}
\eeq
From Eqs.~(\ref{eq:a-coeff}) and (\ref{eq:b-coeff}) one notices
that the $a_2$ contribution cancels out the gauge-dependent
part of $b$, giving for the remaining contribution
$ b - \half a_2 = \twoth \cdot 19 $.
Therefore the gauge dependence has, as one would have hoped on physical
grounds, disappeared from the final answer.
It is easy to see that the same result would have been obtained
if the {\it scaled} cosmological constant $G \lambda_0$
had been held constant,
instead of $\lambda_0$ as in Eq.~(\ref{eq:scaled-cosm}).
One important aspect of the result of Eq.~(\ref{eq:g-ren}) is
that the quantum correction is negative, meaning that the
strength of $G$ is effectively increased by the lowest order
radiative correction.

In the presence of an explicit renormalization scale parameter
$\mu$ the $\beta$-function for pure gravity is obtained by
requiring the independence of the quantity $G_e$ (here
identified as an effective coupling constant, with lowest order
radiative corrections included) from the original
renormalization scale $\mu$,
\bea
\mu \, { d \over d \mu } \, G_e & = & 0
\nonumber \\
{ 1 \over G_e} & \equiv & { \mu^\epsilon \over G (\mu) }
\left [ 1 - { 1 \over \epsilon } ( b - \half a_2 ) \, G (\mu) \right ]
\label{eq:g-oneloop}
\eea
To zero-th order in $G$, the renormalization group $\beta$-function entering
the renormalization group equation
\beq
\mu \, { \partial \over \partial \mu } \, G \, = \, \beta (G)
\label{eq:rg-equation}
\eeq
is just given by
\beq
\beta (G) = \epsilon \, G \, + \, \dots
\eeq
The above result just follows from the trivial scale dependence
of the classical,
dimensionful gravitational coupling: to achieve a fixed given
$G_e$, the dimensionless quantity $G(\mu)$ itself
has to scale like $\mu^\epsilon$.
Next, to first order in $G$, one has from Eq.~(\ref{eq:g-oneloop}) 
\beq
\mu { \partial \over \partial \mu } \, G =
\beta (G) = \epsilon \, G \, - \, \beta_0 \, G^2 \, + \,
O( G^3, G^2 \epsilon )
\label{eq:beta-oneloop}
\eeq 
with $\beta_0 = \twoth \cdot 19 $.
From the procedure outlined above it is clear that $G$ is
the only coupling that is scale-dependent in pure gravity.
As will be appreciated further below, the importance of
the gravitational $\beta$-function $\beta (G)$
lies in the fact that it can be used 
either to determine the ultraviolet cutoff dependence of the bare
coupling needed to keep the effective coupling fixed (as in 
Eq.~(\ref{eq:g-oneloop}), {\it or} to determine the momentum
dependence of the physical coupling $G (k)$ 
for a {\it fixed} cutoff. 

Matter fields can be included as well.
When $N_S$ scalar fields and $N_F$ Majorana fermion fields
are added, the results of Eqs.~(\ref{eq:b-coeff}), (\ref{eq:g-ren})
and (\ref{eq:g-oneloop}) are modified to
\beq
b \rightarrow b - \twoth c
\eeq
with $c=N_S + \half N_F$ (the central charge of the Virasoro
algebra in two dimensions), and therefore for the combined
$\beta$-function of Eq.~(\ref{eq:beta-oneloop}) to one-loop
order one has $\beta_0 = \twoth (19 - c) $.
Of course one noteworthy aspect of the perturbative calculation is
the appearance of a non-trivial ultraviolet fixed point at 
$G_c=(d-2)/\beta_0$ for which $\beta (G_c)=0$,
whose physical significance will be discussed further below.

To check their consistency, the above one-loop calculations 
have been repeated by performing a number of natural variations.
One modification consists in using the Thirring interaction
\beq
{\cal L} = - { \mu^\epsilon \over 16 \pi G} \, \sqrt{g} \, R
\, + \, e \bar \psi i \gamma^\mu D_\mu \psi 
\, - \, k e \bar \psi \gamma^a \psi \bar \psi \gamma_a \psi
\eeq
instead of the cosmological constant term to set the scale
for the metric.
This results in a $\beta$-function still of the form
of Eq.~(\ref{eq:beta-oneloop}) but with 
$\beta_0 = \twoth (25 - c) $.
The slight discrepancy between the two results was initially
attributed (Kawai and Ninomiya, 1990) to the well known problems
related to the kinematic singularities of the graviton propagator
in $2d$ discussed previously.
To address this issue, a new perturbative expansion can be
performed with the metric parametrized as
\beq
g_{\mu\nu} \, \rightarrow \, \bar g_{\mu\nu} \, = \, 
g_{\mu\rho} \, \left ( e^h \right )^{\rho}_{\;\; \nu}
\, e^{-\phi}
\eeq
where the conformal mode $\phi$ (responsible for the kinematic
singularity in the second term on the r.h.s. of
Eq.~(\ref{eq:prop-2d})) is explicitly separated out, and
$h^\mu_{\;\;\nu}$ is now taken to be traceless, $h^\mu_{\;\;\mu}=0$.
Furthermore the conformal mode is made massive by adding a cosmological
constant term $\lambda_0 \sqrt{g}$, which again acts
as an infrared regulator.
Repeating the calculation for the one loop divergences
(Kawai, Kitazawa and Ninomiya, 1993) one now finds
$\beta_0 = \twoth (25 - c) $ which is consistent with the
above quoted Thirring result
\footnote{
For a while there was considerable uncertainty about the magnitude
of the graviton contribution to $\beta_0$, which was quoted originally 
as $38/3$ (Tsao 1977), later as $2/3$ (Gastman et al 1977, Christensen
and Duff 1978; Weinberg, 1977), and more recently
as $50/3$ (Kawai, Kitazawa and Ninomiya, 1992).
As discussed in (Weinberg, 1977), the original expectation was that the graviton
contribution should be $d(d-3)/2 =-1$ times the scalar contribution close to $d=2$, which would suggest for gravity the value $2/3$.
Direct numerical estimates of the scaling exponent $\nu$ in the
lattice theory for $d=3$ (Hamber and Williams, 1993) give, using
Eq.~(\ref{eq:nueps}), a value $\beta_0 \approx 44/3$ and are therefore
in much better agreement with the larger, more recent values.}.

In the meantime the calculations have been laboriously extended 
to two loops (Aida and Kitazawa, 1997), with the result
\beq
\mu { \partial \over \partial \mu } G  =
\beta (G) = \epsilon \, G \, - \, \beta_0 \, G^2 
\, - \, \beta_1 \, G^3 \, + \, O( G^4, G^3 \epsilon , G^2 \epsilon^2 )
\label{eq:beta-twoloop}
\eeq 
with $\beta_0 = \twoth \, (25 - c) $ and
$ \beta_1 = { 20 \over 3 } \, (25 - c) $.

\subsubsection{Phases of Gravity in $2+\epsilon$ Dimensions}
\label{sec:phaseseps}

\begin{figure}[h]
\epsfig{file=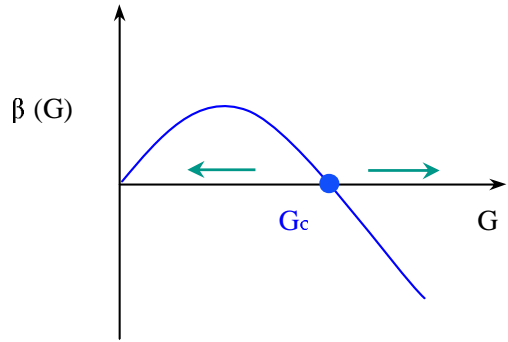,width=8cm}
\caption{The $\beta$-function for gravity in $2+\epsilon$ dimensions. The arrows indicate the coupling constant flow as one approaches increasingly larger
distance scales.}
\label{fig:beta-g-eps}
\end{figure}

The gravitational $\beta$-function of Eqs.~(\ref{eq:beta-oneloop})
and (\ref{eq:beta-twoloop}) determines the scale dependence
of Newton's constant $G$ for $d$ close to two.
It has the general shape shown in Fig.~\ref{fig:beta-g-eps}.
Because one is left, for the reasons described above, with a
single coupling constant in the pure gravity case, the discussion
becomes in fact quite similar to the non-linear $\sigma$-model case.

For a qualitative discussion of the physics it will be simpler in the
following to just focus on the one loop result of Eq.~(\ref{eq:beta-oneloop}); 
the inclusion of the two-loop correction does not alter the qualitative
conclusions by much, as it has the same sign as the lower order, one-loop term.
Depending on whether one is on the right ($G>G_c$) or on the left 
($G<G_c$) of the non-trivial ultraviolet fixed point at
\beq
G_c = { d - 2 \over \beta_0 } + O((d-2)^2 )
\eeq
(with $G_c$ positive provided one has $c<25$)
the coupling will either flow to increasingly larger values of $G$,
or flow towards the Gaussian fixed point at $G=0$, respectively.
In the following we will refer to the two phases as the strong
coupling and weak coupling phase, respectively.
Perturbatively one only has control on the small $G$ regime. 
When one then sets $d=2$ only the strong coupling phase survives,
so two-dimensional gravity is always strongly coupled
within this picture.

The running of $G$ as a function of a sliding momentum scale $\mu=k$
in pure gravity can be obtained by integrating 
Eq.~(\ref{eq:beta-oneloop}), and one finds
\beq
G(k^2) \; = \; { G_c \over 1 \, \pm \, a_0 \, (m^2 / k^2 )^{(d-2)/2} } 
\label{eq:grun-cont} 
\eeq
with $a_0$ a positive constant and $m$ a mass scale.
The choice of $+$ or $-$ sign is determined from whether one is
to the left (+), or to right (-) of $G_c$, in which case
the effective $G(k^2)$ decreases or, respectively, increases as one flows away
from the ultraviolet fixed point towards lower momenta, or larger distances.
Physically the two solutions represent a screening ($G<G_c$) and an 
anti-screening ($G>G_c$) situation.
The renormalization group invariant mass scale $\sim m$
arises here as an arbitrary
integration constant of the renormalization group equations
(one could have absorbed the constant $a_0$ in $m$, but we will not
do so here for reasons that will become clearer later).

While in the continuum perturbative calculation both phases,
and therefore both signs, seem acceptable, the Euclidean lattice
results on the other hand seem to rule out the weak coupling
phase as pathological, in the sense that the lattice collapses into a
two-dimensional branched polymer, as will be discussed later
in this review in Sec.~\ref{sec:phases}.

At the opposite end, at energies sufficiently high to become comparable
to the ultraviolet cutoff (the inverse lattice spacing in a lattice
discretization), the gravitational coupling $G$ flows towards the
ultraviolet fixed point,
\beq
G(k^2) \, \mathrel{\mathop\sim_{ k^2 \rightarrow \Lambda^2 }} \, G (\Lambda)
\eeq
where $G(\Lambda)$ is the coupling at the cutoff scale $\Lambda$,
to be identified with the bare (or lattice) coupling. 
Note that it would seem meaningless to consider, within this framework, momenta
which are larger than the ultraviolet cutoff $\Lambda$.
At such energies higher dimension operators (such as higher derivative
curvature terms) are expected to become important and should
therefore be included in the microscopic action.

Note that the result of Eq.~(\ref{eq:grun-cont}) is quite different
from the naive expectation based on straight perturbation
theory in $d>2$ dimensions (where the theory is not perturbatively
renormalizable)
\beq
{ G(k^2) \over G } \sim 1 + {\rm const.} \; G \, k^{d-2} + O(G^2)
\eeq
which gives a much worse ultraviolet behavior.
The existence of a non-trivial ultraviolet fixed point alters
the naive picture and drastically improves the ultraviolet behavior.

The $k^2$-dependent contribution in the denominator of 
Eq.~(\ref{eq:grun-cont}) is the quantum correction, which at
least within a perturbative framework is assumed to be small.
In the weak coupling phase $G<G_c$ the coupling then flows towards 
the origin corresponding to a gravitational screening solution, 
which sounds a bit odd as one would not expect gravity to be screened.
On the other hand the infrared growth of the coupling in the strong
coupling phase $G>G_c$ can be written equivalently as 
\beq
G(k^2) \; \simeq \; G_c \, \left [ 
\, 1 \, + \, a_0 \, \left ( {m^2 \over k^2 } \right )^{(d-2)/2}
\, + \, \dots \right ]
\label{eq:grun-cont1} 
\eeq
where the dots indicate higher order radiative corrections, and
which exhibits a number of interesting features.
Firstly the fractional power suggests new non-trivial gravitational 
scaling dimensions, just as in the case of the non-linear $\sigma$-model.
Furthermore, the quantum correction
involves a new physical, renormalization group invariant scale $\xi=1/m$
which cannot be fixed perturbatively, and whose size determines the 
scale for the quantum effects.
In terms of the bare coupling $G(\Lambda)$, it is given by
\beq
m \, = \, A_m \cdot \Lambda \, 
\exp \left ( { - \int^{G(\Lambda)} \, {d G' \over \beta (G') } }
\right )
\label{eq:m-cont}
\eeq
which just follows from integrating 
$ \mu { \partial \over \partial \mu }G = \beta (G)$
and then setting as the arbitrary scale $\mu \rightarrow \Lambda$.
Conversely, since $m$ is an invariant, one has 
$ \Lambda { d \over d \Lambda } m =0 $;
the running of $G(\mu)$ in accordance with the renormalization
group equation of Eq.~(\ref{eq:rg-equation}) ensures that the l.h.s. 
is indeed a renormalization group invariant.
The constant $A_m$ on the r.h.s. of Eq.~(\ref{eq:m-cont})
cannot be determined perturbatively,
it needs to be computed by non-perturbative (lattice) methods,
for example by evaluating invariant correlations at fixed
geodesic distances.
It is related to the constant $a_0$ in Eq.~(\ref{eq:grun-cont1})
by $a_0 = 1/ (A_m^{1/\nu} G_c )$.

At the fixed point $G=G_c$ the theory is scale invariant by definition.
In statistical field theory language the fixed point corresponds to a phase transition, where the correlation length $\xi=1/m$ diverges 
and the theory becomes scale (conformally) invariant.
In general in the vicinity of the fixed point, for which $\beta(G)=0$,
one can write
\beq
\beta (G) \, \mathrel{\mathop\sim_{ G \rightarrow G_c }} \, 
\beta' (G_c) \, (G-G_c) \, + \, O ((G-G_c)^2 )
\label{eq:beta-lin-g}
\eeq
If one then defines the exponent $\nu$ by
\beq
\beta ' (G_c) \, = \, - 1/ \nu
\label{eq:nu-def-beta}
\eeq
then from Eq.~(\ref{eq:m-cont}) one has by integration in the vicinity
of the fixed point
\beq
m \, \mathrel{\mathop\sim_{G \rightarrow G_c }} \,
\Lambda \cdot A_m \, | \, G (\Lambda) - G_c |^{\nu} \;\;\; .
\label{eq:m-cont1}
\eeq
which is why $\nu$ is often referred to as the mass gap exponent.
Solving the above equation (with $\Lambda \rightarrow k$)
for $G(k)$ one obtains back Eq.~(\ref{eq:grun-cont1}), with 
the constant $a_0$ there related to
$A_m$ in Eq.~(\ref{eq:m-cont1}) by $a_0 = 1/ (A_m^{1/\nu} G_c )$
and $\nu = 1 / (d-2)$.

That $m$ is a renormalization group invariant is seen from
\beq
\mu \, { d \over d \mu } \, m \, \equiv \, 
\mu \, { d \over d \mu } \,
\Bigl [ \, A_m \, \mu \, | \, G (\mu) - G_c |^{ \nu } \, \Bigr ]
\, = \, 0 
\label{eq:m-rginv}
\eeq
provided $G$ runs in accordance with Eq.~(\ref{eq:grun-cont1}).
To one-loop order one has from Eqs.~(\ref{eq:beta-oneloop}) 
and (\ref{eq:nu-def-beta}) $\nu=1/(d-2)$.
When the bare (lattice) coupling $G(\Lambda)=G_c$ one has achieved
criticality, $m=0$.
How far the bare theory is from the critical point is determined by
the choice of $G(\Lambda)$, the distance from criticality being
measured by the deviation $\Delta G = G(\Lambda)-G_c$.

Furthermore Eq.~(\ref{eq:m-cont1}) shows how the (lattice)
continuum limit is to be taken.
In order to reach the continuum limit $a\equiv 1/\Lambda \rightarrow 0$ for
fixed physical correlation $\xi=1/m$, the bare coupling $G(\Lambda)$ needs to be
tuned so as to approach the ultraviolet fixed point at $G_c$,
\beq
\Lambda \rightarrow \infty \; , 
\;\;\; m \; {\rm fixed } \; ,
\;\;\; G \rightarrow G_c \;\; .
\eeq
The fixed point at $G_c$ thus plays a central role in the cutoff
theory: together with
the universal scaling exponent $\nu$ it determines the correct
unique quantum continuum limit in the presence of an ultraviolet
cutoff $\Lambda$.
Sometimes it can be convenient to measure all quantities in units
of the cutoff and set $\Lambda=1/a=1$.
In this case the quantity $m$ measured in units of the cutoff
(i.e. $m/\Lambda$) has to be tuned to zero in order to construct the
lattice continuum limit: for a fixed lattice cutoff, 
the continuum limit is approached by tuning the bare lattice
$G(\Lambda)$ to $G_c$.
In other words, the lattice continuum limit has to be taken in
the vicinity of the non-trivial ultraviolet point.

The discussion given above is not altered significantly, at least
in its qualitative aspects, by the inclusion of the
two-loop correction of Eq.~(\ref{eq:beta-twoloop}).
From the expression for the two-loop $\beta$-function 
\beq
\mu { \partial \over \partial \mu } G  =
\beta (G) = \epsilon \, G \, - \, { 2 \over 3 } \, (25 - c) \, G^2 \,
- \, { 20 \over 3 } \, (25 - c) \, G^3 \, + \dots \;\; ,
\label{eq:beta-twoloops1}
\eeq
for $c$ massless real scalar fields minimally coupled to gravity,
one computes the roots $\beta (G_c) \, = \, 0 $ to obtain the location
of the ultraviolet fixed point, and from it
on can then determine the universal exponent $\nu = -1 /\beta'(G_c)$.
One finds
\bea
G_c & = & {3 \over 2 \, ( 25 - c ) } \, \epsilon
\, - \, {45 \over 2 ( 25 - c )^2 } \, \epsilon^2 \, + \, \dots
\nonumber \\
\nu^{-1} & = & 
\epsilon \, + \, {15 \over 25 - c } \, \epsilon^2 \, + \, \dots
\label{eq:nueps}
\eea
which gives, for pure gravity without matter ($c=0$) in four dimensions, to lowest order
$\nu^{-1} = 2$, and $\nu^{-1} \approx 4.4 $ at the next order.

Also, in general higher order corrections 
to the results of the linearized renormalization group equations
of Eq.~(\ref{eq:beta-lin-g}) are present, which affect the
scaling away from the fixed point.
Let us assume that close to the ultraviolet fixed point at
$G_c$ one can write for the $\beta$-function the following expansion
\beq
\beta (G) \, = \, - {\textstyle { 1 \over \nu } \displaystyle }
\, ( G - G_c ) \, - \, \omega \,  ( G -G_c )^2  
+ {\cal O} ( ( G -G_c )^3 ) \;\; ,
\eeq
After integrating $\mu { \partial \over \partial \mu } G = \beta(G)$
as before, one finds for the structure of the correction
to $m$ [see for comparison Eq.~(\ref{eq:m-cont1})]
\beq
\Bigl ( { m \over \Lambda } \Bigr )^{1 / \nu }
\, = \, A_m \, \Bigl [
( G (\Lambda) - G_c ) \, - \, \omega \, \nu \, ( G (\Lambda) -G_c )^2
\, + \, \dots \Bigr ] \;\; .
\label{eq:m-cont2}
\eeq
The hope of course is that these corrections to scaling 
are small, $(\omega \ll 1)$; in the vicinity of the fixed point
the higher order term becomes unimportant when 
$ | G-G_c | \ll 1 / (\omega \nu ) $.
For the effective running  coupling one then has 
\beq
{ G(\mu) \over G_c } = 1 \, + \, 
a_0 \Bigl ( { m \over \mu} \Bigr )^{1 /\nu }
+  a_0 \, \omega \, \nu \, \Bigl ( { m \over \mu } \Bigr )^{2 / \nu }
+ {\cal O} \Bigl ( \Bigl ( { m \over \mu } \Bigr )^{3 / \nu } \Bigr ) 
\;\; .
\label{eq:grun-cont2} 
\eeq
which gives an estimate for the size of the modifications to 
Eq.~(\ref{eq:grun-cont1}).

Finally, as a word of caution, one should mention that
in general the convergence properties of the $2+\epsilon$
expansion are not well understood.
The poor convergence found in some better known cases is 
usually ascribed to the suspected existence of infrared renormalon-type
singularities $ \sim e^{-c/G}$ close to two dimensions, and
which could possibly arise in gravity as well.
At the quantitative level, the results of the $2+\epsilon$ expansion for gravity
therefore remain somewhat limited, and obtaining the three- or four-loop
term still represents a daunting task.
Nevertheless they provide, through Eqs.~(\ref{eq:grun-cont1})
and (\ref{eq:m-cont1}), an analytical insight into the scaling properties
of quantum gravity close and above two dimensions, including
the suggestion of a non-trivial phase structure and an
estimate for the non-trivial universal scaling exponents
(Eq.~(\ref{eq:nueps})).
The key question raised by the perturbative calculations is
therefore: what remains of the above phase transition in four dimensions,
how are the two phases of gravity characterized there non-perturbatively, 
and what is the value of the exponent $\nu$ determining
the running of $G$ in the vicinity of the fixed point 
{\it in four dimensions}.

Finally we should mention that there are other continuum renormalization
group methods which can be used to estimate the scaling exponents.
An approach which is closely related to the $2+\epsilon$ expansion
for gravity is the derivation of approximate flow equations
from the changes of the Legendre effective action with respect
to a suitably introduced infrared cutoff $\mu$.
The method can be regarded as a variation on Wilson's original
momentum slicing technique for obtaining approximate
renormalization group equations for lattice couplings.
In the simplest case of a scalar field theory (Morris, 1994)
one starts from the partition function
\beq
\exp ( W[J] ) = \int [ d \phi ] \, \exp \left \{
- \half \phi \cdot C^{-1} \cdot \phi 
- I_{\Lambda} [ \phi ] + J \cdot \phi \right \}
\eeq
The $C \equiv C ( k, \mu )$ term is taken to be
an 'additive infrared cutoff term'. 
For it to be an infrared cutoff it needs to be small
for $k<\mu$, ideally tending to zero as $k \rightarrow 0$,
and such that $k^2 C(k,\mu)$ is large when $k > \mu$.
Since the method is only ultimately applied to the vicinity of
the fixed point, for which all physical relevant scales are
much smaller than the ultraviolet cutoff $\Lambda$, it is argued that
the specific nature of this cutoff is not really relevant
in the following.
Taking a derivative of $W[J]$ with respect to the scale $\mu$ gives
\beq
{ \partial W[J] \over \partial \mu } =
- \half \left [ \,
{ \delta W \over \delta J } \cdot { \partial C^{-1} \over \partial \mu }
\cdot  { \delta W \over \delta J }
+ \tr \left ( 
{ \partial C^{-1} \over \partial \mu }
{ \delta^2 W \over \delta J \, \delta J }
\right ) \right ]
\eeq
which can be re-written in terms of the Legendre transform
$\Gamma [ \phi ] = - W [J] - \half \phi \cdot C^{-1} \cdot \phi + J \cdot \phi $
as
\beq
{ \partial \, \Gamma [\phi] \over \partial \mu } =
- \half \tr \left [ \,
{ 1 \over C } \,  
{ \partial C \over \partial \mu }
\cdot  \left ( 1 + C \cdot 
{ \delta^2 \Gamma \over \delta \phi \, \delta \phi }
\right )^{-1} \right ] 
\label{eq:gamma-rg}
\eeq
where now $\phi = \delta W / \delta J$ is regarded as the classical field.
The traces can then be simplified by writing them in momentum space.
What remains to be done is first settle on a suitable cutoff function 
$C(k, \mu)$, and subsequently compute the effective action 
$\Gamma [\phi]$ in a derivative expansion, thus involving terms
of the type $\partial^n \phi^m$, with $\mu$ dependent coefficients.

As far as the cutoff function is concerned, it is first written
as $C(k,\mu)= \mu^{\eta-2} C(k^2 / \mu^2 )$ so as to include
the expected anomalous dimensions of the $\phi$ propagator.
To simplify things further, it is then assumed for the remaining
function of a single variables that
$C(q^2)=q^{2p}$ with $p$ a non-negative integer (Morris, 1994). 
The subsequent derivative expansion gives for example for
the $O(N)$ model in $d=3$ to lowest order $O(\partial^0)$
an anomalous dimensions $\eta=0$
for all $N$, and $\nu=0.73$ for $N=2$.
At the next order $O(\partial^2)$ in the derivative expansion
the method gives $\nu=0.65$, compared to the best
theoretical and experimental value $\nu=0.67$ (Morris and Turner 1997).

In the gravitational case one can proceed in a similar way.
First the gravity analog of Eq.~(\ref{eq:gamma-rg}) is
clearly
\beq
{ \partial \, \Gamma [g] \over \partial \mu } =
- \half \tr \left [ \,
{ 1 \over C }  
{ \partial C \over \partial \mu }
\cdot  \left ( 1 + C \cdot 
{ \delta^2 \Gamma \over \delta g \, \delta g }
\right )^{-1} \right ] 
\label{eq:gamma-rg-grav}
\eeq
where now $g_{\mu\nu} = \delta W / \delta J_{\mu\nu}$ corresponds
to the classical metric.
The effective action itself contains the Einstein and cosmological terms
\beq
\Gamma_\mu [g] = - { 1 \over 16 \pi G(\mu) } \int d^d x \sqrt{g} 
\, [ \, R(g) - 2 \lambda ( \mu ) \, ] + \dots
\eeq
as well as gauge fixing and possibly higher derivative terms (Reuter 1998). 
After the addition of a background harmonic gauge fixing term 
with gauge parameter $\alpha$, the choice of a suitable (scalar) cutoff function
is required, $C^{-1} (k,\mu) = (\mu^2 - k^2) \theta ( \mu^2 - k^2 )$
(Litim 2004), which is inserted into
\beq
\int [ d h ] \, \exp \left \{
- \half h \cdot C^{-1} \cdot h
- I_{\Lambda} [ g ] + J \cdot h \right \}
\eeq
Note that this added momentum-dependent cutoff term violates both the
weak field general coordinate
invariance [see for instance Eq.~(\ref{eq:h-gauge})], as well as the general
rescaling invariance of Eq.~(\ref{eq:rescale}).

The solution of the resulting renormalization group
equation for the two couplings $G(\mu)$ and $\lambda (\mu)$ 
is then truncated to the Einstein
and cosmological term, a procedure which is equivalent to the
derivative expansion discussed previously.
A nontrivial fixed point in both couplings
($G^{*}, \lambda^{*}$) is then found
in four dimensions, with complex eigenvalues
$\nu^{-1} = 1.667 \pm 4.308 i $ for a gauge parameter
choice $\alpha \rightarrow \infty$ [for general gauge
parameter the exponents can vary by as much as seventy percent
(Lauscher and Reuter 2002)].
In the special limit of vanishing cosmological constant
the equations simplify further and one finds a trivial Gaussian
fixed point at $G=0$, as well as a non-trivial ultraviolet 
fixed point with $\nu^{-1} = 2 d (d-2)/(d+2)$, which in $d=4$ gives
now $\nu^{-1} = 2.667 $.
So in spite of the apparent crudeness of the lowest order approximation,
an ultraviolet fixed point similar to the one found in the
$2+\epsilon$ expansion is recovered.

\subsubsection{Running of $\alpha (\mu)$ in $QED$ and $QCD$}
\label{sec:qcd}

$QED$ and $QCD$ provide two invaluable illustrative cases where the running
of the gauge coupling with energy is not only theoretically well understood, but also verified experimentally.
This section is intended to provide analogies and distinctions between the
two theories, 
in a way later suitable for a comparison with the gravitational case.
Most of the results found in this section are well known 
(see, for example, Frampton, 2000), but the purpose
here is to provide some contrast (and in some instances, a relationship)
with the gravitational case.

In $QED$ the non-relativistic static Coulomb potential is affected by the vacuum
polarization contribution due to electrons (and positrons) of mass $m$.
To lowest order in the fine structure constant, the contribution is
from a single Feynman diagram involving a fermion loop.
One finds for the vacuum polarization contribution $\omega_R ({\bf k^2}) $
at small ${\bf k^2}$ the well known result (Itzykson and Zuber, 1980)
\beq
{ e^2 \over {\bf k^2} } \; \rightarrow \;
{ e^2 \over {\bf k^2} [ \, 1 + \, \omega_R ({\bf k^2}) ] }
\; \sim \; 
{ e^2 \over {\bf k^2} } \, \left [ \, 1 \, + \, {\alpha \over 15 \, \pi } \,
{ {\bf k^2} \over m^2 } \, + \, O(\alpha^2) \, \right ] 
\label{eq:alpha_qed}
\eeq
which, for a Coulomb potential with a charge centered at the origin of strength
$-Ze$ leads to well-known Uehling $\delta$-function correction 
\beq
V(r) \; = \; \left ( \, 1 \, - \, {\alpha \over 15 \, \pi } \,
{ \Delta \over m^2 } \, \right ) \, { - Z \, e^2 \over 4 \, \pi \, r} \; = \;
{ - Z \, e^2 \over 4 \, \pi \, r} \, - \, {\alpha \over 15 \, \pi } \, { - Z \, e^2 \over m^2 } 
\, \delta^{(3)} ({\bf x})
\label{eq:pot_uel}
\eeq
It is not necessary though to resort to the small-${\bf k^2}$ approximation,
and in general a static charge of strength $e$ at the origin will
give rise to a modified potential
\beq
{ e \over 4 \, \pi \, r } \; \rightarrow \;
{ e \over 4 \, \pi \, r } \, Q(r)
\eeq
with 
\beq
Q(r) \; = \; 1 \, + \, {\alpha \over 3 \, \pi } \, \ln { 1 \over m^2 \, r^2 } \, + \, \dots 
\;\;\;\;\; m \, r \ll 1
\label{eq:qed_s}
\eeq
for small $r$, and
\beq
Q(r) \; = \; 1 \, + \, {\alpha \over 4 \, \sqrt{\pi} \, (m r)^{3/2} } \, 
e^{- \, 2 \, m \, r  } \, + \, \dots 
\;\;\;\;\; m \, r \gg 1
\label{eq:qed_l}
\eeq
for large $r$.
Here the normalization is such that the potential at infinity has $Q(\infty)=1$
\footnote{
The running of the fine structure constant has recently been verified
experimentally at $LEP$.
The scale dependence of the vacuum polarization effects gives
a fine structure constant changing from $\alpha (0) \sim 1/137.036$
at atomic distances
to about $ \alpha (m_{Z_0}) \sim 1/128.978 $ at energies comparable to
the $Z^0$ boson mass, in good agreement with the theoretical 
renormalization group prediction.}.
The reason we have belabored this example is to show that the screening
vacuum polarization contribution would have dramatic effects in QED if for some
reason the particle running through the fermion loop diagram had a much
smaller (or even close to zero) mass.
There are two interesting aspects of the (one-loop) result of
Eqs.~(\ref{eq:qed_s}) and (\ref{eq:qed_l}).
The first one is that the exponentially small size of the correction at large $r$
is linked with the fact that the electron mass $m_e$ is not too small:
the range of the correction term is 
$\xi = 2 \hbar / m c = 0.78 \times 10^{-10} cm$, but would have been much
larger if the electron mass had been a lot smaller.

\begin{figure}[h]
\epsfig{file=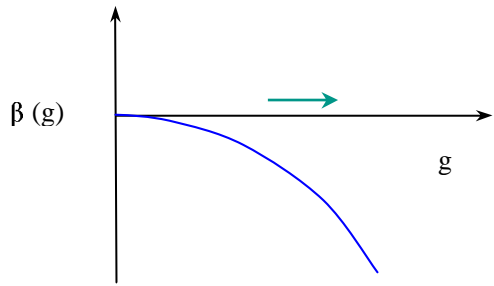,width=8cm}
\caption{The $QCD$ $\beta$-function in four dimensions, with an
ultraviolet stable fixed point at $g=0$.}
\label{fig:beta-qcd}
\end{figure}

In $QCD$ (and related Yang-Mills theories) radiative corrections are also known to alter 
significantly the behavior of the static potential at short distances.
The changes in the potential are best expressed in terms 
of the running strong coupling constant $\alpha_S (\mu) $, whose scale
dependence is determined by the celebrated beta function of $SU(3)$ $QCD$
with $n_f$ light fermion flavors
\beq
\mu \, { \partial \, \alpha_S \over \partial \, \mu } \; = \; 
2 \, \beta ( \alpha_S ) \; = \; 
- \, { \beta_0 \over 2 \, \pi } \, \alpha_S^2 
\, - \, { \beta_1 \over 4 \, \pi^2 } \, \alpha_S^3 
\, - \, { \beta_2 \over 64 \, \pi^3 } \, \alpha_S^4 \, - \, \dots
\label{eq:beta_qcd}
\eeq
with $\beta_0 = 11 - {2 \over 3} n_f $, $\beta_1 = 51 - {19 \over 3} n_f $, 
and $\beta_2 = 2857 - {5033 \over 9} n_f + {325 \over 27} n_f^2 $.
The solution of the renormalization group equation  
Eq.~(\ref{eq:beta_qcd}) then gives for the running of $\alpha_S (\mu )$
\beq
\alpha_S (\mu ) \; = \; 
{ 4 \, \pi \over \beta_0 \ln { \mu^2 / \Lambda_{\overline{MS}}^2  } }
\, \left [ 1 \, - \, { 2 \beta_1 \over \beta_0^2 } \,
{ \ln \, [ \ln { \mu^2 / \Lambda_{\overline{MS}}^2 ] } 
\over \ln { \mu^2 / \Lambda_{\overline{MS}}^2  } }
\, + \, \dots \right ]
\label{eq:alpha_qcd}
\eeq
(see Fig.~\ref{fig:beta-qcd}).
The non-perturbative scale $ \Lambda_{\overline{MS}} $ appears as an
integration constant of the renormalization group equations, and
is therefore - by construction - scale independent.
The physical value of $ \Lambda_{\overline{MS}} $
cannot be fixed from perturbation theory alone, and needs to be determined
by experiment, giving $ \Lambda_{\overline{MS}} \simeq 220 MeV$.

In principle one can solve for $ \Lambda_{\overline{MS}} $ in terms of the 
coupling at any scale, and in particular at the cutoff scale $\Lambda$,
obtaining 
\beq
\Lambda_{\overline{MS}} \; = \; 
\Lambda \, \exp \left ( 
{ - \int^{\alpha_S (\Lambda)} \, {d \alpha_S' \over 2 \, \beta ( \alpha_S') } }
\right )
\; = \;
\Lambda \, \left ( { \beta_0 \, \alpha_S ( \Lambda ) \over 4 \, \pi } 
\right )^{\beta_1 / \beta_0^2 }
\, e^{ - { 2 \, \pi \over \beta_0 \, \alpha_S (\Lambda ) } }
\, \left [ \, 1 \, + \, O( \alpha_S ( \Lambda) ) \, \right ] 
\label{eq:lambda_qcd}
\eeq
In lattice QCD this is usually taken as the definition of
the running strong coupling constant $\alpha_S (\mu) $.
It then leads to an effective potential between quarks and anti-quarks of the form
\beq
V ( {\bf k^2} ) \; = \; - \, {4 \over 3} \, { \alpha_S ( {\bf k^2} ) 
\over {\bf k^2} }
\label{eq:qcd_pot}
\eeq
and the leading logarithmic correction makes the potential appear
softer close to the origin, $V(r) \sim 1 / ( r \ln r) $.

When the $QCD$ result is contrasted with the $QED$ answer of
Eqs.~(\ref{eq:alpha_qed}) and (\ref{eq:pot_uel}) it appears
that the infrared small ${\bf k^2}$ singularity in 
Eqs.~(\ref{eq:qcd_pot}) is quite serious.
An analogous conclusion is reached when examining 
Eqs.~(\ref{eq:alpha_qcd}): the coupling strength
$ \alpha_S ( {\bf k^2} )$ diverges in the infrared due to the 
singularity at $k^2=0$.
In phenomenological approaches to low energy $QCD$ (Richardson, 1979)
the uncontrolled growth in $ \alpha_S ( {\bf k^2} )$ due to
the spurious small-$k^2$ divergence is regulated by the 
dynamically generated $QCD$ infrared cutoff $\Lambda_{\overline{MS}}$,
which can then be shown to give a confining linear potential at
large distances.

Not all physical properties can be computed reliably in weak coupling
perturbation theory.
In non-Abelian gauge theories a confining potential 
is found at strong coupling by examining the behavior of the Wilson loop (Wilson, 1973), defined for a large closed loop $C$ as
\beq
W( C ) \, = \, 
< \tr {\cal P} \exp \Bigl \{ i g \oint_{C} A_{\mu} (x) dx^{\mu} 
\Bigr \} > \;\; ,
\label{eq:wloop_sun}
\eeq
with $A_\mu \equiv t_a A_\mu^a $ and the $t_a$'s the group
generators of $SU(N)$ in the fundamental representation. 
In the pure gauge theory at strong coupling, the leading contribution to
the Wilson loop can be shown to follow an area law for sufficiently
large loops
\beq
W( C )
\; \mathrel{\mathop\sim_{ A \, \rightarrow \, \infty  }} \;
\exp ( - A(C) / \xi^2 ) 
\label{eq:wloop_sun1}
\eeq
where $A(C)$ is the minimal area spanned by the planar loop $C$.
The quantity $\xi $ is the gauge field correlation length, defined 
for example from the exponential decay of the Euclidean correlation
function of two infinitesimal loops separated by a distance $|x|$,
\beq
G_{\Box} ( x ) \, = \, 
< \tr {\cal P} \exp \Bigl \{ i g \oint_{C_\epsilon} A_{\mu} (x') dx'^{\mu} 
\Bigr \} (x)
\, 
\tr {\cal P} \exp \Bigl \{ i g \oint_{C_\epsilon} A_{\mu} (x'') dx''^{\mu} 
\Bigr \} (0)
>_c \;\; ,
\label{eq:box_sun}
\eeq
Here the $C_\epsilon$'s are two infinitesimal loops centered around $x$ ands $0$
respectively, suitably defined on the lattice as elementary square loops, 
and for which one has at sufficiently large separations
\beq
G_{\Box} ( x ) 
\; \mathrel{\mathop\sim_{ |x|  \, \rightarrow \, \infty  }} \;
\exp ( - |x| / \xi ) 
\label{eq:box_sun1}
\eeq
The inverse of the correlation length $\xi$ corresponds to the lowest
mass excitation in the gauge theory, the scalar glueball,
$m_0 = 1 / \xi$.
Notice that since the glueball mass $m_0$ is expected to be proportional
to the parameter $\Lambda_{\overline{MS}}$ of Eq.~(\ref{eq:lambda_qcd}) 
for small $g$, it is non-analytic in the gauge coupling.

\section{LATTICE REGULARIZED QUANTUM GRAVITY}

\label{sec:lattice}

The following sections are based on the lattice discretized description
of gravity known as Regge calculus, where the Einstein theory is expressed in terms of a simplicial decomposition of space-time manifolds.
Its use in quantum gravity is prompted by the desire to make use of techniques
developed in lattice gauge theories (Wilson, 1973)
\footnote{
As an example of a state-of-the-art calculation of hadron properties in
the lattice formulation of $SU(3)$ $QCD$ see (Aoki et al., 2003).
},
but with a lattice  which reflects the
structure of space-time rather than just providing a flat passive background (Regge, 1961).
It also allows one to use powerful nonperturbative analytical techniques
of statistical mechanics as well as numerical methods.
A regularized lattice version of the continuum field theory is also usually
perceived as a necessary prerequisite for a rigorous study of the latter.
  
In Regge gravity the infinite number of degrees of freedom in the continuum
is restricted by considering Riemannian spaces described
by only a finite number of variables, the geodesic distances between
neighboring points.
Such spaces are taken to be flat almost everywhere and are called
{\it piecewise linear} (Singer and Thorpe, 1967).
The elementary building blocks for $d$-dimensional space-time are
{\it simplices} of dimension $d$.
A 0-simplex is a point, a 1-simplex is an edge, a 2-simplex is a triangle, a
3-simplex is a tetrahedron.
A $d$-simplex is a $d$-dimensional object with $d+1$ vertices and
$d(d+1)/2$ edges connecting them.
It has the important property that the values of its edge lengths specify
the shape, and therefore the relative angles, uniquely.
  
\begin{figure}[h]
\epsfig{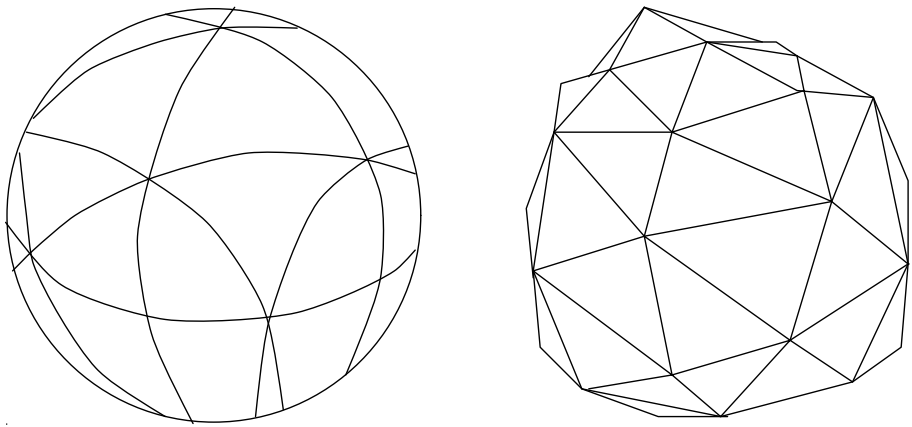}
\caption[Polyhedral approximation]{Polydedral approximation to a sphere.}
\label{fig:polyhedron}
\end{figure}

A simplicial complex can be viewed as a set of simplices glued together
in such a way that either two simplices are
disjoint or they touch at a common face.
The relative position of points on the lattice is thus completely specified
by the {\it incidence matrix} (it tells which point is next to which) and the
{\it edge lengths},
and this in turn induces a metric structure on the piecewise linear space.
Finally the polyhedron constituting the union of all the simplices of
dimension $d$ is called a geometrical complex or {\it skeleton}.
The transition from a smooth triangulation of a sphere to the corresponding
secant approximation is illustrated in Fig.~\ref{fig:polyhedron}.

A manifold can then be defined by its relationship to a piecewise linear
space: a topological space is called a closed $d$-dimensional manifold if
it is homeomorphic to a connected polyhedron, and furthermore, if its
points possess neighborhoods which are homeomorphic to the interior of
the d-dimensional sphere.

\subsection{General Formulation}

\label{sec:lattgen}

We will consider here a general simplicial lattice in $d$ dimensions, made out
of a collection of flat $d$-simplices glued together at their common faces 
so as to constitute a triangulation of a smooth continuum manifold,
such as the $d$-torus or the surface of a sphere.
If we focus on one such $d$-simplex, it will itself contain sub-simplices of smaller dimensions; as an example in four dimensions a given 4-simplex will contain 5 tetrahedra, 10 triangles (also referred to as hinges in
four dimensions), 10 edges and 5 vertices.
In general, an $n$-simplex will contain $ { n+1 \choose k+1 } $ $k$-simplices in its boundary.
It will be natural in the following to label simplices by the letter
$s$, faces by $f$ and hinges by $h$.
A general connected, oriented simplicial manifold consisting of $N_s$ $d-$simplices will also be characterized by an incidence matrix
$I_{s,s'}$, whose matrix element $I_{s,s'}$ is chosen to be equal
to one if the two simplices labeled by $s$ and $s'$ share a common face,
and zero otherwise. 

The geometry of the interior of a $d$-simplex is assumed to be flat,
and is therefore completely specified by the lengths of its $d(d+1)/2$ edges.
Let $x^\mu (i)$ be the $\mu$-th coordinate of the $i$-th site.
For each pair of neighboring sites $i$ and $j$ the link length
squared is given by the usual expression
\beq
l^2_{ij} \; = \; \eta_{\mu\nu} \, \left [ x(i)-x(j) \right ]^\mu
\, \left [ x(i)-x(j) \right ]^\nu
\eeq
with $\eta_{\mu\nu}$ the flat metric.
It is therefore natural to associate, within a given
simplex $s$, an edge vector $l_{ij}^{\mu} (s) $
with the edge connecting site $i$ to site $j$.

\begin{figure}[h]
\epsfig{file=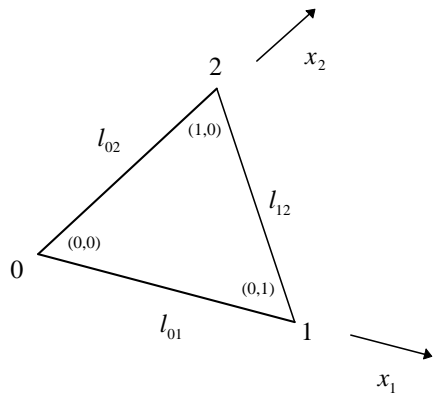,width=8cm}
\caption[Coordinates in simplex]{Coordinates chosen along edges of a simplex, here a triangle.}
\label{fig:coord}
\end{figure}

\subsubsection{Volumes and Angles}

\label{sec:volumes}

When focusing on one such $n$-simplex it will be convenient to label
the vertices by $0,1, 2, 3, \dots , n$ and denote the 
square edge lengths by $l _ {01}^2 = l _ {10}^2$, ... , $l _ {0n}^2 $.
The simplex can then be spanned by the set of $n$ vectors 
$e_1$, ... $e_n$ connecting the vertex $0$ to the 
other vertices.
To the remaining edges within the simplex one then assigns
vectors $e_{ij} = e_i-e_j$ with $1 \le i < j \le n$.
One has therefore $n$ independent vectors, but $\half n (n+1)$
invariants given by all the edge lengths squared within $s$. 

In the interior of a given $n-$simplex one can also assign a second, 
orthonormal (Lorentz) frame, which we will denote in the following 
by $\Sigma (s)$.
The expansion coefficients relating this orthonormal frame to the one specified
by the $n$ directed edges of the simplex associated with the vectors
$e_i $ is the lattice analog of the $n$-bein or tetrad $e_{\mu}^a$.

\begin{figure}[h]
\epsfig{file=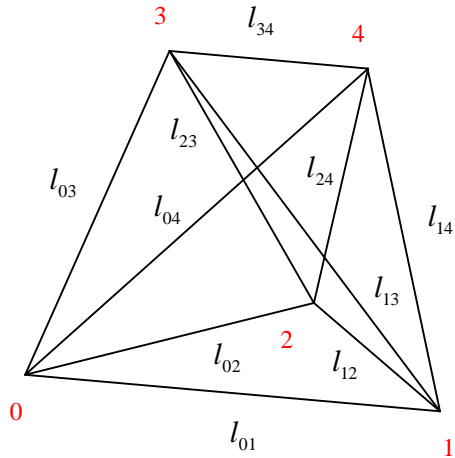,width=8cm}
\caption{A four-simplex, the four-dimensional analog of a tetrahedron.
It contains five vertices, ten edges, ten triangles and five tetrahedra.}
\label{fig:simplex}
\end{figure}

Within each $n$-simplex one can define a metric
\beq
g_{ij} (s) \; = \; e_i \cdot e_j \;\; , 
\eeq
with $1 \leq i,j \leq n $, and which in the Euclidean case is positive definite.
In components one has $g_{ij} = \eta_{ab} e_i^a e_j^b$.
In terms of the edge lengths
$l_{ij} \, = \, | e_i - e_ j | $, the metric is given by
\beq
g_{ij} (s) \; = \; \half \, 
\left ( l_{0i}^2 + l_{0j}^2 - l_{ij}^2 \right ) \;\; .
\label{eq:latmet}
\eeq
Comparison with the standard expression for the invariant interval
$ds^2 = g_{\mu\nu} dx^\mu dx^\nu$ confirms that for the metric
in Eq.~(\ref{eq:latmet}) coordinates have been chosen along the
$n$ $e_i$ directions.

The volume of a general $n$-simplex is given by the $n$-dimensional
generalization of the well-known formula for a tetrahedron, namely
\beq
V_n (s) \; = \; {1 \over n ! }  \sqrt { \det  g_{ij} (s) } \;\; .
\label{eq:vol-met}
\eeq
An equivalent, but more symmetric, form for the volume of an
$n$-simplex can be given in terms of the bordered determinant of an
$(n+2) \times (n+2)$ matrix (Wheeler, 1964)
\beq
V_n (s) \; = \; {(-1)^{n+1 \over 2 } \over n! \, 2^{n/2} } \,
\left|\matrix{
0      &    1     &    1     & \ldots \cr
1      &    0     & l_{01}^2 & \ldots \cr 
1      & l_{10}^2 &    0     & \ldots \cr
1      & l_{20}^2 & l_{21}^2 & \ldots \cr
\ldots &  \ldots  &  \ldots  & \ldots \cr
1      & l_{n,0}^2 & l_{n,1}^2 & \ldots \cr
}\right| ^{1/2}  .
\label{eq:vol}
\eeq
It is possible to associate $p$-forms with lower dimensional
objects within a simplex, which will become useful later (Hartle, 1984).
With each face $f$ of an $n$-simplex (in the shape of a tetrahedron
in four dimensions) one can associate
a vector perpendicular to the face
\beq
\omega (f)_\alpha \; = \; \epsilon_{\alpha \beta_1 \dots \beta_{n-1}} \, 
e^{\beta_1}_{(1)} \dots e^{\beta_{n-1}}_{(n-1)}
\label{eq:omega-face}
\eeq
where $ e_{(1)} \dots e_{(n-1)} $ are a set of oriented edges
belonging to the face $f$, and
$ \epsilon_{\alpha_1 \dots \alpha_n}$ is the sign of the permutation
$ ( \alpha_1 \dots \alpha_n )$.

The volume of the face $f$ is then given by
\beq
V_{n-1} (f) \; = \; \left ( \sum_{\alpha=1}^n  
\omega_\alpha^2 (f) \right )^{1/2}
\eeq
Similarly, one can consider a hinge (a triangle in four dimensions)
spanned by edges $ e_{(1)}$,$\dots$, $ e_{(n-2)}$.
One defines the (un-normalized) hinge bivector
\beq
\omega (h)_{\alpha\beta} \; = \; 
\epsilon_{\alpha \beta \gamma_1 \dots \gamma_{n-2}} \, 
e^{\gamma_1}_{(1)} \dots e^{\gamma_{n-2}}_{(n-2)}
\label{eq:omega-hinge}
\eeq
with the area of the hinge then given by
\beq
V_{n-2} (h) \; = \; {1 \over (n-2)! } 
\left ( \sum_{\alpha < \beta }  \omega_{\alpha\beta}^2 (h) \right )^{1/2}
\eeq
Next, in order to introduce curvature, one needs to define
the {\it dihedral angle} between faces in an $n$-simplex.
In an $n$-simplex $s$ two $n-1$-simplices $f$ and $f'$ will intersect
on a common $n-2$-simplex $h$, and the dihedral angle at the specified hinge $h$ is defined as
\beq
\cos \theta (f,f') \; = \; { \omega (f) _{n-1} \cdot \omega (f')_{n-1} \over
V_{n-1}(f) \, V_{n-1} (f') }
\label{eq:dihedralcos}
\eeq
where the scalar product appearing on the r.h.s. can be re-written in terms
of squared edge lengths using
\beq
\omega_{n} \cdot \omega_{n}' \; = \; 
{1 \over (n!)^2 } \, \det ( e_i \cdot e_j' )
\label{eq:omega-dot}
\eeq
and $e_i \cdot e_j'$ in turn expressed in terms of squared edge lengths
by the use of Eq.~(\ref{eq:latmet}).
(Note that the dihedral angle $\theta$ would have to be defined as $\pi$ minus
the arccosine of the expression on the r.h.s. if the orientation
for the $e$'s had been chosen in such a way that the $\omega$'s would
all point from the face $f$ inward into the simplex $s$).  
As an example, in two dimensions and within a given triangle, two edges
will intersect at a vertex, giving $\theta$ as the angle between
the two edges. 
In three dimensions within a given simplex two triangles will intersect
at a given edge, while in four dimension two tetrahedra will
meet at a triangle.
For the special case of an equilateral $n$-simplex, one has simply
$\theta = \arccos {1 \over n}$. 
A related and often used formula for the sine of the dihedral
angle $\theta$ is
\beq
\sin \theta (f,f') \; = \; { n \over n-1 } \, 
{ V_n (s) \, V_{n-2} (h) \over V_{n-1} (f) \, V_{n-1} (f') } 
\label{eq:dihedral}
\eeq
but is less useful for practical calculations,
as the sine of the angle does not unambiguously determine
the angle itself, which is needed in order
to compute the local curvature.

\begin{figure}[h]
\epsfig{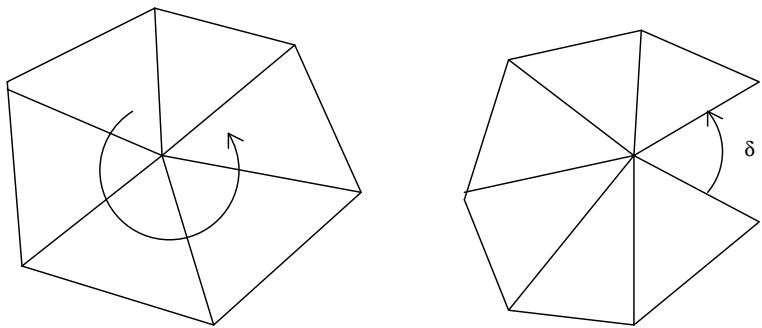}
\caption{Illustration of the deficit angle $\delta$ in two
dimensions, where several flat triangles meet at a vertex.}
\label{fig:2d-deficit}
\end{figure}

\begin{figure}[h]
\epsfig{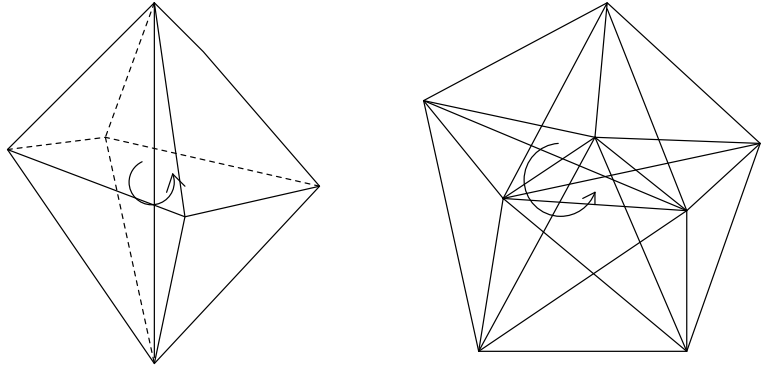}
\caption{Deficit angle in three dimensions 
(where several flat tetrahedra meet at an edge), and in four dimensions
(where several flat four-simplices meet at a triangle).}
\label{fig:3d-deficit}
\end{figure}

In a piecewise linear space curvature is detected by going around
elementary loops which are dual to a ($d-2$)-dimensional subspace.
From the dihedral angles associated with the faces of the simplices meeting
at a given hinge $h$ one can compute the {\it deficit angle} $\delta (h)$,
defined as
\beq
\delta (h) \; = \; 2 \pi \, - \, \sum_{ s \supset h } \; \theta (s,h)
\label{eq:deficit}
\eeq
where the sum extends over all simplices $s$ meeting on $h$.
It then follows that the deficit angle $\delta$ is a measure
of the curvature at $h$.
The two-dimensional case is illustrated in Fig.~\ref{fig:2d-deficit},
while the three- and four-dimensional cases are shown in
Fig.~\ref{fig:3d-deficit}.

\subsubsection{Rotations, Parallel Transports and Voronoi Loops}

\label{sec:rotations}

Since the interior of each simplex $s$ is assumed to be flat, one can assign
to it a Lorentz frame $\Sigma (s)$.
Furthermore inside $s$ one can define a $d$-component vector 
$\phi (s) = ( \phi_0 \dots \phi_{d-1} )$.
Under a Lorentz transformation of $\Sigma (s)$, described by the
$d \times d$ matrix $\Lambda (s)$ satisfying the usual relation
for Lorentz transformation matrices
\beq
\Lambda^{T} \, \eta \, \Lambda \; = \; \eta
\eeq
the vector $\phi (s)$ will rotate to 
\beq
\phi ' (s) \; = \; \Lambda (s) \, \phi (s)
\eeq
The base edge vectors $e_i^{\mu} = l_{0i}^{\mu} (s)$ themselves
are of course an example of such a vector. 

Next consider two $d$-simplices, individually labeled by $s$ and $s'$,
sharing a common face $f (s,s')$ of dimensionality $d-1$.
It will be convenient to label the $d$ edges residing in the common face
$f$ by indices $i,j=1 \dots d$.
Within the first simplex $s$ one can then assign a Lorentz frame $ \Sigma (s)$,
and similarly within the second $s'$ one can assign the frame $\Sigma (s')$.
The $\half d (d-1) $ edge vectors on the common interface
$f(s,s')$ (corresponding physically to the same edges, viewed from
two different coordinate systems)
are expected to be related to each other by a Lorentz rotation $\bf R$,
\beq
l_{ij}^\mu (s') \; = \; R_{\;\;\nu}^{\mu} (s',s) \; l_{ij}^{\nu} (s) 
\eeq
Under individual Lorentz rotations in $s$ and $s'$ one has of course a 
corresponding change in $\bf R$, namely 
${\bf R} \rightarrow \Lambda (s') \, {\bf R} (s',s) \, \Lambda (s)$.
In the Euclidean $d$-dimensional case $\bf R$ is an orthogonal matrix,
element of the group $SO(d)$.

In the absence of torsion, one can use the matrix ${\bf R}(s',s)$ to describes
the parallel transport of any vector $\phi$ from simplex $s$ to a
neighboring simplex $s'$,
\beq
\phi^\mu (s') \; = \; R_{\; \; \nu}^{\mu} (s',s) \, \phi^{\nu} (s) 
\eeq
${\bf R}$ therefore describes a lattice version of the connection (Lee, 1983).
Indeed in the continuum such a rotation would be described by the matrix
\beq
R_{\;\;\nu}^{\mu} \; = \; \left ( e^{\Gamma \cdot dx} \right )_{\;\;\nu}^{\mu}
\eeq
with $\Gamma^{\lambda}_{\mu\nu}$ the affine connection.
The coordinate increment $dx$ is interpreted as joining
the center of $s$ to the center of $s'$, thereby intersecting
the face $f(s,s')$. 
On the other hand, in terms of the Lorentz frames 
$\Sigma (s)$ and $\Sigma (s')$ defined within the two
adjacent simplices, the rotation matrix is given instead by 
\beq
R^a_{\;\;b} (s',s) \; = \; e^a_{\;\;\mu} (s') \, e^{\nu}_{\;\;b} (s)
\; R_{\;\;\nu}^{\mu} (s',s)
\eeq
(this last matrix reduces to the identity if the two orthonormal bases
$\Sigma (s)$ and $\Sigma (s')$ are chosen to be the same,
in which case the connection is simply given by 
$ R(s',s)_{\mu}^{\;\; \nu} = e_{\mu}^{\;\;a} \, e^{\nu}_{\;\;a} $).
Note that it is possible to choose coordinates so that
$ {\bf R} (s,s')$ is
the unit matrix for one pair of simplices, but it will not then be unity for
all other pairs.

This last set of results will be useful later when discussing lattice Fermions.
Let us consider here briefly the problem of how to introduce lattice
{\it spin rotations}.
Given in $d$ dimensions the above rotation matrix $ {\bf R} (s',s) $, 
the spin connection ${\bf S}(s,s')$ between two neighboring simplices 
$s$ and $s'$ is defined as follows.
Consider $\bf S$ to be an element of the $2^\nu$-dimensional representation
of the covering group of $SO(n)$, $Spin(d)$, with $d=2 \nu$ or $d=2 \nu+1$, and
for which $S$ is a matrix of dimension $2^\nu \times 2^\nu$.
Then $\bf R$ can be written in general as
\beq
{\bf R} \; = \; \exp \left [ \, 
\half \, \sigma^{\alpha\beta} \theta_{\alpha\beta} \right ]
\eeq
where $\theta_{\alpha\beta}$ is an antisymmetric matrix
The $\sigma$'s are $\half d(d-1)$ $d \times d$ matrices, 
generators of the Lorentz group 
($SO(d)$ in the Euclidean case, and $SO(d-1,1)$ in the Lorentzian case),
whose explicit form is 
\beq
\left [ \sigma_{\alpha\beta} \right ]^{\gamma}_{\;\; \delta}
\; = \; \delta_{\;\; \alpha}^{\gamma} \, \eta_{\beta\delta} \, - \, 
\delta_{\;\; \beta}^{\gamma} \, \eta_{\alpha\delta}
\eeq
so that, for example,
\beq
\sigma_{12} \; = \; 
\left ( \matrix{
0  & 1 & 0 & 0 \cr
-1 \;\;  & 0 & 0 & 0 \cr 
0  & 0 & 0 & 0 \cr
0  & 0 & 0 & 0 } \right ) \;\; .
\eeq
For Fermions the corresponding spin rotation matrix is then obtained from
\beq
{\bf S} \; = \; \exp \left [ \, {\textstyle {i\over4} \displaystyle} \,
\gamma^{\alpha\beta} \theta_{\alpha\beta} \right ]
\eeq
with generators
$ \gamma^{\alpha\beta} = { 1 \over 2 i } [ \gamma^\alpha , \gamma^\beta ] $,
and with the Dirac matrices $\gamma^\alpha$ satisfying as usual  
$\gamma^\alpha \gamma^\beta + \gamma^\beta \gamma^\alpha = 2 \,\eta^{\alpha\beta}$.
Taking appropriate traces, one can obtain a direct relationship
between the original rotation matrix ${\bf R} (s,s')$ and the corresponding
spin rotation matrix ${\bf S}(s,s')$
\beq
R_{\alpha\beta} \; = \; \tr \left ( 
{\bf S}^\dagger \, \gamma_\alpha \, {\bf S} \, \gamma_\beta \right )
/ \tr {\bf 1 }
\label{eq:spinrot}
\eeq
which determines the spin rotation matrix up to a sign. 

One can consider a sequence of rotations along an arbitrary
path $P (s_1, \dots , s_{n+1})$ going through simplices 
$s_1 \dots s_{n+1}$, whose combined rotation matrix is given by
\beq
{\bf R} (P) \; = \; {\bf R} (s_{n+1}, s_n ) \cdots {\bf R} (s_2, s_1 )
\eeq
and which describes the parallel transport of an arbitrary vector
from the interior of simplex $s_1$ to the interior of simplex $s_{n+1}$,
\beq
\phi^\mu (s_{n+1}) \; = \; R_{\; \; \nu}^{\mu} (P) \, \phi^{\nu} (s_1) 
\eeq
If the initial and final simplices $s_{n+1}$ and $s_1$ coincide,
one obtains a closed path $C (s_1, \dots , s_n)$, for which the
associated expectation value can be considered as the gravitational
analog of the Wilson loop.
Its combined rotation is given by
\beq
{\bf R} (C) \; = \; {\bf R} (s_1, s_n ) \cdots {\bf R} (s_2, s_1 )
\label{eq:loop-rot}
\eeq
Under Lorentz transformations within each simplex $s_i$ along
the path one has a pairwise cancellation of the $\Lambda (s_i)$ matrices
except at the endpoints, giving in the closed loop case
\beq
{\bf R} (C) \; \rightarrow \; \Lambda ( s_1 ) \, {\bf R} ( C )
\, \Lambda^{T} ( s_1 )
\eeq
Clearly the deviation of the matrix ${\bf R} (C)$ from unity is a measure of curvature.
Also, the trace $\tr {\bf R} (C)$ is independent of the choice
of Lorentz frames.

\begin{figure}[h]
\epsfig{file=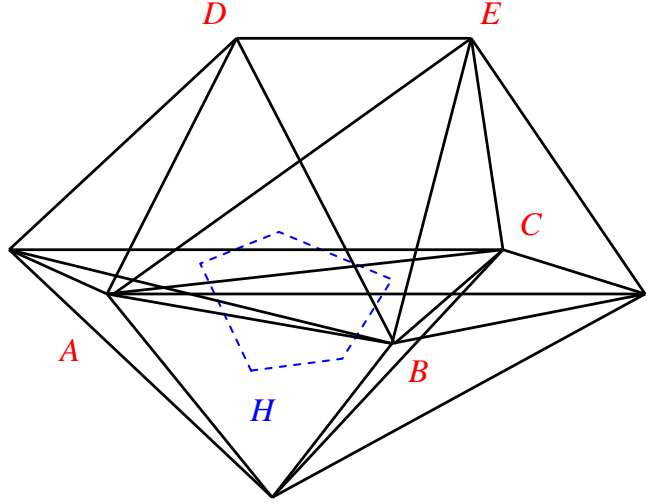,width=10cm}
\caption{Elementary polygonal path around a hinge
(triangle) in four dimensions. The hinge $ABC$, contained in the simplex
$ABCDE$, is encircled by the polygonal path $H$ connecting the
surrounding vertices which reside in the dual lattice.
One such vertex is contained within the simplex $ABCDE$.}
\label{fig:hinge-path}
\end{figure}

Of particular interest is the elementary loop associated with
the smallest non-trivial, segmented parallel transport path one can
build on the lattice.
One such polygonal path in four dimensions is shown in 
Fig.~\ref{fig:hinge-path}.
In general consider a $(d-2)$-dimensional simplex (hinge) $h$, which
will be shared by a certain number $m$ of $d$-simplices, 
sequentially labeled by $s_1 \dots s_m$, and whose common faces 
$f(s_1,s_2) \dots f( s_{m-1}, s_m ) $ will also contain the hinge $h$.
Thus in four dimensions several four-simplices will contain,
and therefore encircle, a given triangle (hinge).
In three dimensions the path will encircle an edge, while in two
dimensions it will encircle a site. 
Thus for each hinge $h$ there is a unique elementary closed path $C_h$
for which one again can define the ordered product
\beq
{\bf R} (C_h) \; = \; {\bf R} (s_1, s_m ) \cdots {\bf R} (s_2, s_1 )
\label{eq:loop-rot1}
\eeq
The hinge $h$, being geometrically an object of dimension $(d-2)$, is naturally
represented by a tensor of rank $(d-2)$, referred to a coordinate system
in $h$: an edge vector $l_h^\mu$ in $d=3$, and an area bi-vector 
$\half ( l_h^\mu l_h^{'\nu} - l_h^\nu l_h^{'\mu} ) $ in $d=4$ etc.
Following Eq.~(\ref{eq:omega-hinge}) it will therefore be convenient
to define a hinge bi-vector $U$ in any dimension as
\beq
U_{\mu\nu} (h) \; = \; {\cal N} \, \epsilon_{\mu\nu \alpha_1 \alpha_{d-2}} \, l_{(1)}^{\alpha_1} \dots l_{(d-2)}^{\alpha_{d-2}} \;\; ,
\label{eq:bivector-d}
\eeq
normalized, by the choice of the constant ${\cal N}$, in such a way that 
$U_{\mu\nu} U^{\mu\nu} =2$. 
In four dimensions
\beq
U_{\mu\nu} (h) \; = \; { 1 \over 2 A_h } \;
\epsilon_{\mu\nu\alpha\beta} \, l_1^{\alpha} \, l_2^{\beta}
\label{eq:bivector}
\eeq
where $l_1 (h)$ and $l_2 (h)$ two independent edge vectors
associated with the hinge $h$, 
and $A_h$ the area of the hinge.

An important aspect related to the rotation of an 
arbitrary vector, when parallel transported around a hinge $h$,
is the fact that, due to the hinge's intrinsic orientation,
only components of the vector in the plane perpendicular
to the hinge are affected.
Since the direction of the hinge $h$ is specified locally by
the bivector $U_{\mu\nu}$ of Eq.~(\ref{eq:bivector}), 
one can write for the loop rotation matrix $\bf R$
\beq
R_{\;\;\nu}^{\mu} (C) \; = \; 
\left ( e^{ \delta \, U } \right )_{\;\;\nu}^{\mu}
\label{eq:rot-hinge}
\eeq
where $C$ is now the small polygonal loop entangling the hinge $h$,
and $\delta$ the deficit angle at $h$, previously
defined in Eq.~(\ref{eq:deficit}).
One particularly noteworthy aspect of this last result is the fact that the area
of the loop $C$ does not enter in the expression for the
rotation matrix, only the deficit angle and the hinge direction.
 
At the  same time, in the continuum a vector $V$ carried around
an infinitesimal loop of area $A_C$ will change by 
\beq
\Delta V^{\mu} \; = \; \, \half \, R^{\mu}_{ \;\; \nu \lambda \sigma }
\, A^{ \lambda \sigma } \, V^\nu
\eeq
where $A^{ \lambda \sigma }$ is an area bivector in the plane of $C$,
with squared magnitude $ A_{ \lambda \sigma } A^{ \lambda \sigma } = 2 A_C^2$.
Since the change in the vector $V$ is given by
$ \delta V^\alpha = ({\bf{R-1}})^{\alpha}_{\;\;\beta} \, V^\beta $
one is led to the identification
\beq
\half \; R^{\alpha}_{\;\;\beta\mu\nu} \, A^{\mu\nu} \; = \;
({\bf {R-1}})^{\alpha}_{\;\;\beta} \;\; .
\label{eq:riemrot}
\eeq
Thus the above change in $V$ can equivalently be re-written in terms of the
infinitesimal rotation matrix
\beq
R_{\;\;\nu}^{\mu} (C) \; = \; \left ( e^{ \, \half \, R \cdot A } \right )_{\;\;\nu}^{\mu}
\label{eq:rot-cont}
\eeq
(where the Riemann tensor appearing in the exponent on the r.h.s. should not
be confused with the rotation matrix $\bf R$ on the l.h.s.).

It is then immediate to see that the two expressions for the rotation
matrix $\bf R$ in Eqs.~(\ref{eq:rot-hinge}) and (\ref{eq:rot-cont})
will be compatible provided one uses for the Riemann tensor
at a hinge $h$ the expression 
\beq
R_{\mu\nu\lambda\sigma} (h) \; = \; {\delta (h) \over A_C (h) } 
\, U_{\mu\nu} (h) \, U_{\lambda\sigma} (h)
\label{eq:riem-hinge}
\eeq
expected to be valid in the limit of small curvatures,
with $A_C (h) $ the area of the loop entangling the hinge $h$.
Here use has been made of the geometric relationship
$U_{\mu\nu} \, A^{\mu\nu} = 2 A_C$.
Note that the bivector $U$ has been defined to be perpendicular
to the $(d-2)$ edge vectors spanning the hinge $h$, 
and lies therefore in the same plane as the loop $C$.
Furthermore, the expression of Eq.~(\ref{eq:riem-hinge}) for the 
{\it Riemann
tensor at a hinge} has the correct algebraic symmetry
properties, such as the antisymmetry in the first and
second pair of indices, as well as the swap symmetry
between first and second pair, and is linear in the curvature,
with the correct dimensions of one over length squared.

The area $A_C$ is most suitably defined by introducing the
notion of a {\it dual lattice},
i.e. a lattice constructed by assigning centers to the simplices,
with the polygonal curve $C$ connecting these centers sequentially,
and then assigning an area to the interior of this curve.
One possible way of assigning such centers is by introducing
perpendicular bisectors to the faces of a simplex, and locate
the vertices of the dual lattice at their common intersection,
a construction originally discussed in (Voronoi, 1908) and in (Meijering, 1953).
Another, and perhaps even simpler, possibility is to use a barycentric
subdivision (Singer and Thorpe, 1967).

\subsubsection{Invariant Lattice Action}

\label{sec:regge}

The first step in writing down an invariant lattice action,
analogous to the continuum Einstein-Hilbert action, 
is to find the lattice analog of the Ricci scalar.
From the expression for the Riemann tensor at a hinge
given in Eq.~(\ref{eq:riem-hinge}) one obtains by contraction
\beq
R (h) \; = \; 2 \, { \delta (h) \over A_C (h) }
\eeq
The continuum expression $\sqrt{g} \, R$ is then obtained
by multiplication with the volume element $V (h) $ associated with
a hinge.
The latter is defined by first joining the vertices of the
polyhedron $C$, whose vertices lie in the dual lattice,
with the vertices of the hinge $h$, and then computing its volume.

By {\it defining} the polygonal area $A_C$ as 
$A_C (h) = d \, V (h) / V^{(d-2)} (h) $, where $V^{(d-2)} (h)$
is the volume of the hinge (an area in four dimensions),
one finally obtains for the Euclidean lattice action for pure gravity
\beq
I_{R} (l^2) \; = \; - \; k \, \sum_{\rm hinges \; h}
\, \delta (h) \, V^{(d-2)} (h) \;\; ,
\label{eq:regge-d}
\eeq
with the constant $k=1/(8 \pi G)$. 
One would have obtained the same result for the single-hinge
contribution to the lattice action
if one had contracted the infinitesimal form of the rotation
matrix $R(h)$ in Eq.~(\ref{eq:rot-hinge}) with the hinge bivector 
$\omega_{\alpha\beta}$ of Eq.~(\ref{eq:omega-hinge}) (or equivalently
with the bivector $U_{\alpha\beta}$ of Eq.~(\ref{eq:bivector})
which differs from $\omega_{\alpha\beta}$ by a constant).
The fact that the lattice action only involves the content of
the hinge $ V^{(d-2)} (h)$ (the area of a triangle in four dimensions)
is quite natural in view of the fact that the rotation matrix
at a hinge in Eq.~(\ref{eq:rot-hinge}) only involves the deficit angle, 
and not the polygonal area $A_C (h)$.

An alternative form for the lattice action (Fr\"ohlich, 1981) can be obtained
instead by contracting the elementary rotation matrix ${\bf R}(C)$ 
of Eq.~(\ref{eq:rot-hinge}), and not just its infinitesimal form,
with the hinge bivector of Eq.~(\ref{eq:omega-hinge}),
\beq
I_{\rm com} (l^2) \; = \; - \; k \, \sum_{\rm hinges \; h} \,  \half \, 
\omega_{\alpha\beta}(h) \, R^{\alpha\beta} (h) 
\label{eq:regge-compact}
\eeq
The above construction can be regarded as analogous to Wilson's lattice
gauge theory, for which the action also involves traces of products of
$SU(N)$ color rotation matrices (Wilson, 1973).
For small deficit angles one can of course use
$\omega_{\alpha\beta} = (d-2)! \, V^{(d-2)} \, U_{\alpha\beta} $
to show the equivalence of the two lattice actions.

But in general, away from a situation of small curvatures,
the two lattice action are not equivalent,
as can be seen already in two dimensions.
Writing the rotation matrix at a hinge as
${\bf R}(h)= \left ( \matrix { 
\cos \delta & \sin \delta \cr
- \sin \delta & \cos \delta } \right ) $, expressed 
for example in terms of Pauli matrices, and taking the appropriate trace 
($\omega_{\alpha\beta}=\epsilon_{\alpha\beta}$ in two dimensions) one finds
\beq
\tr \left [ 
\half ( - i \sigma_y ) ( \cos \delta_p + i \sigma_y \sin \delta_p ) \right ] 
\; = \; \sin \delta_p
\eeq
and therefore $I_{\rm com} \; = \; - \; k \, \sum_p \,  \sin \delta_p $.
In general one can show that the compact action $I_{com}$ in $d$
dimensions involves the sine of the deficit angle, instead of just 
the angle itself as in the Regge case.
In the weak field limit the two actions should lead to similar
expansions, while away from the weak field limit one would
have to verify that the same universal long
distance properties are recovered.

The preceding observations can in fact be developed into a consistent
first order (Palatini) formulation of Regge gravity, with suitably chosen
independent transformation matrices and metrics, related to each other
by a set of appropriate lattice equations of motion 
(Caselle, d'Adda and Magnea, 1986). 
Ultimately one would expect the first and second order formulations to
describe the same quantum theory, with common universal
long-distance properties.
How to consistently define finite rotations, frames
and connections in Regge gravity was first discussed
systematically in (Fr\"ohlich, 1981).

One important result that should be mentioned 
in this context is the rigorous proof of
convergence in the sense of measures of
the Regge lattice action towards the continuum Einstein-Hilbert action
(Cheeger, M\"uller and Schrader, 1984).
Some general aspects of this result have recently been reviewed from a
mathematical point of view in (Lafontaine, 1986).
A derivation of the Regge action from its continuum
counterpart was later presented in (Lee, Feinberg and Friedberg, 1984).

Other terms can be added to the lattice action.
Consider for example a cosmological constant term, which in the continuum theory
takes the form $ \lambda_0 \int d^d x \sqrt g $.
The expression for the cosmological constant term on the lattice involves
the total volume of the simplicial complex.
This may be written as
\beq
V_{{\bf total}} = \sum_{\rm simplices \; s} V_s
\label{eq:totlatvol}
\eeq
or equivalently as
\beq
V_{{\bf total}} = \sum_{\rm hinges \; h} V_h 
\label{eq:totlatvol1}
\eeq
where $V_h$ is the volume associated with each hinge via the construction
of a dual lattice, as described above.
Thus one may regard the local volume element $ \sqrt g \, d^d x $ as
being represented by either $ V_h $ (centered on $h$) or $ V_s$
(centered on $s$).

The Regge and cosmological constant term then lead to the combined action
\beq
I_{\rm latt} (l^2) \; = \; \lambda_0 \sum_{\rm simplices \; s} \, V^{(d)}_s
\, - \, \sum_{\rm hinges \; h} \,  \delta_h \, V^{(d-2)}_h
\label{eq:latac}
\eeq
One would then write for the lattice regularized version of the Euclidean Feynman path integral
\beq
Z_{\rm latt} ( \lambda_0 , \, k ) \; = \; 
\int [ d \, l^2 ] \, \exp \left ( - I_{\rm latt} (l^2)
\right ) \;\; ,
\label{eq:zdef}  
\eeq
where $ [ d \, l^2 ] $ is an appropriate functional integration
measure over the edge lengths, to be discussed later.

The structure of the gravitational action of Eq.~(\ref{eq:latac}) leads
naturally to some rather general observations, which we will pursue here.
The first, cosmological constant, term represents the total four-volume
of space-time.
As such, it does not contain any derivatives (or finite differences) of the metric and is completely local; 
it does not contribute to the propagation of gravitational degrees of
freedom and is more akin to a mass term (as is already clear
from the weak field expansion of $\int \sqrt{g}$ in the continuum).
In an ensemble in which the total four-volume is fixed in the
thermodynamic limit (number of simplices tending to infinity)
one might in
fact take the lattice coupling $\lambda_0 =1$, since different values of
$\lambda_0$ just correspond to a trivial rescaling of the overall four-volume
(of course in a traditional renormalization group approach to field
theory, the overall four-volume is always kept fixed while the scale
or $q^2$ dependence of the action and couplings are investigated).
Alternatively, one might even want to choose directly an ensemble for
which the probability distribution in the total four-volume $V$ is
\beq
{\cal P} (V) \; \propto \; \delta ( V - V_0 )
\eeq
in analogy with the microcanonical ensemble of statistical mechanics.

The second, curvature contribution to the action contains,
as in the continuum, the proper kinetic term.
This should already be clear from the derivation of the lattice action
given above, and will
be made even more explicit in the section dedicated to the lattice weak
field expansion.
Such a term now provides the necessary coupling between neighboring lattice
metrics, but the coupling still remains local.
Geometrically, it can be described as a sum of elementary 
loop contributions, as it contains as its primary ingredient the deficit
angle associated with an elementary parallel transport loop
around the hinge $h$.
When $k=0$ one resides in the extreme strong coupling regime
There the fluctuations in the metric are completely unconstrained
by the action, insofar as only the total four-volume of
the manifold is kept constant.

At this point it might be useful to examine some specific cases
with regards to the overall dimensionality of the simplicial complex.
In {\it two dimensions} the Regge action reduces to a sum over lattice
sites $p$ of the $2-d$ deficit angle, giving the discrete analog
of the Gauss-Bonnet theorem
\beq
\sum_{\rm sites \; p} \,  \delta_p \; = \; 2 \, \pi \, \chi
\label{eq:regge-2d}
\eeq
where $\chi = 2-2g$ is the Euler characteristic of the surface,
and $g$ the genus (the number of handles). 
In this case the action is therefore a topological invariant, and
the above lattice expression is therefore completely
analogous to the well known continuum result
\beq
\half \int d^2 x \sqrt g \, R \; = \; 2 \, \pi \, \chi
\eeq
This remarkable identity ensures that two-dimensional lattice $R$-gravity is
as trivial as in the continuum, since the variation of the local action density
under a small variation of an edge length $l_{ij}$ is still zero.
Of course there is a much simpler formula for the Euler characteristic of a
simplicial complex, namely
\beq
\chi = \sum_{i=0}^d (-1)^i N_i
\eeq
where $ N_i $ is the number of simplices of dimension $i$.
Also it should be noted that in two dimensions the compact action
$I_{com} = \sum_p \,  \sin \delta_p$ does not satisfy the Gauss-Bonnet
relation.

In {\it three dimensions} the Regge lattice action reads
\beq
I_{R}  =  - \, k \, \sum_ {\rm edges \; h } \; l_h \, \delta_h
\label{eq:regge-3d}
\eeq
where $\delta_h$ is the deficit angle around the edge labeled by $h$. 
Variation with respect to an edge length $l_h$ gives two
terms, of which only the term involving the variation of the
edge is non-zero
\beq
\delta \, I_{R}  =  - \, k \, \sum_ {\rm edges \; h } \; 
\delta l_h  \cdot \delta_h .
\eeq
In fact it was shown by Regge that for any $d>2$ the term involving the 
variation of the deficit angle does not contribute to the equations
of motion (just as in the continuum the variation of the Ricci tensor
does not contribute to the equations of motion either).
Therefore in three dimensions the lattice equations of motion, in the
absence of sources and cosmological constant term, reduce to
\beq
\delta_h \; = \; 0
\eeq
implying that all deficit angles have to vanish, i.e. a flat space.

In four dimensions variation of $I_R$ with respect to the edge lengths
gives the simplicial analog of Einstein's field equations, whose
derivation is again, as mentioned, simplified by the fact that the
contribution from the variation of the deficit angle is zero
\beq
\delta \, I_R = \sum_{\rm triangles \; h } \delta ( A_h ) \cdot \delta_h
\eeq
In the discrete case the field equations reduce therefore to
\beq
\lambda_0 \, \sum_s { \partial V_s \over \partial l_{ij} }
- \, k \, \sum_h \delta_h \, { \partial A_h \over \partial l_{ij} }
\; = \; 0
\label{eq:regge-eq}
\eeq
and the derivatives can then be worked out for example from Eq.~(\ref{eq:vol}).
Alternatively, a rather convenient and compact expression can be given (Hartle, 1984) for the derivative of the squared volume $V_n^2$ of an arbitrary
$n$-simplex with respect to one of its squared edge lengths 
\beq
{ \partial V^2_n \over \partial l^2_{ij} } \; = \; { 1 \over n^2 } 
\, \omega_{n-1} \cdot \omega_{n-1}'
\eeq
where the $\omega_{(n-1)}^\alpha$'s (here referring to two $(n-1)$ simplices
part of the same $n$-simplex)
are given in Eqs.~(\ref{eq:omega-face}) and (\ref{eq:omega-dot}).
In the above expression the $\omega$'s are meant to be 
associated with vertex labels 
$0, \dots , i-1,i+1,\dots,n$ for $\omega_{n-1}$, and
$0, \dots , j-1,j+1, \dots , n$ for $\omega_{n-1}'$
respectively.

\begin{figure}[h]
\epsfig{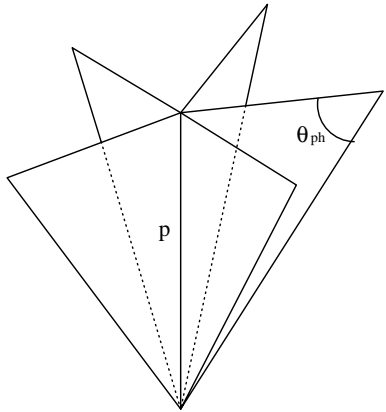}
\caption{Angles appearing in the Regge equations.}
\label{fig:regge-eq}
\end{figure}

Then in the absence of a cosmological term one finds the remarkably
simple expression for the lattice field equations
\beq
\half \; l_p
\sum_ { h \supset l_p } \delta_h \, \cot \theta_{ph} \; = \; 0
\eeq
where the sum is over hinges (triangles) labeled by $h$ meeting on the
common edge $l_p$, and $\theta_{ph} $ is the angle in the hinge $h$
opposite to the edge $l_p$.
This is illustrated in Fig.~\ref{fig:regge-eq}.

The discrete equations given above represent the lattice analogues of
the Einstein field equations in a vacuum, for which suitable
solutions can be searched for by adjusting the edge lengths. 
Since the equations are in general non-linear, 
the existence of multiple solutions cannot in general be ruled out
(Misner, Thorne and Wheeler, 1973).
A number of papers have adressed the general issue of convergence
to the continuum in the framework of the classical formulation
(Brewin and Gentle, 2001).
Several authors have discussed non-trivial applications of the Regge
equations to problems in classical general relativity such as the
Schwarzschild and Reissner-Nordstrom geometries (Wong, 1971), the
Friedmann and Tolman universes (Collins and Williams, 1974), 
and the problem of radial motion and circular (actually polygonal) orbits (Williams and Ellis, 1980).
Spherically symmetric, as well as more generally inhomogeneous, vacuum
spacetimes were studied using a discrete $3+1$ formulation with a variety
of time-slicing prescriptions in (Porter, 1987),
and later extended (Dubal, 1989) to a systematic investigation of 
the axis-symmetric non-rotating vacuum solutions and to 
the problem of relativistic spherical collapse for polytropic
perfect fluids.

In classical gravity the general time evolution problem plays of
course a central role.
The $3+1$ time evolution problem in Regge gravity was discussed originally 
in (Sorkin, 1975) and later re-examined from a numerical,
practical prespective in 
(Barrett, Galassi, Miller, Sorkin, Tuckey and Williams, 1994) 
using a discrete time step formulation, whereas
in (Piran and Williams, 1986) a continuous time fomalism was proposed.
The choice of lapse and shift functions in Regge gravity were discussed
further in (Tuckey, 1989; Galassi, 1993) and in (Gentle and Miller, 1998), and
applied to the Kasner cosmology in the last reference.
An alternative so-called null-strut approach was proposed in 
(Miller and Wheeler, 1985) which builds up a spacelike-foliated spacetime
with a maximal number of null edges, but seems difficult to implement in
practice. 
Finally in (Khatsymovsky, 1991) and (Immirzi,1996) a continuous time
Regge gravity formalism in the tetrad-connection variables was developed,
in part targeted towards quantum gravity calculations.
A recent comprehensive review of classical applications of Regge gravity
can be found for example in (Gentle, 2002), as well as a more complete
set of references.

\subsubsection{Lattice Diffeomorphism Invariance}

\label{sec:diffeos}

\begin{figure}[h]
\epsfig{file=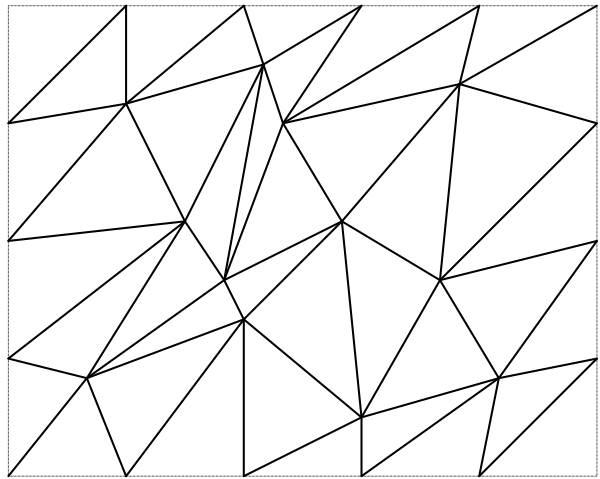,width=10cm}
\caption{On a random simplicial lattice there are in general
no preferred directions.}
\label{fig:random}
\end{figure}

Consider the two-dimensional flat skeleton shown in Fig.~\ref{fig:random}.
It is clear that one can move around a point on the surface, keeping
all the neighbors fixed, without violating the triangle inequalities and
leave all curvature invariants unchanged.

\begin{figure}[h]
\epsfig{file=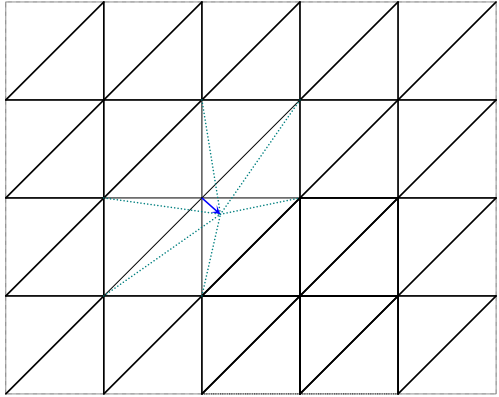,width=8cm}
\caption{Example of a lattice diffeomorphism, the local gauge transform 
of a flat lattice, corresponding to a $d$-parameter local deformation
of the edges.}
\label{fig:latt-diffeo}
\end{figure}

\begin{figure}[h]
\epsfig{file=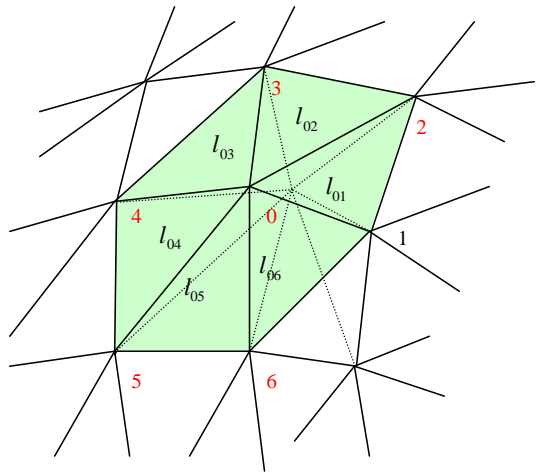,width=8cm}
\caption{Another example of a lattice diffeomorphism, the gauge deformation
of a lattice around a vertex $0$, leaving the local action contribution from that vertex invariant.}
\label{fig:gauge}
\end{figure}

In $d$ dimensions this transformation has $d$ parameters and is an exact
invariance of the action.
When space is slightly curved, the invariance is in general only an approximate
one, even though for piecewise linear spaces piecewise diffeomorphisms
can still be defined as the
set of local motions of points that leave the local contribution to the action,
the measure and the lattice analogues of the continuum curvature invariants unchanged (Hamber and Williams, 1998).
Note that in general the gauge deformations of the edges are still constrained
by the triangle inequalities.
The general situation is illustrated in Figs.~\ref{fig:random},
\ref{fig:latt-diffeo} and \ref{fig:gauge}.
In the limit when the number of edges becomes very large, the full continuum
diffeomorphism group should be recovered.

In general the structure of lattice local gauge transformations
is rather complicated and will not be given here; it can be found in
the above quoted reference. 
These are defined as transformations acting locally on a given
set of edges which leave the local lattice curvature invariant. 
The simplest context in which this local invariance can be exhibited
explicitly is
the lattice weak field expansion, which will be discussed later
in Sec.~\ref{sec:latticewfe}.
The local gauge invariance corresponding to continuum diffeomorphism
is given there in Eq.~(\ref{eq:eigenmodes}).
From the transformation properties of the edge lenghts it is clear
that their transformation properties are related to those of
the local metric, as already suggested for example by
the identification of Eqs.~(\ref{eq:latmet}) and (\ref{eq:latmet1}).

\subsubsection{Lattice Bianchi Identities}

\label{sec:bianchi}

Consider therefore a closed path 
$C_h$ encircling a hinge $h$ and passing through
each of the simplices that meet at that hinge.
In particular one may take $C_h$ to be the boundary of the polyhedral
dual (or Voronoi) area surrounding the hinge.
We recall that the Voronoi polyhedron dual to a vertex $P$ is the set
of all points on the lattice which are closer to $P$ than any other vertex;
the corresponding new vertices then represent the sites on the dual lattice.
A unique closed parallel transport path can then be assigned to each hinge,
by suitably connecting sites in the dual lattice.

With each neighboring pair of simplices $s,s+1$ one associates a
Lorentz transformation ${\bf R}^{\alpha}_{\;\; \beta} (s,s+1)$, which describes
how a given vector $ V_\mu $ transforms between the local coordinate
systems in these two simplices
As discussed previously,
the above transformation is directly related to the continuum
path-ordered ($P$) exponential of the integral of the local affine connection
$ \Gamma^{\lambda}_{\mu \nu}$ via
\beq
R^\mu_{\;\; \nu} \; = \; \Bigl [ {\cal P} \; e^{\int_
{{\bf path \atop between \; simplices}}
\Gamma^\lambda d x_\lambda} \Bigr ]^\mu_{\;\; \nu}  \;\; .
\eeq
The connection here has support only on the common interface between the two
simplices.

Just as in the continuum, where the affine connection and therefore
the infinitesimal rotation matrix is determined by the metric
and its first derivatives, on the lattice one expects that
the elementary rotation matrix between simplices ${\bf R}_{s,s+1}$
is fixed by the difference between the $g_{ij}$'s of 
Eq.~(\ref{eq:latmet}) within neighboring simplices.

For a vector $V$ transported once around a Voronoi loop,
i.e. a loop formed by Voronoi edges surrounding a chosen hinge,
the change in the vector $V$ is given by
\beq
\delta V^\alpha = ({\bf{R-1}})^{\alpha}_{\;\;\beta}\, V^\beta \;\; ,
\eeq
where ${\bf R} \equiv \prod_s {\bf R}_{s,s+1} $
is now the total rotation matrix associated with the given hinge,
given by 
\beq
\Bigl [ \prod_s {\bf R}_{s,s+1}   \Bigr ]^{\mu}_{\;\; \nu} \; = \;
\Bigl [ \, e^{\delta (h) U (h)} \Bigr ]^{\mu}_{\;\; \nu}  \;\; .
\eeq
It is these lattice parallel transporters around closed elementary loops 
that satisfy the lattice analogues of the Bianchi identities.
These are derived by considering paths which encircle more than one
hinge and yet are topologically trivial, in the sense that
they can be shrunk to a point without entangling any hinge (Regge, 1961). 

\begin{figure}[h]
\epsfig{file=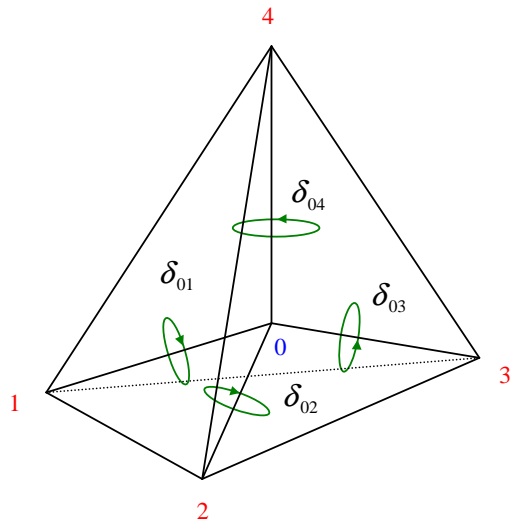,width=8cm}
\caption{Illustration of the lattice Bianchi identity in the case of three
dimensions, where several hinges (edges) meet on a vertex. The combined
rotation for a path that sequentially encircles several hinges 
and which can be shrunk to a point is given by the identity matrix.}
\label{fig:bianchi1}
\end{figure}

\begin{figure}[h]
\epsfig{file=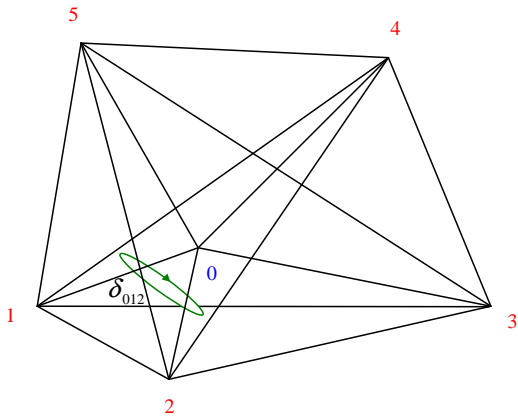,width=8cm}
\caption{Another illustration of the lattice Bianchi identity, now in four
dimensions. Here several hinges (triangles) meet at the vertex labelled by 0.
Around each hinge one has a corresponding rotation and therefore
a deficit angle $\delta$.
The product of rotation matrices that sequentially encircle several hinges
and is topologically trivial gives the identity matrix.}
\label{fig:bianchi2}
\end{figure}

Thus, for example, the ordered product of rotation matrices associated with the
triangles meeting on a given edge has to give one, since a single path can be
constructed which sequentially encircles all the triangles and is
topologically trivial
\beq
\prod_{ {\bf hinges \; h \atop meeting \; on \; edge \; p } } 
\left [ \, e^{ \delta (h) U (h) } \right ]^{\mu}_{\;\; \nu } \; = \; 1
\label{eq:bianchi}
\eeq
Other identities might be derived by considering paths that encircle 
several hinges meeting on one point.
Regge has shown that the above lattice relations correspond precisely
to the continuum Bianchi identities.
One can therefore explicitly construct exact lattice analogues of the continuum
uncontracted, partially contracted, and fully contracted Bianchi
identities (Hamber and Kagel, 2004).
The lattice Bianchi identities are illustrated in Figs.~ \ref{fig:bianchi1}
for the three-dimensional case, 
and \ref{fig:bianchi2} for the four-dimensional case.

The resulting lattice equations are quite similar in structure to the Bianchi identities in $SU(N)$ lattice gauge theories, where one considers identities
arising from the multiplication of group elements associated with the
square faces of a single cube part of a hypercubic lattice (Wilson, 1973).
The motivation there was the possible replacement of the integration over
the group elements by an integration over the ``plaquette variables'' associated
with an elementary square (thereby involving the ordered product of four
group elements), provided the Bianchi identity constraint is included
as well in the lattice path integral.

\subsubsection{Gravitational Wilson Loop}

\label{sec:loop}


We have seen that with each neighboring pair of simplices $s,s+1$ one 
can associate a
Lorentz transformation ${\bf R}^{\mu}_{\;\; \nu} (s,s+1)$, which describes
how a given vector $ V^\mu $ transforms between the local coordinate
systems in these two simplices, and
that the above transformation is directly related to the continuum
path-ordered ($P$) exponential of the integral of the local affine connection
$ \Gamma^{\lambda}_{\mu \nu}(x)$ via
\beq
R^\mu_{\;\; \nu} \; = \; \Bigl [ P \; e^{\int_
{{\bf path \atop between \; simplices}}
\Gamma_\lambda d x^\lambda} \Bigr ]^\mu_{\;\; \nu}  \;\; .
\eeq
with the connection having support only on the common interface between the two
simplices.
Also, for a closed elementary path 
$C_h$ encircling a hinge $h$ and passing through
each of the simplices that meet at that hinge one has
for the total rotation matrix ${\bf R} \equiv \prod_s {\bf R}_{s,s+1} $
associated with the given hinge 
\beq
\Bigl [ \prod_s {\bf R}_{s,s+1}   \Bigr ]^{\mu}_{\;\; \nu} \; = \;
\Bigl [ \, e^{\delta (h) U (h)} \Bigr ]^{\mu}_{\;\; \nu}  \;\; .
\eeq
Equivalently, this last expression can be re-written in terms of a surface
integral of the Riemann tensor, projected along the surface area element
bivector $A^{\alpha\beta}(C_h)$ associated with the loop,
\beq
\Bigl [ \prod_s {\bf R}_{s,s+1} \Bigr ]^{\mu}_{\;\; \nu} \; \approx \;
\Bigl [ \, e^{\half \int_S 
R^{\, \cdot}_{\;\; \cdot \, \alpha\beta} \, A^{\alpha\beta} (C_h) } 
\Bigr ]^{\mu}_{\;\; \nu}  \;\; .
\eeq

\begin{figure}[h]
\epsfig{file=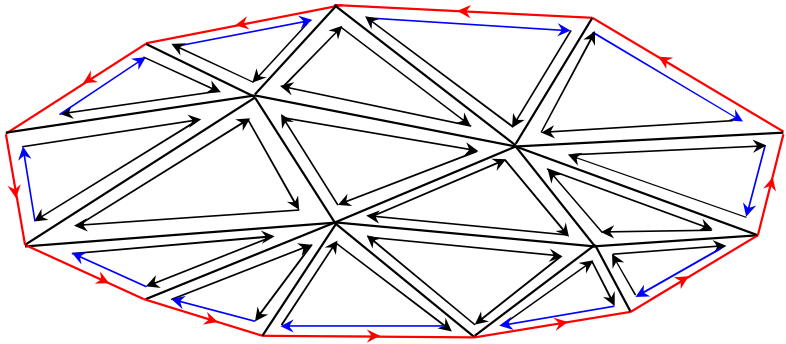,width=12cm}
\caption{Gravitational analog of the Wilson loop.
A vector is parallel-transported along the larger outer loop.
The enclosed minimal surface is tiled with parallel
transport polygons, here chosen to be triangles for illustrative
purposes.
For each link of the dual lattice, the elementary parallel transport
matrices ${\bf R}(s,s')$ are represented by arrows. 
In spite of the fact that the (Lorentz) matrices ${\bf R}$ can fluctuate 
strongly in accordance with the local geometry, two contiguous, oppositely
oriented arrows always give ${\bf R} {\bf R}^{-1} = 1$.}
\label{fig:wilson}
\end{figure}

More generally one might want to consider a near-planar, but non-infinitesimal,
closed loop $C$, as shown in Fig.~\ref{fig:wilson}.
Along this closed loop the overall rotation matrix will still be given by 
\beq
R^{\mu}_{\;\; \nu} (C) \; = \;
\Bigl [ \prod_{s \, \subset C} {\bf R}_{s,s+1} \Bigr ]^{\mu}_{\;\; \nu} 
\eeq
In analogy with the infinitesimal loop case,
one would like to state that for the overall rotation matrix one has
\beq
R^{\mu}_{\;\; \nu} (C) \; \approx \; 
\Bigl [ \, e^{\delta (C) U (C))} \Bigr ]^{\mu}_{\;\; \nu}  \;\; ,
\label{eq:latt-wloop}
\eeq
where $U_{\mu\nu} (C)$ is now an area bivector perpendicular to the
loop - which will work only if the loop is close to planar so
that $U_{\mu\nu}$ can be taken to be approximately constant
along the path $C$.

If that is true, then one can define, again in analogy with
the infinitesimal loop case, an appropriate coordinate scalar 
by contracting the above rotation matrix ${\bf R}(C)$ 
with the bivector of Eq.~(\ref{eq:omega-hinge}), namely
\beq
W ( C ) \; = \; \omega_{\alpha\beta}(C) \, R^{\alpha\beta} (C) 
\label{eq:latt-wloop1}
\eeq
where the loop bivector,
$\omega_{\alpha\beta} (C ) = (d-2)! \, V^{(d-2)} \, U_{\alpha\beta} $
$ = 2 \, A_C \, U_{\alpha\beta} ( C ) $ in four dimensions,
is now intended as being representative of the overall
geometric features of the loop.
For example, it can be taken as an average of the hinge bivector
$\omega_{\alpha\beta} (h)$ along the loop.

In the quantum theory one is of course interested in the average
of the above loop operator $W (C)$, as in
Eq.~(\ref{eq:wloop_sun}).
The previous construction is indeed quite analogous to the
Wilson loop definition in ordinary lattice gauge theories
(Wilson, 1973), where it is defined
via the trace of path ordered products of $SU(N)$ color rotation matrices.
In gravity though the Wilson loop does not give any information
about the static potential (Modanese, 1993; Hamber, 1993).
It seems that the Wilson loop in gravity provides instead
some insight
into the large-scale curvature of the manifold, just
as the infinitesimal loop contribution entering the lattice action
of Eqs.~(\ref{eq:regge-d}) and (\ref{eq:regge-compact})
provides, through its averages, insight into the very short distance,
local curvature.
Of course for any continuum manifold one can define locally
the parallel transport of a vector around a near-planar loop
$C$.
Indeed parallel transporting a vector around a closed loop represents
a suitable operational way of detecting curvature locally.
If the curvature of the manifold is small, one can treat the larger loop
the same way as the small one; then the expression of Eq.~(\ref{eq:latt-wloop})
for the rotation matrix ${\bf R} (C )$ associated with a near-planar
loop can be re-written in terms of a surface
integral of the large-scale Riemann tensor, projected along the surface
area element bivector $A^{\alpha\beta} (C )$ associated with the loop,
\beq
R^{\mu}_{\;\; \nu} (C) \; \approx \; 
\Bigl [ \, e^{\half \int_S 
R^{\, \cdot}_{\;\; \cdot \, \alpha\beta} \, A^{\alpha\beta} ( C )} 
\Bigr ]^{\mu}_{\;\; \nu}  \;\; .
\eeq
Thus a direct calculation of the Wilson loop provides a way of determining
the {\it effective} curvature at large distance scales, even in the case
where short distance fluctuations in the metric may be significant.
Conversely, the rotation matrix appearing in the elementary Wilson loop of Eqs.~(\ref{eq:loop-rot1}) and (\ref{eq:rot-hinge}) only provides 
information about the parallel transport of vectors around
{\it infinitesimal} loops, with size comparable to the ultraviolet cutoff.

One would expect that for a geometry fluctuating strongly
at short distances (corresponding therefore to the small $k$ limit) the
infinitesimal parallel transport matrices
${\bf R} (s,s')$ should be distributed close to randomly, with a measure close
to the uniform Haar measure, and with little correlation between
neighboring hinges.
In such instance one would have for the local quantum averages of the
infinitesimal lattice parallel transports $< {\bf R} > = 0$, but 
$< {\bf R} {\bf R}^{-1} > \neq 0$, which would require, for a non-vanishing
lowest order contribution to the Wilson loop, that the loop at least
be tiled by elementary action contributions from Eqs.~(\ref{eq:regge-d}) 
or (\ref{eq:regge-compact}), thus forming a minimal surface spanning the loop.
Then, in close analogy to the Yang-Mills case of Eq.~(\ref{eq:wloop_sun1})
(see for example Peskin and Schroeder, 1995),
the leading contribution to the gravitational Wilson loop would be expected to follow an area law,
\beq
W( C ) \; \sim \; 
{\rm const.} \, k^{A (C)} \; \sim \; \exp ( - A(C) / \xi^2 ) 
\label{eq:wloop_curv}
\eeq
where $A(C))$ is the minimal physical area spanned by the near-planar loop $C$,
and $\xi = 1 /\sqrt{ \vert \ln k \vert }$ the gravitational correlation
length at small $k$.
Thus for a close-to-circular loop of perimeter $P$ one would use 
$A (C) \approx P^2 / 4 \pi$.

The rapid decay of the gravitational Wilson loop as a function of the
area is seen here simply as a general and direct consequence of the
disorder in the fluctuations of the parallel transport matrices
${\bf R}(s,s')$ at strong coupling.
It should then be clear from the above discussion
that the gravitational Wilson loop provides in a sense 
a measure of the magnitude of the large-scale, averaged curvature, where
the latter is most suitably defined by
the process of parallel-transporting test vectors around very large loops,
and is therefore, from the above expression, computed to be of the 
order $R \sim 1 / \xi^2 $, at least in the small $k$ limit.
A direct calculation of the Wilson loop should therefore
Provide, among other things, a direct insight into whether the manifold is de Sitter or anti-de Sitter at large distances.
More details on the lattice construction of the gravitational Wilson loop,
the various issues that arise in its precise definition, 
and a sample calculation in the strong coupling limit of lattice
gravity, can be found in (Hamber and Williams, 2007). 

Finally we note that the definition of the gravitational Wilson loop is 
based on a surface with a given boundary $C$,
in the simplest case the minimal surface spanning the loop.
It is possible though to consider other surfaces built out of elementary
parallel transport loops.
An example of such a generic closed surface tiled with elementary parallel
transport polygons (here chosen for illustrative purposes to be triangles) will
be given later in Fig.~\ref{fig:icos}.

Later similar surfaces will arise naturally in the context of the strong
coupling (small $k$) expansion for gravity, as
well as in the high dimension (large $d$) expansion.

\subsubsection{Lattice Regularized Path Integral}

\label{sec:lattmeas}

As the edge lengths $l_{ij}$ play the role of the continuum metric
$g_{\mu\nu}(x)$, one would expect the discrete measure to involve an
integration over the squared edge lengths 
(Hamber, 1984; Hartle, 1984; Hamber and Williams, 1999).
Indeed the induced metric at a simplex is related to the squared edge
lengths within that simplex, via the expression for the
invariant line element $ds^2 = g_{\mu \nu} dx^\mu dx^\nu$.
After choosing coordinates along the edges emanating from a vertex,
the relation between metric perturbations and squared edge
length variations for a given simplex based at 0 in $d$ dimensions is
\beq
\delta g_{ij} (l^2) \; = \; \half \;
( \delta l_{0i}^2 + \delta l_{0j}^2 - \delta l_{ij}^2 ) \;\; .
\label{eq:latmet1}
\eeq
For one $d$-dimensional simplex labeled by $s$
the integration over the metric is thus equivalent to an 
integration over the edge lengths, and one has the identity
\beq
\left ( {1 \over d ! } \sqrt { \det g_{ij}(s) } \right )^{\sigma} \! \!
\prod_{ i \geq j } \, d g_{i j} (s) = 
{\textstyle \left ( - { 1 \over 2 } \right ) \displaystyle}^{ d(d-1) \over 2 }
\left [ V_d (l^2) \right ]^{\sigma} \!
\prod_{ k = 1 }^{ d(d+1)/2 } \! \! dl_{k}^2
\label{eq:simpmeas}
\eeq
There are $d(d+1)/2$ edges for each
simplex, just as there are $d(d+1)/2$ independent components for the metric
tensor in $d$ dimensions (Cheeger, M\"uller and Schrader, 1981).
Here one is ignoring temporarily the triangle inequality constraints,
which will further require all sub-determinants of $g_{ij}$ to be
positive, including the obvious restriction $l_k^2 >0$.

Let us discuss here briefly the simplicial inequalities that
need to be imposed on the edge lengths (Wheeler 1964).
These are conditions on the edge lengths $l_{ij}$ 
such that the sites $i$ can then be considered the vertices of a
$d$-simplex embedded in flat $d$-dimensional Euclidean space.
In one dimension, $d=1$, one requires trivially for all edge lengths
\beq
l_{ij}^2 > 0 \;\; .
\label{eq:tieq-1d}
\eeq
In two dimensions, $d=2$, the conditions on the edge lengths 
are again $l_{ij}^2 > 0$ as in one dimensions, as well as
\beq
A_\Delta^2 = \left ( {1 \over 2 ! } \right )^2  \det g_{ij}^{(2)}(s) \, > 0 
\label{eq:tieq-2d}
\eeq
which is equivalent, by virtue of Heron's
formula for the area
of a triangle $A^2_\Delta = s (s-l_{ij})(s-l_{jk})(s-l_{ki})$
where $s$ is the semi-perimeter $s=\half (l_{ij}+l_{jk}+l_{ki})$,
to the requirement that the area of the triangle be positive.
In turn Eq.~(\ref{eq:tieq-2d}) implies that the triangle inequalities
must be satisfied for all three edges,
\bea
l_{ij} + l_{jk} & > & l_{ik}
\nonumber \\
l_{jk} + l_{ki} & > & l_{ji}
\nonumber \\
l_{ki} + l_{ij} & > & l_{kj}
\eea
In three dimensions, $d=3$, the conditions on the edge
lengths are again such that one recovers a physical tetrahedron.
One therefore requires for the
individual edge lengths the condition of Eq.~(\ref{eq:tieq-1d}), 
the reality and positivity of all four triangle areas as in
Eq.~(\ref{eq:tieq-2d}),
as well as the requirement that the volume of the tetrahedron
be real and positive,
\beq
V_{\rm tetrahedron}^2 = \left ( {1 \over 3 ! } \right )^2  
\det g_{ij}^{(3)}(s) \, > 0 
\label{eq:tieq-3d}
\eeq
The generalization to higher dimensions is such that one requires
all triangle inequalities and their higher dimensional analogues
to be satisfied, 
\bea
l_{ij}^2 & > & 0
\nonumber \\
V_{k}^2 & = & \left ( {1 \over k ! } \right )^2  \det g_{ij}^{(k)}(s) \, > 0 
\label{eq:tieq-d}
\eea
with $k=2 \dots d$ for every possible choice of sub-simplex
(and therefore sub-determinant) within the original simplex $s$.

The extension of the measure to many simplices glued together
at their common faces is then immediate. 
For this purpose one first needs to identify edges
$ l_k (s) $ and $ l_{k'} (s') $ which are shared between
simplices $s$ and $s'$,
\beq
\int_0^\infty d l^2_k (s) \, \int_0^\infty d l^2_{k'} (s') \;
\delta \left [ l^2_k (s) - l^2_{k'} (s') \right ]
\, = \, \int_0^\infty d l^2_k (s) \;\; .
\eeq
After summing over all simplices one derives,
up to an irrelevant numerical constant, the unique functional measure
for simplicial geometries
\beq
\int [ d l^2] \; = \; 
\int_\epsilon^\infty \; \prod_s \; \left [ V_d (s) \right ]^{\sigma} \;
\prod_{ ij } \, dl_{ij}^2 \; \Theta [l_{ij}^2]  \;\; .
\label{eq:lattmeas}
\eeq
Here $ \Theta [l_{ij}^2] $ is
a (step) function of the edge lengths, with the property
that it is equal to one whenever the triangle inequalities and their
higher dimensional analogues are satisfied,
and zero otherwise.
The quantity $\epsilon$ has been introduced as a cutoff at small
edge lengths.
If the measure is non-singular for small edges, one can safely
take the limit $\epsilon \rightarrow 0$.
In four dimensions the lattice analog of the DeWitt measure
($\sigma=0$) takes on a particularly simple form, namely
\beq
\int [ d l^2] \; = \; \int_0^\infty \prod_{ ij } \, dl_{ij}^2 
\; \Theta [ l_{ij}^2 ] \;\; .
\label{eq:dewlattmeas}
\eeq
Lattice measures over the space of squared edge lengths
have been used extensively in numerical simulations of simplicial
quantum gravity (Hamber and Williams, 1984; Hamber, 1984; Berg 1985).
The derivation of the above lattice measure
closely parallels the analogous procedure in the
continuum.

There is no obstacle in defining a discrete analog of the
supermetric, as a way of introducing an invariant notion of distance
between simplicial manifolds, as proposed
in (Cheeger, M\"uller and Schrader, 1984).
It leads to an alternative way of deriving the lattice measure in
Eq.~(\ref{eq:dewlattmeas}), by
considering the discretized distance between induced metrics
$g_{ij}(s)$ 
\beq
\Vert \, \delta g (s) \, \Vert^2 \; = \; \sum_{s} \;
G^{ i j k l } \left [ g(s) \right ] \; 
\delta g_{i j} (s) \, \delta g_{k l} (s) \;\; ,
\eeq
with the inverse of the lattice DeWitt supermetric now given by
the expression
\beq
G^{ i j k l } [ g(s) ] =
\half \sqrt{g(s)} \; \left [ \,
g^{i k} (s) g^{j l} (s) +
g^{i l} (s) g^{j k} (s) + \lambda \,
g^{i j} (s) g^{k l} (s) \right ]
\label{eq:dewittsuperl}
\eeq
and with again $\lambda \neq - 2 / d $.
This procedure defines a metric on the tangent space of positive real
symmetric matrices $g_{ij}(s)$. After computing the determinant
of $G$, the resulting functional measure is
\beq
\int d \mu [l^2] = \int
\prod_{s} \, \left [ \; \det G(g(s)) \; \right ]^{\half}
\prod_{i \geq j} d g_{i j} (s) \;\; ,
\eeq
with the determinant of the super-metric $G^{i j k l} (g(s))$
given by the local expression
\beq
\det G [ g(s)] \; \propto \; 
(1 + \half d \lambda ) \; \left [ g(s) \right ]^{ (d-4)(d+1)/4 } \;\; ,
\eeq
Using Eq.~(\ref{eq:simpmeas}), and up to irrelevant constants,
one obtains again the standard lattice measure
of Eq.~(\ref{eq:lattmeas}). 
Of course the same procedure can be followed for the Misner-like
measure, leading to a similar result for the lattice measure,
but with a different power $\sigma$. 

One might be tempted to try to find alternative lattice
measures by looking directly at the discrete form for the supermetric,
written as a quadratic form in the squared edge lengths (instead of
the metric components), and then evaluating the resulting determinant.
The main idea, inspired by work described in a paper
(Lund and Regge, 1974) on the $3+1$ formulation of simplicial
gravity, is as follows.
First one considers a lattice analog of the DeWitt supermetric by writing
\beq
\Vert \delta l^2 \Vert^2 \; = \; \sum_{ij} \; G_{ij} (l^2)
\; \delta l^2_i \; \delta l^2_j \; \; ,
\label{eq:lund}
\eeq
with $G_{ij} (l^2)$ now defined on the space of squared edge lengths
(Hartle, Miller and Williams, 1997).
The next step is to find an appropriate form for $G_{ij} (l^2)$ expressed in
terms of known geometric objects.
One simple way of constructing the explicit form for $G_{ij} (l^2)$, in any
dimension, is to first focus on one simplex, and
write the squared volume of a given simplex in terms
of the induced metric components within the {\it same} simplex $s$,
\beq
V^2 ( s ) \; = \; {\textstyle \left ( { 1 \over d! } \right )^2 \displaystyle}
\det g_{ij}(l^2(s)) \;\; .
\eeq
One computes to linear order
\beq
{1 \over V (l^2)} \; \sum_{i} 
{\partial V^2 (l^2) \over \partial l^2_i} \; \delta l^2_i
\; = \; {\textstyle { 1 \over d! } \displaystyle} \sqrt{ \det ( g_{ij} ) }
\; g^{ij} \; \delta g_{ij} \;\; ,
\eeq
and to quadratic order
\beq
{1 \over V (l^2)} \; 
\sum_{ij} { \partial^2 V^2 (l^2) \over \partial l^2_i \partial l^2_j }
\; \delta l^2_i \; \delta l^2_j \; = \;
{\textstyle { 1 \over d! } \displaystyle} \sqrt{ \det ( g_{ij} ) }
\left [ \; g^{ij} g^{kl} \delta g_{ij} \delta g_{kl}
- g^{ij} g^{kl} \delta g_{jk} \delta g_{li} \; \right ] \;\; .
\eeq
The right hand side of this equation contains precisely
the expression appearing in the continuum supermetric
of Eq.~(\ref{eq:dw-def}), for the specific choice of the
parameter $\lambda = -2$.
One is led therefore to the identification
\beq
G_{ij} (l^2) \; = \; - \; d! \; \sum_{s} \;
{1 \over V (s)} \; 
{ \partial^2 \; V^2 (s) \over \partial l^2_i \; \partial l^2_j } \;\; ,
\eeq
and therefore for the norm
\beq
\Vert \delta l^2 \Vert^2 \; = \; \sum_{s} \; V (s) \;
\left \{
\; - \; {d! \over V^2 (s)} \; \sum_{ij} \;
{ \partial^2 \; V^2 (s) \over \partial l^2_i \; \partial l^2_j }
\; \delta l^2_i \; \delta l^2_j \; \right \} \;\; .
\eeq
One could be tempted at this point
to write down a lattice measure,
in parallel with Eq.~(\ref{eq:dw-det0}), and write
\beq
\int \; [ d l^2 ] \; = \; \int \; \prod_i \;
\sqrt{ \det G_{ij}^{( \omega ' )} (l^2) } \, dl^2_i \;\; .
\label{eq:lundmeas}
\eeq
with 
\beq
G^{(\omega')}_{ij} (l^2) \; = \; - \; d! \; \sum_{s} \;
{ 1 \over [ V(s) ]^{1+{\omega'}} } \;
{ \partial^2 \; V^2 (s) \over \partial l^2_i \; \partial l^2_j } \;\; ,
\label{eq:gmatrix}
\eeq
where one has allowed for a parameter $\omega'$, possibly
different from zero, interpolating between apparently equally
acceptable measures.
The reasoning here is that, as in the continuum, different edge
length measures, here parametrized by
$\omega$', are obtained, depending on whether the local volume
factor $V(s)$ is included in the supermetric or not.

One rather undesirable, and puzzling, feature of the lattice measure of 
Eq.~(\ref{eq:lundmeas}) is that in 
general it is non-local, in spite of the fact that the original
continuum measure of Eq.~(\ref{eq:dw-dewitt}) is completely local
(although it is clear that for some special choices
of $\omega'$ and $d$, one does recover a local measure;
thus in two dimensions and for ${\omega'} = -1$ one obtains again
the simple result
$\int \, [ d l^2 ] = \int_0^\infty \, \prod_i \, dl^2_i $).
Unfortunately irrespective of the value chosen for $\omega'$, one
can show (Hamber and Williams, 1999) that the measure 
of Eq.~(\ref{eq:lundmeas}) disagrees with the continuum measure
of Eq.~(\ref{eq:dw-dewitt}) already to lowest
order in the weak field expansion,
and does not therefore describe an acceptable lattice measure.

The lattice action for pure four-dimensional Euclidean 
gravity contains
a cosmological constant and Regge scalar curvature term
\beq 
I_{latt} \; = \;  \lambda_0 \, \sum_h V_h (l^2) \, - \, 
k \sum_h \delta_h (l^2 ) \, A_h (l^2) \;\; , 
\label{eq:ilatt} 
\eeq
with $k=1/(8 \pi G)$, as well as possibly higher derivative terms.
It only couples edges which belong either to
the same simplex or to a set of neighboring simplices, and can therefore
be considered as {\it local}, just like the continuum action, and 
leads to the regularized lattice functional integral
\beq 
Z_{latt} \; = \;  \int [ d \, l^2 ] \; e^{ 
- \lambda_0 \sum_h V_h \, + \, k \sum_h \delta_h A_h } \;\; ,
\label{eq:zlatt} 
\eeq
where, as customary, the lattice ultraviolet cutoff is set equal to one
(i.e. all length scales are measured in units of the lattice cutoff).

The lattice partition function $Z_{latt}$ should then be compared to the
continuum Euclidean Feynman path integral of Eq.~(\ref{eq:zcont}),
\beq
Z_{cont} \; = \; \int [ d \, g_{\mu\nu} ] \; e^{ 
- \lambda_0 \, \int d x \, \sqrt g \, + \, 
{ 1 \over 16 \pi G } \int d x \sqrt g \, R} \;\; ,
\eeq
Occasionally it can be convenient to include the
$\lambda_0$-term in the measure. 
For this purpose one defines 
\beq 
d \mu (l^2) \; \equiv \; [ d \, l^2 ] \; e^{- \lambda_0 \sum_h V_h }
\;\; .
\label{eq:mulatt} 
\eeq
It should be clear that this last expression represents a fairly non-trivial 
quantity, both in view of the relative complexity
of the expression for the volume of a simplex, Eq.~(\ref{eq:vol}),
and because of the generalized triangle inequality constraints 
already implicit in $[d\,l^2]$.
But, like the continuum functional measure, it is certainly {\it local},
to the extent that each edge length
appears only in the expression for the volume of those simplices
which explicitly contain it.
Furthermore, $\lambda_0$ sets the overall scale and can therefore be set 
equal to one without any loss of generality.

\subsubsection{An Elementary Example}

\label{sec:oned-meas}

In the very simple case of one dimension ($d=1$) one can work out 
explicitly a number of details, and see how potential problems
with the functional measure arise, and how they are resolved.

In one dimension one discretizes the line by introducing $N$ points,
with lengths $l_n$ associated with the edges, and periodic
boundary conditions, $l_{N+1} = l_1 $. 
Here $l_n$ is the distance between points $n$ and $n+1$.
The only surviving invariant term in one dimension is then the 
overall length of a curve,
\beq
L (l) \; = \; \sum_{n=1}^N l_n  \;\; ,
\label{eq:length}
\eeq
which corresponds to
\beq
\int dx {\textstyle \sqrt{g(x)} \displaystyle} \; = \;
\int dx \; e(x)
\label{eq:acl}
\eeq
(with $g(x) \equiv g_{00}(x)$) in the continuum.
Here $e(x)$ is the ``einbein'', and satisfies
the obvious constraint $ \sqrt{g(x)} = e(x) > 0 $.
In this context the discrete action is unique, preserving
the geometric properties of the continuum definition.
From the expression for the invariant line element, $ds^2 = g dx^2$,
one associates $g(x)$ with $l_n^2$ (and therefore $e(x)$ with $l_n$).
One can further take the view that distances can only be assigned
between vertices which appear on some lattice in the ensemble, although
this is not strictly necessary, as distances can also be defined for
locations that do not coincide with any specific vertex.

The gravitational measure then contains an integration over
the elementary lattice degrees of freedom, the
lattice edge lengths.
For the edges one writes the lattice integration measure
as
\beq
\int d \mu [ l ] \; = \;
\prod_{n=1}^N  \int_0^\infty \, d l_n^2  \; l_n^\sigma  \;\; ,
\label{eq:meas} 
\eeq
where $\sigma$ is a parameter interpolating between different
local measures. The positivity of the edge lengths
is all that remains of the triangle inequality constraints
in one dimension.
The factor $l_n^\sigma$ plays a role analogous to the 
$ g^{\sigma/2} $ which appears for continuum
measures in the Euclidean functional integral.

The functional measure does not have compact support, and the cosmological
term (with a coefficient $\lambda_0 > 0 $)
is therefore necessary to obtain convergence of the functional integral,
as can be seen for example from the expression for the average edge
length,
\beq
\langle L(l) \rangle  \; = \;
\langle \sum_{n=1}^N l_n \rangle  \; = \; {\cal Z}_N^{-1} \; 
\prod_{n=1}^N  \int_0^\infty \, d l_n^2 \; l_n^\sigma \,
\exp \left ( - \lambda_0 \sum_{n=1}^N l_n \right ) \; \sum_{n=1}^N l_n 
\; = \; { 2 + \sigma \over \lambda_0 } \; N
\label{eq:avel} 
\eeq
with
\beq
{\cal Z}_N (\lambda_0) \; = \;
\prod_{n=1}^N  \int_0^\infty \, d l_n^2 \; l_n^\sigma \,
\exp \left ( - \lambda_0 \sum_{n=1}^N l_n \right )  \; = \;
\left [ { 2 \; \Gamma ( 2 + \sigma ) \over \lambda_0^{ 2 + \sigma } } \right ]^N
\label{eq:avel1} 
\eeq
Similarly one finds for the fluctuation in the total length
$ \Delta L / L = 1 / \sqrt{ (2 + \sigma) N } $, which requires
$\sigma > -2 $.
Different choices for $\lambda_0$ then correspond to trivial rescalings of
the average lattice spacing,
$l_0 \equiv \langle l \rangle  = (2 + \sigma ) / \lambda_0 $.

In the continuum, the action of Eq.~(\ref{eq:acl}) is invariant
under continuous reparametrizations
\beq
 x \rightarrow x' (x) = x - \epsilon(x)
\eeq
\beq
 g(x) \rightarrow g'(x') = \left ( { dx \over dx' } \right )^2 g(x)
= g(x) + 2 \; g(x) \left ( { d \epsilon  \over dx } \right ) + O(\epsilon^2)
 \;\; ,
\eeq
or equivalently
\beq
\delta g(x) \equiv g'(x')- g(x) =  2 g \partial \epsilon  \;\; ,
\eeq
and we have set $ \partial \equiv d / dx $.
A gauge can then be chosen by imposing $g'(x')=1$, which can be achieved
by the choice of coordinates $x'= \int dx \sqrt{ g(x) } $.

The discrete analog of the transformation rule is
\beq
\delta l_n = \epsilon_{n+1} - \epsilon_n  \;\; ,
\label{eq:gauge} 
\eeq
where the $\epsilon_n$'s represent continuous gauge transformations defined on
the lattice vertices. In order for the edge lengths to remain positive,
one needs to require $\epsilon_{n} - \epsilon_{n+1}  < l_n $, which
is certainly satisfied for sufficiently small $\epsilon$'s.
The above continuous symmetry is an exact invariance of the lattice
action of Eq.~(\ref{eq:length}), since
\beq
\delta L \; = \; \sum_{n=1}^N \delta l_n = \sum_{n=1}^{N} \epsilon_{n+1} 
- \sum_{n=1}^N \epsilon_{n} = 0  \;\; ,
\eeq
and we have used $\epsilon_{N+1} = \epsilon_1 $.
Moreover, it is the only local symmetry of the action of Eq.~(\ref{eq:length}).

The infinitesimal local
invariance property defined in  Eq.~(\ref{eq:gauge}) formally selects a
unique measure over the edge lengths, corresponding to $\prod_n dl_n$
($\sigma = -1 $ in Eq.~(\ref{eq:meas})), as long as we ignore the effects
of the lower limit of integration.
On the other hand for sufficiently large lattice diffeomorphisms,
the lower limit of integration comes into play (since we require
$ l_n > 0 $ always) and the measure is
no longer invariant.
Thus a measure $\int_{-\infty}^{\infty} \prod dl_n $ would not be
acceptable on physical grounds; it would violate the
constraint $\sqrt{g}>0$ or $e>0$.

The same functional
measure can be obtained from the following physical consideration.
Define the gauge invariant distance $d$ between two configurations 
of edge lengths $\{ l_n \}$ and $\{ l'_n \}$ by
\beq
d^2 (l,l') \equiv \left [ L(l) - L'(l') \right ]^2 \; = \;
\left ( \sum_{n=1}^N l_n - \sum_{n'=1}^N l'_{n'} \right )^2 
=  \sum_{n=1}^N \sum_{n'=1}^N \delta l_n \; M_{n,n'} \; \delta l_{n'}  \;\; ,
\eeq
with $M_{n,n'} =1 $. Since $M$ is independent of $l_n$ and $ l'_{n'}$,
the ensuing measure is again simply proportional to $\prod dl_n$.
Note that the above metric over edge length deformations $\delta l$
is non-local.

In the continuum, the functional measure is usually determined by
considering the following (local) norm in function space,
\beq
|| \delta g ||^2 \; = \; \int dx \; {\textstyle \sqrt{g(x)} \displaystyle} \;
\delta g(x) \; G(x) \; \delta g(x)  \;\; ,
\eeq
and diffeomorphism invariance would seem to require $G(x) = 1/g^2 (x) $.
The volume element in function space is then an ultraviolet 
regulated version of  $\prod_x \sqrt{G(x)} \; dg(x) 
= \prod_x dg(x)/g(x)$, which is the Misner measure in one dimension.
Its naive discrete counterpart would be $\prod dl_n/l_n$, which is 
not invariant under the transformation of Eq.~(\ref{eq:gauge}) 
(it is invariant under $ \delta l_n = l_n ( \epsilon_{n+1}  -
\epsilon_{n} )$, which is not an invariance of the action).

The point of the discussion of the one-dimensional case
is to bring to the surface the several non-trivial
issues that arise when defining a properly regulated
version of the continuum Feynman functional measure $[dg_{\mu\nu}]$,
and how they can be systematically resolved.

\subsubsection{Lattice Higher Derivative Terms}

\label{sec:higher}

So far only the gravitational Einstein-Hilbert contribution to
the lattice action and the cosmological constant term
have been considered.
There are several motivations for extending the discussion
to lattice higher derivative terms, which would include the
fact that these terms a) might appear in the original microscopic
action, or might have to be included to cure the classical
unboundedness problem of the Euclidean Einstein-Hilbert action,
b) that they are in any case generated by radiative corrections,
and c) that on a more formal level they may shed new light on the relationship
between the lattice and continuum expressions for curvature
terms as well as quantities such as the Riemann tensor on a hinge,
Eq.~(\ref{eq:riem-hinge}).

For these reasons we will discuss here a generalization
of the Regge gravity equivalent of the Einstein action
to curvature squared terms.
When considering contributions quadratic in the
curvature there are overall six possibilities, listed in Eq.~(\ref{eq:curv2}).
Among the two topological invariants, the Euler characteristic $ \chi $
for a simplicial decomposition may be
obtained from a particular case of the general formula for the analog of the
Lipschitz-Killing curvatures of smooth Riemannian manifolds for piecewise
flat spaces.
The formula of (Cheeger, M\"uller and Schrader, 1984) reduces in four dimensions to
\beq
\chi = \sum_ { \sigma^0 } \Bigl [ 1 - \sum_ { \sigma^2 \supset
\sigma^0 }
(0,2)- \sum_ { \sigma^4 \supset \sigma^0 } (0,4)+ \sum_ { \sigma^4
\supset \sigma^2 \supset \sigma^0 } (0,2)(2,4) \Bigr ]
\eeq
where $ \sigma^i $ denotes an i-dimensional simplex and $(i,j)$ denotes the
(internal) dihedral angle at an i-dimensional face of a j-dimensional simplex.
Thus, for example, $(0,2)$ is the angle at the vertex of a triangle and $(2,4)$
is the dihedral angle at a triangle in a 4-simplex (The normalization
of the angles is such that the volume of a sphere in any dimension is one;
thus planar angles are divided by $2 \pi $, 3-dimensional solid angles by
$4 \pi$ and so on).
 
Of course, as noted before, there is a much simpler formula for the Euler characteristic of a simplicial complex
\beq
\chi = \sum_{i=0}^d (-1)^i N_i
\eeq
where $ N_i $ is the number of simplices of dimension $i$.
However, it may turn out to be useful in quantum gravity calculations to have
a formula for $\chi$ in terms of the angles, and hence of the edge lengths, of
the simplicial decomposition.
These expression are interesting and useful, but do not shed much light
on how the other curvature squared terms should be constructed.

In a piecewise linear space curvature is detected by going around
elementary loops which are dual to a ($d-2$)-dimensional subspace.
The area of the loop itself is not well defined, since any loop inside the
$d$-dimensional simplices bordering the hinge will give the same result
for the deficit angle.
On the other hand the hinge has a content (the length of the edge in $d=3$ and
the area of the triangle in $d=4$), and there is a natural volume
associated with each hinge, defined by dividing the volume of each simplex
touching the hinge into a contribution belonging to that hinge, and other
contributions belonging to the other hinges on that simplex 
(Hamber and Williams, 1984).
The contribution belonging to that simplex will be called
{\it dihedral volume} $V_d $.
The volume $V_h $ associated with the hinge $h$ is then naturally the sum
of the dihedral volumes $V_d $ belonging to each simplex 
\beq
V_h = \sum_ { {\bf d-simplices \atop meeting \; on \; h }} V_d
\label{eq:vol-hinge}
\eeq
The dihedral volume associated with each hinge in a simplex can be defined
using dual volumes, a barycentric subdivision, or some other natural
way of dividing the volume of a $d$-simplex into $d(d+1)/2$ parts.
If the theory has some reasonable continuum limit, then the final results
should not depend on the detailed choice of volume type.

\begin{figure}[h]
\epsfig{file=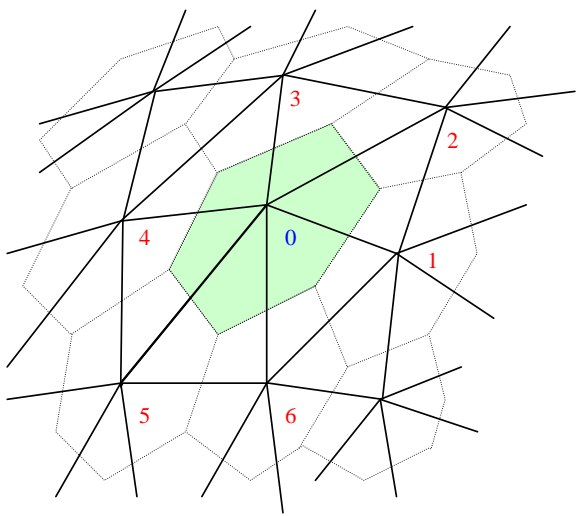,width=8cm}
\caption{ Illustration of dual volumes in two dimensions. 
The vertices of the polygons reside in the dual lattice. 
The shaded region describes the dual area
associated with the vertex 0.}
\label{fig:dual2d}
\end{figure}

As mentioned previously, there is a well-established procedure for
constructing a dual lattice for any given lattice.
This involves constructing polyhedral cells, known in the literature as 
Voronoi polyhedra, around each vertex, in such a way that the cell around
each particular vertex contains all points which are nearer to that
vertex than to any other vertex.
Thus the cell is made up from ($d-1$)-dimensional subspaces which are the 
perpendicular bisectors of the edges in the original lattice,
($d-2$)-dimensional
subspaces which are orthogonal to the 2-dimensional subspaces of the
original lattice, and so on.
General formulas for dual volumes are given in (Hamber and Williams, 1986).
In the case of the barycentric subdivision, the dihedral volume is just
$2/d(d+1)$ times the volume of the simplex.
This leads one to conclude that there is a natural area
$ A_{ C_h }$ associated with each hinge
\beq
A_{ C_h } \; = \; { V_h \over A_h^{(d-2)} }
\label{eq:area-hinge}
\eeq
obtained by dividing
the volume per hinge (which is $d$-dimensional) by
the volume of the hinge (which is ($d-2$)-dimensional).

The next step is to find terms equivalent to the continuum expression
of Eq.~(\ref{eq:curv2}), and the remainder of this
section will be devoted to this problem.
It may be objected that since in Regge calculus where the curvature is
restricted to the hinges which are subspaces of dimension 2 less than that
of the space considered, then the curvature tensor involves $\delta$-functions
with support on the hinges, and so higher powers of the
curvature tensor are not defined.
But this argument clearly does not apply to the Euler characteristic 
and the Hirzebruch signature of Eq. Eq.~(\ref{eq:curv2}), 
which are both integrals of 4-forms.
However it is a common procedure in lattice field theory to take powers of
fields defined at the same point, and there is no reason why one should not
consider similar terms in lattice gravity.
Of course one would like the expressions to correspond to the continuum
ones as the edge lengths of the simplicial lattice become smaller and
smaller.
 
Since the curvature is restricted to the hinges, it is natural that expressions
for curvature integrals should involve sums over hinges as in Eq.~(\ref{eq:regge-d}).
The curvature tensor, which involves second derivatives of the metric, is
of dimension $L^{-2} $.
Therefore $ { 1 \over 4 } \int d^d x \sqrt g R^n $ is of dimension
$ L^{d-2n} $.
Thus if one postulates that an $ R^2 $-type term will involve the square of
$ A_h \delta_h $, which is of dimension $L^{2(d-2)} $, then one
will need to divide by some $d$-dimensional volume to obtain the correct
dimension for the extra term in the action.
Now any hinge is surrounded by a number of $d$-dimensional simplices, so the
procedure of dividing by a $d$-dimensional volume seems ambiguous.
The crucial step is to realize that there {\it is} a unique $d$-dimensional
volume associated with each hinge, as described above.
 
If one regards the invariant volume element $ \sqrt g d^d x $ as
being represented by $ V_h $ 
of Eq.~(\ref{eq:vol-hinge}) when one performs the sum over hinges
as in Eq.~(\ref{eq:regge-d}), then
this means that one may regard the scalar curvature $R$ contribution
as being represented at each hinge by $ 2 A_h \delta_h / V_h $
\beq
\half \, \int d^d x \sqrt g \; R \;\; \rightarrow \;\; 
\sum_ {{\rm hinges \; h }} V_h \; { A_h \delta_h \over V_h }
\equiv \sum_ {{\rm hinges \; h }}  A_h \delta_h
\label{eq:r-latt}
\eeq
It is then straightforward to see that a candidate curvature
squared term is
\beq
\sum_ {{\rm hinges \; h }}  V_h \Bigl ( { A_h \delta_h
\over V_h } \Bigr )^2 \equiv 
\sum_ {{\rm hinges \; h }}  V_h \Bigl ( { \delta_h 
\over A_{ C_h } } \Bigr )^2 
\label{eq:riem2-latt}
\eeq
where $A_{ C_h }$ was defined in Eq.~\ref{eq:area-hinge}).
Since the expression in Eq.~(\ref{eq:riem2-latt}) vanishes if
and only if all deficit angles are zero, it is naturally identified
with the continuum Riemann squared term,
\beq
\quarter \int d^d x \sqrt g \; 
R_{ \mu \nu \lambda \sigma }  R^{ \mu \nu \lambda \sigma } 
\;\; \rightarrow \;\;
\sum_ {{\rm hinges \; h }}  V_h \Bigl ( { \delta_h 
\over A_{ C_h } } \bigr )^2 
\label{eq:riem2-latt1}
\eeq
The above construction then leaves open the question of how to construct
the remaining curvature squared terms in four dimensions.
If one takes the form given previously in Eq.~(\ref{eq:riem-hinge}) 
for the Riemann tensor on a hinge and contracts one obtains
\beq
R (h) = 2 \, { \delta_h \over A_{ C_h } } 
\eeq
which agrees with the form used in the Regge action for $R$.
But one also finds readily that with this choice the higher derivative
terms are all proportional to each other (Hamber and Williams, 1986),
\beq
\quarter \,
{ R }_{ \mu \nu \rho \sigma } (h)
{ R }^{ \mu \nu \rho \sigma } (h)
= \half \,
{ R }_{ \mu \nu } (h) { R }^{ \mu \nu } (h)
= \quarter \, R(h)^2
= \Bigl ( { \delta_h \over A_{C_h} } \Bigr )^2
\eeq
Furthermore if one uses the above expression for the Riemann tensor
to evaluate the contribution to the Euler
characteristic on each hinge one obtains zero, and becomes clear
that at least in this case one needs cross terms involving
contributions from different hinges.

The next step is therefore to embark on a slightly more sophisticated approach,
and construct the full Riemann tensor by considering more
than one hinge.
Define the Riemann tensor for a simplex $s$ as a weighted sum of hinge
contributions
\beq
\Bigl [ R_{ \mu \nu \rho \sigma }  \Bigr ]_{s} =
\sum_ { h \subset s } \omega_{h}
 \Bigl [ 
{\delta \over A_C } U_{ \mu \nu } U_{ \rho \sigma } \Bigr ]_{h}
\label{eq:riem-latnew}
\eeq
where the $\omega_{h} $ are dimensionless weights, to be
determined later.
After squaring one obtains
\beq
\Bigl [ R_{ \mu \nu \rho \sigma }  
R^{ \mu \nu \rho \sigma }  \Bigr ]_{s} =
\sum_ { h , h' \subset s } 
\omega_{h} \omega_{h'}
 \Bigl [ 
{\delta \over A_C } U_{ \mu \nu } U_{ \rho \sigma } \Bigr ]_{h}
 \Bigl [ 
{\delta \over A_C } U^{ \mu \nu } U^{ \rho \sigma } \Bigr ]_{h'}
\eeq
The question of the weights $ \omega_{ h } $
introduced in Eq.~(\ref{eq:riem-latnew}) will now be addressed.
Consider the expression for the scalar curvature of a simplex defined as
\beq
\Bigl [ R \Bigr ]_{s} =
\sum_ { h  \subset s }
\omega_{ h }
\Bigl [ 
2 \, { \delta \over A_C } \Bigr ]_{ h }
\eeq
It is clear from the formulae given above for the lattice curvature
invariants
(constructed in a simplex by summing over hinge contributions) that there is
again a natural volume associated with them :
the sum of the volumes of the hinges in the simplex
\beq
V_{ s } = \sum_ { h  \subset s } V_{ h }
\eeq
where $ V_h $ is the volume around the hinge,
as defined in Eq.~(\ref{eq:vol-hinge}).
Summing the scalar curvature
over all simplices, one should recover Regge's expression
\beq
\sum_ { s } V_{ s } \Bigl [ R \Bigr ]_{ s } = \sum_ { s } 
\sum_ { h \subset s } \omega_{ h }  \Bigl [ 
2 \, { \delta \over A_C } \Bigr ]_{ h}
= \sum_ { h } \delta_{ h } A_{ h }
\eeq
which implies
\beq
N_{2,4} V_{ s } \, \omega_{ h } { \delta_{ h } \over A_{ C_{h } } }
\equiv 
N_{2,4} V_{ s } \, \omega_{ h } { \delta_{ h } A_{ h } \over V_{ h } } 
= \delta_{h } A_{h }
\eeq
where $ N_{2,4} $ is the number of simplices
meeting on that hinge.
Therefore the correct choice for the weights is
\beq
\omega_{h} =
{ V_{h} \over N_{2,4} \, V_{s } }
= { V_{h } \over N_{2,4}
\sum_{ h  \subset s } V_{ h } }
\eeq
Thus the weighting factors that reproduce Regge's formula for the Einstein
action are just the volume fractions occupied by the various hinges in a
simplex, which is not surprising (of course the above formulae are not quite unique, since one might have done
the above construction of higher derivative terms by considering a point
$ p $ instead of a four-simplex $ s $).

In particular the following form for the Weyl tensor squared
was given in (Hamber and Williams, 1986)  
\bea
& \int d^d x \, \sqrt g \; C_{\mu \nu \lambda \sigma} C^{\mu \nu \lambda \sigma}
\sim &
\nonumber \\ 
& \;\;\;\;\;\;\;\;\;\;\;\;\;\; \twoth
\sum_ { s } V_{ s} 
\sum_ { h , h' \subset s } 
\epsilon_{ h, h'} \;
\Bigl ( \omega_{ h } 
\Bigl [ { \delta \over A_C } \Bigr ]_{ h }
- \omega_{ h' } \Bigl [ 
{\delta \over A_C } \Bigr ]_{ h' } \Bigr )^2 &
\label{eq:weyl2-latt}
\eea
which introduces a short range coupling between deficit angles.
The numerical factor $ \epsilon_{ h, h^\prime } $ is equal to $1$ if
the two hinges $ h,h' $ have one edge in common and
$-2$ if they do not.
Note that this particular interaction term has the property
that it requires neighboring deficit angles to have similar values, but
it does not require them to be small, which is
a key property one would expect from the Weyl curvature squared term.

In conclusion the formulas given above allow one to construct
the remaining curvature squared terms in four dimensions, 
and in particular to write, for example, the
lattice analog of the continuum curvature squared action of Eq.~(\ref{eq:hdqg})
\bea
I = \int d^4 x \, \sqrt g \; \Bigl [ & \lambda_0 & - k \; R - 
b \; R_{ \mu \nu \rho \sigma } R^{ \mu \nu \rho \sigma }
\nonumber \\
& + \half & ( a + 4 b ) \;
C_{ \mu \nu \rho \sigma } C^{ \mu \nu \rho \sigma } \Bigr ]
\label{eq:hd-action}
\eea
To compare with the form of Eq.~(\ref{eq:hdqg}) use has been made of
$R^2 = 3 R_{\mu\nu\rho\sigma}^2 - 6 C^2 $
and $ R_{\mu\nu}^2 = R_{\mu\nu\rho\sigma}^2 - {3\over2} C^2$, up
to additive constant contributions.
A special case corresponds to $b=-a/4$, which gives
a pure $ R_{\mu\nu\rho\sigma}^2 $ contribution. 
The latter vanishes if and only if the curvature is locally zero,
which is not true of any of the other curvature squared terms.

\subsubsection{Scalar Matter Fields}

\label{sec:scalars}

In the previous section we have discussed the construction
and the invariance properties of a lattice action for pure gravity.
Next a scalar field can be introduced as the simplest type of dynamical
matter that can be coupled invariantly to gravity.
In the continuum the scalar action for a single component field $\phi(x)$
is usually written as 
\beq
I [ g, \phi ] = \half \int d x \, \sqrt g \, [ \,
g^{ \mu \nu } \, \partial_\mu \phi \, \partial_\nu \phi
+ ( m^2 + \xi R ) \phi^2 ] + \dots
\label{eq:scalar}
\eeq
where the dots denote scalar self-interaction terms.
Thus for example a scalar field potential $U(\phi)$ could
be added containing quartic field terms,
whose effects could be of interest in the context of cosmological
models where spontaneously broken symmetries play an important role.
The dimensionless coupling $\xi$ is arbitrary;
two special cases are the minimal ($\xi = 0$) and the conformal
($\xi = \sixth $) coupling case.
In the following we shall mostly consider the case $\xi=0$.
Also, it will be straightforward to extend later the treatment to the
case of an $N_s$-component scalar field $\phi_i^a$ with $a=1,...,N_s$.

One way to proceed is to introduce a lattice scalar $\phi_i$
defined at the vertices of the simplices.
The corresponding lattice action can then be obtained
through a procedure by which the original continuum metric
is replaced by the induced lattice metric, with the latter
written in terms of squared edge lengths as in Eq.~(\ref{eq:latmet}).
For illustrative purposes only the two-dimensional case
will be worked out explicitly here 
(Christ, Friedberg and Lee, 1982; Itzykson, 1983; Bander and Itzykson, 1983;
Jevicki and Ninomiya, 1984).
The generalization to higher dimensions is
straightforward, and in the end the final answer for the lattice scalar
action is almost identical to the two dimensional form.
Furthermore in two dimensions it leads to a natural dicretization of the
bosonic string action (Polyakov, 1989).

\begin{figure}[h]
\epsfig{file=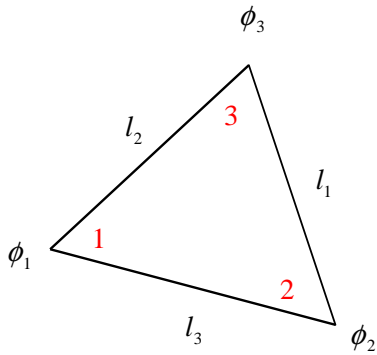,width=8cm}
\caption{Labeling of edges and fields for the construction of the scalar
field action.}
\label{fig:phi}
\end{figure}

In two dimensions the simplicial lattice is built out of
triangles.
For a given triangle it will be convenient to use the
notation of Fig.~\ref{fig:phi}, 
which will display more readily the symmetries of the resulting
scalar lattice action.
Here coordinates will be picked in each triangle along the
(1,2) and (1,3) directions.

To construct a lattice action for the scalar field, one performs in two
dimensions the replacement
\bea
g_{\mu\nu} (x) & \longrightarrow & g_{ij} (\Delta)
\nonumber \\
\det g_{\mu\nu} (x) & \longrightarrow & \det g_{ij} (\Delta)
\nonumber \\
g^{\mu\nu} (x) & \longrightarrow & g^{ij} (\Delta)
\nonumber \\
\partial_\mu \phi \, \partial_\nu \phi & \longrightarrow & 
\Delta_{i} \phi \, \Delta_{j} \phi
\eea
with the following definitions
\beq
g_{ij} (\Delta) = \left( \begin{array}{cc}
l_{3}^2 & \half ( - l_1^2 + l_2^2 + l_{3}^2 ) \cr
\half ( - l_1^2 + l_2^2 + l_{3}^2 ) & l_2^2 \cr
\end{array} \right) \;\; ,
\eeq
\beq
\det g_{ij} (\Delta) = 
\quarter \left [
2 ( l_1^2 l_2^2 + l_2^2 l_{3}^2 + l_{3}^2 l_1^2 ) -
l_1^4 - l_2^4 - l_{3}^4 \right ] \, \equiv \, 4 A_{\Delta}^2 \;\; ,
\eeq
\beq
g^{ij} (\Delta) = 
{ 1 \over \det g (\Delta) } \, \left( \begin{array}{cc}
l_2^2 & \half (l_1^2 - l_2^2 - l_{3}^2 ) \cr
\half (l_1^2 - l_2^2 - l_{3}^2 ) & l_{3}^2 \cr
\end{array} \right) \;\; .
\eeq
The scalar field derivatives get replaced as usual by finite differences
\beq
\partial_\mu \phi \, \longrightarrow \, ( \Delta_\mu \phi )_i \; = \;
\phi_{ i + \mu } - \phi_ i \;\; .
\eeq
where the index $\mu$ labels the possible directions in which one
can move away from a vertex within a given triangle.
Then
\beq
\Delta_{i} \phi \, \Delta_{j} \phi = 
\left( \begin{array}{cc}
( \phi_2 - \phi_1 )^2 & ( \phi_2 - \phi_1 ) ( \phi_3 - \phi_1 ) \cr
( \phi_2 - \phi_1 ) ( \phi_3 - \phi_1 ) & ( \phi_3 - \phi_1 )^2 \cr
\end{array} \right) \;\; ,
\eeq
Then the discrete scalar field action takes the form
\beq
I = {\textstyle {1\over16} \displaystyle}
\sum_{\Delta} { 1 \over A_{\Delta} } \left [
l_1^2 ( \phi_2 - \phi_1 ) ( \phi_3 - \phi_1 ) +
l_2^2 ( \phi_3 - \phi_2 ) ( \phi_1 - \phi_2 ) +
l_3^2 ( \phi_1 - \phi_3 ) ( \phi_2 - \phi_3 ) \right ] \;\; .
\label{eq:jnac}
\eeq
where the sum is over all triangles on the lattice.
Using the identity
\beq
( \phi_i - \phi_j ) ( \phi_i - \phi_k ) \; = \; \half \left [
( \phi_i - \phi_j )^2 + ( \phi_i - \phi_k )^2 -
( \phi_j - \phi_k )^2 \right ] \;\; ,
\eeq
one obtains after some re-arrangements the slightly more appealing
expression for the action of a massless scalar field (Bander and Itzykson, 1984)
\beq
I (l^2, \phi) \; = \; \half \sum_{<ij>} A_{ij} \,
\Bigl ( { \phi_i - \phi_j \over l_{ij} } \Bigr )^2 \;\; ,
\label{eq:acdual} 
\eeq
$A_{ij}$ is the dual (Voronoi) area associated with the edge $ij$,
and the symbol $< \! ij \! >$ denotes a sum over nearest neighbor lattice
vertices.
It is immediate to generalize the action of Eq.~(\ref{eq:acdual}) to
higher dimensions, with the two-dimensional Voronoi volumes
replaced by their higher dimensional analogues, leading to
\beq
I (l^2, \phi) \; = \; \half \sum_{<ij>} V_{ij}^{(d)} \,
\Bigl ( { \phi_i - \phi_j \over l_{ij} } \Bigr )^2 \;\; ,
\label{eq:acdual-d} 
\eeq
Here $V_{ij}^{(d)}$ is the dual (Voronoi) volume associated with the edge $ij$,
and the sum is over all links on the lattice.

\begin{figure}[h]
\epsfig{file=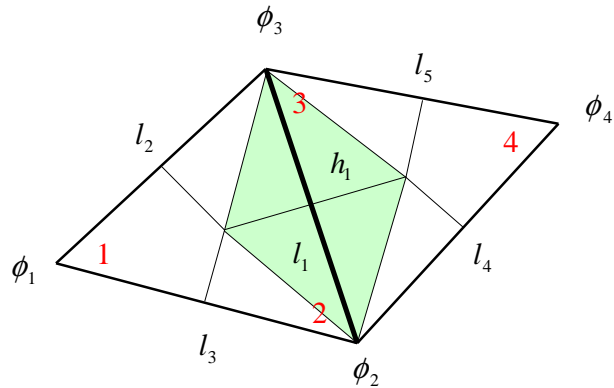,width=10cm}
\caption{Dual area associated with the edge $l_1$ (shaded area),
and the corresponding dual link $h_1$.}
\label{fig:dual-ed}
\end{figure}

In two dimensions, in terms of the edge length $l_{ij}$ and 
the dual edge length $h_{ij}$,
connecting neighboring vertices in the dual lattice,
one has $A_{ij} = \half h_{ij} l_{ij}$
(see Fig.~\ref{fig:dual-ed}).
Other choices for the lattice subdivision will lead to a similar
formula for the lattice action, with the Voronoi dual volumes replaced
by their appropriate counterparts for the new lattice.
Explicitly, for an edge of length $l_1$ the dihedral dual volume contribution is given by
\beq
A_{l_1} =
{ l_1^2 ( l_2^2 + l_{3}^2 - l_1^2 ) \over 16 A_{123} } +
{ l_1^2 ( l_{4}^2 + l_{5}^2 - l_1^2 ) \over 16 A_{234} }
\; = \; \half \, l_1 h_1 \;\; ,
\label{eq:dual_vol_edge}
\eeq
with $h_1$ is the length of the edge dual to $l_1$.

\begin{figure}[h]
\epsfig{file=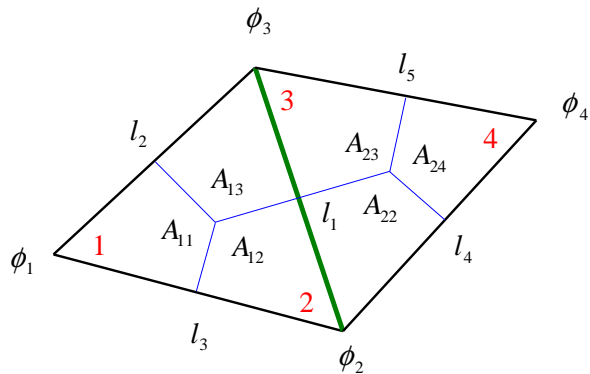,width=10cm}
\caption{More dual areas appearing in the scalar field action.}
\label{fig:phi3}
\end{figure}

On the other hand the barycentric dihedral area for the same edge
would be simply
\beq
A_{l_1} = ( A_{123} + A_{234} ) / 3 \;\; .
\eeq
It is well known that one of the disadvantages of the Voronoi construction
is the lack of positivity of the dual volumes, as pointed out
in (Hamber and Williams, 1984).
Thus some of the weights appearing in Eq.~(\ref{eq:acdual}) can be negative
for such an action. 
For the barycentric subdivision this
problem does not arise, as the areas $A_{ij}$ are always positive due
to the enforcement of the triangle inequalities.
Thus from a practical point of view the barycentric volume subdivision is the simplest to deal with.

The scalar action of Eq.~(\ref{eq:acdual-d}) has a very natural form:
it involves the squared difference of fields at neighboring points
divided by their invariant distances $(\phi_i - \phi_j) / l_{ij}$, 
weighted by the appropriate space-time volume element
$ V_{ij}^{(d)} $ associated with the lattice link $ij$.
This suggests that one could just as well define the scalar
fields on the vertices of the dual lattice, and write
\beq
I (l^2, \phi) \; = \; \half \sum_{<rs>} V_{rs}^{(d)} \,
\Bigl ( { \phi_r - \phi_s \over l_{rs} } \Bigr )^2 \;\; ,
\label{eq:acdual-d2} 
\eeq
with $l_{rs}$ the length of the edge connecting the dual lattice
vertices $r$ and $s$, and consequently
$V_{rs}^{(d)}$ the spacetime volume fraction associated with
the dual lattice edge $rs$.
One would expect both forms to be equivalent in the continuum limit.

Continuing on with the two-dimensional case, mass and curvature terms
such as the ones appearing in Eq.~(\ref{eq:scalar}) can be added to the action,
so that the total scalar lattice action contribution becomes
\beq
I \; = \; \half \sum_{<ij>} A_{ij} \,
\Bigl ( { \phi_i - \phi_j \over l_{ij} } \Bigr )^2 \, +
\half \sum_{i} A_{i} \, (m^2 + \xi R_i ) \, \phi_i^2 \;\; .
\label{eq:acp} 
\eeq
The term containing the discrete analog of the scalar curvature involves
the quantity
\beq
A_{i} R_i \equiv \sum_{ h \supset i } \delta_h \; \sim \; \sqrt{g} \, R \;\; .
\eeq
In the above expression for the scalar action,
$A_{ij}$ is the area associated with the edge $l_{ij}$,
while $A_i$ is associated with the site $i$.
Again there is more than one way to define the volume element $A_i$,
(Hamber and Williams, 1986), but under reasonable assumptions,
such as positivity, one expects to get equivalent results in
the lattice continuum limit, if it exists.

In higher dimensions one would use
\beq
I = \half \sum_{<ij>} V_{ij}^{(d)} \,
\Bigl ( { \phi_i - \phi_j \over l_{ij} } \Bigr )^2 \, +
\half \sum_{i} V_{i}^{(d)} \, (m^2 + \xi R_i ) \phi_i^2  
\label{eq:acp-d} 
\eeq
where the term containing the discrete analog of the scalar curvature involves
\beq
V_{i}^{(d)} R_i \equiv \sum_{ h \supset i } \delta_h V_h^{(d-2)} \sim \sqrt{g} R 
\eeq
In the expression for the scalar action,
$V_{ij}^{(d)}$ is the (dual) volume associated with the edge $l_{ij}$,
while $V_i^{(d)}$ is the (dual) volume associated with the site $i$.

The lattice scalar action contains a mass parameter $m$, which has
to be tuned to zero in lattice units to achieve the lattice continuum
limit for scalar correlations.
The agreement between different lattice actions in the smooth limit
can be shown explicitly in the lattice weak field expansion.
But in general, as is already the case for the purely gravitational action,
the correspondence between lattice and continuum operators is true
classically only up to higher derivative corrections.
But such higher derivative corrections in the scalar
field action are expected to be irrelevant when looking at the large
distance limit, and they will not be considered here any further.

As an extreme case one could even consider a situation in which
the matter action by itself is the only action contribution, without
any additional term for the gravitational field, but still with a
non-trivial gravitational measure; integration over the scalar field would
then give rise to an effective non-local gravitational action.

Finally let us take notice here of the fact that if an $N_s$-component
scalar field is coupled to gravity, the power $\sigma$ appearing in the 
gravitational functional measure has to be modified
to include an additional factor of $\prod_{x} ( \sqrt{g} )^{N_s/2}$.
The additional measure factor insures that the integral
\beq
\int \prod_x \left [ d \phi \, ( \sqrt{g} )^{N_s \over 2} \right ] \,
\exp \left ( - \half m^2 \int \sqrt{g} \, \phi^2 \right ) = 
\left ( \frac{ 2 \pi}{m^2} \right )^{N_s V \over 2}
\eeq
evaluates to a constant.
Thus for large mass $m$ the scalar field completely decouples, leaving
only the dynamics of the pure gravitational field.

The quadratic scalar field action of Eq.~(\ref{eq:acdual-d}) can be 
written in terms of a matrix $\Delta_{ij} (l^2)$
\beq
I (l^2, \phi) \; = \; 
- \half \sum_{<ij>} \, \phi_i \, \Delta_{ij} (l^2) \, \phi_j 
\eeq
The  matrix $\Delta_{ij} (l^2)$ can then be regarded as a lattice version of the 
continuum scalar Laplacian,
\beq
\Delta (g) = {1 \over \sqrt{g} } \partial_\mu \sqrt{g} g^{\mu\nu} \partial_\nu
\eeq
for a given background metric.
This then allows one to define the massless lattice scalar propagator
as the inverse of the above matrix, $G_{ij} (l^2) = \Delta^{-1}_{ij} (l^2)$.
The continuum scalar propagator for a finite scalar mass $m$ and in a given
background geometry, evaluated for large separations $d (x,y) \gg m^{-1}$,
\bea
G ( x,y | g )  & = &
< \! x \, \vert { 1 \over - \Delta (g) \, + \, m^2 } \vert \, y \! >
\nonumber \\
&& \;\;\;\; \, \mathrel{\mathop\sim_{ d(x,y) \rightarrow \infty }} \,
d^{- (d-1)/2 } (x,y) \, \exp \bigl \{ -  m \, d(x,y) \bigr \}
\label{eq:scalar-prop}
\eea
involves the geodesic distance $d(x,y)$ between points $x$ and $y$,
\beq
d(x,y \vert g ) = \int_{\tau (x)}^{\tau (y)} d \tau \sqrt{ g_{\mu\nu} (\tau) \,
{ d x^{\mu} \over d \tau } { d x^{\nu} \over d \tau } }
\eeq
Analogously, one can define the discrete massive lattice scalar propagator
\bea
G_{ij} (l^2 ) & = &
\left [ { 1 \over - \Delta (l^2) \, + \, m^2 } \right ]_{ij} 
\nonumber \\
&& \;\;\;\; \mathrel{\mathop\sim_{ d(i,j) \rightarrow \infty }} \,
d^{- (d-1)/2 } (i,j) \, \exp \bigl \{ -  m \, d(i,j) \bigr \}
\label{eq:latt-scalar-prop}
\eea
where $ d(i,j)$ is the lattice geodesic distance between vertex
$i$ and vertex $j$.
The inverse can be computed, for example, via the recursive expansion
(valid for $m^2>0$ to avoid the zero eigenvalue of the Laplacian)
\beq
{ 1 \over - \Delta (l^2) \, + \, m^2 } = {1 \over m^2 } \sum_{n=0}^\infty
\left ( { 1 \over m^2 } \, \Delta (l^2) \right )^n \;\; .
\eeq
The large distance behavior of the Euclidean (flat space) massive free
field propagator in $d$ dimensions is known in the statistical mechanics
literature as the Ornstein-Zernike result. 

As a consequence, the lattice propagator $ G_{ij} (l^2 )$ can be used to estimate the
lattice geodesic distance $d(i,j | l^2 )$ between
any two lattice points $i$ and $j$ in a fixed background lattice geometry
(provided again that their mutual separation is such that $d(i,j) \gg m^{-1} $).
\beq
d(i,j) \, \mathrel{\mathop\sim_{ d(i,j) \rightarrow \infty }} \,
 - \, { 1 \over m } \, \ln G_{ij} (l^2 ) 
\label{eq:latt-dist}
\eeq

\subsubsection{Invariance Properties of the Scalar Action}

\label{sec:oned}

In the very simple case of one dimension ($d=1$) one can work out the details
to any degree of accuracy, and see how potential problems
arise and how they are resolved.
Introduce a scalar field $\phi_n$ defined on the sites, with action
\beq
I(\phi) \; = \; \half \; \sum_{n=1}^N V_1 (l_n) 
\left ( { \phi_{n+1} - \phi_n \over l_n } \right )^2
+ \half \; \omega \sum_{n=1}^N V_0 (l_n) \; \phi_n^2 \;\; ,
\label{eq:acp-1}  
\eeq
with $\phi(N+1)=\phi(1)$. It is natural in one dimension to take for
the ``volume per edge'' $V_1(l_n)=l_n$, and for the ``volume per site''
$V_0(l_n)=(l_n+l_{n-1})/2$. Here $\omega$ plays the role
of a mass for the scalar field, $\omega=m^2$.
In addition one needs a term
\beq
\lambda \; L(l) \; = \; \lambda_0 \sum_{n=1}^N l_n
\label{eq:acl1} 
\eeq
which is necessary in order to make the $dl_n$ integration
convergent at large $l$.
Varying the action with respect to $\phi_n$ gives
\beq
{ 2 \over l_{n-1} + l_n } \left [
{ \phi_{n+1} - \phi_n \over l_n } - { \phi_{n} - \phi_{n-1} \over l_{n-1} }
\right ] \; = \; \omega \; \phi_n  \;\; .
\label{eq:lapl} 
\eeq
This is the discrete analog of the equation
$ g^{-1/2} \partial g^{-1/2} \partial \phi = \omega \phi $.
The spectrum of the Laplacian of Eq.~(\ref{eq:lapl}) corresponds
to $\Omega \equiv - \omega > 0 $.
Variation with respect to $l_n$ gives instead
\beq
{1 \over 2 l_n^2 } \; ( \phi_{n+1} - \phi_n )^2 \; = \; \lambda_0 +
{1 \over 4} \; \omega \; ( \phi_n^2 + \phi_{n+1}^2 )  \;\; .
\eeq
For $\omega=0$ it suggests the well-known interpretation of the fields $\phi_n$
as coordinates in embedding space.
In the following we shall only consider the case $\omega=0$,
corresponding to a massless scalar field.

It is instructive to look at the invariance properties of the
scalar field action under the continuous lattice 
diffeomorphisms defined in Eq.~(\ref{eq:gauge}).
Physically, these local gauge transformations, which act on the vertices,
correspond to re-assignments of edge lengths which
leave the distance between two fixed points unchanged.
In the simplest case, only two neighboring edge lengths are changed, leaving
the total distance between the end points unchanged.
On physical grounds one would like to maintain such an invariance also
in the case of coupling to matter, just as is done in the continuum.

The scalar nature of the field requires that in the continuum under a 
change of coordinates $x \rightarrow x'$,
\beq
 \phi' (x') = \phi (x)  \;\; ,
\eeq
where $x$ and $x'$ refer to the same physical point in the
two coordinate systems.
On the lattice, as discussed previously, diffeomorphisms
move the points around, and at the same vertex labeled by $n$ we expect
\beq
\phi_n \; \rightarrow \; \phi'_n \approx \phi_n + 
\left ( { \phi_{n+1} - \phi_n \over l_n } \right ) \epsilon_n  \;\; ,
\eeq
One can determine the exact form of the change needed
in $\phi_n$ by requiring that the local variation of the scalar field
action be zero.
Solving the resulting quadratic equation for $\Delta \phi_n$ one
obtains a rather unwieldy expression, given to lowest order by
\bea
\Delta \phi_n  & = & { \epsilon_n \over 2 } \left [
{  \phi_{n} - \phi_{n-1} \over l_{n-1} } +
{ \phi_{n+1} - \phi_n  \over l_{n} }
\right ]  \nonumber \\
& & + { \epsilon_n^2 \over 8 } \left [ 
- {  \phi_{n} - \phi_{n-1} \over l_{n-1}^2 } +
{ \phi_{n+1} - \phi_n  \over l_{n}^2 } +
{ \phi_{n+1} - 2 \phi_n + \phi_{n-1} \over l_{n-1} l_{n} }
\right ] + O( \epsilon_n^3 )  \;\; ,
\\ \nonumber
\label{eq:sgauge} 
\eea
and which is indeed of the expected form (as well as symmetric in the
vertices $n-1$ and $n+1$).
For fields which are reasonably smooth, this correction
is suppressed if $ | \phi_{n+1} - \phi_n | /  l_{n} \ll 1 $.
On the other hand it should be clear that
the measure $d \phi_n $ is no longer manifestly
invariant, due to the rather involved transformation property of the scalar
field.

The full functional integral for $N$ sites then reads
\beq
{\cal Z}_N \; = \; \prod_{n=1}^N  \int_0^\infty \, d l_n^2 \; l_n^\sigma  \,
\int_{-\infty}^\infty \, d \phi_n \,
\exp \left \{ - \lambda_0 \sum_{n=1}^N l_n -
\half \, \sum_{n=1}^N { 1 \over l_n } \;
( \phi_{n+1} - \phi_n )^2 \right \}  \;\; .
\label{eq:z} 
\eeq
In the absence of the scalar field one just has the ${\cal Z}_N (\lambda_0)$
of Eq.~(\ref{eq:avel1}).
The trivial translational mode in $\phi$ can be eliminated for example
by setting $\sum_{n=1}^N  \phi_n = 0$.

It is possible to further constrain the measure over the edge lengths
by examining some local averages.
Under a rescaling of the edge lengths $l_n \rightarrow \alpha \, l_n $
one can derive the following identity for ${\cal Z}_N$
\beq
{\cal Z}_N ( \lambda_0, z ) \; = \; \lambda_0^{ - ( 5/2 + \sigma ) N }
z^{ - N/2 } {\cal Z}_N (1,1)  \;\; ,
\eeq
where we have replaced the coefficient $1/2$ of the scalar kinetic term
by $ z / 2 $.
It follows then that
\beq
\langle l \rangle  \; \equiv { 1 \over N } \langle \sum_{n=1}^N l_n \rangle
\; = \; ( {5 \over 2} + \sigma ) \; \lambda_0^{-1}
\label{eq:srule} 
\eeq
and 
\beq
{ 1 \over N } \langle \sum_{n=1}^N { 1 \over l_n }  \;
( \phi_{n+1} - \phi_n )^2  \rangle  \; = \; 1  \;\; .
\eeq
Without loss of generality we can fix the average edge 
length to be equal to one, $\langle l \rangle  \; =1$, which then requires
$ \lambda_0 = {5 \over 2} + \sigma $. 
In order for the model to be meaningful, the measure
parameter is constrained by $\sigma > -5/2$, i.e. the measure
over the edges cannot be too singular.

\subsubsection{Lattice Fermions, Tetrads and Spin Rotations}

\label{sec:fermions}

On a simplicial manifold spinor fields $\psi_s$ 
and $\bar \psi_s $ are most naturally placed
at the center of each d-simplex $s$.
In the following we will restrict our discussion for
simplicity to the four-dimensional case, and largely
follow the original discussion given by (Fr\"ohlich, 1981) and
(Drummond 1986).
As in the continuum (see for example Veltman, 1974), the
construction of a suitable lattice action requires the
introduction of the Lorentz group and its associated
tetrad fields $e_\mu^a (s) $ within each simplex labeled
by $s$.  

Within each simplex one can choose a representation of the
Dirac gamma matrices, denoted here by $ \gamma^\mu (s)$,
such that in the local coordinate basis
\beq
\left \{ \gamma^\mu (s) , \gamma^\nu (s) \right \} \; = \; 2 \, g^{\mu\nu} (s)
\eeq
These in turn are related to the ordinary Dirac gamma
matrices $\gamma^a$, which obey
\beq
\left \{ \gamma^a , \gamma^b \right \} \; = \; 2 \, \eta^{ab} \;\; ,
\eeq
by
\beq
\gamma^\mu (s) \; = \; e^\mu_a (s) \, \gamma^a
\eeq
so that within each simplex the tetrads $e_\mu^a (s) $
satisfy the usual relation
\beq
e_a^\mu (s) \; e_b^\nu (s) \; \eta^{ab} \; = \; g^{\mu\nu} (s)
\eeq 
In general the tetrads are not fixed uniquely within a simplex,
being invariant under the local Lorentz transformations
discussed earlier in Sec.~\ref{sec:rotations}.

In the continuum the action for a massless spinor field is
given by
\beq
I \; = \; \int d x \sqrt{g} \; \bar \psi (x) \, 
\gamma^\mu \, D_\mu \, \psi (x)
\label{eq:fermac}
\eeq
where $D_\mu = \partial_\mu + \half \omega_{\mu ab} \sigma^{ab}$
is the spinorial covariant derivative containing the spin connection 
$\omega_{\mu ab}$.
It will be convenient to first consider only two neighboring
simplices $s_1$ and $s_2$, covered by a common coordinate
system $x^{\mu}$.
When the two tetrads in $s_1$ and $s_2$ are made to coincide,
one can then use a common set of gamma matrices $\gamma^\mu$
within both simplices.
Then in the absence of torsion the covariant derivative $D_\mu$
in Eq.~(\ref{eq:fermac}) reduces to just an ordinary derivative.
The fermion field $\psi(x)$ within the two simplices can then be
suitably interpolated, by writing for example
\beq
\psi (x) \; = \; \theta (n \cdot x) \, \psi (s_1) \, + \,
(1 - \theta (n \cdot x) ) \, \psi (s_2)
\label{eq:psi-int}
\eeq 
where $n_\mu$ is the common normal to the face $f(s_1,s_2)$
shared by the simplices $s_1$ and $s_2$, and chosen to point into $s_1$.
Inserting the expression for $\psi (x) $ from Eq.~(\ref{eq:psi-int})
into Eq.~(\ref{eq:fermac}) and applying the divergence theorem
(or equivalently using the fact that the derivative of a step function
only has support at the origin) one obtains
\beq
I \; = \; \half \, V^{(d-1)} (f) \; ( \bar \psi_1 + \bar \psi_2 ) \,
\gamma^\mu \, n_\mu \; (\psi_1 - \psi_2 )
\eeq
where $ V^{(d-1)} (f)$ represents the volume of the $(d-1)$-dimensional
common interface $f$, a tetrahedron in four dimensions.
But the contributions from the diagonal terms containing 
$\bar \psi_1 \psi_1 $ and $\bar \psi_2 \psi_2 $ vanish when summed
over the faces of an $n$-simplex, by virtue of the useful identity
\beq
\sum_{p=1}^{n+1} \, V( f^{(p)} ) \, n_\mu^{(p)} \; = \; 0
\eeq
where $ V( f^{(p)} )$ are the volumes of the $p$ faces of
a given simplex, and $ n_\mu^{(p)} $ the inward pointing unit
normals to those faces.

So far the above partial expression for the lattice spinor action was
obtained by assuming that the tetrads $e^\mu_a (s_1)$
and $e^\mu_a (s_2)$ in the two simplices coincide.
If they do not, then they will be related by a matrix 
${\bf R} (s_2, s_1)$ such that
\beq
e_a^\mu (s_2) \; = \; R^\mu_{\;\;\nu} (s_2,s_1) \; e_a^\nu (s_1)
\eeq
and whose spinorial representation $\bf S$
was given previously for example in Eq.~(\ref{eq:spinrot}).
Such a matrix ${\bf S}(s_2,s_1)$ is now needed to additionally
parallel transport the spinor $\psi$ from a simplex $s_1$
to the neighboring simplex $s_2$. 

The invariant lattice action for a massless spinor takes therefore the form
\beq
I \; = \; \half \sum_{\rm faces \; f(s s')} \, V( f(s,s')) \,
\bar \psi_s \, {\bf S} ( {\bf R} (s,s') ) \, \gamma^\mu (s') \,
n_\mu (s,s') \, \psi_{s'}
\eeq
where the sum extends over all interfaces $f(s,s')$ connecting one
simplex $s$ to a neighboring simplex $s'$.
As shown in (Drummond 1986) it can be further extended to include a
dynamical torsion field.

The above spinorial action can be considered analogous to the lattice
Fermion action proposed originally in (Wilson 1973) for non-Abelian
gauge theories.
It is possible that it still suffers from the
Fermion doubling problem, although the situation is less clear
for a dynamical lattice (Nielsen and Ninomiya, 1981; Christ and Lee, 1982).

\subsubsection{Alternate Discrete Formulations}

\label{sec:otherlatt}

The simplicial lattice formulation offers a natural way of representing
gravitational degrees in a discrete framework by employing
inherently geometric concepts such as areas, volumes and angles. 
It is possible though to formulate quantum gravity on a flat
hypercubic lattice, in analogy to Wilson's discrete formulation
for gauge theories, by putting the connection center stage.
In this new set of theories the natural variables are then lattice
versions of the spin connection and the vierbein.
Also, because the spin connection variables appear 
from the very beginning, it is much easier to incorporate fermions later.
Some lattice models have been based on the pure Einstein theory
(Smolin, 1979; Das, Kaku and Townsend, 1979;
Mannion and Taylor, 1981; Caracciolo and Pelissetto, 1987),
while others attempt to incorporate higher derivative terms
(Tomboulis 1984; Kondo 1984).

Difficult arise when attempting to put quantum gravity on a flat hypercubic
lattice a la Wilson, since it is not entirely clear what the gravity analog
of the Yang-Mills connection is.
In continuum formulations invariant under the Poincar\'{e} or de Sitter group
the action is invariant under a local extension of the Lorentz transformations,
but not under local translations (Kibble, 1961).
Local translations are replaced by diffeomorphisms which
have a different nature. 
One set of lattice discretizations starts from the action of (MacDowell and Mansouri, 1977; West, 1978)
whose local invariance group is the de Sitter group $Sp(4)$.
In the lattice formulation of (Smolin, 1979; Das, Kaku and Townsend, 1979)
the lattice variables are gauge potentials $e_{a\mu} (n)$ and 
$\omega_{\mu ab}(n)$
defined on lattice sites $n$, generating local $Sp(4)$ matrix transformations
with the aid of the de Sitter generators $P_a$ and $M_{ab}$.
The resulting lattice action reduces classically
to the Einstein action with cosmological term in first order form in the limit of the lattice spacing $a \rightarrow 0$; 
to demonstrate the quantum equivalence one needs an additional zero torsion
constraint.
In the end the issue of lattice diffeomorphism invariance remains somewhat open,
with the hope that such an invariance will be restored in the full quantum theory.

As an example, we will discuss here the approach of (Mannion and Taylor, 1981)
which relies on a four-dimensional lattice discretization of the Einstein-Cartan
theory with gauge group $SL(2,C)$, and does not initially require the presence
of a cosmological constant, as would be the case if one had started out
with the de Sitter group $Sp(4)$.
On a lattice of spacing $a$ with vertices labelled by $n$ and directions
by $\mu$ one relates
the relative orientations of nearest-neighbor local $SL(2,C)$ frames by
\beq
U_\mu (n) = \Bigr [ U_{-\mu} ( n+ \mu ) \Bigl ]^{-1} 
\, = \, \exp [ \, i \, B_\mu (n) ]
\eeq
with $B_\mu = \half a B_\mu^{ab} (n) J_{ba}$, $J_{ba}$ being the set of six
generators of $SL(2,C)$, the covering group of the Lorentz group $SO(3,1)$,
usually taken to be 
\beq
\sigma_{ab} = { 1 \over 2 i} [ \gamma_a , \gamma_b ]
\eeq
with $\gamma_a$'s the Dirac gamma matrices.
The local lattice curvature is then obtained in the usual way
by computing the product of four parallel transport matrices
around an elementary lattice square,
\beq
U_\mu (n) \, U_\nu (n + \mu) \, U_{-\mu} (n+\mu+\nu) \, U_{-\nu} (n+\nu)
\eeq
giving in the limit of small $a$ by the Baker-Hausdorff formula
the value $\exp [ i a R_{\mu\nu} (n) ] $, where $R_{\mu\nu}$
is the Riemann tensor defined in terms of the spin connection $B_\mu$
\beq
R_{\mu\nu} = \partial_\mu B_\nu - \partial_\nu B_\mu + i [ B_\mu, B_\nu ]
\eeq
If one were to write for the action the usual Wilson lattice gauge form
\beq
\sum_{n, \mu, \nu} \tr [ \,
U_\mu (n) \, U_\nu (n + \mu) \, U_{-\mu} (n+\mu+\nu) \, U_{-\nu} (n+\nu) \, ]
\eeq
then one would obtain a curvature squared action proportional to
$\sim \int R_{\mu\nu}^{\;\;ab} R^{\mu\nu}_{\;\;ab} $
instead of the Einstein-Hilbert one.
One needs therefore to introduce lattice vierbeins $e_\mu^{\;\;b} (n)$
on the sites by defining the matrices
\beq
E_\mu (n) = a \, e_\mu^{\;\;a} \, \gamma_a 
\eeq
Then a suitable lattice action is given by
\beq
I = { i \over 16 \kappa^2 }
\sum_{n, \mu, \nu, \lambda, \sigma} \tr [ \, \gamma_5 \,
U_\mu (n) \, U_\nu (n + \mu) \, U_{-\mu} (n+\mu+\nu) \, U_{-\nu} (n+\nu) \,
E_\sigma (n) \, E_\lambda (n) \, ]
\label{eq:lagr-mt}
\eeq
The latter is invariant under local $SL(2,C)$ transformations $\Lambda (n)$
defined on the lattice vertices
\beq
U_\mu \rightarrow \Lambda (n) \, U_\mu (n) \, \Lambda^{-1} (n+\mu)
\eeq
for which the curvature transforms as
\bea
&& U_\mu (n) \, U_\nu (n + \mu) \, U_{-\mu} (n+\mu+\nu) \, U_{-\nu} (n+\nu)
\nonumber \\
&& \;\;\;\; \rightarrow \;
\Lambda (n) \,
U_\mu (n) \, U_\nu (n + \mu) \, U_{-\mu} (n+\mu+\nu) \, U_{-\nu} (n+\nu) \,
\Lambda^{-1} (n)
\eea
and the vierbein matrices as
\beq
E_\mu (n) \rightarrow \Lambda (n) \, E_\mu (n) \, \Lambda^{-1} (n)
\eeq
Since $\Lambda (n) $ commutes with $\gamma_5$, the expression in
Eq.~(\ref{eq:lagr-mt}) is invariant.
The metric is then obtained as usual by
\beq
g_{\mu\nu} (n) = \quarter \tr [ E_\mu (n) \, E_\nu (n) ] \;\; .
\eeq
From the expression for the lattice curvature
$ R_{\mu\nu}^{\;\;ab} $ given above if follows immediately that the lattice action
in the continuum limit becomes
\beq
I \, = \, { a^4 \over 4 \kappa^2 }
\sum_n \epsilon^{\mu\nu\lambda\sigma} \, \epsilon_{abcd} \,
R_{\mu\nu}^{\;\;\;\;ab} (n) \, e^{\;\;c}_\lambda (n) \, e^{\;\;d}_\sigma (n)
\, + \, O( a^6)
\eeq
which is the Einstein action in Cartan form
\beq
I \, = \, { 1 \over 4 \kappa^2 } \int d^4 x \,
\epsilon^{\mu\nu\lambda\sigma} \, \epsilon_{abcd} \,
R_{\mu\nu}^{\;\;\;\;ab} \, e^{\;\;c}_\lambda \, e^{\;\;d}_\sigma
\eeq
with the parameter $\kappa$ identified with the Planck length.
One can add more terms to the action;
in this theory a cosmological term can be represented  by
\beq
\lambda_0 \sum_n \epsilon^{\mu\nu\lambda\sigma}
\tr [ \, \gamma_5 \,
E_\mu (n) \, E_\nu (n) \, E_\sigma (n) \, E_\lambda (n) \, ]
\label{eq:lagr-mt1}
\eeq
Both Eqs.~(\ref{eq:lagr-mt}) and Eq.~(\ref{eq:lagr-mt1}) are locally
$SL(2,C)$ invariant.
The functional integral is then given by
\beq
Z \, = \, \int \prod_{n, \mu}  d B_\mu (n) \, \prod_{n, \sigma}  d E_\sigma (n)
\, \exp \, \Bigl \{ - I ( B,E) \Bigr \}
\eeq
and from it one can then compute suitable quantum averages.
Here $ d B_\mu (n)$ is the Haar measure for $SL(2,C)$;
it is less clear how to choose the integration measure over the
$E_\sigma$'s, and how it should suitably constrained, which
obscure the issue of diffeomorphism invariance in this theory.

In these theories it is possible to formulate curvature squared terms as well.
In general for a hypercubic lattices 
the formulation of $ R^2 $-type terms in four dimensions
involves constraints between the connections and the tetrads, which are
a bit difficult to handle.
Also there is no simple way of writing down topological invariants, which
are either related to the Einstein action (in two dimensions),
or are candidates for extra terms to be included in the action.
A flat hypercubic lattice action has been written with higher derivative
terms which appears to be reflection positive but has a very cumbersome form.
These difficulties need not be present on a simplicial
lattice (except that it is not known how to write an exact expression for
the Hirzebruch signature in lattice terms).

There is another way of discretizing gravity, still using largely
geometric concepts as is done in the Regge theory.
In the dynamical triangulation approach one fixes the edge lengths to
unity, and varies the incidence matrix. 
As a result the volume of each simplex is fixed at
\beq
V_d \; = \; { 1 \over d! } \sqrt{ d+1 \over 2^d } \;\; ,
\eeq
and all dihedral angles are given by the constant value
\beq
\cos \theta_d \; = \; { 1 \over d } 
\eeq
so that for example in four dimensions one has 
$\theta_d = \arccos (1/4) \approx 75.5^{o}$.
Local curvatures are then determined by how many simplices $n_s (h)$
meet on a given hinge,
\beq
\delta (h)  \; = \; 2 \, \pi - n_s (h ) \, \theta_d
\eeq
The action contribution from a single hinge is
therefore from Eq.~(\ref{eq:regge-d}) 
$\delta (h) A(h) = \quarter \sqrt{3} [ 2 \, \pi - n_s (h ) \, \theta_d ]$
with $n_s$ a positive integer.
The local curvatures are inherently discrete, and there
is no equivalent lattice notion of continuous diffeomorphisms, 
or for that matter of
continuous local deformations corresponding for example to shear waves.
The hope is that for lattices made of some large number of simplices
one recovers some sort of discrete version of diffeomorphism invariance.
The Euclidean dynamical triangulation approach has been reviewed recently in
(Ambj{\o}rn, Carfora and Marzuoli, 1999), and we refer the reader to
further references therein.
A recent discussion of attempts at simulating the Lorentzian case,
which leads to complex weights in the functional integral which
are difficult to handle,
can be found in (Loll et al, 2006).
  
Another lattice approach closely related to the Regge theory 
described in this review is based on the so-called
spin foam models, which have their origin in an observation found in
(Ponzano and Regge, 1968) relating the geometry of simplicial lattices
to the asymptotics of Racah angular momentum addition coefficients.
The original Regge-Ponzano concepts were later developed into a 
spin model for gravity (Hasslacher
and Perry, 1981) based on quantum spin variables attached to lattice links.
In these models representations of $SU(2)$ label edges.
One natural underlying framework for such theories is the
canonical $3+1$ approach to quantum gravity, wherein quantum spin variables
are naturally related to $SU(2)$ spin connections.
Extensions to four dimensions have been attempted, and we
refer the reader to the recent review of spin foam models in (Perez, 2003).

\subsection{Analytical Expansion Methods}

\label{sec:analytical}

The following sections will discuss a number of instances
where the lattice theory of quantum gravity can be investigated
analytically, subject to some simplifying assumptions.

The first problem is the lattice weak field expansion about
a flat background.
It will be shown that in this case the relevant modes are
the lattice analogues of transverse-traceless deformations.

The second problem is the strong coupling (large $G$) expansion,
where the weight factor in the path integral is expanded
in powers of $1/G$.
The domain of validity of this expansion can be regarded as
somewhat complementary to the weak field limit.

The third case to be discussed is what happens in lattice gravity in
the limit of large dimensions $d$, which formally is similar in some
ways to the large-$N$ expansion discussed previously in this review.
In this limit one can derive exact estimates for the phase transition
point and for the scaling dimensions.

\subsubsection{Lattice Weak Field Expansion and Transverse-Traceless Modes}

\label{sec:latticewfe}

One of the simplest possible problems that can be treated in quantum Regge
calculus is the analysis of small fluctuations about a fixed flat Euclidean simplicial background (Ro\v cek and Williams, 1981).
In this case one finds that the lattice graviton propagator in a 
De Donder-like gauge is precisely analogous to the continuum expression.

\begin{figure}[h]
\epsfig{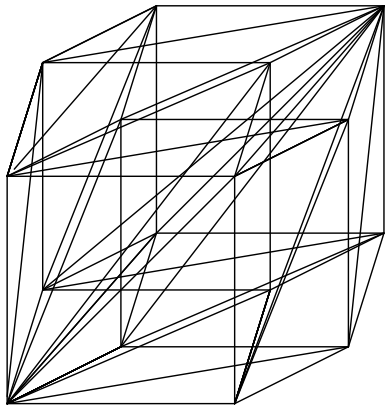}
\caption{A four-dimensional hypercube divided up into four-simplices.}
\label{fig:4d-cube}
\end{figure}

To compute an expansion of the lattice Regge action 
\beq
I_R \; \propto \; \sum_{\rm hinges} \, \delta (l) \; A (l) 
\eeq
to quadratic order in the lattice weak fields one needs first and second variations with respect to the edge lengths.
In four dimensions the first variation of the lattice Regge action is given by
\beq
\delta I_R \; \propto \; \sum_{\rm hinges} \, \delta \cdot 
\left ( \sum_{\rm edges} { \partial A \over \partial l } \, \delta l \, \right )
\eeq
since Regge has shown that the term involving the variation of the deficit
angle $\delta$ vanishes (here the variation symbol should obviously not be confused with the deficit angle).
Furthermore in flat space all the deficit angles vanish, so that the
second variation is given simply by  
\beq
\delta^2 I_R \; \propto \; 
\sum_{\rm hinges} \,
\left ( \sum_{\rm edges} { \partial \delta \over \partial l }
\, \delta l \, \right ) \cdot
\left ( \sum_{\rm edges} { \partial A \over \partial l } \, \delta l \, \right )
\eeq
Next a specific lattice structure needs to be chosen as a background geometry.
A natural choice is to use a flat hypercubic lattice, made
rigid by introducing face diagonals, body diagonals and hyperbody diagonals,
which results into a subdivision of each hypercube into $d!$ (here
4!=24) simplices.
This subdivision is shown in Fig.~\ref{fig:4d-cube}.

By a simple translation, the whole lattice can then be constructed from
this one elemental hypercube.
Consequently there will be $2^d - 1 = 15$ lattice fields per point,
corresponding to all the edge lengths emanating in the positive lattice
directions from any one vertex. 
Note that the number of degrees per lattice point is slightly
larger than what one would have in the continuum, where the
metric $g_{\mu\nu}(x)$ has $d(d+1)/2=10$ degrees of freedom per
spacetime point $x$ in four dimensions 
(perturbatively, the physical degrees of freedom in the continuum are much less:
$ {1\over 2} d (d+1)-1-d-(d-1)$ = ${1 \over 2} d (d-3) $, for a traceless symmetric tensor, and after imposing gauge conditions).
Thus in four dimensions each lattice hypercube will contain 4 body principals, 
6 face diagonals, 4 body diagonals and one hyperbody diagonal.
Within a given hypercube it is quite convenient to label the coordinates of the
vertices using a binary notation, so that the four body principals
with coordinates $(1,0,0,0)$ … $(0,0,0,1)$ will be labeled
by integers 1,2,4,8, and similarly for the other vertices  
(thus for example the vertex $(0,1,1,0)$, corresponding to a face diagonal
along the second and third Cartesian direction, will be labeled by the
integer 6).

For a given lattice of fixed connectivity, the edge lengths are then
allowed to fluctuate around an equilibrium value $l_i^0 $
\beq
l_i = l_i^0 \;( 1 + \epsilon_i ) 
\eeq
In the case of the hypercubic lattice subdivided into simplices,
the unperturbed edge lengths $ l_i^0 $ take on the values $1,\sqrt{2},\sqrt{3},2$, depending on edge type.
The second variation of the action then reduces to a quadratic form in 
the 15-component small fluctuation vector $ {\bf \epsilon}_n $
\beq
\delta^2 I_R \; \propto \;  \sum_ { m n } \; {\bf \epsilon }_m^T \;
M_{ m n } \; {\bf \epsilon }_n 
\eeq
Here $M$ is the small fluctuation matrix, whose inverse determines the
free lattice graviton propagator, and the indices $m$ and $n$ 
label the sites on the lattice.
But just as in the continuum, $M$ has zero
eigenvalues and cannot therefore be inverted until one supplies an appropriate
gauge condition. 
Specifically, one finds that the matrix $M$ in four dimensions has
four zero modes corresponding to periodic translations of the lattice,
and a fifth zero mode corresponding to periodic fluctuations in the
hyperbody diagonal.
After block-diagonalization it is found that 4 modes completely decouple
and are constrained to vanish, and thus the remaining degrees of freedom
are 10, as in the continuum, where the metric has 10 independent components.
The wrong sign for the conformal mode, which is present in the
continuum, is also reproduced by the lattice propagator.

Due to the locality of the original lattice action, the matrix $M$ can be considered local as well, since it only couples edge fluctuations on neighboring lattice sites. 
In Fourier space one can write for
each of the fifteen displacements $\epsilon_n^{i+j+k+l}$,
defined at the vertex of the hypercube with labels $(i,j,k,l)$,
\beq
\epsilon_n^{i+j+k+l} \; = \;  
( \omega_1 )^i ( \omega_2 )^j ( \omega_4 )^k ( \omega_8 )^l \; \epsilon_n^0
\label{eq:trans}
\eeq
with $\omega_1 = e^{i k_1} $, $\omega_2 = e^{i k_2} $,
$\omega_4 = e^{i k_3} $ and $\omega_8 = e^{i k_4} $
(it will be convenient in the following to use binary notation for
$\omega$ and $\epsilon$, but the regular notation for $k_i$).
Here and in the following we have set the lattice spacing $a$
equal to one.

In this basis the matrix $M$ reduces to a block-diagonal form,
with entries given by the $15\times15$ dimensional matrices
\beq
M_{\omega} \; = \; \left ( \matrix{ 
A_{10} & B & 0 \cr B^{\dagger} & 18 I_4 & 0 \cr 0 & 0 & 0 } \right ) ,
\eeq
where $A_{10}$ is a $10 \times 10$ dimensional matrix, $B$ a $10 \times 4$ dimensional matrix and $I_{4}$ is the $4 \times 4$ dimensional identity
matrix.
Explicitly the above cited authors find
\bea
( M_\omega )_{1,1} & = & \; 6
\nonumber \\
( M_\omega )_{1,2} & = & \; \omega_1 ( \omega_4  + \omega_8 )+ \bar\omega_2 ( \bar\omega_4 + \bar\omega_8 )
\nonumber \\
( M_\omega )_{1,3} & = & \; 2 + 2 \bar\omega_2
\nonumber \\
( M_\omega )_{1,6} & = & \; 2 \omega_1 + 2 \bar\omega_2 \bar\omega_4
\nonumber \\
( M_\omega )_{1,7} & = & \; \bar\omega_2 + \bar\omega_4
\nonumber \\
( M_\omega )_{1,14} & = & \; 0
\nonumber \\
( M_\omega )_{3,3} & = & \; 4
\nonumber \\
( M_\omega )_{3,5} & = & \omega_2 + \bar\omega_4
\nonumber \\
 ( M_\omega )_{3,12} & = & \; 0
\nonumber \\
 ( M_\omega )_{3,7} & = & \; 1 + \bar\omega_4
\nonumber \\
( M_\omega )_{3,13} & = & \; 0
\eea
where the remaining non-vanishing matrix elements can be
obtained either by permuting appropriate indices, or by complex conjugation.

Besides one obvious zero eigenvalue, corresponding to a periodic
fluctuation in $\epsilon_{15}$, the matrix $M_\omega$ exhibits
four additional zero modes corresponding to the four-parameter group
of translations in flat space.
An explicit form for these eigenmodes is 
\bea
\epsilon_i & = & ( 1 - \omega_i ) \, x_i
\nonumber \\
\epsilon_{i+j} & = & \half ( 1 - \omega_i \omega_j ) \, ( x_i + x_j )
\nonumber \\
\epsilon_{i+j+k} & = & \third ( 1 - \omega_i \omega_j \omega_k ) 
\, ( x_i + x_j + x_k )
\nonumber \\
\epsilon_{i+j+k+l} & = & \quarter ( 1 - \omega_i \omega_j \omega_k \omega_l ) 
\, ( x_i + x_j + x_k + x_l )
\nonumber \\
\label{eq:eigenmodes}
\eea
with $i,j,k,l=1,2,4,8$ and $i \neq j \neq k \neq l $.

The next step consists in transforming the lattice action $M_{\omega}$ into 
a form more suitable for comparison with the continuum action.
To this end a set of transformations are performed sequentially, the
first of which involves the matrix
\beq
S \; = \; \left ( \matrix{ 
I_{10} & 0 & 0 \cr - {1 \over 18 } B^{\dagger} & I_4 & 0 \cr 0 & 0 & 1
} \right ) ,
\label{eq:s-matrix}
\eeq
which rotates $M_{\omega}$ into
\beq
M_{\omega}' \; = \; S^{\dagger} \, M_\omega \, S \; = \; 
\left ( \matrix{ 
A_{10}-{1 \over 18} B B^\dagger & 0 & 0 \cr 0 & 18 \, I_4 & 0 \cr 0 & 0 & 0 } \right ) ,
\eeq
thus completely decoupling the (body diagonal) fluctuations
$\epsilon_7, \epsilon_{11}, \epsilon_{13}, \epsilon_{14}$.
These in turn can now be integrated out, as they appear
in the action with no $\omega$ (i.e. derivative) term.
As a result the number of dynamical degrees of freedom has been reduced from
15 to 10, the same number as in the continuum.

The remaining dynamics is thus encoded in the $10 \times 10$ dimensional
matrix $L_\omega \; = \; A_{10}-{1 \over 18} B B^\dagger $.
By a second rotation, here affected by the matrix $T$, it can finally
be brought into the form
\beq
\tilde L_\omega \; = \; T^{\dagger} \, L_\omega \, T \; = \; 
\left [ 8-( \Sigma + \bar \Sigma ) \right ] 
\left ( \matrix{ \half \beta & 0 \cr 0 & I_6 } \right ) \, - \, C^{\dagger} C
\label{eq:lattwfe}
\eeq
with the matrix $\beta$ given by
\beq
\beta \; = \; \half \left ( \matrix{ 
1  & -1 & -1 & -1 \cr
-1 & 1  & -1 & -1 \cr 
-1 & -1 & 1  & -1 \cr 
-1 & -1 & -1 &  1 \cr 
} \right ) 
\label{eq:beta-mat}
\eeq
The other matrix $C$ appearing in the second term is given by
\beq
C = \left ( \matrix{ 
f_1 & 0 & 0 & 0 & \tilde f_2 & \tilde f_4 & 0 & \tilde f_8 & 0 & 0 \cr
0 & f_2 & 0 & 0 & \tilde f_1 & 0 & \tilde f_4 & 0 & \tilde f_8 & 0 \cr
0 & 0 & f_4 & 0 & 0 & \tilde f_1 & \tilde f_2 & 0 & 0 & \tilde f_8 \cr
0 & 0 & 0 & f_8 & 0 & 0 & 0 & \tilde f_1 & \tilde f_2 & \tilde f_4 \cr
} \right )
\label{eq:c-matrix}
\eeq
with $f_i \equiv \omega_i -1 $ and $\tilde f_i \equiv 1- \bar \omega_i $.
Furthermore $\Sigma = \sum_i \omega_i $, and for small momenta one finds
\beq
8-( \Sigma + \bar \Sigma ) \; = \; 8- \sum_{i=1}^4 ( e^{i k_i} + e^{- i k_i } )
\; \sim \; k^2 + O(k^4)
\eeq
which shows that the surviving terms in the lattice action are indeed
quadratic in $k$. 
The rotation matrix $T$ involved in the last transformation is given by
\beq
T \; = \; 
\left ( \matrix{ \Omega_4 \beta & 0 \cr 0 & I_6 } \right )
\left ( \matrix{ I_4 & 0 \cr \Omega_6 \gamma & I_6 } \right )
\label{eq:t-matrix}
\eeq
with $\Omega_4 = {\rm diag} (\omega_1,\omega_2,\omega_4,\omega_8)$
and $\Omega_6 = {\rm diag} (\omega_1 \omega_2, \omega_1 \omega_4 ,
\omega_2 \omega_4 , \omega_1 \omega_8 , \omega_2 \omega_8 ,
\omega_4 \omega_8 )$, and 
\beq
\gamma \; = \; - \half 
\left ( \matrix{ 
0 & 0 & 1 & 1 \cr
0 & 1 & 0 & 1 \cr 
1 & 0 & 0 & 1 \cr 
0 & 1 & 1 & 0 \cr
1 & 0 & 1 & 0 \cr 
1 & 1 & 0 & 0 \cr 
} \right ) 
\eeq
At this point one is finally ready for a comparison with the continuum
result, namely with the Lagrangian for pure gravity in the weak field limit
as given in Eq.~(\ref{eq:h-action})
\bea
{\cal L}_{sym} \; = \; &-&
\half \partial_\lambda \, h_{\lambda \mu} \, \partial_\mu h_{\nu\nu}
+ \half \partial_\lambda \, h_{\lambda \mu} \, \partial_\nu h_{\nu\mu}
\nonumber \\
&-& \quarter \partial_\lambda \, h_{\mu \nu} \, \partial_\lambda h_{\mu\nu}
+ \quarter \partial_\lambda \, h_{\mu \mu} \, \partial_\lambda h_{\nu\nu}
\eea
The latter can be conveniently split into two parts, as was done already
in Eq.~(\ref{eq:h-quadr-tv}), as follows
\beq
{\cal L}_{sym} \; = \; - \half \partial_\lambda \, h_{\alpha\beta}
V_{\alpha\beta\mu\nu} \, \partial_\lambda h_{\mu\nu} \, + \, \half C^2
\eeq
with
\beq
V_{\alpha\beta\mu\nu} \; = \; \half \, \eta_{\alpha\mu} \eta_{\beta\nu}
-\quarter \, \eta_{\alpha\beta} \eta_{\mu\nu}
\eeq
or as a matrix,
\beq
V = \left (
  \matrix{ 
  {1\over 4} & -{1\over 4} & -{1\over 4} & -{1\over 4} &
  0 & 0 & 0 & 0 & 0 & 0 \cr
  -{1\over 4} & {1\over 4} & -{1\over 4} & -{1\over 4} &
  0 & 0 & 0 & 0 & 0 & 0 \cr
  -{1\over 4} & -{1\over 4} & {1\over 4} & -{1\over 4} &
  0 & 0 & 0 & 0 & 0 & 0 \cr
  -{1\over 4} & -{1\over 4} & -{1\over 4} & {1\over 4} &
  0 & 0 & 0 & 0 & 0 & 0 \cr
  0 & 0 & 0 & 0 & 
  1 & 0 & 0 & 0 & 0 & 0 \cr
  0 & 0 & 0 & 0 & 
  0 & 1 & 0 & 0 & 0 & 0 \cr
  0 & 0 & 0 & 0 & 
  0 & 0 & 1 & 0 & 0 & 0 \cr
  0 & 0 & 0 & 0 & 
  0 & 0 & 0 & 1 & 0 & 0 \cr
  0 & 0 & 0 & 0 & 
  0 & 0 & 0 & 0 & 1 & 0 \cr
  0 & 0 & 0 & 0 & 
  0 & 0 & 0 & 0 & 0 & 1 \cr }  \right ) ,
\label{eq:v-matrix}
\eeq
with metric components $11,22,33,44,12,13,14,23,24,34$ more conveniently 
labeled sequentially by integers $1 \dots 10$,
and the gauge fixing term $C_\mu$ given by the term
in Eq.~(\ref{eq:gauge-fix})
\beq
C_\mu \; = \; \partial_\nu h_{\mu\nu} - \half \partial_\mu h_{\nu\nu}
\label{eq:gauge-fix-1}
\eeq
The above expression is still not quite the same as the lattice weak
field action, but a simple transformation to trace reversed variables
$\bar h_{\mu\nu} \equiv h_{\mu\nu} - \half \delta_{\mu\nu} h_{\lambda\lambda}$
leads to
\beq
{\cal L}_{sym} \; = \;
\half k_\lambda \bar h_i V_{ij} k_\lambda \bar h_j
\, - \, \half \bar h_i ( C^{\dagger} C )_{ij} \bar h_j
\eeq
with the matrix $V$ given by 
\beq
V_{ij} \; = \; 
\left ( \matrix{ \half \beta & 0 \cr 0 & I_6 } \right )
\eeq
with $k = i \partial $. 
Now $\beta$ is the same as the matrix in Eq.~(\ref{eq:beta-mat}),
and $C$ is nothing but the small $k$ limit of the matrix by the same name
in Eq.~(\ref{eq:c-matrix}), for which one needs to set 
$\omega_i -1 \simeq i \, k_i$.
The resulting continuum expression is then recognized to be
identical to the lattice weak field results of Eq.~(\ref{eq:lattwfe}).

This concludes the outline of the proof of equivalence of the lattice weak
field expansion of the Regge action to the corresponding continuum expression.
To summarize, there are several ingredients to this proof,
the first of which is a relatively straightforward weak field expansion of
both actions, and the second of which is the correct identification of the lattice degrees of freedom $\epsilon_i (n) $ with their continuum counterparts $h_{\mu\nu}(x)$, which involves a sequence of non-trivial 
$\omega$-dependent transformations, expressed by the matrices $S$ and $T$. 
One more important aspect of the process is the disappearance of 
redundant lattice variables (five in the case of the hypercubic lattice),
whose dynamics turns out to be trivial, in the
sense that the associated degrees of freedom are non-propagating. 

It is easy to see that the sequence of transformations expressed by 
the matrices $S$ of Eq.~(\ref{eq:s-matrix}) and $T$ of Eq.~(\ref{eq:t-matrix}), and therefore ultimately relating the lattice fluctuations 
$\epsilon_i (n) $ to their continuum counterparts $h_{\mu\nu}(x)$,
just reproduces the expected relationship between lattice and continuum
fields (Hamber and Williams, 1993).
On the one hand one has $ g_{\mu\nu} = \eta_{\mu\nu} + h_{\mu\nu} $, where
$ \eta_{\mu\nu} $ is the flat metric.
At the same time one has from Eq.~(\ref{eq:latmet}) for each simplex
within a given hypercube
\beq
g_{ij} \; = \; \half ( l_{0i}^2 + l_{0j}^2-l_{ij}^2 )
\eeq
By inserting $l_i \; = \; l_i^0 \; ( 1 + \epsilon_i )$,
with $l_i^0 = 1, \sqrt{2}, \sqrt{3},2 $ for the body principal
($i=1,2,4,8$), face diagonal ($i=3,5,6,9,10,12$), 
body diagonal ($i=7,11,13,14$) and hyperbody diagonal ($i=15$),
respectively, one gets for example $( 1 + \epsilon_1 )^2 = 1 + h_{11}$,
$ ( 1 + \epsilon_3 )^2 = 1 + \half \, ( h_{11} + h_{22} ) + h_{12}$ etc.,
which in turn can then be solved for the $\epsilon$'s in terms
of the $h_{\mu\nu}$'s.
One would then obtain
\bea
\epsilon_1 & = -1 + [ 1 & + h_{11} ]^{1/2} 
\nonumber \\
\epsilon_3 & = -1 + [ 1 & + 
\half (h_{11} + h_{22}) + h_{12} ) ]^{1/2}
\nonumber \\
\epsilon_7 & = -1 + [ 1 & +  
\third ( h_{11} + h_{22} + h_{33} ) 
\nonumber \\
&& + \twoth ( h_{12} + h_{23} + h_{13} ) ]^{1/2} 
\nonumber \\
\epsilon_{15} & = -1 + [ 1 & + 
\quarter ( h_{11} + h_{22} + h_{33} + h_{44} )
\nonumber \\
&& + \thrqu ( h_{12} + h_{13} + h_{14} + h_{23} + h_{24} + h_{34} )
]^{1/2}
\nonumber \\
\label{eq:etoh}
\eea
and so on for the other edges, by suitably permuting indices.
These relations can then be expanded out for weak $h$, giving for example
\bea
\epsilon_1 & = & \half \, h_{11} + O( h^2) 
\nonumber \\
\epsilon_3 & = & \half \, h_{12} + \quarter \, ( h_{11} +  h_{22} ) + O( h^2)  
\nonumber \\
\epsilon_7 & = & \sixth \, ( h_{12} + h_{13} + h_{23} ) +
\sixth \, ( h_{23} +  h_{13} + h_{12} )
\nonumber \\
&& + \sixth \, ( h_{11} +  h_{22} + h_{33} ) + O( h^2)  
\nonumber \\
\eea
and so on.
The above correspondence between the $\epsilon$'s and the $h_{\mu\nu}$
are the underlying reason for the existence of the rotation matrices
$S$ and $T$ of Eqs.~(\ref{eq:s-matrix}) and (\ref{eq:t-matrix}), with one further
important amendment: on the hypercubic lattice four edges within a given simplex 
are assigned to one vertex, while the remaining six edges are
assigned to neighboring vertices, and require therefore a translation
back to the base vertex of the hypercube, using the result of Eq.~(\ref{eq:trans}).
This explains the additional factors of $\omega$ appearing in the
rotation matrices $S$ and $T$.
More importantly, one would expect such a combined rotation to be
independent of what particular term in the lattice action one is considering, implying that it can be used to
relate other lattice gravity contributions, such as the cosmological term
and higher derivative terms, to their continuum counterparts
(Hamber and Williams, 1993).

The choice of gauge in Eq.~(\ref{eq:gauge-fix-1}) is of course 
not the only possible one.
Another possible choice is the so-called vacuum gauge 
for which in the continuum $ h_{ik,k}=0 $, $ h_{00}=h_{0i}=0 $.
Expressed in terms of the lattice small fluctuation variables
such a condition reads in momentum space 
\bea
e_8  & = & 0 \nonumber \\
e_9  & = &  \half  \; \omega_8 \; e_1 \nonumber \\
e_{10} & = & \half \; \omega_8 \; e_2 \nonumber \\
e_{12} & = & \half \; \omega_8 \; e_4 \nonumber \\
e_{11} & = & \third \;  (1 + \omega_8 ) e_3 - 
\; \sixth \; (1 - \omega_8 ) ( \omega_2 e_1 + \omega_1 e_2 ) \nonumber \\
e_{13} & = & \third \; (1 + \omega_8 ) e_5 - 
\; \sixth \; (1 - \omega_8 ) ( \omega_4 e_1 + \omega_1 e_4 ) \nonumber \\
e_{14} & = & \third \; (1 + \omega_8 ) e_6 - 
 \; \sixth \; (1 - \omega_8 ) ( \omega_2 e_4 + \omega_4 e_2 )
\label{eq:vac-gauge}
\eea
One can then evaluate the lattice action in such a gauge and
again compare to the continuum expression.
First one expands again the $e_i$'s in terms of the $h_{ij}$'s, as given in 
Eq.~(\ref{eq:etoh}), and then expand out the $\omega$'s in powers of $k$. 
If one then sets $k_4=0$ one finds that the resulting contribution
can be re-written as the sum of two parts
(Hamber and Williams, 2005), the first part being
the transverse-traceless contribution
\beq
\quarter {\bf k}^2 \,
\Tr [ \; {}^3 h (P \; {}^3 h P - \half P \Tr (P \; {}^3 h )) ]
= \quarter {\bf k}^2 \, \bar h_{ij}^{TT} ({\bf k}) \; h^{TT}_{ij} ({\bf k})
\eeq
\beq
\bar h^{TT}_{ij} h^{TT}_{ij}  \equiv
\Tr [ \; {}^3 h (P \; {}^3 h P - \half P \Tr (P \; {}^3 h )) ]
\eeq
with $P_{ij} = \delta_{ij} - k_i k_j / {\bf k}^2 $ acting on 
the three-metric ${}^3 h_{ij}$,
and the second part arising due to the trace component of the metric
\beq
-  \quarter {\bf k}^2 \,
\Tr [ P \, \Tr (P \; {}^3 h) P \Tr (P \; {}^3 h ) ]
\; = \; {\bf k}^2 \, \bar h_{ij}^{T} ({\bf k}) \; h^{T}_{ij} ({\bf k})
\eeq
with $h^T= \half P \, \Tr (P \; {}^3 h) $. 
In the vacuum gauge $ h_{ik,k}=0 $, $ h_{ii}=0 $, $ h_{0i}=0 $ 
one can further solve for the metric components 
$h_{12}$, $h_{13}$, $h_{23}$ and $h_{33}$
in terms of the two remaining degrees of freedom, $h_{11}$ and $h_{22}$, 
\bea
h_{12} & = & - { 1 \over  2 k_1 k_2 } 
( h_{11} k_1^2 + h_{22} k_2^2 + h_{11} k_3^2 + h_{22} k_3^2 ) \nonumber \\
h_{13} & = & - { 1 \over  2 k_1 k_3 } 
( h_{11} k_1^2 - h_{22} k_2^2 - h_{11} k_3^2 - h_{22} k_3^2 ) \nonumber \\
h_{23} & = & - { 1 \over  2 k_2 k_3 } 
( - h_{11} k_1^2 + h_{22} k_2^2 - h_{11} k_3^2 - h_{22} k_3^2 ) \nonumber \\
h_{33} & = & - h_{11} - h_{22} 
\eea
and show that the second (trace) part vanishes.

The above manipulations underscore the fact that the lattice action,
in the weak field limit and for small momenta, 
only propagates transverse-traceless modes, as for linearized
gravity in the continuum.
They can be used to derive an expression for the lattice analog 
of the result given in (Kuchar, 1972) and (Hartle, 1982) for the
vacuum wave functional of linearized gravity, which
gives therefore a suitable starting point for a lattice
candidate for the same functional.

A cosmological constant term can be analyzed in the lattice
weak field expansion along similar lines.
According to Eqs.~(\ref{eq:totlatvol}) or (\ref{eq:totlatvol1}) it is given on the lattice by the total 
space-time volume, so that the action contribution is given by
\beq
I_V \; = \; \lambda_0 \sum_ {\rm edges \; h } \; V_h ,
\eeq
where $V_h$ is defined to be the volume associated with an edge $h$.
The latter is obtained by subdividing the volume of each four-simplex 
into contributions 
associated with each hinge (here via a barycentric subdivision),
and then adding up the contributions from each four-simplex touched
by the given hinge.
Expanding out in the small edge fluctuations one has
\beq
I_{V} \; \sim \; \sum_{n} \;
( \epsilon_1^{(n)} + \epsilon_2^{(n)} + \epsilon_4^{(n)} + \epsilon_8^{(n)})
+ \half \, \sum_{m n, i j} \;
\epsilon_i^{(m) \; T} ~ M_{i,j}^{(m,n)} ~ \epsilon_j^{(n)}
\eeq
One needs to be careful since the expansion of $\epsilon_i$ in
terms of $h_{\mu\nu}$ contains terms quadratic in $h_{\mu\nu}$, so
that there are additional diagonal contributions to the small fluctuation
matrix $L_\omega$,
\beq
\epsilon_1 + \epsilon_2 + \epsilon_4 + \epsilon_8
\, = \,
\half \, ( h_{11} + h_{22} + h_{33} + h_{44}) -
\eigth \, ( h_{11}^2 + h_{22}^2 + h_{33}^2 + h_{44}^2) + \cdots
\eeq
These additional contributions are required for the volume term
to reduce to the continuum form
of Eq.~(\ref{eq:h-cosm}) for small momenta
and to quadratic order in the weak field expansion.

Next the same set of rotations needs to be performed as for the Einstein
term, in order to go from the lattice variables $\epsilon_i$ 
to the continuum variables $\bar h_{\mu\nu}$.
After the combined $S_\omega$- and $T_{\omega}$-matrix rotations 
of Eqs.~(\ref{eq:s-matrix}) and (\ref{eq:t-matrix}) one obtains
for the small fluctuation matrix $L_{\omega}$ arising from
the gauge-fixed lattice Einstein-Regge term [see Eq.~(\ref{eq:lattwfe})]
\beq
L_{\omega} \; = \; - \half \; k^2 \; V ,
\eeq
with the matrix $V$ given by Eq.~(\ref{eq:v-matrix}).
Since the lattice cosmological term can also be expressed in terms of the
matrix $V$,
\beq
\sqrt{g} \, = \, 1 + \half h_{\mu\mu} -
\half h_{\alpha\beta} V^{\alpha\beta\mu\nu} h_{\mu\nu} + O(h^3) ,
\eeq
one finds, as in the continuum, for the combined Einstein and 
cosmological constant terms
\beq
\lambda_0 \; ( 1 + \half \, h_{\mu\mu} )
+ \half \cdot {k \over 2} \;
h_{\alpha\beta} V^{\alpha\beta\mu\nu} \; 
( \partial^2 + {2 \lambda_0 \over k} )
\; h_{\mu\nu} + O(h^3)  \;\; ,
\eeq
corresponding in this gauge to the exchange of a particle 
of "mass" $\mu^2 = - 2 \lambda_0 / k$, in agreement with the
continuum weak field result of Eq.~(\ref{eq:h-quadr-gf-tv2}).
As for the Regge-Einstein term, there are higher order
lattice corrections to the cosmological
constant term of $O(k)$ (which are completely absent in the continuum,
since no derivatives are present there).
These should be irrelevant in the lattice continuum limit.

\subsubsection{Strong Coupling Expansion}

\label{sec:strong}

In this section the strong coupling
(large $G$ or small $k=1/(8 \pi G)$)
expansion of the lattice gravitational functional integral will be discussed.
The resulting series is expected to be useful up to some $k=k_c$,
where $k_c$ is the lattice critical point,
at which the partition function develops a singularity.

There will be two main aspects to the following discussion.
The first aspect will be the development of a systematic expansion
for the partition function and the correlation functions
in powers of $k$, and a number of rather general considerations that follow
from it.
The second main aspect will be a detailed analysis and
interpretation of the individual terms which appear order by order 
in the strong coupling expansion. 
This second part will lead to a later discussion of what
happens for large $d$.

One starts from the lattice regularized
path integral with action Eq.~(\ref{eq:latac}) and 
measure Eq.~(\ref{eq:lattmeas}).
In the following we will focus at first on the four-dimensional
case. 
Then the four-dimensional Euclidean lattice action contains
the usual cosmological constant and Regge scalar curvature terms
of Eq.~(\ref{eq:ilatt})
\beq 
I_{latt} \; = \;  \lambda \, \sum_h V_h (l^2) \, - \, 
k \sum_h \delta_h (l^2 ) \, A_h (l^2) \;\; , 
\label{eq:ilatt1} 
\eeq
with $k=1/(8 \pi G)$, and possibly additional higher derivative terms
as well.
The action only couples edges which belong either to
the same simplex or to a set of neighboring simplices, and can therefore
be considered as {\it local}, just like the continuum action. 
It leads to a lattice partition function defined in Eq.~(\ref{eq:zlatt})
\beq 
Z_{latt} \; = \;  \int [ d \, l^2 ] \; e^{ 
- \lambda_0 \sum_h V_h \, + \, k \sum_h \delta_h A_h } \;\; ,
\label{eq:zlatt-1} 
\eeq
where, as customary, the lattice ultraviolet cutoff is set equal to one
(i.e. all length scales are measured in units of the lattice cutoff).
For definiteness the measure will be of the form 
\beq
\int [ d \, l^2 ] \; = \;
\int_0^\infty \; \prod_s \; \left ( V_d (s) \right )^{\sigma} \;
\prod_{ ij } \, dl_{ij}^2 \; \Theta [l_{ij}^2] \;\; .
\eeq
The lattice partition function $Z_{latt}$ should be compared to the
continuum Euclidean Feynman path integral of Eq.~(\ref{eq:zcont}),
\beq
Z_{cont} \; = \; \int [ d \, g_{\mu\nu} ] \; e^{ 
- \lambda \, \int d x \, \sqrt g \, + \, 
{ 1 \over 16 \pi G } \int d x \sqrt g \, R} \;\; .
\eeq
When doing an expansion in the kinetic term
proportional to $k$, it will be convenient to include the
$\lambda$-term in the measure. 
We will set therefore in this Section as in Eq.~(\ref{eq:mulatt})
\beq 
d \mu (l^2) \; \equiv \; [ d \, l^2 ] \, e^{- \lambda_0 \sum_h V_h }
\;\; .
\label{eq:mulatt1} 
\eeq
It should be clear that this last expression represents a fairly non-trivial 
quantity, both in view of the relative complexity
of the expression for the volume of a simplex, Eq.~(\ref{eq:vol}),
and because of the generalized triangle inequality constraints 
already implicit in $[d\,l^2]$.
But, like the continuum functional measure, it is certainly {\it local},
to the extent that each edge length
appears only in the expression for the volume of those simplices
which explicitly contain it.
Also, we note that in general the integral $\int d \mu$ can only be
evaluated numerically; nevertheless this can be done, at least in principle, 
to arbitrary precision.
Furthermore, $\lambda_0$ sets the overall scale and can therefore be set 
equal to one without any loss of generality.

Thus the effective strong coupling measure of 
Eq.~(\ref{eq:mulatt1}) has the properties
that (a) it is local in the lattice metric of Eq.~(\ref{eq:latmet}),
to the same extent that the continuum measure is ultra-local, 
(b) it restricts all edge lengths to be positive, and (c) it imposes
a soft cutoff on large simplices due to the $\lambda_0$-term and the
generalized triangle inequalities.
Apart from these constraints, it 
does {\it not} significantly restrict the fluctuations
in the lattice metric field at short distances.
It will be the effect of the curvature term to restrict such
fluctuation, by coupling the metric field between simplices, in
the same way as the derivatives appearing in the continuum Einstein
term couple the metric between infinitesimally close space-time points.

As a next step, $Z_{latt}$ is expanded in powers of $k$,
\beq 
Z_{latt}(k) \; = \;  \int d \mu (l^2) \, \; e^{k \sum_h \delta_h \, A_h } 
\; = \;  \sum_{n=0}^{\infty} \, { 1 \over n!} \, k^n \, 
\int d \mu (l^2) \, \left ( \sum_h \delta_h \, A_h \right )^n \;\; .
\label{eq:zlatt-k}
\eeq
It is easy to show that $Z (k) \, = \, \sum_{n=0}^{\infty} a_n \, k^n $
is analytic at $k=0$, so this expansion should be well defined up to the
nearest singularity in the complex $k$ plane.
A quantitative estimate for the expected location of such a singularity
in the large-$d$ limit will be given later in Sec.~\ref{sec:larged}.
Beyond this singularity $Z(k)$ can sometimes be extended, for example, 
via Pad\'e or differential approximants
\footnote{
A first order transition cannot affect
the singularity structure of $Z(k)$ as viewed from the strong coupling phase,
as the free energy is $C_\infty$ at a first order transition. 
}.
The above expansion is of course analogous to the high temperature expansion
in statistical mechanics systems,
where the on-site terms are treated exactly and
the kinetic or hopping term is treated as a perturbation.
Singularities in the free energy or its derivatives can usually
be pinned down with the knowledge of a large enough number of terms in the
relevant expansion (Domb and Green, 1973).

Next consider a fixed, arbitrary hinge on the lattice, and
call the corresponding curvature term in the action $\delta A $.
Such a contribution will be denoted in the following, as is customary
in lattice gauge theories, a {\it plaquette} contribution.
For the average curvature on that hinge one has
\beq 
< \delta A > \; = \; 
{\displaystyle \sum_{n=0}^{\infty} \, { 1 \over n!} \, k^n \, 
\int d \mu (l^2) \, \delta A \, \left ( \sum_h \delta_h \, A_h \right )^n
\over
\displaystyle \sum_{n=0}^{\infty} \, { 1 \over n!} \, k^n \, 
\int d \mu (l^2) \, \left ( \sum_h \delta_h \, A_h \right )^n } \;\; .
\label{eq:rseries}
\eeq
After expanding out in $k$ the resulting expression, one
obtains for the cumulants
\beq
< \delta \, A > \; = \; \sum_{n=0}^{\infty} c_n \, k^n \;\; ,
\label{eq:rseries2}
\eeq
with 
\beq
c_0 \; = \; 
{\displaystyle 
\int d \mu (l^2) \, \delta \, A 
\over
\displaystyle \int d \mu (l^2) } \;\; ,
\label{eq:k0}
\eeq
whereas to first order in $k$ one has
\beq
c_1 \; = \; 
{\displaystyle 
\int d \mu (l^2) \, \delta \, A \, \left ( \sum_{h} \delta_{h} \, A_{h} \right )
\over
\displaystyle \int d \mu (l^2) }
\; - \;
{\displaystyle 
\int d \mu (l^2) \, \delta \, A \, \cdot 
\int d \mu (l^2) \, \sum_{h} \delta_{h} \, A_{h}
\over
\displaystyle \left ( \int d \mu (l^2) \right )^2 } \;\; .
\label{eq:k1}
\eeq
This last expression clearly represents a measure of the fluctuation
in $\delta \, A$, namely 
$[ \langle ( \sum_{h} \delta_{h} \, A_{h} )^2 \rangle - 
\langle \sum_{h} \delta_{h} \, A_{h} \rangle^2 ] / N_{h} $, 
using the homogeneity properties of the lattice
$\delta A \rightarrow \sum_{h} \delta_{h} A_{h} / N_{h} $.
Here $N_h$ is the number of hinges in the lattice. 
Equivalently, it can be written in an even more compact way as
$N_{h} [ \langle (\delta A)^2 \rangle - \langle \delta A \rangle^2 ] $. 

To second order in $k$ one has
$c_2 = N_{h}^2 \, [ \langle (\delta A)^3 \rangle - \, 3 \langle \delta A \rangle \langle ( \delta A )^2 \rangle
+ \,  2 \langle \delta A \rangle^3 ]/2 $.
At the next order one has 
$c_3 = N_{h}^3 \, [ \langle (\delta A)^4 \rangle - \, 4 \langle \delta A \rangle \langle ( \delta A )^3 \rangle
- \, 3 \langle (\delta A)^2 \rangle^2 
+ \, 12 \langle (\delta A)^2 \rangle \langle \delta A \rangle^2
- \,  6 \langle \delta A \rangle^4 ]/6 $, 
and so on.
Note that the expressions in square parentheses become rapidly quite small,
$O(1/N_h^n)$ with increasing order $n$, as a result of large 
cancellations that must arise eventually between individual terms
inside the square parentheses.
In principle, a careful and systematic numerical evaluation of the above
integrals (which is quite feasible in practice) would allow the
determination of the expansion coefficients in $k$ for the average
curvature $<\delta A>$ to rather high order.

As an example, consider a non-analyticity in the average scalar curvature
\beq
{\cal R} (k) \; = \; 
{ < \int d x \, {\textstyle {\sqrt{g(x)}} \displaystyle} \, R(x) >
\over < \int d x \, {\textstyle {\sqrt{g(x)}} \displaystyle} > } 
\, \approx \,
{ < \sum_h \delta_h \, A_h  >
\over <  \sum_h V_h  > }  \; ,
\label{eq:rave}
\eeq
assumed for concreteness to be of the form of an algebraic singularity at
$k_c$, namely
\beq
{\cal R} (k) \; \mathrel{\mathop\sim_{ k \rightarrow k_c}} \;
A_{\cal R} \, ( k_c - k )^\delta
\label{eq:r-nonan}
\eeq
with $\delta$ some exponent.
It will lead to a behavior, for the general term in the series in $k$,
of the type
\beq
(-1)^n \, A_{\cal R} \, { (\delta -n +1) (\delta -n +2) \dots \delta 
\over n! \, k_c^{n - \delta} } \, k^n \;\; .
\eeq
Given enough terms in the series, the singularity structure can
then be investigated using a variety of increasingly sophisticated
series analysis methods.

It can be advantageous to isolate in the above expressions the {\it local}
fluctuation term, from those terms that involve {\it correlations} between
different hinges.
To see this, one needs to go back, for example,
to the first order expression in Eq.~(\ref{eq:k1})
and isolate in the sum $\sum_{h}$ the contribution which
contains the selected hinge with value $\delta A$, namely
\beq
\sum_{h} \delta_{h} \, A_{h} \; = \;
\delta \, A \, + \, \sum_{h}\, ' \, \delta_{h} \, A_{h} \;\; ,
\eeq
where the primed sum indicates that the term containing $\delta A$
is {\it not} included.
The result is
\bea
c_1 \; = \; && {\displaystyle 
\int d \mu (l^2) \, ( \delta \, A )^2 
\over
\displaystyle \int d \mu (l^2) }
\; - \;
{\displaystyle 
\left ( \int d \mu (l^2) \, \delta \, A \, \right )^2 
\over
\displaystyle \left ( \int d \mu (l^2) \right )^2 }
\nonumber \\
&& \; + \;
{\displaystyle 
\int d \mu (l^2) \, \delta \, A \, 
\sum_{h}\, ' \, \delta_{h} \, A_{h}
\over
\displaystyle \int d \mu (l^2) }
\; - \;
{\displaystyle 
\left ( \int d \mu (l^2) \, \delta \, A \right ) \, \left (
\int d \mu (l^2) \, \sum_{h}\, ' \, \delta_{h} \, A_{h} \right )
\over
\displaystyle \left ( \int d \mu (l^2) \right )^2 } \;\; .
\label{eq:k1ex}
\eea
One then observes the following: the first two terms describe the
{\it local} fluctuation of $\delta A$ on a given hinge;
the third and fourth terms describe {\it correlations} between $\delta A$
terms on {\it different} hinges.
But because the action is {\it local}, the only non-vanishing
contribution to the last two terms comes from edges and hinges 
which are in the immediate vicinity of the hinge in question.
For hinges located further apart (indicated below by ``$not \, nn$'')
one has that their fluctuations
remain uncorrelated, leading to a vanishing variance 
\beq
{\displaystyle 
\int d \mu (l^2) \, \delta \, A \, 
\sum_{h \, {\rm not} \, nn}\, ' \, \delta_{h} \, A_{h}
\over
\displaystyle \int d \mu (l^2) }
\; - \;
{\displaystyle 
\left ( \int d \mu (l^2) \, \delta \, A \right ) \, \left (
\int d \mu (l^2) \, \sum_{h \, {\rm not} \, nn}\, ' \, 
\delta_{h} \, A_{h} \right )
\over
\displaystyle \left ( \int d \mu (l^2) \right )^2 } \; = \; 0 \;\; ,
\label{eq:variance}
\eeq
since for uncorrelated random variables $X_n$'s, $<X_n X_m>-<X_n><X_m>=0$.
Therefore the only non-vanishing contributions in the last two terms
in Eq.~(\ref{eq:k1ex}) come from hinges which are {\it close} to each other.

The above discussion makes it clear that a key quantity is the
{\it correlation} between different plaquettes,
\beq
< ( \delta \, A )_{h} \, ( \delta \, A)_{h'} > \; = \; 
{\displaystyle 
\int d \mu (l^2) \,
( \delta \, A)_{h} \, ( \delta \, A)_{h'} \,
e^{k \sum_h \delta_h \, A_h }
\over
\displaystyle \int d \mu (l^2) \, e^{k \sum_h \delta_h \, A_h } } \;\;
,
\label{eq:corr}
\eeq
or, better, its {\it connected} part (denoted here by $< \dots >_C$)
\beq
< ( \delta \, A )_{h} \, ( \delta \, A)_{h'} >_C \;\; \equiv \;\;
< ( \delta \, A )_{h} \, ( \delta \, A)_{h'} > \; - \;
< ( \delta \, A )_{h} > \, < ( \delta \, A)_{h'} > \;\; ,
\label{eq:corrconn}
\eeq
which subtracts out the trivial part of the correlation.
Here again the exponentials in the numerator and denominator can be
expanded out in powers of $k$, as in Eq.~(\ref{eq:rseries}).
The lowest order term in $k$ will involve the correlation
\beq
\int d \mu (l^2) \, ( \delta \, A)_{h} \, ( \delta \, A)_{h'} \;\; .
\eeq
But unless the two hinges are close to each other, they will fluctuate
in an uncorrelated manner, with
$< ( \delta \, A )_{h} \, ( \delta \, A)_{h'} > -
< ( \delta \, A )_{h} > < ( \delta \, A)_{h'} > \, = \, 0 $.
In order to achieve a non-trivial correlation, the path between
the two hinges 
$h$ and $h'$ needs to be tiled by at least as many terms from
the product $ ( \sum_h \delta_h \, A_h )^n $ in
\beq
\int d \mu (l^2) \, ( \delta \, A)_{h} \, ( \delta \, A)_{h'} \,
\left ( \sum_h \delta_h \, A_h \right )^n
\eeq
as are needed to cover the distance $l$ between the two hinges.
One then has
\beq
< ( \delta \, A )_{h} \, ( \delta \, A)_{h'} >_C \; \sim \; 
k^l \; \sim \; e^{- l / \xi } \;\; ,
\label{eq:kxi}
\eeq
with the correlation length $ \xi = 1 / | \log k | \rightarrow 0 $
to lowest order as $k \rightarrow 0 $
(here we have used the usual definition of the correlation length $\xi$,
namely that a generic correlation function
is expected to decay as $ \exp (- {\rm distance} / \xi) $ for
large separations)
\footnote{This statement, taken literally, oversimplifies the situation a bit, 
as depending on the spin (or tensor structure) of the operator appearing in the
correlation function, the large distance decay of the corresponding
correlator is determined by the lightest excitation in that specific channel.
But in the gravitational context one is mostly concerned with correlators
involving spin two (transverse-traceless) objects, evaluated at fixed
geodesic distance.
}.
This last result is quite general, and holds for example irrespective
of the boundary conditions (unless of course $\xi \sim L$, where $L$ is the
linear size of the system, in which case a path can be found
which wraps around the lattice).

\begin{figure}[h]
\epsfig{file=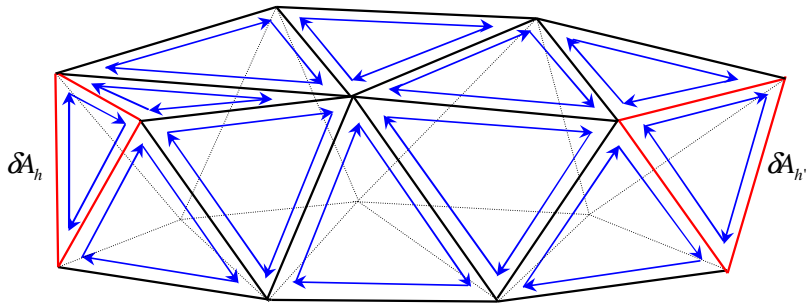,width=12cm}
\caption{Correlations between action contributions on hinge $h$
and hinge $h'$ arise to lowest order in the strong coupling expansions
from diagrams describing a narrow tube connecting the two hinges.
Here vertices represent points in the dual lattice, with
the tube-like closed surface tiled with parallel transport polygons.
For each link of the dual lattice, the $SO(4)$ parallel transport matrices
${\bf R}$ of Sec.~\ref{sec:rotations} are represented by an arrow.}
\label{fig:tube}
\end{figure}

But further thought reveals that the above result is in fact not
completely correct, due to the fact that in order to achieve
a non-vanishing correlation one needs, at least to lowest order,
to connect the two hinges by a narrow tube (Hamber and Williams, 2006).
The previous result should then read correctly as 
\beq
< ( \delta \, A )_{h} \, ( \delta \, A)_{h'} >_C \; \sim \; 
\left ( k^{n_d} \right )^l \;\; ,
\label{eq:nd}
\eeq
where $n_d \, l$ represents the minimal number of dual
lattice polygons needed to form a closed surface connecting
the hinges $h$ and $h'$, with $l$ the actual distance (in lattice units) 
between the two hinges.
Fig.~\ref{fig:tube} provides an illustration of the situation.

With some additional effort many additional terms can be computed in the strong
coupling expansion.
In practice the method is generally not really competitive with direct
numerical evaluation of the path  integral via Monte Carlo methods.
But it does provide a new way of looking at the functional
integral, and provide the basis for new approaches, such
as the large $d$ limit to be discussed in the
second half of the next section.

\subsubsection{Discrete Gravity in the Large-d Limit}

\label{sec:larged}

In the large-$d$ limit the geometric expressions for volume, areas and
angles simplify considerably, and as will be shown below one can obtain
a number of interesting results for lattice gravity.
These can then be compared to earlier investigations of continuum
Einstein gravity in the same limit (Strominger, 1981).

Here we will consider a general simplicial lattice in $d$ dimensions, made out
of a collection of flat $d$-simplices glued together at their common faces 
so as to constitute a triangulation of a smooth continuum manifold,
such as the $d$-torus or the surface of a sphere.
Each simplex is endowed with $d+1$ vertices, and its geometry
is completely specified by assigning the lengths of its $d(d+1)/2$ edges.
We will label the vertices by $1, 2, 3, \dots , d+1$ and denote the 
square edge lengths by $l _ {12}^2 = l _ {21}^2$, ... , $l _ {1,d+1}^2 $.

As discussed in Sec.~\ref{sec:volumes}, the volume of a $d$-simplex 
can be computed from the determinant of a $(d+2) \times (d+2)$ matrix,
\beq
V_d \; = \; {(-1)^{d+1 \over 2 } \over d! \, 2^{d/2} } \,
\left|\matrix{
0      &    1     &    1     & \ldots \cr
1      &    0     & l_{12}^2 & \ldots \cr 
1      & l_{21}^2 &    0     & \ldots \cr
1      & l_{31}^2 & l_{32}^2 & \ldots \cr
\ldots &  \ldots  &  \ldots  & \ldots \cr
1      & l_{d+1,1}^2 & l_{d+1,2}^2 & \ldots \cr
}\right| ^{1/2}  .
\eeq
If one calls the above matrix $M_d$ then
previous expression can the re-written as
\beq
V_d \; = \; {(-1)^{d+1 \over 2 } \over d! \, 2^{d/2} } 
\, \sqrt{ \det  M_d } \;\; ,
\eeq
In general the formulae for volumes and angles are quite complicated
and therefore of limited use in large dimensions.
The next step consists in expanding them out in 
terms of small edge length variations, by setting
\beq
l_{ij}^2 \; = \; l_{ij}^{(0)\, 2} \; + \; \delta \, l_{ij}^2 \;\; .
\eeq
From now on we will set $ \delta \, l_{ij}^2 \, = \, \epsilon_{ij} $.
Unless stated otherwise, we will be considering the expansion about
the equilateral case, and set $l_{ij}^{(0)} = 1$; later on this
restriction will be relaxed.
In the equilateral case one has for the volume of a simplex
\beq
V_d \; = \; { 1 \over d! } \sqrt{ d+1 \over 2^d } \;\; ,
\label{eq:vequilat}
\eeq
From the well-known expansion for determinants
\bea
\det ( 1 + M ) \, & = & \, e^{ \tr \ln (1+M) }  
\nonumber \\
& = & \, 1 \, + \, \tr M \, + \, 
{1 \over 2!} \left [ ( \tr M )^2 \, - \, \tr M^2 \right ]
\, + \, \dots \; .
\nonumber \\
\eea
one finds after a little algebra
\beq
V_d \; \mathrel{\mathop\sim_{ d \rightarrow \infty }} \;
{ \sqrt{d} \over d! \, 2^{d/2} } \, \left \{ 
1 \, - \, 
\half \, \epsilon_{12}^2 \, + \, \dots \, + \,
{\textstyle {1 \over d } \displaystyle} \, 
( \epsilon_{12} \, + \, \dots \, + \, \epsilon_{12} \, \epsilon_{13} 
\, + \, \dots ) \, + \, O ( d^{-2} )  \right \} \;\; .
\label{eq:volex}
\eeq
Note that the terms linear in $\epsilon$, which would have required a
shift in the ground state value of $\epsilon$ for non-vanishing
cosmological constant $\lambda_0$, vanish to leading order in $1/d$.
The complete volume term $\lambda_0 \sum V_d $ appearing in the action can then
be easily written down using the above expressions.

In $d$ dimensions the dihedral angle in a $d$-dimensional simplex of volume
$V_d$, between faces of volume $V_{d-1}$ and $V_{d-1}^{'}$, is obtained from
Eq.~(\ref{eq:dihedral})
\beq
\sin \theta_d \; = \; { d \over d-1 } \, 
{ V_d \, V_{d-2} \over V_{d-1} \, V_{d-1}^{'} } \;\; .
\label{eq:dihedral-d}
\eeq
In the equilateral case one has for the dihedral angle
\beq
\theta_d \; = \; \arcsin { \sqrt{d^2 -1} \over d } \; 
\mathrel{\mathop\sim_{ d \rightarrow \infty } } \;
{ \pi \over 2 } \, - \, { 1 \over d } \, - \, { 1 \over 6 \, d^3 } 
\, + \, \dots \;\; ,
\label{eq:arcsin}
\eeq
which will require {\it four} simplices to meet on a hinge, to give
a deficit angle of $ 2 \pi - 4 \times {\pi \over 2 } \approx 0 $ in large
dimensions.
One notes that in large dimensions the simplices look locally
(i.e. at a vertex) more like hypercubes.
Several $d$-dimensional simplices will meet on a $(d-2)$-dimensional
hinge, sharing a common face of dimension $d-1$ between adjacent simplices.
Each simplex has $(d-2)(d-1)/2$ edges ``on'' the hinge, 
some more edges are then situated on the two ``interfaces'' between
neighboring simplices meeting at the hinge, and finally one edge 
lies ``opposite'' to the hinge in question.

In the large $d$ limit one then obtains, to leading order for the dihedral
angle at the hinge with vertices labelled by $1 \dots d-1$
\bea
\theta_d & \; \mathrel{\mathop\sim_{ d \rightarrow \infty }} \; &
\arcsin { \sqrt{d^2 -1} \over d } \, + \,
\epsilon_{d,d+1} \, + \, \epsilon_{1,d} \, \epsilon_{1,d+1} \, + \, \dots 
\nonumber \\
\, && + \, {\textstyle {1 \over d } \displaystyle} \, \left ( 
- \, \epsilon_{1,d} \, + \, \dots \, - \, 
\half \, \epsilon_{1,d}^2 \, + \, \dots \, - \, 
\half \, \epsilon_{d,d+1}^2 
\, - \, \epsilon_{12} \, \epsilon_{1,d+1} 
\, - \, \epsilon_{1,d} \, \epsilon_{3,d+1} 
\, - \, \epsilon_{1,d} \, \epsilon_{d,d+1}  \, + \, \dots \right )
\nonumber \\
&& \, + \, O ( d^{-2} ) \;\; .
\label{eq:dihedr}
\eea

From the expressions in Eq.~(\ref{eq:volex}) for the volume
and (\ref{eq:dihedr}) for the dihedral angle one can then
evaluate the $d$-dimensional Euclidean lattice action, 
involving cosmological constant and scalar curvature terms
as in Eq.~(\ref{eq:latac})
\beq
I (l^2) \; = \; 
\lambda_0 \, \sum \, V_d \; - \; k \, \sum \,  \delta_d \, V_{d-2} \;\; ,
\label{eq:latac1}
\eeq
where $\delta_d$ is the $d$-dimensional deficit angle,
$\delta_d \; = \; 2 \, \pi - \sum_{\rm simplices} \theta_d $.
The lattice functional integral is then
\beq
Z ( \lambda_0, \, k ) \; = \; \int [ d \, l^2 ] \, \exp \left ( - I(l^2)
\right ) \;\; .
\label{eq:zdef1}  
\eeq
To evaluate the curvature term $- k \sum \delta_d V_{d-2} $
appearing in the gravitational lattice action one needs the
hinge volume $V_{d-2}$, which is easily obtained from Eq.~(\ref{eq:volex}), 
by reducing $d \rightarrow d-2$.

We now specialize to the case where four simplices meet at a hinge.
When expanded out in terms of the $\epsilon$'s one obtains for the
deficit angle
\bea
\delta_d \; & = & \; 2 \, \pi - 4 \cdot {\pi \over 2} \, + \,
\sum_{\rm simplices} {\textstyle {1 \over d } \displaystyle} 
\, - \, \epsilon_{d,d+1} \, + \, \dots 
\, - \, \epsilon_{1,d} \, \epsilon_{1,d+1} \, + \, \dots 
\nonumber \\
\, && - \, {\textstyle {1 \over d } \displaystyle} \, \left ( 
- \, \epsilon_{1,d} \, - \, \half \, \epsilon_{1,d}^2
\, - \, \half \, \epsilon_{d,d+1}^2 \, - \, 
\epsilon_{12} \, \epsilon_{1,d+1} \, - \, 
\epsilon_{1,d} \, \epsilon_{3,d+1} \, - \, 
\epsilon_{1,d} \, \epsilon_{d,d+1} \, + \, \dots \right ) 
\, + \, O( {1 \over d^2} ) .
\nonumber \\
\label{eq:deficit-d}
\eea
The action contribution involving the deficit angle is then, for a single hinge,
\beq
- \, k \, \delta_d \, V_{d-2} \, = \, ( - \, k ) \, 
{ 2 \, d^{3/2} \, (d-1) \over d! \, 2^{d/2} } \, \left (
\, - \, \epsilon_{d,d+1} \, + \, \dots \, - \, 
\epsilon_{1,d} \, \epsilon_{1,d+1} \, + \, \dots \, \right ) \;\; .
\label{eq:curvterm}
\eeq
It involves two types of terms: one linear in the (single) edge
opposite to the hinge, as well as a term involving a product
of two distinct edges, connecting any hinge vertex to the two vertices
opposite to the given hinge.
Since there are four simplices meeting on one hinge, one will
have 4 terms of the first type, and $4(d-1)$ terms of the second type.

To obtain the total action, a sum over all simplices, resp. hinges, has
still to be performed.
Dropping the irrelevant constant term and summing over edges
one obtains for the total action 
$\lambda_0 \sum V_d - k \sum \delta_d \, V_{d-2}$
in the large $d$ limit
\beq
\lambda_0 \, \left ( \, - \, \half \, \sum \epsilon_{ij}^2 \,  \right )
\, - \, 2 \, k \, d^2 \, \left ( 
\, - \, \sum \epsilon_{jk} \, - \, \sum \epsilon_{ij} \, \epsilon_{ik} \, 
\right ) \;\; ,
\label{eq:totact}
\eeq
up to an overall multiplicative factor $ \sqrt{d} / d! \, 2^{d/2} $,
which will play no essential role in the following.

The next step involves the choice of a specific lattice. 
Here we will evaluate the action for the cross polytope $\beta_{d+1}$.
The cross polytope $\beta_n$
is the regular polytope in $n$ dimensions corresponding
to the convex hull of the points formed by permuting the coordinates 
$(\pm 1, 0, 0, ..., 0)$, and has therefore $2n$ vertices.
It is named so because its  vertices are located equidistant
from the origin, along the Cartesian axes in $n$-space.
The cross polytope in $n$ dimensions
is bounded by $2^n$ $(n-1)$-simplices, has $2n$ vertices
and $2n(n-1)$ edges.

\begin{figure}[h]
\epsfig{file=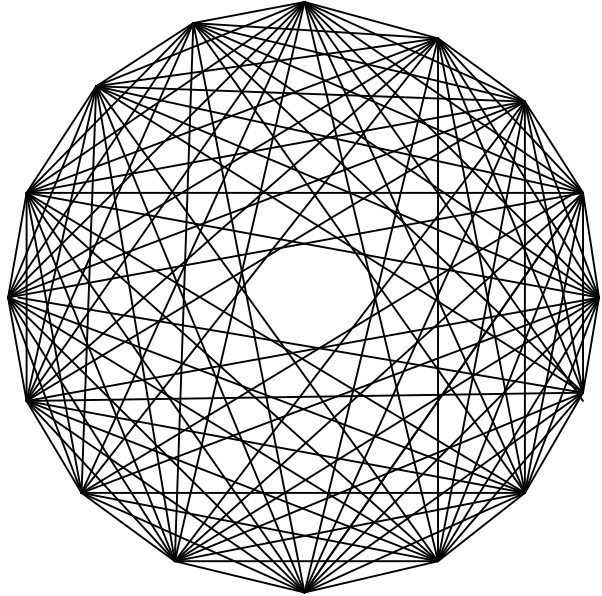,width=8cm}
\caption{Cross polytope $\beta_n$ with $n=8$ and $2n=16$
vertices, whose surface can be used to define a simplicial manifold of
dimension $d=n-1=7$. For general $d$, the cross polytope $\beta_{d+1}$
will have $2(d+1)$ vertices, connected to each other by $2d(d+1)$ edges.}
\label{fig:polytope}
\end{figure}

In three dimensions, it represents the convex hull of the
octahedron, while in four dimensions the cross polytope  is the 16-cell
(Coxeter 1948; Coxeter 1974).
In the general case it is dual to a hypercube in $n$ dimensions, with
the `dual' of a regular polytope being another regular
polytope having one vertex in the center of each cell of the polytope 
one started with.
Fig.~\ref{fig:polytope} shows as an example the polytope $\beta_8$.

When we consider the surface of the cross polytope in $d+1$ dimensions,
we have an object of dimension $n-1=d$, which 
corresponds to a triangulated manifold with no boundary, homeomorphic
to the sphere.
From Eq.~(\ref{eq:deficit-d}) the deficit angle is then given to 
leading order by
\beq
\delta_d \; = \; 0 \, + \,  { 4 \over d } \, - \, \left (
\epsilon_{d,d+1} \, + \, {\rm 3 \, terms} \, + \, 
\epsilon_{1,d} \, \epsilon_{1,d+1} \, + \, \dots \right ) \, + \, \dots 
\eeq
and therefore close to flat in the large $d$ limit (due to our
choice of an equilateral starting configuration).
Indeed if the choice of triangulation is such that the deficit angle is not
close to zero, then the discrete model leads to an average curvature
whose magnitude is comparable to the lattice spacing or ultraviolet cutoff,
which from a physical point of view does not seem very attractive:
one obtains a space-time with curvature radius comparable to the
Planck length.

When evaluated on such a manifold the lattice action becomes
\beq
{ \sqrt{d} \, 2^{d/2} \over d! } \, 2 \,
\left ( \lambda_0 \, - \, k \, d^3 \right ) \, \left [ 
1 \, - \, {1 \over 8} \, \sum \epsilon_{ij}^2 \, + \, 
{ 1 \over d } \left ( { 1 \over 4 } \sum \epsilon_{ij} \, + \, 
{ 1 \over 8 } \sum \epsilon_{ij} \, \epsilon_{ik} \right ) \, + \, O(1/d^2) 
\right ] \;\; .
\label{eq:beta-pol}
\eeq
Dropping the $1/d$ correction the action is proportional to
\beq
- \, \half \, \left ( \lambda_0 \, - \, k \, d^3 \right ) \, 
\sum \epsilon_{ij}^2 \;\; .
\label{eq:beta2}
\eeq
Since there are $2 d (d+1)$ edges in the cross polytope, one finds
therefore that, at the critical point $k d^3 = \lambda_0$, the
quadratic form in $\epsilon$, defined by the above action,
develops $2 d (d+1) \sim 2 d^2 $ zero eigenvalues.

This result is quite close to the $d^2/2$ zero eigenvalues expected
in the continuum for large $d$, with the factor of four discrepancy
presumably attributed to an underlying intrinsic ambiguity that
arises when trying to identify lattice points with points in the continuum.

It is worth noting here 
that the competing curvature ($k$) and cosmological constant 
($\lambda_0$) terms will have comparable magnitude when
\beq
k_c \; = \; { \lambda_0 \, l_0^2 \over \, d^3 } \;\; .
\label{eq:kc4}
\eeq
Here we have further allowed for the possibility that the
average lattice spacing $l_0 = \langle l^2 \rangle^{1/2}$ is not equal to one
(in other words, we have restored the appropriate overall scale
for the average edge length, which is in fact largely determined by the
value of $\lambda_0$).

The average lattice spacing $l_0$ can easily be estimated from the
following argument. 
The volume of a general equilateral simplex is given by
Eq.~(\ref{eq:vequilat}), multiplied by an additional factor of $l_0^d$.
In the limit of small $k$ the average volume of a simplex
is largely determined by the cosmological term, 
and can therefore be computed from
\beq
< V > \, = \, - \, {\partial \over \partial \lambda_0 } \,
\log \, \int [dl^2] \, e^{- \lambda_0 V (l^2) } \;\; ,
\label{eq:vd}
\eeq
with $V(l^2) = ( \sqrt{d+1} / d! \, 2^{d/2} ) \, l^d \equiv c_d l^d $.
After doing the integral over $l^2$ with measure $dl^2$ 
and solving this last expression for $l_0^2$ one obtains 
\beq
l_0^2 = { 1 \over \lambda_0^{2/d} } \left [ 
{2 \over d } \, { d! \, 2^{d/2} \over \sqrt{d+1} } \right ]^{2/d}
\label{eq:l2d}
\eeq
(which, for example, gives $l_0 = 2.153$ for $\lambda_0=1$ in four dimensions,
in reasonable agreement with the actual value 
$l_0 \approx 2.43$ found near the transition point).

This then gives for $\lambda_0=1$ the estimate 
$k_c = \sqrt{3} / (16 \cdot 5^{1/4} ) = 0.0724$ in $d=4$,
to be compared with $k_c = 0.0636(11)$ obtained in (Hamber, 2000)
by direct numerical simulation in four dimensions.
Even in $d=3$ one finds again for $\lambda_0=1$, from Eqs.~(\ref{eq:kc4}) 
and (\ref{eq:l2d}), $k_c= 2^{5/3} / 27 = 0.118$, to be compared with 
$k_c=0.112(5)$ obtained in (Hamber and Williams, 1993)
by direct numerical simulation.

Using Eq.~(\ref{eq:l2d}) inserted into Eq.~(\ref{eq:kc4})
one obtains in the large $d$ limit for the dimensionless
combination $k / \lambda_0^{(d-2)/d}$ 
\beq
{ k_c \over \lambda_0^{1 - 2/d} } \; = \; 
{ 2^{1+ 2/d } \over d^3 } \, \left [ 
{ \Gamma (d) \over \sqrt{d+1} } \right ]^{2/d}  \;\; .
\label{eq:kcd1}
\eeq

To summarize, an expansion in powers of $1/d$ can be developed,
which relies on a combined use of the weak field expansion.
It can be regarded therefore as a double expansion in $1/d$
and $\epsilon$, valid wherever the fields are smooth enough and the
geometry is close to flat, which presumably is the
case in the vicinity of the lattice critical point at $k_c$.


A somewhat complementary $1/d$ expansion can be set up, which does not require
weak fields, but relies 
instead on the strong coupling (small $k=1/8 \pi G$, or large $G$) limit.
As such it will be a double expansion in $1/d$ and $k$.
Its validity will be in a regime where the fields are not smooth, and
in fact will involve lattice field configurations
which are very far from smooth at short distances.

The general framework for the strong coupling expansion for pure 
quantum gravity was outlined in the previous section, and
is quite analogous to what one does in gauge theories
(Balian, Drouffe and Itzykson, 1975).
One expands $Z_{latt}$ in powers of $k$ as in Eq.~(\ref{eq:zlatt-k})
\beq 
Z_{latt}(k) \; = \;  \int d \mu (l^2) \, \; e^{k \sum_h \delta_h \, A_h } 
\; = \;  \sum_{n=0}^{\infty} \, { 1 \over n!} \, k^n \, 
\int d \mu (l^2) \, \left ( \sum_h \delta_h \, A_h \right )^n \;\; .
\eeq
Then one can show that dominant diagrams contributing
to $Z_{latt}$ correspond to closed
surfaces tiled with elementary transport loops.
In the case of the hinge-hinge connected correlation function
the leading contribution at strong coupling
come from closed surfaces anchored on the two hinges,
as in Eq.~(\ref{eq:nd}).

It will be advantageous to focus on general properties of the 
parallel transport matrices ${\bf R}$, discussed previously
in Sec.~\ref{sec:rotations}.
For smooth enough geometries, with small curvatures, these 
rotation matrices can be chosen to be close to the identity.
Small fluctuations in the geometry will then imply small deviations
in the ${\bf R}$'s from the identity matrix.
But for strong coupling ($k \rightarrow 0$) 
the measure $\int d \mu (l^2)$ does not significantly restrict fluctuations
in the lattice metric field.
As a result we will assume that these fields can be regarded, 
in this regime, as basically unconstrained random variables, only subject
to the relatively mild constraints implicit in the measure $d \mu$.
The geometry is generally far from smooth
since there is no coupling term to enforce long range
order (the coefficient of the lattice Einstein term is zero),
and one has as a consequence large local fluctuations in the geometry.
The matrices ${\bf R}$ will therefore fluctuate with the local
geometry, and average out to zero, or a value close to
zero.
In the sense that, for example, the $SO(4)$ rotation
\beq
{\bf R}_\theta = \left ( \matrix{ 
\cos \theta & - \sin \theta & 0 & 0 \cr
\sin \theta &   \cos \theta & 0 & 0 \cr 
0 & 0 & 1 & 0 \cr     0 & 0 & 0 & 1 \cr 
} \right )
\eeq 
averages out to zero when integrated over $\theta$. 
In general an element of $SO(n)$ is described by $n(n-1)/2$ independent 
parameters, which in the case at hand can be conveniently chosen as the six
$SO(4)$ Euler angles.
The uniform (Haar) measure over the group is then
\beq
d \mu_H ( {\bf R} ) = {1 \over 32 \pi^9 }
\int_0^{2 \pi} d \theta_1 \int_0^{\pi} d \theta_2 \int_0^{\pi} d \theta_3
\int_0^{\pi} d \theta_4 \sin \theta_4 \int_0^{\pi} d \theta_5 \sin \theta_5
\int_0^{\pi} d \theta_6 \sin^2 \theta_6
\eeq
This is just a special case of the general $n$ result, which reads
\beq
d \mu_H ( {\bf R} ) = 
\left ( \prod_{i=1}^n \Gamma (i/2) / 2^n \, \pi^{n(n+1)/2} \right )
\prod_{i=1}^{n-1} \prod_{j=1}^i 
\sin^{j-1} \theta_{\; j}^i \, d \theta_{\; j}^i
\eeq
with $ 0 \le \theta_{\; k}^1 < 2 \pi $, $ 0 \le \theta_{\; k}^j < \pi $.

\begin{figure}[h]
\epsfig{file=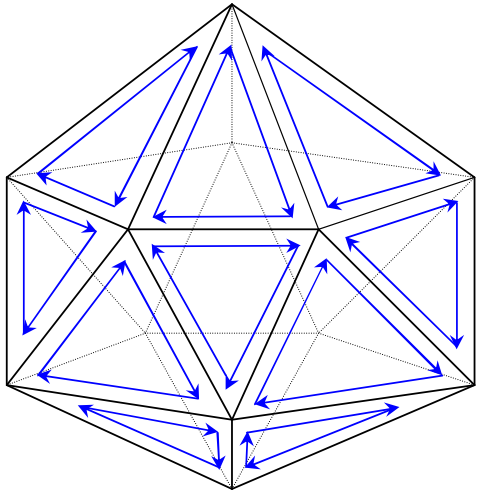,width=8cm}
\caption{Elementary closed surface tiled with parallel
transport polygons, here chosen to be triangles for illustrative
purposes. For each link of the dual lattice, the elementary parallel transport
matrices ${\bf R}(s,s')$ are represented by an arrow. 
In spite of the fact that the (Lorentz) matrices ${\bf R}$ fluctuate with
the local geometry, two contiguous, oppositely oriented arrows always
give ${\bf R} {\bf R}^{-1} = 1$.}
\label{fig:icos}
\end{figure}

These averaging properties of rotations are quite similar of course to what happens in $SU(N)$ Yang-Mills theories,
or even more simply in (compact) QED, where the analogues of the $SO(d)$ rotation
matrices ${\bf R}$ are phase factors $U_{\mu}(x)=e^{iaA_{\mu}(x)}$.
There one has $\int { d A_{\mu} \over 2 \, \pi } \, U_{\mu}(x) = 0 $
and 
$\int { d A_{\mu} \over 2 \, \pi } \, U_{\mu}(x) \, U^{\dagger}_{\mu}(x) = 1 $.
In addition, for two contiguous closed paths $C_1$ and $C_2$
sharing a common side one has
\beq
e^{ i \oint_{C_1} {\bf A \cdot dl } } \, e^{ i \oint_{C_2} {\bf A \cdot dl } }
\; = \; e^{ i \oint_{C} {\bf A \cdot dl } }
\; = \; e^{ i \int_S {\bf B \cdot n } \, dA } \;\; ,
\eeq
with $C$ the slightly larger path encircling the two loops.
For a closed surface tiled with many contiguous infinitesimal closed loops
the last expression evaluates to $1$, due to the
divergence theorem. 
In the lattice gravity case the discrete analog of this last result
is considerably more involved, and ultimately represents the (exact) lattice
analog of the contracted Bianchi identities.
An example of a closed surface tiled with parallel transport polygons
(here chosen for simplicity to be triangles) is shown in 
Fig.~\ref{fig:icos}.

As one approaches the critical point, $k \rightarrow k_c$, one is interested
in random surfaces which are of very large extent.
Let $n_p$ be the number of polygons in the surface, and set $n_p = T^2$
since after all one is describing a surface.
The critical point then naturally corresponds to the appearance
of surfaces of infinite extent,
\beq
n_p \; = \; T^{2} \; 
\sim \; 
{ 1 \over  k_c - k }  \; \rightarrow \; \infty \;\; .
\label{eq:T}
\eeq
A legitimate parallel is to the simpler case of scalar field theories, 
where random walks of length $T$ describing particle paths become of
infinite extent at the critical point, situated where the inverse of
the (renormalized) mass $\xi=m^{-1}$, expressed in units of the 
ultraviolet cutoff, diverges.

In the present case of polygonal random surfaces, one can provide the
following concise argument in support of the identification
in Eq.~(\ref{eq:T}).
First approximate the discrete sums over $n$, as they appear for example
in the strong coupling expansion for the average curvature, 
Eq.~(\ref{eq:rseries}) or
its correlation, Eq.~(\ref{eq:corr}), by continuous integrals over
areas
\beq
\sum_{n=0}^\infty \, c_n \, \left ( { k \over k_c } \right )^n  \;
\rightarrow \;
\int_0^\infty dA \, A^{\gamma-1} \left ( { k \over k_c } \right )^A
\; = \; 
\Gamma ( \gamma ) \, \left ( \log { k_c \over k } \right )^{-\gamma}
\;\; ,
\label{eq:areasum}
\eeq
where $A \equiv T^2$ is the area of a given surface.
The $A^{\gamma-1}$ term can be regarded as counting the
multiplicity of the surface (its entropy, in statistical mechanics terms).
The exponent $\gamma$ depends on the specific quantity one is looking at.
For the average curvature one has from Eq.~(\ref{eq:r-nonan}) $\gamma=-\delta$, 
while for its derivative,
the curvature fluctuation (the curvature correlation function at
zero momentum), one expects $\gamma=1-\delta$.
The saddle point is located at
\beq
A \; = \; { ( \gamma - 1 ) \over \log { k_c \over k } } 
\; \mathrel{\mathop\sim_{ k \rightarrow k_c }} \;
{ ( \gamma - 1 ) \, k_c \over k_c - k } \;\; .
\eeq
From this discussion one then concludes that close to the critical point
very large areas dominate, as claimed in Eq.~(\ref{eq:T}).

Furthermore, one would expect that the universal geometric scaling properties
of such a (closed) surface would not depend on its short distance
details, such as whether it is constructed out
of say triangles or more complex polygons.
In general excluded volume effects at finite $d$ will provide constraints on
the detailed geometry of the surface, but as $d \rightarrow \infty$ these
constraints can presumably be neglected and one is dealing then with a
more or less unconstrained random surface.
This should be regarded as a direct consequence of the fact that
as $d \rightarrow \infty$ there are infinitely many dimensions
for the random surface to twist and fold into, giving a negligible
contribution from unallowed (by interactions) directions.
In the following we will assume that this is indeed the case, and that
no special pathologies arise, such as the collapse of the
random surface into narrow tube-like, lower dimensional geometric
configurations.
Then in the large $d$ limit the problem simplifies considerably.

Related examples for what is meant in this
context are the simpler cases of random walks in infinite
dimensions, random polymers and random surfaces
in gauge theories (Drouffe, Parisi and Sourlas, 1979),
which have been analysed in detail in the large-$d$ limit.
There too the problem simplifies considerably in such a limit since
excluded volume effects (self-intersections) can be neglected there as well.
A summary of these results, with a short derivation, is given
in the appendices of (Hamber and Williams, 2006).

Following (Gross, 1984) one can define the partition function
for such an ensemble of unconstrained random surfaces, and one finds that
the mean square size of the surface increases logarithmically with
the intrinsic area of the surface.
This last result is usually interpreted as the statement that an
unconstrained random surface has infinite fractal (or Hausdorff) dimension.
Although made of very many triangles (or polygons), the random
surface remains quite compact in overall size, as viewed
from the original embedding space.
In a sense, an unconstrained random surface is a much more compact object
than an unconstrained random walk, for which $< {\bf X}^2 > \sim T $.
Identifying the size of the random surface with the gravitational
correlation length $\xi$ then gives
\beq
\xi \; \sim \; \sqrt{ \log T} 
\; \mathrel{\mathop\sim_{ k \rightarrow k_c }} \;
\vert \log ( k_c - k ) \vert^{1/2} \;\; .
\label{eq:xilog}
\eeq
From the definition of the exponent $\nu$, namely
$\xi \sim (k_c - k)^{-\nu}$, the above result then implies
$\nu = 0$ (i.e. a weak logarithmic singularity) at $d=\infty$.

It is of interest to contrast the result $\nu \sim 0$ for gravity in
large dimensions with what one finds for scalar (Wilson and Fisher, 1972; Wilson 1973) and gauge (Drouffe, Parisi and Sourlas, 1979) fields, in the same limit $d=\infty$.
So far, known results can be summarized as follows
\bea
& {\rm scalar \; field } \;\;\;\;\;\;\;\;\;\;\;\; & \nu \; = \; \half 
\nonumber \\
& {\rm lattice \; gauge \; field } \;\; & \nu \; = \; \quarter
\nonumber \\
& {\rm lattice \; gravity } \;\;\;\;\;\;\; & \nu \; = \; 0 \;\; . 
\eea
It should be regarded as encouraging that the new value obtained here,
namely $\nu=0$ for gravitation, appears to some extent to be consistent with
the general trend observed for lower spin, at least at infinite dimension.
What happens in finite dimensions? 
The situation becomes much more complicated since the self-intersection
properties of the surface have to be taken into account.
But a simple geometric argument then suggests in finite but large
dimensions $\nu=1/(d-1)$ (Hamber and Williams, 2004).

\section{NUMERICAL STUDIES IN FOUR DIMENSIONS}

\label{sec:numerical}

The exact evaluation of the lattice functional integral
for quantum gravity by numerical methods
allows one to investigate a regime
which is generally inaccessible by perturbation theory, where
the coupling $G$ is strong and quantum fluctuations
are expected to be large.

The hope in the end is to make contact with the analytic
results obtained, for example, in the $2+\epsilon$
expansion, and determine which scenarios are physically
realized in the lattice regularized model, and then
perhaps even in the real world.

Specifically, one can enumerate several major questions
that one would like to get at least partially answered.

\begin{enumerate}

\item[$\circ$]
The first one is: which scenarios suggested by perturbation
theory are realized in the lattice theory?
Perhaps a stable ground state for the quantum theory cannot be found, which
would imply that the regulated theory is still inherently
pathological.

\item[$\circ$]
Furthermore, if a stable ground state exists for some range
of bare parameters, does it require the
inclusion of higher derivative couplings in an essential
way, or is the minimal theory, with an
Einstein and a cosmological term, sufficient? 

\item[$\circ$]
Does the presence of dynamical matter, say in the form
of a massless scalar field, play an important role, or
is the non-perturbative dynamics of gravity largely
determined by the pure gravity sector (as in Yang-Mills
theories)?

\item[$\circ$]
Is there any indication that the non-trivial ultraviolet fixed
point scenario is realized
in the lattice theory in four dimensions? This would imply, 
as in the non-linear sigma model, the existence of at least
two physically distinct phases and non-trivial exponents.
Which quantity can be used as an order parameter to physically 
describe, in a {\it qualitative}, way the two phases?

\item[$\circ$]
A clear physical characterization of the two phases would
allow one, at least in principle, to decide which phase, 
if any, could be realized in nature.
Ultimately this might or might not be possible based on
purely qualitative aspects.
As will discussed below, the lattice continuum limit
is taken in the vicinity of the fixed point,
so close to it is the physically most relevant regime.

\item[$\circ$]
At the next level one would hope to be able to establish
a {\it quantitative} connection with those continuum
perturbative results
which are not affected by uncontrollable errors, such
as for example the $2+\epsilon$ expansion of Sec.~\ref{sec:graveps}.
Since the lattice cutoff and the method of dimensional regularization
cut the theory off in the ultraviolet in rather different
ways, one needs to compare universal quantities
which are {\it cutoff-independent}.
One example is the critical exponent $\nu$, as well as any other
non-trivial scaling dimension that might arise.
Within the $2+\epsilon$ expansion only {\it one}
such exponent appears, to {\it all} orders in the loop
expansion, as $ \nu^{-1} = - \beta ' (G_c) $.
Therefore one central issue in the lattice regularized theory 
is the value of the universal exponent $\nu$.

\end{enumerate}

Knowledge of $\nu$ would allow one to be more
specific about the running of the gravitational
coupling.
One purpose of the discussion in Sec.~\ref{sec:largen} was
to convince the reader that the exponent $\nu$
determines the renormalization group running
of $G (\mu^2)$ in the vicinity of the fixed
point, as in Eq.~(\ref{eq:grun-nonlin}) for
the non-linear $\sigma$-model, and more appropriately
in Eq.~(\ref{eq:grun-cont1}) for quantized gravity.
From a practical point of view, on the lattice it is 
difficult to determine the running of 
$G(\mu^2)$ directly from correlation functions ,
since the effects from the running of $G$ are
generally small.
Instead one would like to make use of the analog
of Eqs.~(\ref{eq:m-coeff}),
(\ref{eq:m-largen}) and (\ref{eq:m-largen1})
for the non-linear $\sigma$-model, and, again, more appropriately
of Eqs.~(\ref{eq:m-cont1}) and possibly (\ref{eq:m-cont2})
for gravity to determine $\nu$, and from there the running
of $G$.
But the correlation length $\xi = m^{-1}$ is also difficult
to compute, since it enters the curvature 
correlations at fixed geodesic distance, which are hard
to compute for (genuinely geometric) reasons to be discussed later.
Furthermore, these generally decay exponentially in the distance
at strong $G$, and can therefore be difficult to
compute due to the signal to noise problem of
numerical simulations.

Fortunately the exponent $\nu$ can be determined instead, 
and with good accuracy, from singularities of the derivatives
of the path integral $Z$, whose singular part is
expected, on the basis of very general arguments,
to behave in the vicinity of the fixed point
as $F \equiv - {1 \over V} \ln Z \sim \xi^{-d} $
where $\xi$ is the gravitational correlation length.
From Eq.~(\ref{eq:m-cont1}) relating $\xi (G) $ to
$G-G_c$ and $\nu$ one can then determine $\nu$,
as well as the critical coupling $G_c$.

\subsection{Observables, Phase Structure and Critical Exponents}
\label{sec:phases}

The starting point is once again the lattice regularized
path integral with action as in Eq.~(\ref{eq:latac}) and
measure as in Eq.~(\ref{eq:lattmeas}).
Then the lattice action for pure four-dimensional Euclidean 
gravity contains
a cosmological constant and Regge scalar curvature term
as in Eq.~(\ref{eq:ilatt})
\beq 
I_{latt} \; = \;  \lambda_0 \, \sum_h  V_h (l^2) \, - \, 
k \sum_h \delta_h (l^2 ) \, A_h (l^2) \;\; , 
\label{eq:ilatt2} 
\eeq
with $k=1/(8 \pi G)$, 
and leads to the regularized lattice functional integral
\beq 
Z_{latt} \; = \;  \int [ d \, l^2 ] \; e^{ 
- \lambda_0 \sum_h V_h \, + \, k \sum_h \delta_h A_h } \;\; ,
\eeq
where, as customary, the lattice ultraviolet cutoff is set equal to one
(i.e. all length scales are measured in units of the lattice cutoff).
The lattice measure is given in Eq.~(\ref{eq:lattmeas}) and
will be therefore of the form 
\beq
\int [ d \, l^2 ] \; = \;
\int_0^\infty \; \prod_s \; \left ( V_d (s) \right )^{\sigma} \;
\prod_{ ij } \, dl_{ij}^2 \; \Theta [l_{ij}^2] \;\; .
\eeq
with $\sigma$ a real parameter given below.

Ultimately the above lattice partition function $Z_{latt}$ is intended as a
regularized form of the continuum Euclidean Feynman path integral
of Eq.~(\ref{eq:zcont}),
\beq
Z_{cont} \; = \; \int [ d \, g_{\mu\nu} ] \; e^{ 
- \lambda_0 \, \int d x \, \sqrt g \, + \, 
{ 1 \over 16 \pi G } \int d x \sqrt g \, R} \;\; .
\eeq
with functional measure over the $g_{\mu\nu}(x)$'s of the form
\beq
\int [d \, g_{\mu\nu} ] \,  \equiv \, 
\prod_x \, \left [ g(x) \right ]^{\sigma / 2} \, 
\prod_{ \mu \ge \nu } \, d g_{ \mu \nu } (x) \;\; ,
\eeq
where $\sigma$ is a real parameter constrained by the
requirement $\sigma \ge - (d+1)$.
For $\sigma= \half (d-4)(d+1)$ one obtains the De Witt measure
of Eq.~(\ref{eq:dw-dewitt}), while for
$\sigma= - (d+1)$ one recovers the original Misner
measure of
Eq.~(\ref{eq:dw-misn}).
In the following we will mostly be interested in the four-dimensional
case, for which $d=4$ and therefore $\sigma=0$ for the DeWitt measure.

It is possible to add higher derivative terms to the lattice action
and investigate how the results are affected.
The original motivation was that they would improve the
convergence properties of functional integral for the lattice theory, but
extensive numerical studies suggest that they don't seem
to be necessary after all.
In any case, with such terms included the lattice action for pure gravity
aquires the two additional terms whose lattice expressions
can be found in Eqs.~(\ref{eq:riem2-latt}) and (\ref{eq:weyl2-latt}),
\bea
I_{latt} & = &
\sum_{ h } \Bigl [ \lambda_0 \, V_h - k \, \delta_h A_h
- b \, A_h^2 \delta_h^2 / V_h  \Bigr ]
\nonumber \\
&& + \third \, ( a + 4 b ) 
\sum_ { s } \, V_{ s} 
\sum_ { h , h' \subset s } 
\epsilon_{ h, h'} \;
\Bigl ( \omega_{ h } 
\Bigl [ { \delta \over A_C } \Bigr ]_{ h }
- \omega_{ h' } \Bigl [ 
{\delta \over A_C } \Bigr ]_{ h' } \Bigr )^2 
\eea
The above action is intended as a lattice form
for the continuum action
\beq
I = \int d x \, \sqrt g \; \Bigl [ \lambda_0 - \half \, k \, R -
\quarter \, b \, R_{ \mu \nu \rho \sigma } R^{ \mu \nu \rho \sigma } +
\half \, ( a + 4 b ) \;
C_{ \mu \nu \rho \sigma } C^{ \mu \nu \rho \sigma }\Bigr ]
\eeq
and is therefore of the form in Eqs.~(\ref{eq:hdqg})
and (\ref{eq:hd-action}).
Because of its relative complexity, in the following the
Weyl term will not be considered any further, 
and $b$ will chosen so that $b=- \quarter a$.
Thus the only curvature term to be discussed here will be
a Riemann squared contribution, with a (small) positive
coefficient $ + \quarter a \rightarrow 0$.

\subsubsection{Invariant Local Gravitational Averages}
\label{sec:obs}

Among the simplest quantum mechanical averages is the
one associated with the local curvature
\beq
{\cal R} (k) \; \sim \;
{ < \int d x \, \sqrt{ g } \, R(x) >
\over < \int d x \, \sqrt{ g } > } \;\;\; ,
\eeq
The curvature associated with the quantity above is the
one that would be detected when parallel-tranporting 
vectors around infinitesimal loops, with size
comparable to the average lattice spacing $l_0$.
Closely related to it is the fluctuation
in the local curvature 
\beq
\chi_{\cal R}  (k) \; \sim \;
{ < ( \int dx \sqrt{g} \, R )^2 > - < \int dx \sqrt{g} \, R >^2 \;\;\; .
\over < \int dx \sqrt{g} > }
\eeq
The latter is related to the connected curvature correlation at
zero momentum
\beq
\chi_{\cal R} \; 
\sim \; { \int dx \int dy < \sqrt{g(x)} R(x) \; \sqrt{g(y)} R(y) >_c
\over < \int dx \sqrt{g(x)} > } \;\;\; .
\eeq
Both ${\cal R}(k)$ and $\chi_{\cal R}(k)$ are directly
related to derivatives of $Z$ with respect to $k$,
\beq
{\cal R} (k) \, \sim \,
\frac{1}{V} \frac{\partial}{\partial k} \ln Z  \; ,
\label{eq:avrz} 
\eeq
and 
\beq
\chi_{\cal R}  (k) \, \sim \,
\frac{1}{V} \frac{\partial^2}{\partial k^2} \ln Z \; .
\label{eq:chirz} 
\eeq
Thus a divergence or non-analyticity in $Z$, as caused for example
by a phase transition, is expected to show up in these local averages as well.
Note that the above expectation values are manifestly invariant, since
they are related to derivatives of $Z$.

On the lattice one prefers to define quantities in such a way that
variations in the average lattice spacing $\sqrt{<l^2>}$ are compensated by
an appropriate factor determined from dimensional considerations.
In the case of the average curvature one defines therefore 
the lattice quantity ${\cal R}$ as
\beq
{\cal R} (k) \; \equiv \; 
<l^2> { < 2 \; \sum_h \delta_h A_h > \over < \sum_h V_h > } \;\;\; ,
\label{eq:avr} 
\eeq
and similarly for the curvature fluctuation,
\beq
\chi_{\cal R}  (k) \; \equiv \; 
{ < (\sum_h \delta_h A_h)^2 >
- < \sum_h \delta_h A_h >^2 \over < \sum_h V_h > } \;\;\; ,
\label{eq:chir} 
\eeq
Fluctuations in the local curvature probe graviton correlations,
and are expected to be sensitive
to the presence of a massless spin two particle.
Note that both of the above expressions are dimensionless, and
are therefore unaffected by an overall rescaling of the edge lenghts.
As in the continuum, they are proportional to first and second derivatives
of $Z_{latt}$ with respect to $k$.

One can contrast the behavior of the preceding averages,
related to the curvature, with the corresponding quantities
involving the local volumes $V_h$ (the quantity $\sqrt{g} \, dx$
in the continuum).
Consider the average volume per site
\beq
\langle V \rangle  \; \equiv \; { 1 \over N_0 } < \sum_h V_h > \;\;\; ,
\label{eq:avv} 
\eeq
and its fluctuation, defined as
\beq
\chi_V (k) \; \equiv \; 
{ < (\sum_h V_h)^2 >
- < \sum_h V_h >^2 \over < \sum_h V_h > } \;\;\; ,
\label{eq:chiv} 
\eeq
where $V_h$ is the volume associated with the hinge $h$.
The last two quantities 
are again simply related to derivatives of $Z_{latt}$ with respect to the
bare cosmological constant $\lambda_0$, as for example in 
\beq
<\!V\!> \; \sim \; \frac{\partial}{\partial \lambda_0} \ln Z_{latt} \;\;\; ,
\label{eq:avvz} 
\eeq
and 
\beq
\chi_V (k) \; \sim \; \frac{\partial^2}{\partial \lambda_0^2} \ln Z_{latt} \;\;\; .
\label{eq:chivz} 
\eeq

Some useful relations and sum rules can be derived, which follow directly
from the scaling properties of the discrete functional integral.
Thus a simple scaling argument, based on neglecting the effects
of curvature terms entirely (which, as will be seen below,
vanish in the vicinity of the critical point), gives 
an estimate of the average volume per edge
[for example from Eqs.~(\ref{eq:vd}) and (\ref{eq:l2d})]
\beq
< \! V_l \! > \; \sim \; { 2 \,( 1 + \sigma \, d ) \over \lambda_0 \, d }
\; \mathrel{\mathop\sim_{ d=4, \; \sigma=0 }} \;
{ 1 \over 2 \lambda_0 } \;\; .
\eeq
where $\sigma$ is the functional measure parameter in Eqs.~(\ref{eq:gen-meas})
and (\ref{eq:lattmeas}).
In four dimensions direct numerical simulations with $\sigma =0$
(corresponding to the lattice DeWitt measure) agree
quite well with the above formula.

Some exact lattice identities can be obtained from
the scaling properties of the action and measure.
The bare couplings $k$ and $\lambda_0$ in the gravitational action are
dimensionful in four dimensions, but one can define the dimensionless
ratio $k^2 / \lambda_0$, and rescale the edge lengths so as to eliminate
the overall length scale $\sqrt{ k/\lambda_0} $.
As a consequence the path integral for pure gravity,
\beq
Z_{latt} ( \lambda_0, k, a,b) \; = \; 
\int [ d l^2 ] \; e^{ - I (l^2) } ,
\eeq
obeys the scaling law
\beq
Z_{latt} ( \lambda_0, k, a,b) \; = \;
\left ( \lambda_0 \right )^{-N_1 / 2}
Z_{latt} \left ( 1 , { k \over \sqrt{\lambda_0} }, a,b \right ) 
\label{eq:rescale1} 
\eeq
where $N_1$ represents the number of edges in the lattice, and
the $dl^2$ measure ($\sigma=0$) has been selected.
This implies in turn a sum rule for local averages, which
for the $dl^2$ measure reads
\beq
2 \lambda_0 < \sum_h V_h > \; - \; 
k < \sum_h \delta_h A_h > \; - \;  N_1  \; = \; 0 \;\;\; ,
\label{eq:sumr1} 
\eeq
and is easily derived from Eq.~(\ref{eq:rescale1}) and the definitions in
Eqs.~(\ref{eq:avrz}) and (\ref{eq:avvz}).
$N_0$ represents the number of sites in the lattice, and the
averages are defined per site (for the hypercubic lattice divided
up into simplices as in Fig.~\ref{fig:4d-cube},
$N_1 = 15$).
This last formula can be very useful in checking the accuracy of
numerical evaluations of the path integral.
A similar sum rule holds for the fluctuations
\bea
4 & \lambda_0^2 & \left [ < (\sum_h V_h)^2 > - < \sum_h V_h >^2 \right ] 
\nonumber \\
&& \; - \; 
k^2 \left [ < (\sum_h \delta_h A_h)^2 > -  < \sum_h \delta_h A_h >^2 \right ]
\; - \; 2 N_1 \; = \; 0 \;\; .
\label{eq:sumr2} 
\eea
In light of the above discussion one can therefore consider without
loss of generality the case of unit bare cosmological
coupling $\lambda_0=1$ (in units of the cutoff).
Then all lengths are expressed in units of the
fundamental microscopic length scale $\lambda_0^{-1/4}$.

\subsubsection{Invariant Correlations at Fixed Geodesic Distance}
\label{sec:corr}

Compared to ordinary field theories, new issues arise in quantum gravity
due to the fact that the physical distance between any two points $x$ and $y$
\beq
d(x,y \, \vert \, g) \; = \; \min_{\xi} \; \int_{\tau(x)}^{\tau(y)} d \tau \;
\sqrt{ \textstyle g_{\mu\nu} ( \xi )
{d \xi^{\mu} \over d \tau} {d \xi^{\nu} \over d \tau} \displaystyle } \;\; ,
\eeq
is a fluctuating function of the background metric $g_{\mu\nu}(x)$.
In addition, the Lorentz group used to classify spin states is
meaningful only as a local concept. 

In the continuum the shortest distance between two events is determined
by solving the equation of motion (equation of free fall, or geodesic
equation)
\beq
{ d^2 x^{\mu} \over d \tau^2 } \, + \, 
\Gamma^{\mu}_{\lambda\sigma} \, 
{ d x^{\lambda} \over d \tau }
{ d x^{\sigma} \over d \tau } \, = \, 0
\eeq
On the lattice the geodesic distance between two 
lattice vertices $x$ and $y$ requires the determination of
the shortest lattice path connecting several lattice vertices,
and having the two given vertices as endpoints.
This can be done at least in principle
by enumerating all paths connecting the two
points, and then selecting the shortest one.
Equivalently it can be computed from the scalar field
propagator, as in Eq.~(\ref{eq:scalar-prop}).

Consequently physical correlations have to be defined at fixed
geodesic distance $d$, as in the following correlation
between scalar curvatures
\beq
< \int d x \, \int d y \, \sqrt{g} \, R(x) \; \sqrt{g} \, R(y) \;
\delta ( | x - y | - d ) > 
\eeq
Generally these do not go to zero at large separation, so
one needs to define the connected part, by
subtracting out the value at $d=\infty$.
These will be indicated in the following by the connected $<>_c$ average,
and we will write the resulting connected curvature correlation
function at fixed geodesic distance compactly as
\beq
G_R (d) \; \sim \; < \sqrt{g} \; R(x) \; \sqrt{g} \; R(y) \;
\delta ( | x - y | -d ) >_c \; .
\eeq
One can define several more invariant correlation functions at 
fixed geodesic distance for other operators involving curvatures
(Hamber, 1994).
The gravitational correlation function just defined is similar to the one
in non-Abelian gauge theories, Eq.~(\ref{eq:box_sun}).

\begin{figure}[h]
\epsfig{file=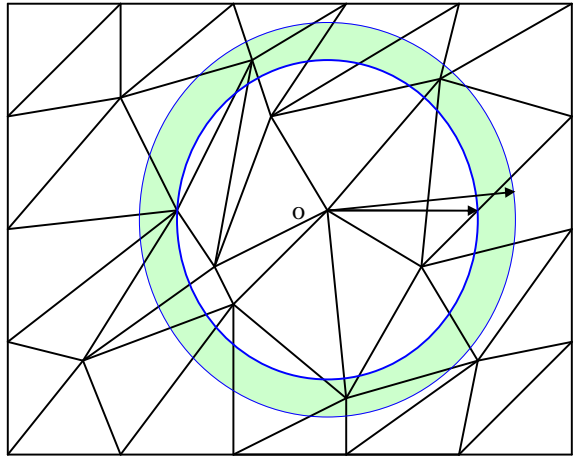,width=10cm}
\caption{Geodesic distance and correlations.
Correlation functions are computed for lattice vertices
in range $[d,d+\Delta d]$.}
\label{fig:geodesic}
\end{figure}

In the lattice regulated theory one can define similar correlations,
using for example the correspondence of Eqs.~(\ref{eq:regge-d})
or (\ref{eq:r-latt}) for the scalar curvature
\beq
\, \sqrt{g} \, R(x) \; \rightarrow \;
2 \, \sum_{ h \supset x } \delta_h A_h 
\eeq
If the deficit angles are averaged over a number of contiguous hinges $h$
sharing a common vertex $x$, one is naturally lead to the
connected correlation function
\beq
G_R (d) \; \equiv \; < \sum_{ h \supset x } \delta_h A_h \;
\sum_{ h' \supset y } \delta_{h'} A_{h'} \;
\delta ( | x - y | -d ) >_c \; ,
\eeq
which probes correlations in the scalar curvatures.
In practice the above lattice correlations have to be computed by
a suitable binning procedure: one averages out all correlations
in a geodesic distance interval $[d,d+\Delta d]$ with
$\Delta d$ comparable to one lattice spacing $l_0$.
See Fig.~\ref{fig:geodesic}.
Similarly one can construct the connected correlation functions
for local volumes at fixed geodesic distance 
\beq
G_V (d) \; \equiv \; < \sum_{ h \supset x } V_h \;
\sum_{ h' \supset y } V_{h'} \;
\delta ( | x - y | -d ) >_c \; ,
\eeq

In general one expects for the curvature
correlation either a power law decay, for distances sufficiently
larger than the lattice spacing $l_0$,
\beq
< \sqrt{g} \; R(x) \; \sqrt{g} \; R(y) \; \delta ( | x - y | -d ) >_c
\;\; \mathrel{\mathop\sim_{d \; \gg \; l_0 }} \;\; 
{ 1 \over d^{2 n} }  \;\;\;\; ,
\label{eq:rr-pow1}
\eeq
with $n$ some exponent characterizing the power law decay,
or at very large distances an exponential decay,
characterized by a correlation length $\xi$,
\beq
< \sqrt{g} \; R(x) \; \sqrt{g} \; R(y) \; \delta ( | x - y | -d ) >_c
\; \; \mathrel{\mathop\sim_{d \; \gg \; \xi }} \;\;
e^{-d / \xi } \;\;\;\; .
\label{eq:rr-exp}
\eeq
In fact the invariant correlation length $\xi$ is generally defined
(in analogy with what one does for other theories)
through the
long-distance decay of the connected, invariant correlations at
fixed geodesic distance $d$.
In the pure power law decay case of Eq.~(\ref{eq:rr-pow1})
the correlation length $\xi$ is of course infinite.
One can show from scaling considerations (see below) that
the power $n$ in Eq.~(\ref{eq:rr-pow1}) is related to
the critical exponent $\nu$ by $n=4-1/\nu$.

In the presence of a finite correlation length $\xi$ one needs
therefore to carefully distinguish between the ''short distance'' regime
\beq
l_0 \, \ll \, d \, \ll \, \xi
\eeq
where Eq.~(\ref{eq:rr-pow1}) is valid, and the
''long distance'' regime 
\beq
\xi \, \ll \, d \ll \, L
\eeq
where Eq.~(\ref{eq:rr-exp}) is appropriate. 
Here $l_0 = \sqrt{<l^2>}$ is the
average lattice spacing, and
$L=V^{1/4}$ the linear size of the system.

Recently the issue of defining diffeomorphism invariant correlations
in quantum gravity has been re-examined from a new perspective
(Giddings, Marolf and Hartle, 2006).

\subsubsection{Wilson Lines and Static Potential}
\label{sec:lattstruct}

In a gauge theory such as QED the static potential can be computed from the
manifestly gauge invariant Wilson loop.
To this end one considers the process where a particle-antiparticle
pair are created at time zero, separated by a fixed distance $R$, and
re-annihilated at a later time $T$.
In QED the amplitude for such a process associated
with the closed loop $\Gamma$ is given by the Wilson loop
\beq
W(\Gamma) \; = \; < \exp \Bigl \{ i e \oint_{\Gamma} A_{\mu} (x) dx^{\mu} 
\Bigr \} > \;\; ,
\eeq
which is a manifestly gauge invariant quantity.
Performing the required Gaussian average using the (Euclidean)
free photon propagator one obtains
\beq
< \exp \Bigl \{ i e \oint_{\Gamma} A_{\mu} dx^{\mu} \Bigr \} > \; = \;
\exp \Bigl \{ 
- \half e^2 \oint_{\Gamma} \oint_{\Gamma} dx^{\mu} dy^{\nu} 
\Delta_{\mu\nu} (x-y) \Bigr \} \;\; .
\eeq
For a rectangular loop of sides $R$ and $T$ one has after a short
calculation
\beq
< \exp \Bigl \{ i e \oint_{\Gamma} A_{\mu} dx^{\mu} \Bigr \} > \; \simeq \;
\exp \Bigl \{ 
- {e^2 \over 4 \pi \epsilon} (T+R) + {e^2 \over 4 \pi } {T \over R}
+ {e^2 \over 2 \pi^2 } \log ( { T \over \epsilon } ) + \cdots \Bigr \}
\eeq
\beq
\mathrel{\mathop\sim_{ T \gg R }} \; \exp \Bigl [ -V(R) \; T ) \Bigr ] \;\; ,
\eeq
where $\epsilon \rightarrow 0$ is an ultraviolet cutoff.
In the last line use has been made of the fact that for large imaginary times
the exponent in the amplitude involves the energy for the
process multiplied by the time $T$.
Thus for $V(R)$ itself one obtains
\beq
V(R) = - \lim_{ T \rightarrow \infty } { 1 \over T }
\log  < \exp \Bigl \{ i e \oint_{\Gamma} A_{\mu} dx^{\mu}  \Bigr \} > \;
\sim \mbox{\rm cst.} - {e^2 \over 4 \pi R }  \;\; ,
\eeq
which is the correct Coulomb potential for two oppositely
charged particles.

To obtain the static potential it is not necessary to consider
closed loops.
Alternatively, in a periodic box of length $T$ one can introduce two long
oppositely oriented parallel lines in the time
direction, separated  by a distance $R$  and closed by the
periodicity of the lattice, and associated with oppositely charged particles,
\beq
< \exp \Bigl \{ i e \int_{\Gamma} A_{\mu} dx^{\mu}  \Bigr \}
\exp \Bigl \{ i e \int_{\Gamma'} A_{\nu} dy^{\nu}  \Bigr \} > 
\eeq
In the large time limit one can then show that the result
for the potential $V(R)$ is the same.

In the gravitational case  there is no notion of ``oppositely charged
particles'', so one cannot use the closed Wilson loop to extract the
potential (Modanese 1995).
One is therefore forced to consider a process in which one introduces
two separate world-lines for the two particles.
It is well known that the equation for free fall 
can be obtained by extremizing the space-time distance travelled.
Thus the quantity
\beq
\mu  \int_{\tau(a)}^{\tau(b)} d \tau 
\sqrt{ \textstyle g_{\mu\nu} ( x )
{d x^{\mu} \over d \tau} {d x^{\nu} \over d \tau} \displaystyle } \;\; ,
\eeq
can be taken as the Euclidean action contribution associated with the
heavy spinless particle of mass $\mu$.

Next consider two particles of mass $M_1$, $M_2$, propagating along
parallel lines in the `time' direction and 
separated by a fixed distance $R$. 
Then the coordinates for the two particles can be chosen
to be $x^{\mu} = (\tau, r,0,0)$ with $r$ either $0$ or $R$.
The amplitude for this process is a product of two factors, one for
each heavy particle.
Each is of the form
\beq
L(0; \; M_1) =   \exp \Bigl \{ - M_1 \int d \tau 
\sqrt{ \textstyle g_{\mu\nu} ( x )
{d x^{\mu} \over d \tau} {d x^{\nu} \over d \tau} \displaystyle } \;
\Bigl \} \;\; .
\eeq
where the first argument indicates the spatial location of the Wilson
line.
For the two particles separated by a distance $R$ the amplitude is
\beq
\mbox{\rm Amp.}
\; \equiv W(0,R; \; M_1,M_2) = L(0; \; M_1) \; L(R; \;M_2) \;\; .
\eeq
For weak fields one sets $g_{\mu\nu} = \delta_{\mu\nu} + h_{\mu\nu}$, 
with $ h_{\mu\nu} \ll 1$, and therefore
$ g_{\mu\nu} ( x ) {d x^{\mu} \over d \tau}
{d x^{\nu} \over d \tau} = 1 + h_{00}(x) $.
Then the amplitude reduces to
\beq
W(M_1,M_2) =
\exp \Bigl \{ - M_1 \int_0^T d \tau \sqrt{ 1 + h_{00}(\tau)} \; \Bigr \} \;
\exp \Bigl \{ - M_2 \int_0^T d \tau' \sqrt{ 1 + h_{00}(\tau')} \;
\Bigr \}
\;\; .
\eeq
In perturbation theory the averaged amplitude can then be easily
evaluated (Hamber and Williams, 1995)
\beq
< W(0,R; \;  M_1,M_2) > \; 
= \exp \Bigl \{ - T \; (M_1+M_2
-  G \; { M_1 M_2 \over R} ) + \cdots
\Bigr \} \;\; .
\eeq
and the static potential has indeed the expected form, 
$ V(R) = -  G \; M_1 M_2 / R  $.
The contribution involving the sum of the two particle masses is
$R$ independent, and 
can therefore be subtracted, if the Wilson line correlation is divided
by the averages of the individual single line contribution, as in
\beq
V(R) = - \lim_{ T \rightarrow \infty } \; {1 \over T } \; \log \;
{ < W(0,R; \;  M_1,M_2) > \over < L(0; \; M_1) > < L(R; \; M_2) > }
\sim  - \; G \; {  M_1 M_2  \over R} \;\; .
\label{eq:wcorrcont}
\eeq
If one is only interested in the spatial dependence of the potential,
one can simplify things further and take $M_1=M_2=M$.
To higher order in the weak field expansion one has to take into
account multiple graviton exchanges,
contributions from graviton loops and
self-energy contributions due to other particles.

\begin{figure}[h]
\epsfig{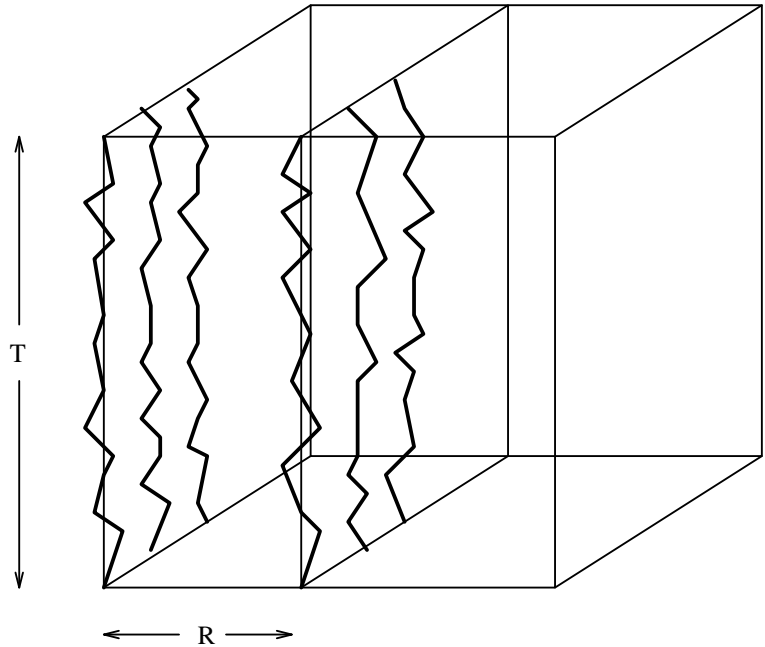}
\caption{Correlations between Wilson lines closed by the lattice
periodicity. }
\label{fig:lines}
\end{figure}

How does all this translate to the lattice theory?
At this point, the prescription for computing the Newtonian
potential for quantum gravity should be clear.
For each metric configuration (which is a given configuration of edge
lengths on the lattice) one chooses a geodesic that closes due
to the lattice periodicity (and there might be many that have
this property for the topology of a four-torus), with length $T$
(see Fig.~\ref{fig:lines}).
One then enumerates all the geodesics that lie at a fixed 
distance $R$ from the original one, and computes the associated
correlation between the Wilson lines. 
After averaging the Wilson
line correlation over many metric configurations, one extracts
the potential from the $R$ dependence of the correlation
of Eq.~(\ref{eq:wcorrcont}). 
In general since two geodesics will not be at a fixed geodesic
distance from each other in the presence of curvature, one needs
to introduce some notion of average distance, which then gives
the spatial separations of the sources $R$.

On the lattice one can construct the analog of the Wilson line for
one heavy particle,
\beq
L(x,y,z) \; = \; \exp \bigl \{ - M \sum_i l_i \bigr \} \;\; ,
\eeq
where edges are summed in the ``$t$'' direction, and the path
is closed by the periodicity of the lattice in the $t$ 
direction. 
One can envision the simplicial lattice as divided
up in hypercubes, in which case
the points $x,y,z$ can be taken as the remaining labels for the Wilson line.

For a single line one expects
\beq
< L(x,y,z) > \; = \; < \exp \bigl \{ - M \sum_i l_i \bigr \} > \;
\sim e^{  - \tilde{M} T } \;\; ,
\eeq
where $T$ is the linear size of lattice in the $t$ direction
and $\tilde M$ the renormalized mass.
The correlation between Wilson lines at average ``distance'' $R$ is
then given by 
\beq
- {1 \over T} \; \log \; \Bigl [
{ < L(x,y,0) \;  L(x,y,R) > \over < L(x,y,0) > < L(x,y,R) > }
\Bigr ] \;
\mathrel{\mathop\sim_{ T \; \gg \; R }}  \; V(R) \;\; .
\label{eq:wcorrlatt}
\eeq
Numerical studies suggest that the correct qualitative features of
the potential emerge close to the critical point.
In particular it was found that the potential is attractive close
to the critical point, and for two equal mass
particles of mass $\mu$ scales, as expected, like the mass squared.
As for any correlation in gravity, the accurate determination
of the potential as a function of distance $R$ is a more difficult task,
since at large distance the correlations are small and the statistical
noise becomes large. 
Still, the first results (Hamber and Williams, 1995) suggest that the
potential has more or less the expected classical form in the vicinity of the
critical point, both as far as the mass dependence and perhaps even the distance
dependence are concerned.
In particular it is attractive.

The correlation of Wilson lines is not the only possible method to determine 
that non-perturbative gravity is still attractive.
One alternate procedure involves the properties of correlation functions
of scalar particles in the presence of gravity (de Bakker and Smit, 1995).

\subsubsection{Scaling in the Vicinity of the Critical Point}
\label{sec:critical}

In practice the correlation functions at fixed geodesic distance
are difficult to compute numerically, and therefore not the best route
to study the critical properties.
But scaling arguments allow one to determine the scaling
behavior of correlation functions from critical exponents
characterizing the singular behavior of the {\it free energy}
and various local averages in the vicinity of the critical point.
In general a divergence of the correlation length $\xi$
\beq
\xi (k) \; \equiv \; 
\mathrel{\mathop\sim_{ k \rightarrow k_c}} \; A_\xi \;
| k_c - k | ^{ -\nu }
\label{eq:xi-k}
\eeq
signals the presence of a phase transition, and leads to the appearance
of a singularity in the free energy $F(k)$.
The scaling assumption for the free energy postulates that a divergent
correlation length in the vicinity of the critical point at $k_c$
leads to non-analyticities of the type
\bea
F \equiv - { 1 \over V } \, \ln Z & = & F_{reg} + F_{sing}
\nonumber \\
&& F_{sing} \sim \xi^{-d}
\label{eq:fsing}
\eea
where the second relationship follows simply from dimensional arguments
(the free energy is an extensive quantity).
The regular part $F_{reg}$ is generally not determined from $\xi$
by purely dimensional considerations, but
as the name implies is a regular function in the vicinity
of the critical point.
Combining the definition of $\nu$ in Eq.~(\ref{eq:xi-k}) with
the scaling assumption of Eq.~(\ref{eq:fsing}) one obtains
\beq
F_{sing} (k) \; 
\mathrel{\mathop\sim_{ k \rightarrow k_c}} \;
({\rm const.} ) \, | k_c - k | ^{ d \nu }
\eeq
The presence of a phase transition can then
be inferred from non-analytic terms
in invariant averages, such as the average curvature
and its fluctuation.
For the average curvature one obtains
\beq
{\cal R} (k) \; \mathrel{\mathop\sim_{ k \rightarrow k_c}} \;
A_{\cal R} \, | k_c - k |^{d \nu -1} \;\; ,
\label{eq:rsing}
\eeq
up to regular contributions (i.e. constant terms in the
vicinity of $k_c$).
An additive constant can be added, but numerical evidence
sor far points to this constant being consistent with zero.
Similarly one has for the curvature fluctuation
\beq
\chi_{\cal R} (k) \; \mathrel{\mathop\sim_{ k \rightarrow k_c}} \;
A_{ \chi_{\cal R} } \; | k_c - k | ^{ -(2- d \nu) } \;\;\;\; .
\label{eq:chising}
\eeq
At a critical point the fluctuation $\chi$ is in general
expected to diverge, corresponding to the
presence of a divergent correlation length.
From such averages one can therefore in principle extract
the correlation length exponent $\nu$ of 
Eq.~(\ref{eq:xi-k}) without having to compute a correlation
function.

An equivalent result, relating the quantum expectation value
of the curvature to the physical correlation length $\xi$ , is
obtained from Eqs.~(\ref{eq:xi-k}) and (\ref{eq:rsing})
\beq
{\cal R} ( \xi ) \; \mathrel{\mathop\sim_{ k \rightarrow k_c}} \;
\xi^{ 1 / \nu - 4 }
\;\; ,
\label{eq:rm}
\eeq
again up to an additive constant.
Matching of dimensionalities in this last equation is restored by
inserting an appropriate power of the Planck length $l_P=\sqrt{G}$
on the r.h.s..

One can relate the critical exponent $\nu$ to the 
scaling behavior of correlations at large distances.
The curvature fluctuation is related to the connected scalar curvature
correlator at zero momentum
\beq
\chi_{\cal R} (k) 
\sim { \int d x \int d y < \sqrt{g} \, R (x) \, \sqrt{g} \, R (y) >_c
\over < \int d x \, \sqrt{g} > }
\label{eq:chir-corr}
\eeq
A divergence in the above fluctuation is then indicative of long range
correlations, corresponding to the presence of a massless particle.
Close to the critical point one expects for large separations
$ l_0 \ll |x-y| \ll \xi$
a power law decay in the geodesic distance, as in Eq.~(\ref{eq:rr-pow1}),
\beq
< \sqrt{g} R (x) \sqrt{g} R (y) > \;
\mathrel{\mathop\sim_{ \vert x - y \vert \rightarrow \infty}} \;
\frac{1}{ \vert x-y \vert^{2n} } \;\;\;\; ,
\label{eq:rcorr}
\eeq
Inserting the above expression in Eq.~(\ref{eq:chir-corr}) and
comparing with Eq.~(\ref{eq:chising}) determines
the $n$ as $ n = d - 1 / \nu  $.
A priori one cannot exclude to possibility that some states
acquire a mass away from the critical point, in which case
the correlation functions would have the behavior of
Eq.~(\ref{eq:rr-exp}) for $ | x-y| \gg \xi$.

\subsubsection{Physical and Unphysical Phases}
\label{sec:phasephys}

An important alternative to the analytic methods in the continuum is an
attempt to solve quantum gravity directly via numerical simulations.
The underlying idea is to evaluate the gravitational functional integral
in the discretized theory $Z$ by summing over a suitable finite set of representative field configurations.
In principle such a method given enough configurations and a fine enough
lattice can provide an arbitrarily accurate solution to the
original quantum gravity theory.

In practice there are several important factors to consider, which effectively
limit the accuracy that can be achieved today in a practical calculation. 
Perhaps the most important one is the enormous amounts of computer time
that such calculations can use up.
This is particularly true when correlations of operators at fixed geodesic
distance are evaluated.
Another practical limitation is that one is mostly interested in the behavior
of the theory in the vicinity of the critical point at $G_c$, where the
correlation length $\xi$ can be quite large and significant correlations
develop both between different lattice regions, as well as among
representative field configurations, an effect known as critical slowing down.
Finally, there are processes which are not well suited to a lattice
study, such as problems with several different length (or energy) scales.
In spite of these limitations, the progress in lattice field theory has
been phenomenal in the last few years, driven in part by
enormous advances in computer technology, and in part by the
development of new techniques relevant to the problems of lattice
field theories. 

The starting point is the generation of a large ensemble of
suitable edge length configurations.
The edge lengths are updated by a
straightforward Monte Carlo algorithm,
generating eventually an ensemble of configurations distributed
according to the action and measure of Eq.~(\ref{eq:zlatt})
(Hamber and Williams, 1984; Hamber, 1984; Berg, 1985); some
more recent references are (Beirl et al, 1994;
Riedler et al, 1999; Bittner et al, 2002).
Further details of the method as applied to pure gravity 
can be found for example in the recent work (Hamber, 2000) and will not be repeated here.

\begin{figure}[h]
\epsfig{file=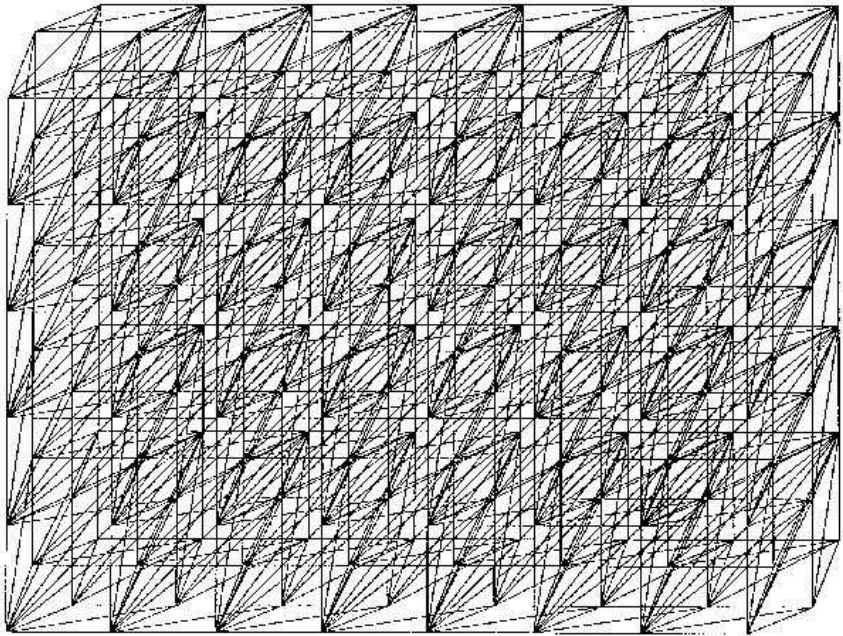,width=12cm}
\caption{Four-dimensional hypercubes divided into simplices and
stacked to form a four-dimensional lattice.}
\label{fig:lat4d}
\end{figure}

As far as the lattice is concerned, one
starts with the 4-d hypercube of Fig.~\ref{fig:4d-cube}
divided into simplices, and then stacks a number of such
cubes in such a way as to construct an arbitrarily large lattice,
as shown in Fig.~\ref{fig:lat4d}.
Other lattice structures are of course possible, including even a
random lattice.
The expectation is that for long range correlations involving
distance scales much larger than the lattice spacing the 
precise structure of the underlying lattice structure
will not matter.

The lattice sizes investigated typically range from $4^4$ sites (3840 edges)
to $32^4$ sites (15,728,640 edges).
On a dedicated massively parallel supercomputer millions of 
consecutive edge length configurations can be generated for tens
of values of $k$ in a few months time.

Even though these lattices are not very large, one should keep in mind
that due to the simplicial nature of the lattice there are many edges
per hypercube with many interaction terms, and as a consequence the
statistical fluctuations can be comparatively small, unless measurements
are taken very close to a critical point, and at rather large separation
in the case of the correlation functions or the potential.
In addition, extrapolations to the infinite volume limit can
be aided by finite size scaling methods, which exploit predictable
renormalization group properties of finite size systems. 

Usually the topology is restricted to a four-torus, corresponding
to periodic boundary conditions. 
One can perform similar calculations with lattices employing different
boundary conditions or topology, but one would expect
the universal scaling properties of the theory to be determined 
exclusively by short-distance renormalization effects.
Indeed the Feynman rules of perturbation theory do not
depend in any way on boundary terms, although some
momentum integrals might require an infrared cutoff.
 
Based on physical considerations it would seem reasonable
to impose the constraint that the scale of the curvature be
much smaller than the inverse of the average lattice spacing, 
but still considerably larger than the inverse of the
overall system size. 
Equivalently, that in momentum space physical scales
should be much smaller that the ultraviolet cutoff, but much larger
than the infrared cutoff.
A typical requirement is therefore that 
\beq
l_0 \; \ll \; \xi \; \ll \; L ,
\label{eq:window}
\eeq
where $L$ is the linear size of the system, $\xi$ the correlation
length related for example to the large scale curvature by 
${\cal R} \sim 1/\xi^2$, and $l_0$ the lattice spacing.
Contrary to ordinary lattice field theories, the lattice spacing
in lattice gravity is a dynamical quantity.
Thus the quantity $l_0 = \sqrt{<\!l^2\!>}$ only
represents an average cutoff parameter.

Furthermore the bare cosmological constant $\lambda_0$ appearing in the
gravitational action of Eq.~(\ref{eq:zlatt}) can be fixed at $1$
in units of the cutoff, since it just sets the overall length
scale in the problem.
The higher derivative coupling $a$ can be set to a value very
close to $0$ since one ultimately is interested in the limit
$a \rightarrow 0$, corresponding to the pure Einstein theory.

\begin{figure}[h]
\epsfig{file=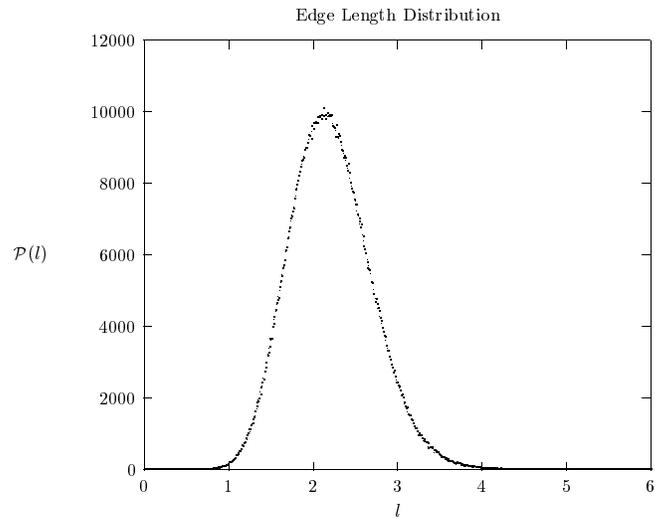,width=10cm}
\caption{A typical edge length distribution in the smooth phase for which
$k<k_c$, or $G>G_c$.
Note that the lattice gravitational measure of Eq.~(\ref{eq:dewlattmeas}) cuts
off the distribution at small edge lengths, while the cosmological
constant term prevents large edge lenghts from appearing.}
\label{fig:dist-ed}
\end{figure}

One finds that for the measure in Eq.~(\ref{eq:dewlattmeas}) this choice of
parameters leads to a well behaved ground state for $k < k_c $
for higher derivative coupling $a \rightarrow 0$.
The system then resides in the `smooth' phase, with an effective 
dimensionality close to four.
On the other hand for $k > k_c$ the curvature becomes very large
and the lattice collapses into degenerate configurations
with very long, elongated simplices.
Fig.~\ref{fig:dist-ed} shows an example of a typical edge length distribution
in the well behaved strong coupling phase close to but below $k_c$.

\begin{figure}[h]
\epsfig{file=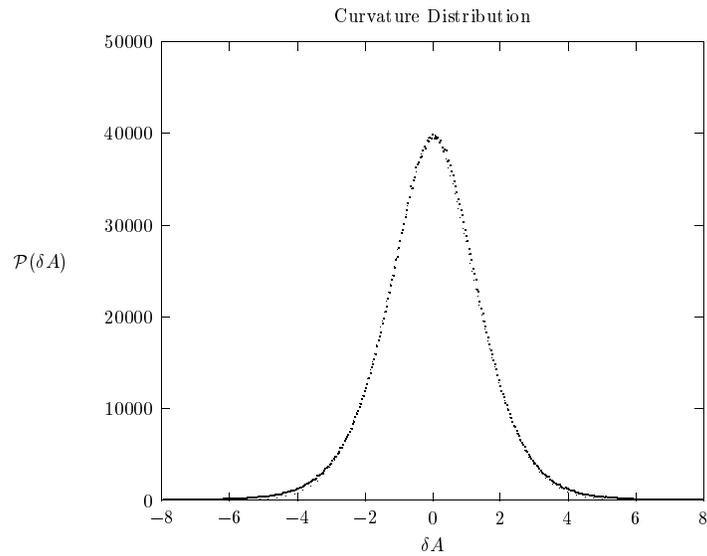,width=10cm}
\caption{A typical curvature distribution in the smooth phase for which
$k<k_c$, or $G>G_c$.
Note that distribution is peaked around close to zero curvature.}
\label{fig:dist-curv}
\end{figure}

Fig.~\ref{fig:dist-curv} shows the corresponding curvature 
($\delta A$ or $\sqrt{g}R$) distribution.

\begin{figure}[h]
\epsfig{file=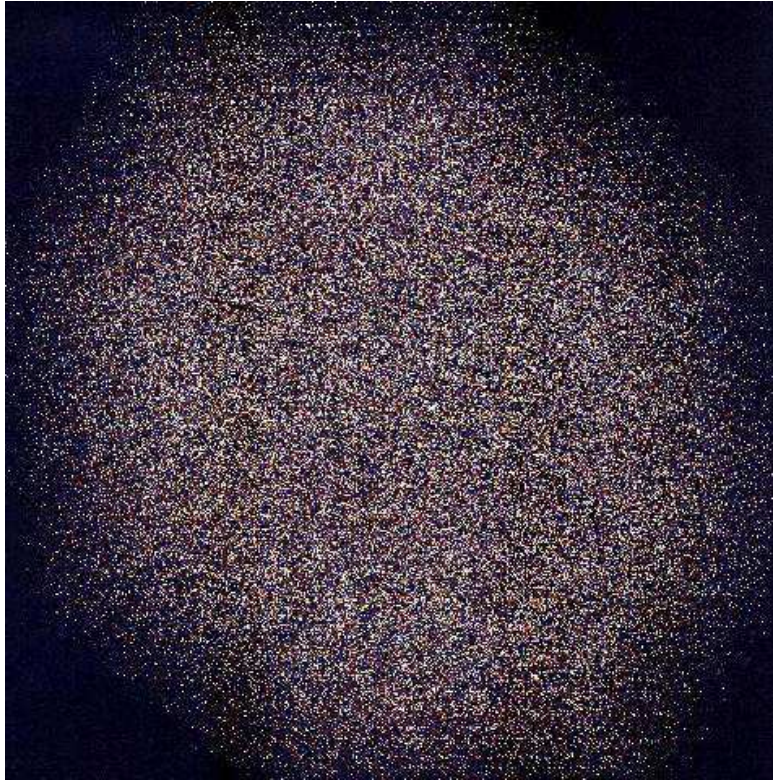,width=10cm}
\caption{Local curvatures projected on a plane, in the smooth phase $G>G_c$.}
\label{fig:r32}
\end{figure}

On one such edge length configuration one can compute the local
curvatures, and then project the result on a two-dimensional
plane.
Using a red color coding for regions of increasingly 
positive curvature, and
a blue color coding for regions of increasingly
negative curvature, one obtains the picture
in Fig.~\ref{fig:r32}.

One finds that as $k$ is varied, the average curvature $\cal R$ is
negative for 
sufficiently small $k$ ('smooth' phase), and appears to go to zero 
continuously at some finite value $k_c$.
For $k > k_c$ the curvature becomes very large, and
the simplices tend to collapse into degenerate configurations
with very small volumes ($ <\!V\!> / <\!l^2\!>^2 \sim 0$).
This 'rough' or 'collapsed' phase 
is the region of the usual weak field expansion ($G \rightarrow 0$).
In this phase the lattice collapses into degenerate configurations
with very long, elongated simplices (Hamber, 1984;
Hamber and Williams, 1985; Berg, 1985).
This phenomenon is usually interpreted as a lattice remnant of the conformal
mode instability of Euclidean gravity discussed in Sec.~(\ref{sec:conformal}).

An elementary argument can be given to explain the fact
that the collapsed phase for $k>k_c$ has an effective dimension of two.
The instability is driven by the
Euclidean Einstein term in the action, and in particular its
unbounded conformal mode contribution.
As the manifold during collapse reaches an effective dimension
of two, the action effectively turns into a topological invariant,
unable to drive the instability further to a still lower dimension
\footnote{
One way of determining coarse aspects of the underlying geometry
is to compute the effective dimension in the scaling regime,
for example by considering how the number
of points within a thin shell of geodesic distance between
$\tau$ and $\tau+\Delta$ scales with the geodesic distance itself.
For distances a few multiples of the average lattice spacing one finds
\beq
N(\tau) \mathrel{\mathop\sim_{\tau \rightarrow \, \infty}} \tau^{\; d_v} ,
\label{fractal}
\eeq
with $d_v =3.1(1)$ for $G>G_c$ (the smooth phase) and $d_v \simeq 1.6(2)$
for $G<G_c$ (the rough phase).
One concludes that in the rough phase the lattice tends to collapse into
a degenerate tree-like configuration,
whereas in the smooth phase the effective dimension of space-time
is consistent with four.
Higher derivative terms affect these results at very short distances,
where they tend to make the geometry smoother. 
}.

Accurate and reproducible curvature data can only be obtained for $k$
below the instability point, since for $k > k_u \approx 0.053$
an instability develops, presumably associated with the unbounded
conformal mode. Its signature is typical of a sharp first order transition,
beyond which the system ventually tunnels into the rough, elongated phase which
is two-dimensional in nature and has no physically acceptable continuum limit.
The instability is caused by the appearance of one or more localized singular
configuration, with a spike-like curvature singularity. 
At strong coupling such singular configurations
are suppressed by a lack of phase space due to the functional measure,
which imposes non-trivial constraints due to the triangle inequalities and
their higher dimensional analogues. In other words, it seems that the measure
regulates the conformal instability at sufficiently strong coupling.

As a consequence, the non-analytic behavior of the free energy (and its
derivatives, which include for example the average curvature) has to be
obtained
by analytic continuation of the Euclidean theory into the metastable
branch.
This procedure, while perhaps unusual, is formally equivalent
to the construction of the continuum theory exclusively from
its strong coupling (small $k$, large $G$) expansion
\beq
{\cal R} (k) \; = \; \sum_{n=0}^{\infty} b_n k^n \;\; .
\label{eq:series}
\eeq
Ultimately it should be kept in mind that one is really only
interested in the {\it pseudo-Riemannian} case, and not the Euclidean
one for which an instability due to the conformal mode is
to be expected. Indeed had such an instability not occurred for small
enough $G$ one might have wondered if the resulting theory still had
any relationship to the original continuum theory.

\subsubsection{Numerical Determination of the Scaling Exponents}
\label{sec:exp}

One way to extract the critical exponent $\nu$ is to fit the average curvature to the form [see Eq.~(\ref{eq:rsing})]
\beq
{\cal R} (k) \; \mathrel{\mathop\sim_{ k \rightarrow k_c}}
- A_{\cal R} \, ( k_c - k )^\delta \;\;\;\; .
\label{eq:rsing1}
\eeq
Using this set of procedures one obtains on lattices of up to $16^4$ sites 
$ k_c = 0.0630(11) $ and $ \nu = 0.330(6) $.
Note that the average curvature is negative at strong coupling up
to the critical point: locally the parallel transport of vectors around
infinitesimal loops seems to be described by a lattice
version of Euclidean anti-de Sitter space.

\begin{figure}[h]
\epsfig{file=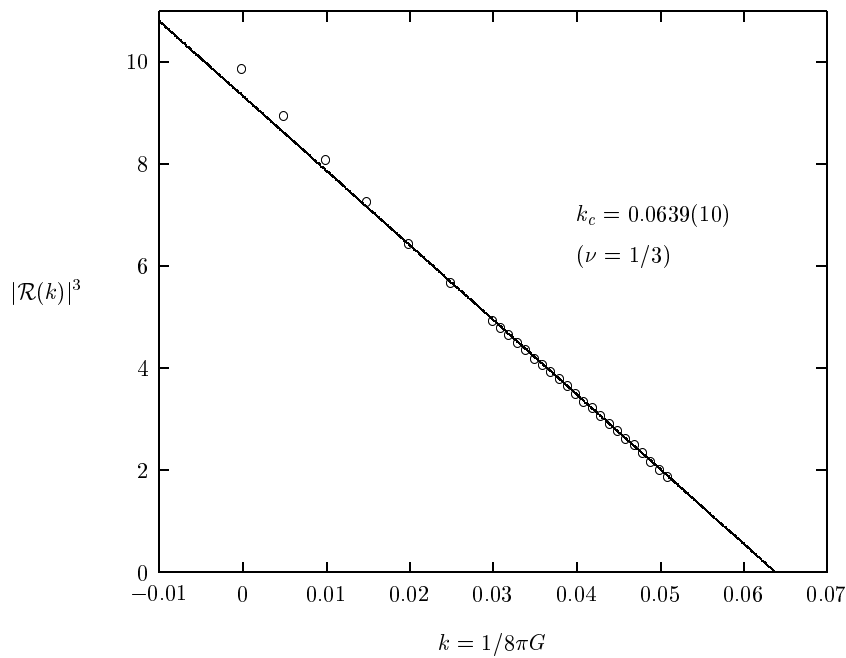,width=12cm}
\caption{ Average curvature ${\cal R}$ on a $16^4$ lattice, raised to the
third power.
If $\nu=1/3$, the data should fall on a straight line.
The continuous line represents a linear fit of the form $A \; (k_c-k)$.
The small deviation from linearity of the transformed data is quite
striking.}
\label{fig:curv3-mc1}
\end{figure}

Fig.~\ref{fig:curv3-mc1} shows as an example a graph of the average curvature
${\cal R}(k)$ raised to the third power.
One would expect to get a straight line close to the critical point if
the exponent for ${\cal R}(k)$ is exactly $1/3$. The numerical data 
indeed supports this assumption, and in fact the linearity of the results
close to $k_c$ is quite striking.
Using this procedure one obtains on the $16^4$-site lattice an
esimate for the critical point, $ k_c = 0.0639(10) $.

Often it can be advantageous to express results obtained in the cutoff
theory in terms of physical (i.e. cutoff independent) quantities.
By the latter one means quantities for which the cutoff dependence has been
re-absorbed, or restored, in the relevant definition.
As an example, an expression equivalent to Eq.~(\ref{eq:rsing}), 
relating the vacuum expectation value of the local scalar curvature 
to the physical correlation length $\xi$ , is
\beq
{ < \int d x \, \sqrt{ g } \, R(x) > \over < \int d x \, \sqrt{ g } > } 
\; \mathrel{\mathop\sim_{ G \rightarrow G_c}} \; {\rm const.} \,
\left ( l_P^2 \right )^{(d -2 - 1 / \nu )/2} \,
\left ( {1 \over \xi^2 } \right )^{ (d - 1 / \nu )/2 }
\;\; ,
\label{eq:curvature}
\eeq
which is obtained by substituting Eq.~(\ref{eq:xi-k}) 
into Eq.~(\ref{eq:rsing}).
The correct dimensions have been restored in this last equation
by supplying appropriate powers of the Planck length $l_P=G_{phys}^{1/(d-2)}$,
which involves the ultraviolet cutoff $\Lambda$.
For $\nu=1/3$ the result of Eq.~(\ref{eq:curvature})
becomes particularly simple
\beq
{ < \int d x \, \sqrt{ g } \, R(x) > \over < \int d x \, \sqrt{ g } > } 
\; \mathrel{\mathop\sim_{ G \rightarrow G_c}} \; 
{\rm const.} \; \, {1 \over l_P \, \xi }  
\label{eq:curvature1}
\eeq
Note that a naive estimate based on dimensional arguments would
have suggested the incorrect result $\sim 1 / l_P^2 $.
Instead the above expression actually vanishes at the critical point.
This shows that $\nu$ plays the role of an anomalous dimension,
determining the magnitude of deviations from naive dimensional
arguments.


Since the critical exponents play such a central role in
determining the existence and nature of the continuum limit,
it appears desirable to have an independent way of estimating
them, which either does not depend on any fitting procedure,
or at least analyzes a different and complementary set of data.
By systematically studying the dependence of averages on the physical size of
the system, one can independently estimate the critical
exponents.

Reliable estimates for the exponents in a lattice field theory
can take advantage of a comprehensive finite-size analysis, a procedure
by which accurate values for the critical exponents are obtained by taking into
account the linear size dependence of the result computed in a finite
volume $V$.

In practice the renormalization group approach is brought in 
by considering the behavior of the system under scale
transformations.
Later the scale dependence is applied to the overall volume
itself.
The usual starting point for the derivation of the scaling properties
is the renormalization group behavior of the free
energy $F\, = \, - { 1 \over V} \log Z $
\beq
F(t, \{ u_j \} ) \; = \; F_{reg} (t, \{ u_j \} ) \; + \;
b^{-d} \; F_{sing} ( b^{y_t} t, \{ b^{y_j} u_j \} ) \;\; ,
\label{eq:fsing-fss}
\eeq
where $F_{sing}$ is the singular, non-analytic part of the free
energy, and $F_{reg}$ is the regular part. 
$b$ is the block size in
the RG transformation, while $y_t$ and $y_j (j \ge 2)$
are the relevant eigenvalues of the RG transformation, and $t$ the reduced
temperature variable that gives the distance from criticality.
One denotes here by $y_t > 0 $ the relevant eigenvalue, while the remaining
eigenvalues $y_j \le 0 $ are associated with either marginal or irrelevant
operators.
Usually the leading critical exponent $y_t^{-1}$ is called $\nu$, while the
next subleading exponent $y_2$ is denoted $-\omega$.
For more details on the method we have to refer to the comprehensive
review in (Cardy, 1988). 

The correlation length $\xi$ determines the asymptotic decay of correlations,
in the sense that one expects for example for the two-point function
at large distances
\beq
< {\cal O} (x) \, {\cal O} (y) > \;
\mathrel{\mathop\sim_{ |x-y| \; \gg \; \xi }} \; e^{-|x-y| / \xi } \;\;\; .
\label{eq:xi}
\eeq
The scaling equation for the correlation length itself
\beq
\xi(t) \; = \; b \; \xi \left ( b^{y_t} t \right )
\eeq
implies for $b = t^{-1/y_t}$ that $ \xi \sim t^{- \nu}$ with a
correlation length exponent
\beq
\nu = 1/y_t \;\; .
\label{eq:nu}
\eeq
Derivatives of the free energy $F$ with respect to $t$ then determine,
after setting the scale factor $b=t^{-1/y_t}$,
the scaling properties of physical observables, including corrections
to scaling.
Thus for example, the second derivative
of the free energy with respect to $t$ yields the specific heat
exponent $\alpha = 2 - d/ y_t = 2-d \nu$,
\beq
{ \partial^2 \over \partial t^2 } \;  F(t, \{ u_j \} )
\; \sim \; t^{-(2-d \nu)} \;\;\;\; .
\label{eq:cv_sing}
\eeq
In the gravitational case one identifies the scaling field $t$
with $k_c - k$, where $k=1/16 \pi G$ involves the bare Newton's constant.
The appearance of singularities in physical averages, obtained from
appropriate derivatives of $F$, is rooted in the fact that
close to the critical point at $t=0$ the correlation length diverges.

The above results can be extended to the case of a finite lattice
of volume $V$ and linear dimension $L=V^{1/d}$.
The volume-dependent free energy is then written as
\beq
F(t, \{ u_j \}, L^{-1} ) \; = \; F_{reg} (t, \{ u_j \} ) \; + \;
b^{-d} \; F_{sing} ( b^{y_t} t, \{ b^{y_j} u_j \}, b/L ) \;\;\; .
\label{eq:fssf}
\eeq
For $b=L$ (a lattice consisting of only one point) one obtains the 
Finite Size Scaling (FSS) form of the free energy
[see for example (Brezin and Zinn-Justin, 1985) for a field-theoretic justification].
After taking derivatives
with respect to the fields $t$ and $ \{ u_j \} $, the FSS
scaling form for physical observables follows,
\beq
O(L,t) \; = \; L^{x_O / \nu} \; \left [ \tilde{f_O}
\left ( L \; t^{\nu} \right ) \; + \; 
{\cal O} ( L^{-\omega}) \right ] \;\;\; ,
\label{eq:fsso}
\eeq
where $x_O$ is the scaling dimension of the operator $O$,
and $ \tilde{f_O} (x)$ an arbitrary function.

As an example, consider the average curvature ${\cal R}$.
From Eq.~(\ref{eq:fsso}), with $t \sim k_c-k$ and $x_O = 1 - 4 \nu$,
one has
\beq
{\cal R} (k,L) \; = \; L^{-(4 - 1 / \nu)} \; \left [ \;
\tilde{ {\cal R} } \left ( (k_c-k) \; L^{1/\nu} \right ) \; + \; 
{\cal O} ( L^{-\omega}) \; \right ]
\label{eq:fss_r}
\eeq
where $\omega>0$ is a correction-to-scaling exponent.
If scaling involving $k$ and $L$ holds according
to Eq.~(\ref{eq:fsso}),
then all points should lie on the same universal curve.

\begin{figure}[h]
\epsfig{file=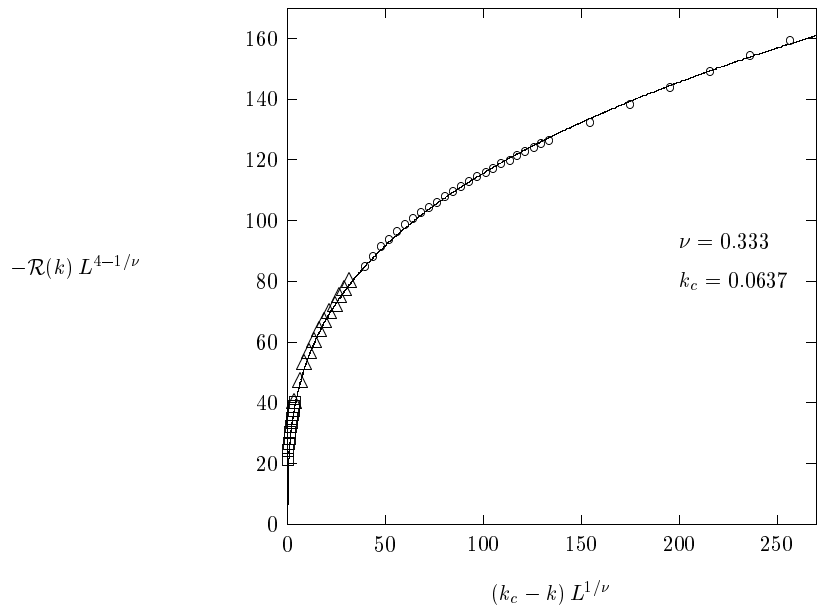,width=12cm}
\caption{Finite size scaling behavior of the scaled curvature
versus the scaled coupling.
Here $L=4$ for the lattice with $4^4$ sites ($\Box$), 
$L=8$ for a lattice with $8^4$ sites ($\triangle$),
and $L=16$ for the lattice with $16^4$ sites ($\circ$).
Statistical errors are comparable to the size of the dots.
The continuous line represents a best fit of the form $ a + b x^c$.
Finite size scaling predicts that all points should lie on the same universal
curve. At $k_c=0.0637$ the scaling plot gives the value $\nu=0.333$.}
\label{fig:curv-fss}
\end{figure}

Fig.~\ref{fig:curv-fss} shows a graph of the scaled curvature 
${\cal R}(k) \; L^{4-1/\nu}$
for different values of $L=4,8,16$, versus the scaled coupling
$(k_c-k) L^{1/\nu}$. 
If scaling involving $k$ and $L$ holds according to Eq.~(\ref{eq:fss_r}),
with $x_{\cal O} = 1 - 4 \nu $ the scaling dimension for the curvature,
then all points should lie on the same universal curve.
The data is in good agreement with such behavior, and provides
a further test on the exponent, which seems consistent
within errors with $\nu = 1/3$.

As a second example consider the curvature fluctuation $\chi_{\cal R}$.
From the general Eq.~(\ref{eq:fsso}) one expects in this case,
for $t \sim k_c-k$ and $x_O = 2 - 4 \nu$,
\beq
\chi_{\cal R} (k,L) \; = \; L^{2 / \nu - 4} \; \left [ \;
\tilde{ \chi_{\cal R} } \left ( (k_c-k) \; L^{1/\nu} \right ) \; + \; 
{\cal O} ( L^{-\omega}) \; \right ] \;\;\;\; ,
\label{eq:fss_c}
\eeq
where $\omega>0$ is again a correction-to-scaling exponent.
If scaling involving $k$ and $L$ holds according to Eq.~(\ref{eq:fsso}),
then all points should lie on the same universal curve.

\begin{figure}[h]
\epsfig{file=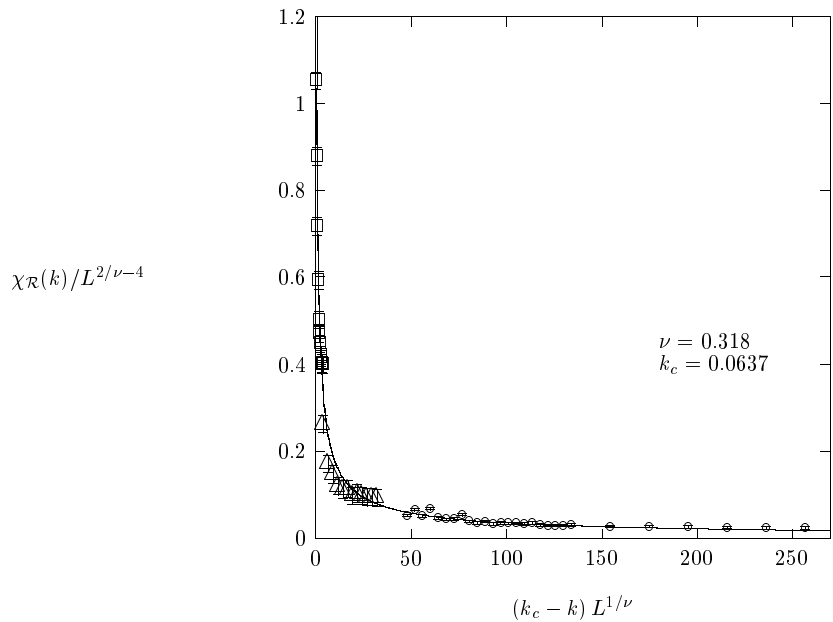,width=12cm}
\caption{Finite size scaling behavior of the scaled curvature
fluctuation versus the scaled coupling.
Here $L=4$ for the lattice with $4^4$ sites ($\Box$), 
$L=8$ for the lattice with $8^4$ sites ($\triangle$),
and $L=16$ for the lattice with $16^4$ sites ($\circ$).
The continuous line represents a best fit of the form $ 1/(a + b x^c)$.
Finite size scaling predicts that all points should lie on the same universal
curve. At $k_c=0.0637$ the scaling plot gives the value $\nu=0.318$.}
\label{fig:chi-fss}
\end{figure}

Fig.~\ref{fig:chi-fss} shows a graph of the scaled curvature fluctuation
$\chi_{\cal R}(k) / L^{2/\nu-4}$ for different values of $L=4,8,16$,
versus the scaled coupling $(k_c-k) L^{1/\nu}$.
If scaling involving $k$ and $L$ holds according to Eq.~(\ref{eq:fss_c}),
then all points should lie on the same universal curve.
Again the data supports such scaling behavior, and provides
a further estimate on the value for $\nu$.

The value of $k_c$ itself should depend on the size of the system.
One expects 
\beq
k_c (L) \mathrel{\mathop\sim_{L \rightarrow \infty}}
k_c (\infty) + c \; L^{-1 / \nu} \; + \; \cdots \;\;\; .
\label{eq:kcl}
\eeq
where $ k_c (\infty)$ is the infinite-volume limit critical point.

The previous discussion applies to continuous, second order phase transitions.
First order phase transitions are driven by instabilities, and are in
general not governed by any renormalization group fixed point. 
The underlying reason
is that the correlation length does not diverge at the first order
transition point,
and thus the system never becomes scale invariant.
In the simplest case, a first order transition develops as the system
tunnels between two neighboring minima of the free energy.
In the metastable branch the free energy acquires a small complex part
with a very weak essential singularity in the coupling at the first order transition point (Langer, 1967; Fisher, 1967; Griffiths, 1969).
As a consequence, such a singularity is not generally visible from the stable
branch, in the sense that a power series expansion in the
temperature is unaffected by such a weak singularity.
Indeed in the infinite volume limit the singularities associated
with a first order transition at $T_u$ become infinitely sharp,
a $\theta$- or $\delta$-function type singularity.
While the singularity in the free energy at the endpoint of the metastable branch (at say $T_c$) cannot be explored directly, it can
be reached by a suitable analytic continuation from the stable side of the free energy branch.
A similar situation arises in the case of lattice $QCD$ with fermions,
where zero fermion mass (chiral) limit is reached by extrapolation 
(Hamber and Parisi, 1983).

From the best data (with the smallest statistical uncertainties
and the least systematic effects) one concludes
\beq
k_c = 0.0636(11) \;\;\;\;\;  \nu = 0.335(9) \;\;\;\; ,
\eeq
which suggests $\nu = 1/3$ for pure quantum gravity.
Note that at the critical point the gravitational coupling
is not weak, $G_c \approx 0.626 $ in units of the ultraviolet
cutoff.
It seems that the value $\nu=1/3$ does not correspond to any
known field theory or statistical mechanics model in four
dimensions. 
For a perhaps related system, namely dilute branched polymers,
it is known that $\nu=1/2$
in three dimensions, and $\nu=1/4$ at the upper critical
dimension $d=8$, so one would expect a value close to
$1/3$ somewhere in between.
On the other hand for a scalar field one would have obtained $\nu=1$ in $d=2$
and $\nu=\half$ for $d \ge 4$, which seems excluded.

\begin{table}

\begin{center}
\begin{tabular}{|l|l|l|}
\hline\hline
& & 
\\
Method & $\nu^{-1}$ in $d=3$ & $\nu^{-1}$ in $d=4$ 
\\ \hline \hline
lattice & 1.67(6) & -
\\ \hline
lattice  & - & 2.98(7) 
\\ \hline
$2+\epsilon$ & 1.6 & 4.4
\\ \hline
truncation & 1.2 & 2.666
\\ \hline \hline
exact ? & 1.5882 & 3 
\\ \hline \hline
\end{tabular}
\end{center}
\label{exp2}

{\small {\it
Table I: Direct determinations of the critical exponent $\nu^{-1}$
for quantum gravitation, using various analytical and numerical
methods in three and four space-time dimensions.
\medskip}}


\end{table}
\vskip 10pt

Table I provides a short summary of the critical
exponents for quantum gravitation as obtained by various perturbative
and non-perturbative methods in three and four dimensions.
The $2+\epsilon$ and the truncation method results were discussed
previously in Secs.~\ref{sec:graveps} and \ref{sec:phaseseps}, respectively.
The lattice model of Eq.~(\ref{eq:zlatt}) in four dimensions
gives for the critical point $G_c \approx 0.626 $ in units of
the ultraviolet cutoff, and $ \nu^{-1} = 2.98(7)$ 
which is used for comparison in Table I.
In three dimensions the numerical results are consistent with the universality
class of the interacting scalar field. 
The same set of results are compared graphically in Fig.~\ref{fig:exp} and
Fig.~\ref{fig:kc} below.

The direct numerical determinations of the critical point
$k_c=1/8 \pi G_c$ in $d=3$ and $d=4$
space-time dimensions are in fact quite close to the analytical prediction
of the lattice $1/d$ expansion given previously in Eq.~(\ref{eq:kcd1}),
\beq
{ k_c \over \lambda_0^{1 - 2/d} } \; = \; 
{ 2^{1+ 2/d } \over d^3 } \, \left [ 
{ \Gamma (d) \over \sqrt{d+1} } \right ]^{2/d}  \;\; .
\eeq
The above expression gives for a bare cosmological constant 
$\lambda_0=1$ the estimate 
$k_c = \sqrt{3} / (16 \cdot 5^{1/4} ) = 0.0724$ in $d=4$,
to be compared with the numerical result 
$k_c = 0.0636(11)$ in (Hamber, 2000).
Even in $d=3$ one has $k_c= 2^{5/3} / 27 = 0.118$, to be compared with 
the direct determination $k_c=0.112(5)$ from (Hamber and Williams, 1993).
These estimates are compared below in Fig.~\ref{fig:kc}.

\begin{figure}[h]
\epsfig{file=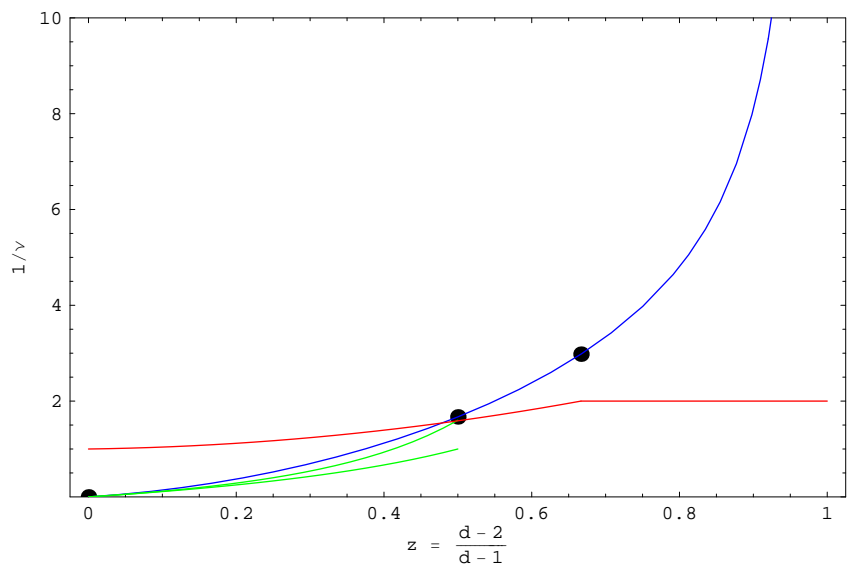,width=12cm}
\caption{Universal gravitational exponent $1/\nu$ as a function of the dimension.
The abscissa is $z=(d-2)/(d-1)$,
which maps $d=2$ to $z=0$ and $d=\infty$ to $z=1$. 
The larger circles at $d=3$ and $d=4$ are the lattice gravity results, interpolated (continuous curve) using the exact
lattice results $1/\nu=0$ in $d=2$, and $\nu=0$ at $d=\infty$
[from Eq.~(\ref{eq:xilog})].
The two curves close to the origin are the $2+\epsilon$ expansion for $1/\nu$
to one loop (lower curve) and two loops (upper curve).
The lower almost horizontal line gives the value for $\nu$ expected for a
scalar field theory, for which it is known that $\nu=1$ in $d=2$
and $\nu=\half$ in $d \ge 4$.}
\label{fig:exp}
\end{figure}

\begin{figure}[h]
\epsfig{file=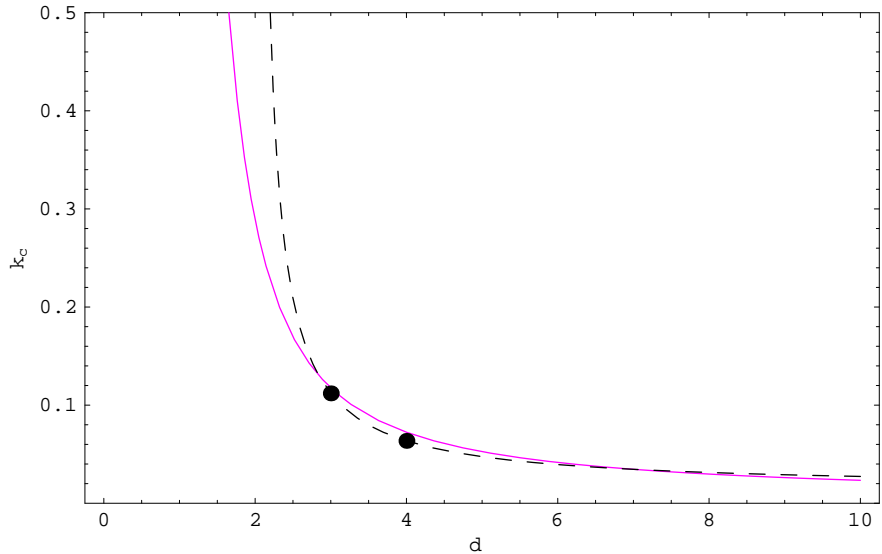,width=12cm}
\caption{Critical point $k_c=1/8 \pi G_c$ in units of the ultraviolet
cutoff as a function of dimension $d$.
The circles at $d=3$ and $d=4$ are the lattice results,
suitably interpolated (dashed curve) using the additional lattice
result $1/k_c=0$ in $d=2$.
The lower continuous curve is the analytical large-$d$ lattice result of
Eq.~(\ref{eq:kcd1}).}
\label{fig:kc}
\end{figure}

\subsubsection{Renormalization Group and Lattice Continuum Limit}
\label{sec:contlim}

The discussion in the previous sections points to the existence
of a phase transition in the lattice gravity theory, with divergent
correlation length in the vicinity of the critical point,
as in Eq.~(\ref{eq:xi-k})
\beq
\xi (k) \; \mathrel{\mathop\sim_{ k \rightarrow k_c}} \; A_\xi \;
| k_c - k | ^{ -\nu }
\label{eq:xi-k1}
\eeq
As described previously, the existence of such a correlation length
is usually inferred indirectly by scaling arguments, from the
presence of singularities in the free energy 
$F_{latt} = - { 1 \over V} \ln Z_{latt}$
as a function of the lattice coupling $k$.
Equivalently, $\xi$ could have been computed directly from
correlation functions at fixed geodesic distance using the
definition in Eq.~(\ref{eq:rr-exp}), or even
from the correlation of Wilson lines associated with the
propagation of two heavy spinless particles.
The outcome of such large scale numerical calculations is 
eventually a determination of the quantities
$\nu$, $k_c=1 /8 \pi G_c$ and $A_\xi$ from first principles,
to some degree of numerical accuracy.

In either case one expects the scaling result of Eq.~(\ref{eq:xi-k1})
close to the fixed point, which we choose to rewrite here
in terms of the inverse correlation length $m \equiv 1 / \xi$
\beq
m \, = \, \Lambda \, A_m \, | \, k \, - \, k_c \, |^{ \nu } \;\; .
\label{eq:m-latt}
\eeq
Note that in the above expression the correct dimension for $m$ (inverse length)
has been restored by inserting explicitly on the r.h.s. the ultraviolet
cutoff $\Lambda$.
Here $k$ and $k_c$ are of course still dimensionless quantities, and
correspond to the bare microscopic couplings at the
cutoff scale, $k \equiv k (\Lambda) \equiv 1/( 8 \pi G(\Lambda) )$.
$A_m$ is a calculable numerical constant, related to $A_\xi$ in 
Eq.~(\ref{eq:xi-k}) by $A_m = A_\xi^{-1}$.
It is worth pointing out that the above expression for $m (k) $ is almost identical in structure
to the one for the non-linear $\sigma$-model in the $2+\epsilon$
expansion, Eq.~(\ref{eq:m-sigma}) and in the
large $N$ limit, Eqs.~(\ref{eq:m-largen}), (\ref{eq:m-largen1})
and (\ref{eq:beta-largen}).
It is of course also quite similar to $2+\epsilon$ result for
continuum gravity, Eq.~(\ref{eq:m-cont1}).

The lattice continuum limit corresponds to the large cutoff limit
taken at {\it fixed} $m$,
\beq
\Lambda \rightarrow \infty \; , 
\;\;\;\; k \rightarrow k_c \; ,
\;\;\;\; m \; {\rm fixed} \; ,
\eeq
which shows that the continuum limit is reached in the
vicinity of the ultraviolet fixed point.
Phrased equivalently, one takes the limit in which the
lattice spacing $a \approx 1/ \Lambda$ is sent to
zero at {\it fixed} $\xi=1/m$, which
requires an approach to the non-trivial UV fixed
point $ k \rightarrow k_c$. 
The quantity $m$ is supposed to be a renormalization group
invariant, a physical scale independent of the scale
at which the theory is probed.
In practice, since the cutoff ultimately determines
the physical value of Newton's constant $G$, $\Lambda$
cannot be taken to $\infty$.
Instead a very large value will suffice, $\Lambda^{-1} \sim 10^{-33} cm$,
for which it will still be true that $ \xi \gg \Lambda$ which
is all that is required for the continuum limit.

For discussing the renormalization group behavior
of the coupling it will be more convenient
to write the result of Eq.~(\ref{eq:m-latt})
directly in terms of Newton's constant $G$ as
\beq
m \, = \,  \Lambda \, \left ( { 1 \over a_0 } \right )^\nu \, 
\left [ { G ( \Lambda ) \over G_c } - 1 \right ]^ \nu 
\; ,
\label{eq:m-latt1}
\eeq
with the dimensionless constant $a_0$ related to
$A_m$ by $A_m = 1 / (a_0 k_c)^\nu $.
Note that the above expression only involves the dimensionless
ratio $G(\Lambda)/G_c$, which is the only relevant
quantity here.
The lattice theory in principle completely determines both the exponent $\nu$
and the amplitude $a_0$ for the quantum correction.
Thus from the knowledge of the dimensionless constant 
$A_m$ in Eq.~(\ref{eq:m-latt})
one can estimate from first principles the value of $a_0$ in 
Eqs.~(\ref{eq:grun-latt}).
Lattice results for the correlation functions at fixed geodesic distance
give a value for $A_m \approx 0.72 $ with a significant uncertainty,
which, when combined with the values $k_c \simeq 0.0636 $ and 
$ \nu \simeq 0.335$ given above, gives 
$a_0 = 1 /( k_c \, A_m^{1/\nu}) \simeq 42$.
The rather surprisingly large value for $a_0$ appears here 
as a consequence of the relatively small value of the
lattice $k_c$ in four dimensions.

The renormalization group invariance of the physical quantity
$m$ requires that the running gravitational
coupling $G(\mu)$ varies in the vicinity of the
fixed point in accordance with the above equation, with
$\Lambda \rightarrow \mu$, where $\mu$ is now an arbitrary scale,
\beq
m \, = \,  \mu \, \left ( { 1 \over a_0 } \right )^\nu \,
\left [ { G ( \mu ) \over G_c } - 1 \right ]^ \nu 
\; ,
\label{eq:m-mu}
\eeq
The latter is equivalent to the renormalization group
invariance requirement
\beq
\mu \, { d \over d \, \mu } \, m ( \mu , G( \mu ) ) \, = \, 0
\label{eq:m-rg-latt}
\eeq
provided $G(\mu)$ is varied in a specific way.
Indeed this type of situation was already encountered before, for example in 
Eqs.~(\ref{eq:mass-indep}) and (\ref{eq:m-rginv}).
Eq.~(\ref{eq:m-rg-latt}) can therefore be used to obtain, if one so wishes,
$a$ $\beta$-function for the coupling $G(\mu)$ in units of the ultraviolet cutoff,
\beq
\mu \, { \partial \over \partial \, \mu } \, G ( \mu ) \; = \; 
\beta ( G ( \mu ) ) \;\; ,
\label{eq:beta-g-mu}
\eeq
with $\beta (G)$ given in the vicinity of the non-trivial fixed point,
using Eq.~(\ref{eq:m-mu}), by
\beq
\beta (G ) \, \equiv \, 
\mu \, { \partial \over \partial \, \mu } \, G( \mu )
\; \mathrel{\mathop\sim_{ G \rightarrow G_c}} \; 
- \, { 1 \over \nu } \, ( G- G_c ) + \dots \;\; .
\label{eq:beta-g-latt}
\eeq
The above procedure is in fact in complete analogy to what was done
for the non-linear $\sigma$-model, for example in Eq.~(\ref{eq:beta-largen}).
But the last two steps are not really necessary, for
one can obtain the scale dependence of the gravitational
coupling directly from Eq.~(\ref{eq:m-mu}), by simply solving for
$G(\mu)$,
\beq
G( \mu ) \; = \;  
G_c \left [ 1 \, + \, a_0 (m^2 / \mu^2 )^{1 / 2 \nu } + \dots \right ]
\label{eq:grun-latt} 
\eeq
This last expression can be compared directly to the $2+\epsilon$
result of Eq.~(\ref{eq:grun-cont1}), as well as to
the $\sigma$-model result of Eq.~(\ref{eq:grun-nonlin}).
The physical dimensions of $G$ can be restored by multiplying the
above expression on both sides by the ultraviolet cutoff $\Lambda$,
if one so desires.
One concludes that the above result physically implies gravitational
anti-screening: the gravitational coupling $G$ increases with distance.

Note that the last equation only involves the dimensionless
ratio $ G(\mu) / G_c$, and is therefore unaffected by
whether the coupling $G$ is dimensionful (after inserting an appropriate
power of the cutoff $\Lambda$) or dimensionless.
It simply relates the gravitational coupling at one scale to the coupling
at a different scale,
\beq
{ G( \mu_1 ) \over G( \mu_2 ) }
\; \approx \;  { 
1 \, + \, a_0 (m / \mu_1 )^{1 / \nu } + \dots
\over
1 \, + \, a_0 (m / \mu_2 )^{1 / \nu } + \dots
}
\label{eq:grun-latt1} 
\eeq
In conclusion, the lattice result for $G(\mu)$ in Eq.~(\ref{eq:grun-latt})
and the $\beta$-function in Eq.~(\ref{eq:beta-g-latt}) are qualitatively
similar to what one finds both in the $2+\epsilon$ expansion
for gravity and in the non-linear $\sigma$-model {\it in the strong coupling
phase}.

But there are also significant differences.
Besides the existence of a phase transition between
two geometrically rather distinct phases, one major new
aspect provided by non-perturbative lattice studies
is the fact that the weakly coupled small $G$ phase 
turns out to be pathological,
in the sense that the theory becomes unstable, with the four-dimensional
lattice collapsing into a tree-like two-dimensional structure
for $G<G_c$.
While in continuum perturbation theory both phases, and therefore both signs
for the coupling constant evolution in Eq.~(\ref{eq:grun-cont}), seem
acceptable (giving rise to both a ``Coulomb'' phase, and
a strong coupling phase), the Euclidean lattice results rule out
the small $G<G_c$ branched polymer phase.
The collapse eventually stops at $d=2$ because the gravitational action then 
becomes a topological invariant.

It appears difficult therefore to physically characterize the weak
coupling phase based on just the lattice results, which only seem
to make sense in the strong coupling phase $G>G_c$.
One could envision an approach wherein such a weak coupling
phase would be discussed in the framework of some sort of analytic continuation
from the strong coupling phase, which would seem possible
at least for some lattice results, such as the gravitational
$\beta$-function of Eq.~(\ref{eq:beta-g-latt}).
The latter clearly makes sense on both sides of the transition,
just as is the case for Eq.~(\ref{eq:beta-largen}) for the
non-linear $\sigma$-model.
In particular Eq.~(\ref{eq:beta-g-latt}) implies that the coupling will
flow towards the Gaussian fixed point $G=0$ for $G< G_c$.
The scale dependence in this phase will be such that one expects
gravitational screening: the coupling $G(\mu)$ gets increasingly weaker at
larger distances.
But how to remove the geometric collapse to a two-dimensional manifold
remains a major hurdle; one could envision an approach
where one introduces one more cutoff on the edges at short
distances, so that each simplex cannot go below a certain
fatness.
But if the results so far can be used as a guide, the gradual removal of
such a cutoff would then plunge the theory back into a 
degenerate two-dimensional, and therefore physically unacceptable, geometry.

\subsubsection{Curvature Scales}
\label{sec:curvature}

As can be seen from Eqs.~(\ref{eq:rescale}) and (\ref{eq:rescale1}) the path
integral for pure quantum gravity can be made to depend on the gravitational
coupling $G$ and the cutoff $\Lambda$ only:
by a suitable rescaling of the metric, or the edge lengths in the discrete case,
one can set the cosmological constant to unity in units of the cutoff.
The remaining coupling $G$ should then be viewed more appropriately
as the gravitational constant
{\it in units of the cosmological constant $\lambda$}.

The renormalization group running of $G (\mu)$ in Eq.~(\ref{eq:grun-latt})
involves an invariant scale $\xi=1/m$.
At first it would seem that this scale could take any value, including
very small ones based on the naive estimate $\xi \sim l_P$, which would
preclude any observable quantum effects in the foreseeable future.
But the result of Eqs.~(\ref{eq:curvature}) and (\ref{eq:curvature1}) suggest
otherwise, namely that the non-perturbative scale $\xi$ is in fact related
to {\it curvature}.
From astrophysical observation the average curvature is very small, 
so one would conclude from Eq.~(\ref{eq:curvature1}) that $\xi$ is very
large, and possibly macroscopic.
But the problem with Eq.~(\ref{eq:curvature1}) is that it involves the 
lattice Ricci scalar, a quantity related curvature probed by parallel
transporting vectors around infinitesimal loops with size comparable
to the lattice cutoff $\Lambda^{-1}$.
What one would like is instead a relationship between $\xi$ and
quantities which describe the geometry on larger scales.

In many ways the quantity $m$ of Eq.~(\ref{eq:m-mu}) behaves as a 
dynamically generated mass scale, quite similar to what
happens in the non-linear $\sigma$-model case [Eq.~(\ref{eq:m-largen1})],
or in the $2+\epsilon$ gravity case [Eq.~(\ref{eq:m-cont})].
Indeed in the weak field expansion, presumably appropriate
for slowly varying fields on very large scales, a mass-like term does appear, 
as in Eq.~(\ref{eq:h-quadr-gf-tv2}),
with $\mu^2 = 16 \pi G \vert \lambda_0 \vert \equiv 2 \vert \lambda \vert $
where $\lambda$ is
the scaled cosmological constant.
From the classical field equation $R=4 \lambda$ one can relate the above $\lambda$, and therefore the mass-like parameter $m$,
to curvature, which leads to the identification
\beq
\lambda_{obs} \; \simeq \; { 1 \over \xi^2 } 
\label{eq:xi_lambda}
\eeq 
with $\lambda_{obs}$ the observed small but non-vanishing cosmological constant.

A further indication that the identification of the observed scaled cosmological
constant with a mass-like - and therefore renormalization group invariant - term 
makes sense beyond the weak field limit can be seen for example
by comparing the structure of the three classical field equations
\bea
R_{\mu\nu} \, - \, \half \, g_{\mu\nu} \, R \, + \, \lambda \, g_{\mu\nu} \; 
& = & \; 8 \pi G  \, T_{\mu\nu}
\nonumber \\
\partial^{\mu} F_{\mu\nu} \, + \, \mu^2 \, A_\nu \, 
& = & \; 4 \pi e \, j_{\nu} 
\nonumber \\
\partial^{\mu} \partial_{\mu} \, \phi \, + \, m^2 \, \phi \; 
& = & \; {g \over 3!} \, \phi^3
\label{eq:masses}
\eea
for gravity, QED (massive via the Higgs mechanism) and a self-interacting scalar
field, respectively.

A third argument suggesting the identification of the scale $\xi$
with large scale curvature and therefore with the observed scaled 
cosmological constant goes as follows.
Observationally the curvature on large scale can be determined by
parallel transporting vectors around very large loops,
with typical size much larger than the lattice cutoff $l_P$.
In gravity, curvature is detected by parallel transporting vectors around
closed loops.
This requires the calculation of a path dependent product of
Lorentz rotations ${\bf R}$, in the Euclidean case elements of $SO(4)$,
as discussed in Sec.~\ref{sec:rotations}.
On the lattice, the above rotation is directly related to the 
path-ordered (${\cal P}$) exponential of the integral of the lattice
affine connection $ \Gamma^{\lambda}_{\mu \nu}$ via
\beq
R^\alpha_{\;\; \beta} \; = \; \Bigl [ \; {\cal P} \, e^{\int_
{{\bf path \atop between \; simplices}}
\Gamma^\lambda d x_\lambda} \; \Bigr ]^\alpha_{\;\; \beta}  \;\; .
\label{eq:rot1}
\eeq
Now, in the strongly coupled gravity regime ($G>G_c$) large fluctuations in the
gravitational field at short distances will be reflected in large fluctuations
of the ${\bf R}$ matrices.
Deep in the strong coupling regime it should be possible to describe these 
fluctuations by a uniform (Haar) measure.
Borrowing from the analogy with Yang-Mills theories, and in particular
non-Abelian lattice gauge theories with compact groups [see Eq.~(\ref{eq:wloop_sun1})], 
one would therefore expect an exponential decay of near-planar Wilson loops with area $A$ of the type
\beq
W(\Gamma) \, \sim \, \tr \exp \, \left [ \;
\int_{S(C)}\, R^{\, \cdot}_{\;\; \cdot \, \mu\nu} \, A^{\mu\nu}_{C} \; \right ]
\; \sim \; \exp ( - A / \xi^2 )
\label{eq:wloop_curv1}
\eeq
where $A$ is the minimal physical area spanned by
the near-planar loop.
A derivation of this standard result for non-Abelian gauge 
theories can be found, for example, in the textbook (Peskin and Schroeder, 1995).

In summary, the Wilson loop in gravity provides potentially
a measure for the magnitude of
the large-scale, averaged curvature, operationally determined by
the process of parallel-transporting test vectors around very large loops,
and which therefore, from the above expression, is computed to be of the 
order $R \sim 1 / \xi^2 $. 
One would expect the power to be universal, but not the amplitude, leaving
open the possibility of having both de Sitter or anti-de Sitter space
at large distances
(as discussed previously in Sec.~\ref{sec:exp}, the average curvature 
describing the parallel transport of vectors around
{\it infinitesimal} loops is described by a lattice
version of Euclidean anti-de Sitter space).
A recent explicit lattice calculation indeed suggests that the de Sitter case is singled out, at least for sufficiently strong copuling (Hamber and Williams, 2007). 
Furthermore one would expect, based on general scaling arguments, that such a behavior would
persists throughout the whole strong coupling phase $G > G_c$, all the
way up to the on-trivial fixed point.
From it then follows the identification of 
the correlation length $\xi$ with a measure of large scale curvature,
the most natural candidate being the scaled cosmological constant
$\lambda_{phys} $, as in Eq.~(\ref{eq:xi_lambda}).
This relationship, taken at face value, implies a very large, cosmological value
for $\xi \sim 10^{28} cm$, given the present bounds on $\lambda_{phys}$.
Other closely related possibilities may exist,
such as an identification of $ \xi $ with the Hubble 
constant (as measured today), determining the macroscopic expansion rate of the
universe via the correspondence
$ \xi \; \simeq \; 1 / H_0  $.
Since this quantity is presumably time-dependent, a possible scenario would
be one in which $\xi^{-1} = H_\infty = \lim_{t \rightarrow \infty} H(t) $,
with $H_\infty^2 = { \lambda \over 3 } $,
for which the horizon radius is $R_{\infty} = H_{\infty}^{-1}$.

Since, as pointed out in Sec.~\ref{sec:graveps} and Sec.~\ref{sec:obs}, the gravitational path integral only depends
in a non-trivial way on the dimensionless combination $G \, \sqrt{\lambda_0}$,
the physical Newton's constant itself $G$ can be decomposed into
non-running and running parts as
\beq
G \; = \; { 1 \over G \, \lambda_0 } \cdot G^2 \, \lambda_0 
\;\; \rightarrow \;\; 
\xi^2 \cdot 
\left [ \, (G \, \sqrt{\lambda_0}) \, (\mu^2) \, \right ]^2
\label{eq:g_decompose}
\eeq
where we have used $1 / G \lambda_0 \sim \xi^2 $.
The running of the second, dimensionless term in square brackets
can be directly deduced from either Eqs.~(\ref{eq:m-mu}) or
(\ref{eq:grun-latt}).
Note that there $\lambda_0$ does not appear there explicitly, since
originally it was set equal to one by scaling the metric (or the edge
lengths).

In conclusion, the modified Einstein equations, incorporating the
quantum running of $G$, read
\beq
R_{\mu\nu} \, - \, \half \, g_{\mu\nu} \, R \, + \, \lambda \, g_{\mu\nu}
\; = \; 8 \pi \, G(\mu)  \, T_{\mu\nu}
\label{eq:field0}
\eeq
with $\lambda \simeq { 1 \over \xi^2 } $,
and only $G(\mu)$ on the r.h.s. scale-dependent in accordance with
Eq.~ (\ref{eq:grun-latt}).
The precise meaning of $G(\mu)$ in a covariant framework
will be given later in Sec.~\ref{sec:effective}.

\subsubsection{Gravitational Condensate}
\label{sec:condensate}

In strongly coupled gravity there appears to be a deep relationship, already encountered previously in non-Abelian gauge theories, between the non-perturbative scale $\xi$ appearing in Eqs.~(\ref{eq:grun-latt}), 
and the non-perturbative vacuum condensate of Eqs.~(\ref{eq:curvature})
and (\ref{eq:xi_lambda}), which is a measure of curvature.
The inescapable conclusion of
the results of Eqs.~(\ref{eq:curvature}) and (\ref{eq:wloop_curv}) is that
the scale $\xi$ appearing in Eq.~(\ref{eq:grun-latt})
is related to curvature, and must be {\it macroscopic} for the lattice 
theory to be consistent.
How can quantum effects propagate to such large distances and give such drastic
modifications to gravity? 
The answer to this paradoxical question presumably lies in the fact
that gravitation is carried by a massless particle whose interactions
cannot be screened, on any length scale.

It is worth pointing out here that the gravitational vacuum condensate,
which only exists in the strong coupling phase $G>G_c$, and which is
proportional to the curvature, is genuinely non-perturbative.
One can summarize the result of Eq.~(\ref{eq:xi_lambda}) as
\beq
{\cal R }_{obs} \; \simeq \; (10^{-30} eV)^2 \, \sim \, \xi^{-2} 
\eeq
where the condensate is, according to Eq.~(\ref{eq:m-latt1}),
non-analytic at $G=G_c$.
A graviton vacuum condensate of order $\xi^{-1} \sim 10^{-30} eV$ is
of course extraordinarily small compared to the QCD color condensate 
($\Lambda_{\overline{MS}} \simeq 220 \, MeV$) and the electro-weak Higgs condensate ($v \simeq 250 \, GeV$).
One can pursue the analogy with non-Abelian gauge theories further
by stating that the quantum gravity theory cannot provide a value for the 
non-perturbative curvature scale $\xi$:
it needs to be fixed by some sort of phenomenological input, either by
Eq.~(\ref{eq:grun-latt}) or by Eq.~(\ref{eq:xi_lambda}).
But one important message is that the scale $\xi$ in those two equations is
one and the same.

Can the above physical picture be used to provide further insight into the
nature of the phase transition, and more specifically the value for $\nu$?
We will mention here a simple geometric argument which can be given to
support the exact value $\nu=1/3$ for pure gravity (Hamber and Williams, 2004).
First one notices that the vacuum polarization induced scale dependence of
the gravitational coupling $G(r)$ as given in Eq.~(\ref{eq:grun-latt}) implies
the following general structure for the quantum corrected static
gravitational potential,
\beq
V(r) \, = \, - \, G(r) \, { m M \over r } \, \approx \, 
- \, G(0) \, { m M \over r } \,
\left [ \, 1 + c \, ( r  / \xi )^{1 / \nu} + {\cal O} ( ( r / \xi)^{2 / \nu} )
\, \right ]
\label{eq:vrun}
\eeq
for a point source of mass $M$ located at the origin
and for intermediate distances $ l_p \ll r \ll \xi $. 
One can visualize the above result by stating that virtual graviton loops
cause an effective anti-screening of the primary gravitational source $M$,
giving rise to a quantum correction to the potential
proportional to $r^{1/ \nu -1}$.
But only for $\nu=1/3$ can the additional contribution 
be interpreted as being due to a close to uniform
mass distribution surrounding the original source,
of strength
\beq
\rho_{\xi} (M) \, = \, { 3 c M  \over 4 \pi \xi^3 } \;\;\; .
\eeq
Such a simple geometric interpretation fails unless
the exponent $\nu$ for gravitation is exactly one third.
In fact in dimensions $d \ge 4$ one would expect based on the geometric argument
that $ \nu = 1/(d-1) $ if the quantum correction to the gravitational
potential arises from such a virtual graviton cloud. 
These arguments rely of course on the lowest order result 
$V(r) \sim \int d^{d-1} k \; e^{i k \cdot x} / k^2
\sim r^{3-d}$ for single graviton exchange in $d>3$ dimensions. 

Equivalently, the running of $G$ can be characterized as being in part
due to a tiny non-vanishing (and positive) 
non-perturbative gravitational vacuum contribution to the cosmological
constant, with
\beq
\lambda_{0} (M) \, = \, { 3 c M \over \xi^3 }
\eeq
and therefore an associated effective classical average curvature of magnitude
$ R_{class} \sim G \lambda_0 \sim G M / \xi^3 $.
It is amusing that for a very large mass distribution $M$,
the above expression for the curvature can only be
reconciled with the naive dimensional estimate $R_{class} \sim 1 / \xi^2 $,
provided for the gravitational coupling $G$ itself one has $G \sim \xi / M $,
which is reminiscent of Mach's principle and its connection with
the Lense-Thirring effect 
(Lense and Thirring, 1918; Sciama, 1953; Feynman, 1962).

\section{SCALE DEPENDENT GRAVITATIONAL COUPLING}
\label{sec:rg}

Non-perturbative studies of quantum gravity suggest
the possibility that gravitational couplings might be weakly scale
dependent due to nontrivial renormalization group effects.
This would introduce a new gravitational scale, unrelated to Newton's
constant, required in order to parametrize the gravitational running
in the infrared region. 
If one is willing to accept such a scenario, then it seems
difficult to find a compelling theoretical
argument for why the non-perturbative scale entering the coupling
evolution equations should be very small, comparable to the Planck length.
One possibility is that
the relevant non-perturbative scale is related to the curvature
and therefore macroscopic in size, which could
have observable consequences. 
One key ingredient in this argument is the relationship, in part supported
by Euclidean lattice results combined with renormalization group arguments,
between the scaling
violation parameter and the scale of the average curvature.

\subsection{Effective Field Equations }
\label{sec:effective}

\subsubsection{Scale Dependence of $G$}
\label{sec:momdep}

To summarize the results of the previous section,
the result of Eq.~(\ref{eq:grun-latt}) implies
for the running gravitational coupling in the vicinity of the
ultraviolet fixed point
\beq
G(k^2) \; = \; G_c \left [ \; 1 \, 
+ \, a_0 \left ( { m^2 \over k^2 } \right )^{1 \over 2 \nu} \, 
+ \, O ( \, ( m^2 / k^2 )^{1 \over \nu} ) \; \right ]
\label{eq:grun-k}
\eeq
with $m=1/\xi$, $a_0 > 0$ and $\nu \simeq 1/3$.
Since $\xi$ is expected to be very large,
the quantity $G_c$ in the above expression should now
be identified with the laboratory scale value 
$ \sqrt{G_c} \sim 1.6 \times 10^{-33} cm$.
Quantum corrections on the r.h.s are therefore quite small as
long as $ k^2 \gg m^2 $, which in real space corresponds to the
``short distance'' regime $ r \ll \xi$.

The interaction in real space is often obtained by Fourier transform,
and the above expression is singular as $k^2 \rightarrow 0$.
The infrared divergence needs to be regulated, which can be achieved
by utilizing as the lower limit of momentum integration
$m=1/\xi$.
Alternatively, as a properly infrared regulated version of the above
expression one can use
\beq
G(k^2) \; \simeq \; G_c \left [ \; 1 \, 
+ \, a_0 \left ( { m^2 \over k^2 \, + \, m^2 } \right )^{1 \over 2 \nu} \, 
+ \, \dots \; \right ]
\label{eq:grun-k-reg}
\eeq
The last form for $G(k^2)$ will only be necessary in the regime where $k$
is small, so that one can avoid unphysical results.
From Eq.~(\ref{eq:grun-k-reg}) the gravitational coupling then approaches
at very large distances $r \gg \xi$ 
the finite value $G_\infty = ( 1 + a_0 + \dots ) \, G_c $.
Note though that in Eqs.~(\ref{eq:grun-k}) or (\ref{eq:grun-k-reg}) the cutoff
no longer appear explicitly, it is absorbed into the definition of $G_c$.
In the follwing we will be mostly interested in the regime
$l_P \ll r \ll \xi$, for which Eq.~(\ref{eq:grun-k}) is completely adequate.

The first step in analyzing the consequences of a running of $G$
is to re-write the expression for $G(k^2)$ in a coordinate-independent
way.
The following methods are not new, and have found over the years their fruitful application in gauge theories and gravity, for example in the discussion
of non-local effective actions 
(Vilkovisky, 1984; Barvinsky and Vilkovisky, 1985).
Since in going from momentum to position space one 
usually employs $k^2 \rightarrow - \Box$,
to obtain a quantum-mechanical running of the gravitational
coupling one should make the replacement
\beq
G  \;\; \rightarrow \;\; G( \Box )
\label{eq:gbox}
\eeq
and therefore from Eq.~(\ref{eq:grun-k})
\beq
G( \Box ) \, = \, G_c \left [ \; 1 \, 
+ \, a_0 \left ( { 1\over \xi^2 \Box  } \right )^{1 \over 2 \nu} \, 
+ \, \dots \, \right ] \; .
\label{eq:grun-box}
\eeq
In general the form of the covariant
d'Alembertian operator $\Box$
depends on the specific tensor nature of the object it is acting on,
\beq
\Box \; T^{\alpha\beta \dots}_{\;\;\;\;\;\;\;\; \gamma \delta \dots}
\; = \; g^{\mu\nu} \nabla_\mu \left( \nabla_\nu \;
T^{\alpha\beta \dots}_{\;\;\;\;\;\;\;\; \gamma \delta \dots} \right)
\eeq
Only on scalar functions one has the fairly simple result
\beq
\Box \, S(x) \; = \; 
{1 \over \sqrt{g} } \, \partial_\mu \, g^{\mu\nu} \sqrt{g} \, \partial_\nu
\, S(x)
\eeq
whereas on second rank tensors one has the already significantly
more complicated expression
$\Box T_{\alpha\beta} \, \equiv \, 
g^{\mu\nu} \nabla_\mu (\nabla_\nu T_{\alpha\beta}) $.

The running of $G$ is expected to lead to
a non-local gravitational action, for example of the form
\beq
I \; = \; { 1 \over 16 \pi G } \int dx \sqrt{g} \,
\left ( 1 \, - \, a_0 \, 
\left ( {1 \over \xi^2 \Box } \right )^{ 1 / 2 \nu} \, 
+ \dots \right ) \, R
\label{eq:ieff_sr}
\eeq
Due to the fractional exponent in general the covariant operator appearing in the above expression, namely
\beq
A (\Box) \; = \; a_0
\left( { 1 \over \xi^2 \Box } \right)^{1/2\nu} 
\label{eq:abox}
\eeq
has to be suitably defined by analytic continuation from positive
integer powers.
The latter can be done for example by computing $\Box^n$ for positive
integer $n$ and then analytically continuing to $n \rightarrow -1/2\nu$.
Alternatively one can make use of the identity
\beq
{ 1 \over {\Box}^n } \; = \; 
{ (-1)^n \over \Gamma (n) } \, \int_0^{\infty} 
ds \, s^{n-1} \exp ( i \, s \, \Box )
\eeq
and later perform the relevant integrals with $n \rightarrow 1/2\nu$.
Other procedures can be used to define $ A (\Box) $, for example based on an
integral representation involving the scalar propagator
(Lopez Nacir and Mazzitelli, 2007).

It should be stressed here that the action in Eq.~(\ref{eq:ieff_sr})
should be treated as a {\it classical} effective action, with dominant
radiative corrections at short distances ($r \ll \xi$) already automatically 
built in, and for which a restriction to generally smooth field
configurations does make some sense. 
In particular one would expect
that in most instances it should be possible, as well as meaningful, 
to neglect terms involving large numbers of derivatives of the
metric in order to compute the effects of the new contributions appearing in
the effective action.

Had one {\it not} considered the action of Eq.~(\ref{eq:ieff_sr})
as a starting point for
constructing the effective theory, one would naturally be led 
(following Eq.~(\ref{eq:gbox}))
to consider the following effective field equations
\beq
R_{\mu\nu} \, - \, \half \, g_{\mu\nu} \, R \, + \, \lambda \, g_{\mu\nu}
\; = \; 8 \pi G  \, \left( 1 + A( \Box ) \right) \, T_{\mu\nu}
\label{eq:field1}
\eeq
the argument again being the replacement 
$G \, \rightarrow \, G(\Box) \equiv G \left( 1 + A( \Box ) \right)$.
Being manifestly covariant, these expressions at least satisfy some
of the requirements for a set of consistent field equations
incorporating the running of $G$.
The above effective field equation can in fact be re-cast in a form
similar to the classical field equations
\beq
R_{\mu\nu} \, - \, \half \, g_{\mu\nu} \, R \, + \, \lambda \, g_{\mu\nu}
\; = \; 8 \pi G  \, {\tilde T_{\mu\nu}}
\eeq
with $ {\tilde T_{\mu\nu}} \, = \, \left( 1 + A( \Box ) \right) \, T_{\mu\nu}$
defined as an effective, or gravitationally dressed, energy-momentum tensor.
Just like the ordinary Einstein gravity case,
in general ${\tilde T_{\mu\nu}}$ might not be covariantly conserved a priori,
$\nabla^\mu \, {\tilde T_{\mu\nu}} \, \neq \, 0 $, but ultimately the
consistency of the effective field equations demands that it
be exactly conserved, in consideration of the Bianchi identity satisfied
by the Riemann tensor (a similar problem arises in other non-local
modifications of gravity (Barvinsky, 2003)).
The ensuing new covariant conservation law
\beq
\nabla^\mu \, {\tilde T_{\mu\nu}} \; \equiv \; 
\nabla^\mu \, \left [ \left ( 1 + A( \Box ) \right ) \, T_{\mu\nu} 
\right ] \;  = \;  0
\label{eq:continuity}
\eeq
can be then be viewed as a {\it constraint}
on ${\tilde T_{\mu\nu}}$ (or $T_{\mu\nu}$) which, for example,
in the specific case of a perfect fluid, 
will imply again a definite relationship between the density $\rho(t) $,
the pressure $p(t)$ and the RW scale factor $a(t)$, just as it does
in the standard case. 
Then the requirement that the bare energy momentum-tensor be conserved would
imply that the quantum contribution $ A( \Box ) \, T_{\mu\nu} $ itself
be separately conserved. 
That this is indeed attainable can be shown in a few simple cases, such
as the static isotropic solution discussed below. 
There a "vacuum fluid" is introduced to account for the vacuum polarization
contribution, whose energy momentum tensor can be shown to be covariantly conserved.
That the procedure is consistent in general is not clear, in which case
the present approach should perhaps be limited to phenomenological considerations.

Let us make a few additional comments regarding the above effective field
equations, in which we will set the cosmological constant $\lambda=0$ 
from now on.
One simple observation is that the trace equation only involves the (simpler)
scalar d'Alembertian, acting on the trace of the energy-momentum tensor
\beq
R \; = \; - \, 8 \pi G  \, \left( 1 + A( \Box ) \right) \, T_{\mu}^{\;\;\mu}
\label{eq:naive_t}
\eeq
Finally, to the order one is working here, the
above effective field equations should be equivalent to
\beq
\left( 1 \, - \, A( \Box )  \, + \, O(A( \Box )^2) \right) 
\left( R_{\mu\nu} - \half \, g_{\mu\nu} \, R \right)
\; = \; 8 \pi G \; T_{\mu\nu}
\label{eq:naive_r}
\eeq
where the running of $G$ has been moved over to the ``gravitational'' side.

\subsubsection{Poisson's Equation and Vacuum Polarization Cloud}
\label{sec:poisson}

One of the simplest cases to analyze is of course the static case.
The non-relativistic, static Newtonian potential is defined as usual as
\beq
\phi (r) \; = \; ( - M ) \int { d^3 {\bf k} \over (2 \pi)^3 } \, 
e^{i \, {\bf k} \cdot {\bf x} } \, G( {\bf k^2} ) \, { 4 \pi \over {\bf k^2} } 
\label{eq:pot_def}
\eeq
and therefore proportional to the $3-d$ Fourier transform of 
\beq
{ 4 \pi \over {\bf k^2} } \; \rightarrow \;
{ 4 \pi \over {\bf k^2} } \left [ \; 1 \, 
+ \, a_0 \left ( { m^2 \over {\bf k^2} } \right )^{1 \over 2 \nu} \, 
+ \, \dots  \; \right ]
\label{eq:pot}
\eeq
But, as already mentioned, for small ${\bf k}$ proper care has to be exercised
in providing a properly infrared regulated version of the above expression,
which, from Eq.~(\ref{eq:grun-k-reg}), reads 
\beq
{ 4 \pi \over ( {\bf k^2} \, + \, \mu^2 ) } \; \rightarrow \; \;
{ 4 \pi \over ( {\bf k^2} \, + \, \mu^2 ) } \left [ \; 1 \, 
+ \, a_0 \left ( { m^2 \over {\bf k^2} \, + \, m^2 } \right )^{1 \over 2 \nu} \, 
+ \, \dots  \; \right ]
\label{eq:pot_m}
\eeq
where the limit $\mu \rightarrow 0$ should be taken at the end of the calculation.

\begin{figure}[h]
\epsfig{file=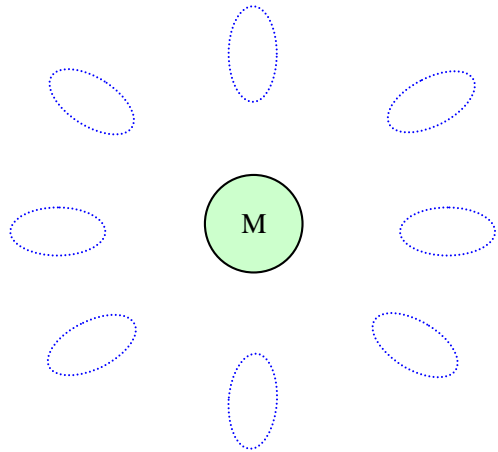,width=8cm}
\caption{A virtual graviton cloud surrounds the point source of mass $M$, leading to an anti-screening modification of the static
gravitational potential.}
\label{fig:polarization}
\end{figure}

\begin{figure}[h]
\epsfig{file=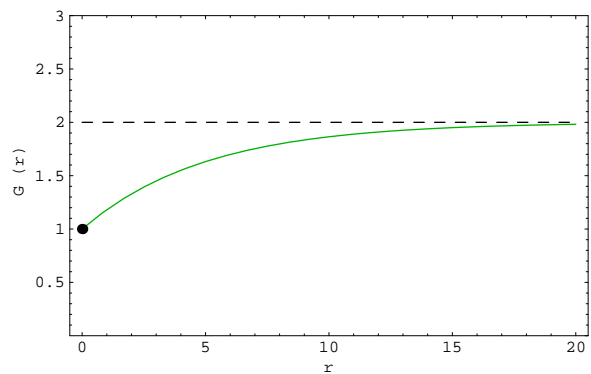,width=8cm}
\caption{Schematic scale dependence of the gravitational coupling
$G(r)$, from Eq.~(\ref{eq:grun-half}) valid for $\nu=1/2$.
The gravitational coupling rises initially like a power of $r$, and later
approaches the asymptotic value $G_{\infty}=(1+a_0) G$ for large $r$.
The behavior for other values of $\nu>1/3$ is similar.}
\label{fig:geff}
\end{figure}

Given the running of $G$ from either Eq.~(\ref{eq:grun-k-reg}), or Eq.~(\ref{eq:grun-k}) 
in the large $\bf k$ limit, the next step is naturally an attempt at
finding a solution to Poisson's equation with a point source at the origin,
so that one can determine the structure of the quantum corrections to
the static gravitational potential in real space.
There are in principle two equivalent ways to compute the potential $\phi(r)$,
either by inverse Fourier transform of Eq.~(\ref{eq:pot}),
or by solving Poisson's equation $\Delta \phi = 4 \pi \rho$ with the source
term $\rho (r)$ given by the inverse Fourier transform of the correction to $G(k^2)$, as given below in Eq.~(\ref{eq:rho_vac}).
The zero-th order term then gives the standard Newtonian $-M G/r$ term, while the correction in general is given by a rather complicated hypergeometric
function.
But for the special case $\nu=1/2$ the Fourier transform of
Eq.~(\ref{eq:pot_m}) is easy to do, the integrals are elementary
and the running of $G(r)$ so obtained is particularly transparent,
\beq
G(r) \; = \; G_{\infty} \left ( 
\, 1 \, - \, { a_0 \over 1 \, + \, a_0 } \, e^{- m r} \, \right )
\label{eq:grun-half}
\eeq
where we have set $G_{\infty} \equiv (1+a_0) \, G $ and $G \equiv G(0)$.
$G$ therefore increases slowly from its value $G$ at small $r$ to the larger
value $(1+a_0)\, G$ at infinity.
Fig.~\ref{fig:polarization} illustrates the anti-screening effect of
the virtual graviton cloud.
Fig.~\ref{fig:geff} gives a schematic illustration of the behavior of $G$
as a function of $r$.

Another possible procedure to obtain the static potential $\phi (r)$
is to solve directly the radial Poisson equation for $\phi(r)$.
This will give a density $\rho(r)$ which can later be used to
generalize to the relativistic case.
In the $a_0 \neq 0$ case one needs to solve $\Delta \phi = 4 \pi \rho$,
or in the radial coordinate for $ r >0 $
\beq
{ 1 \over r^2} \, { d \over d \, r } \, 
\left ( \, r^2 \, {d \, \phi \over d \, r } \, \right ) \; = \;
4 \, \pi \, G \, \rho_m (r)
\eeq
with the source term $\rho_m $ determined from the inverse Fourier
transform of the correction term in Eq.~(\ref{eq:pot_m}).
The latter is given by
\beq
\rho_m (r) \; = \; { 1 \over 8 \pi } \, c_{\nu} \, a_0 \, M \, m^3 \,
( m \, r )^{ - {1 \over 2} (3 - {1 \over \nu}) }  
\, K_{ {1 \over 2} ( 3 - {1 \over \nu} ) } ( m \, r ) 
\label{eq:rho_vac}
\eeq
with the constant
\beq
c_{\nu} \; \equiv \; { 2^{ {1 \over 2} (5 - {1 \over \nu}) }
\over \sqrt{\pi} \, \Gamma( {1 \over 2 \, \nu} ) } \;\; .
\label{eq:rho_vac1}
\eeq
One can verify that the vacuum polarization density $\rho_m$ has the property 
\beq
4 \, \pi \, \int_0^\infty \, r^2 \, d r \, \rho_m (r) \; = \;  a_0 \, M 
\label{eq:rho_vac2}
\eeq
where the standard integral 
$\int_0^\infty dx \, x^{2-n} \, K_n (x) = 
2^{-n} \sqrt{\pi} \Gamma \left ( {3 \over 2} -n \right ) $ has been used. 
Note that the gravitational vacuum polarization distribution is singular close to $r=0$, just as in QED, Eq.~(\ref{eq:qed_s}).

The $r \rightarrow 0 $ result for $\phi (r) $ 
(discussed in the following, as an example, for $\nu=1/3$) can then
be obtained by solving the radial equation for $\phi(r)$,
\beq
{ 1 \over r} \, { d^2 \over d \, r^2 } \, \left [ \, r \, \phi (r) \, \right ] \; = \;
{ 2 \, a_0 \, M \, G \, m^3 \over \pi } \, K_0 ( m \, r ) 
\eeq
where the (modified) Bessel function is expanded out 
to lowest order in $r$, 
$K_0 ( m \, r ) = - \gamma - \ln \, \left ( { m \, r \over 2 } \right ) + O(m^2 \, r^2) $,
giving
\beq
\phi (r) \; = \; - \, { M \, G \over r } \, + \,
a_0 \, M \, G \, m^3 \,  
{ r^2 \over 3 \, \pi } \left [ - \, \ln ( { m \, r \over 2 } ) \, - \, 
\gamma + { 5 \over 6 } \, \right ] \, + \, O(r^3) 
\label{eq:phi_small1}
\eeq
where the two integration constants are matched to the general large $r$
solution
\beq
\phi(r) \; \mathrel{\mathop\sim_{ r  \, \rightarrow \, \infty  }} \; 
- \, { M \, G \over r } \, \left [ 1 \, + \, a_0 \, 
\left ( 1 \, - \, c_l \, ( m \, r )^{ {1 \over 2 \nu} - 1 } \,
e^{- m r } \right ) \, \right ]
\label{eq:phi_large}
\eeq
with $c_l = 1 / ( \nu \, 2^{1 \over 2 \nu} \, \Gamma ( {1 \over 2 \nu } )) $. 
Note again that the vacuum polarization
density $\rho_m (r)$ has the expected normalization property
\beq
4 \, \pi \, \int_0^\infty \, r^2 \, d r \, 
{  a_0 \, M \,  m^3 \over 2 \, \pi^2 } \, K_0 ( m \, r )
\; = \;  
{ 2 \, a_0 \, M \,  m^3 \over \pi } \cdot { \pi \over 2 \, m^3 } \; = \; a_0 \, M 
\eeq
so that the total enclosed additional gravitational charge is indeed just
$a_0 M$, and $G_\infty = G_0 (1 + a_0) $.

\subsubsection{Static Isotropic Solution}
\label{sec:static}

The discussion of the previous section suggests that the quantum correction due
to the running of $G$ can be described, at least in the non-relativistic
limit of Eq.~(\ref{eq:grun-k-reg}) as applied to Poisson's equation,
in terms of a vacuum energy density $\rho_m(r)$, distributed around
the static source of strength $M$ in accordance with the result of 
Eqs.~(\ref{eq:rho_vac}) and Eq.~(\ref{eq:rho_vac2}).

In general a manifestly covariant implementation of the running of $G$, 
via the $G(\Box)$ given in Eq.~(\ref{eq:grun-box}),
will induce a non-vanishing effective pressure term.
It is natural therefore to attempt to represent the vacuum polarization cloud
by a relativistic perfect fluid, with energy-momentum tensor
\beq
T_{\mu\nu} \;  = \; 
\left ( \, p \, + \, \rho \, \right ) \, u_\mu \, u_\nu
\, + \, g_{\mu \nu} \, p 
\label{eq:perfect}
\eeq
which in the static isotropic case reduces to 
\beq
T_{\mu \nu} \; = \; {\rm diag} \,
[ \, B(r) \, \rho(r), \, A(r) \, p(r) ,
\, r^2 \, p(r), \, r^2 \, \sin^2 \theta \, p(r) \, ]
\eeq
and gives a trace $T = 3 \, p - \rho $.
The $tt$, $rr$ and $\theta\theta$ components of the field equations then read 
\beq
-\lambda  B(r)+\frac{A'(r) B(r)}{r A(r)^2}
-\frac{B(r)}{r^2 A(r)}+\frac{B(r)}{r^2} \; = \; 8 \pi G  B(r) \rho (r)
\eeq
\beq
\lambda  A(r)-\frac{A(r)}{r^2}+\frac{B'(r)}{r B(r)}+\frac{1}{r^2}
\; = \; 8 \pi G  A(r) p (r)
\eeq
\beq
-\frac{B'(r)^2 r^2}{4 A(r) B(r)^2}+\lambda  r^2
-\frac{A'(r) B'(r) r^2}{4 A(r)^2 B(r)}
+\frac{B''(r) r^2}{2 A(r) B(r)}
-\frac{A'(r) r}{2 A(r)^2}
+\frac{B'(r) r}{2 A(r) B(r)} \; = \; 8 G \pi  r^2 p(r)
\eeq
with the $\varphi\varphi$ component equal to $\sin^2 \theta$ times the
$\theta\theta$ component.
Covariant energy conservation $\nabla^{\mu} \, T_{\mu\nu} =0 $ implies
\beq
\left [ \, p(r)+\rho (r) \, \right ] \frac{ B'(r)}{2 B(r)} + p'(r) \; = \; 0
\eeq
and forces a definite relationship between $B(r)$, $\rho(r)$ and $p(r)$.
The three field equations and the energy conservation equation
are, as usual, not independent, because of the Bianchi identity.
It seems reasonable to attempt to solve the above equations 
(usually considered in the context of relativistic stellar structure
(Misner Thorne Wheeler 1972))
with the density $\rho(r)$ given by the $\rho_m (r)$ of
Eq.~(\ref{eq:rho_vac}).
This of course raises the question of how the relativistic pressure $p(r)$
should be chosen, an issue that the non-relativistic calculation
did not have to address.
One finds that covariant energy conservation in fact completely
determines the pressure in the static case, leading to
consistent equations and solutions (note that in particular it would
not be consistent to take $p(r)=0$). 

Since the function $B(r)$ drops out of the $tt$ field equation,
the latter can be integrated immediately, giving
\beq
A(r)^{-1} \; = \; 1 \, - \, { 2 M G  \over r } \, - \, { \lambda \over 3 } \, r^2 
\, - \, { 8 \pi G \over r } \, \int_0^r d x \, x^2 \, \rho (x)
\label{eq:ar}
\eeq
It is natural to identify $c_1 = - 2 M G $,
which of course corresponds to the solution with $a_0=0$ ($p=\rho=0$).
Next, the $rr$ field equation can be solved for $B(r)$,
\beq
B(r) \; = \; \exp \left \{ c_2 \, - \, \int_{r_0}^r \, d y \, 
\frac{1 + A(y) \left( \lambda \, y^2 - 8 \pi G \, y^2 \, p(y) - 1 \right)}{y} 
\right \}
\label{eq:br}
\eeq
with the constant $c_2$ again determined by the requirement that the
above expression for $B(r)$ reduce to the standard Schwarzschild solution
for $a_0=0$ ($p=\rho=0$), giving 
$c_2=\ln (1 - 2 M G / r_0 - \lambda r_0^2 / 3 )$.
The last task left therefore is the determination of the pressure $p(r)$.
One needs to solve the equation
\beq
p'(r) + \frac{\left ( 8 \pi G \, r^3 \, p(r) \, + \, 2 M G \, 
- \, { 2 \over 3} \lambda r^3 \, 
+ \, 8 \pi G \, \int_{r_0}^r dx \, x^2 \rho (x) \right ) \, (p(r)+\rho (r))}
{2 \, r \left( r \, - \, 2 M G \, - \, {\lambda \over 3}\, r^3  
- 8 \pi G \, \int_0^r dx \, x^2 \, \rho (x) \right) } \; = \; 0
\label{eq:pres1}
\eeq
which is usually referred to as the equation of hydrostatic equilibrium.
From now on we will focus only the case $\lambda=0$.
The last differential equation can be solved for $p(r)$,
\beq
p_m (r) \; = \; {1 \over \sqrt{1 - \frac{2 M G}{r} } } \,
\left ( c_3 - \int_{r_0}^r dz \, 
\frac{M G \rho (z)} {z^2 \, \sqrt{1-\frac{2 M G}{z}} }
\right )
\label{eq:pres0_sol}
\eeq
where the constant of integration has to be chosen so that
when $\rho(r)=0$ (no quantum correction) one has $p(r)=0$ as well.
Because of the singularity in the integrand at $r= 2 M G$, we will
take the lower limit in the integral to be $r_0 = 2 M G + \epsilon$,
with $\epsilon \rightarrow 0$.

To proceed further, one needs the explicit form for $\rho_m(r)$, which
was given in Eq.~(\ref{eq:rho_vac}),
\beq
\rho_m (r) \; = \; {1 \over 8 \pi } \, c_{\nu} \, a_0 \, M \, m^3 \,
( m \, r )^{ - {1 \over 2} (3 - {1 \over \nu}) }  
\, K_{ {1 \over 2} ( 3 - {1 \over \nu} ) } ( m \, r ) 
\eeq
The required integrands involve for general $\nu$ the modified Bessel function $K_n(x)$, which can lead to rather complicated expressions for the
general $\nu$ case.
To determine the pressure, one supposes that it as well has a power
dependence on $r$ in the regime under consideration, 
$p_m (r) = c_p \, A_0 \, r^\gamma $, where $c_p$ is a numerical constant,
and then substitute $p_m (r)$ into the pressure equation Eq.~(\ref{eq:pres1}).
This gives, past the horizon $r \gg 2 M G$ the algebraic condition
\beq
( 2 \gamma \, - \,1 ) \, c_p \, M \, G \, r^{\gamma - 1} \, 
- \, c_p \, \gamma \, r^\gamma \, - \, M \, G \, r^{1/\nu -4} \, \simeq 0
\eeq
giving the same power $\gamma=1/\nu-3$ as for $\rho(r)$, $c_p=-1$
and surprisingly also $\gamma=0$, implying that in this regime
only $\nu=1/3$ gives a consistent solution.

The case $\nu=1/3$ can be dealt with separately,
starting from the expression for $\rho_m(r)$ for $\nu=1/3$
\beq
\rho_m (r) \; = \; {1 \over 2 \pi^2 } \, a_0 \, M \, m^3 \, 
\, K_0 ( m \, r ) 
\eeq
One has for small $r$ 
\beq
\rho_m (r) \; = \; - \, {a_0 \over 2 \pi^2 } \, M \, m^3 \, 
\left ( \ln { m \, r \over 2 } \, + \, \gamma \, \right ) \, + \, \dots
\label{eq:rho_vac_3_s}
\eeq
and consequently
\beq
A^{-1} (r) \; = \; 1 \, - { 2 \, M \, G \over r } \, + \, 
{4 \, a_0 \, M \, G \, m^3 \over 3 \, \pi } \, r^2 \, \ln \, ( m \, r ) 
\, + \, \dots
\label{eq:a_small_r3}
\eeq
From Eq.~(\ref{eq:pres1}) one can then obtain an expression for the
pressure $p_m(r)$, and one finds again in the limit $r \gg 2 M G $ 
\beq
p_m (r) \; = \; {a_0 \over 2 \pi^2 } \, M \, m^3 \, \ln \, (  m \, r ) 
\, + \, \dots
\label{eq:p_vac_3}
\eeq
After performing the required $r$ integral in Eq.~(\ref{eq:br}),
and evaluating the resulting expression in the limit $r \gg 2 M G$, one obtains
\beq
B(r) \; = \; 1 \, - \, { 2 \, M \, G \over r} \, + \, 
{4 \, a_0 \, M \, G \, m^3 \over 3 \, \pi } \, r^2 \, \ln \, ( m \, r ) 
\, + \, \dots
\label{eq:b_small_r3}
\eeq
It is encouraging to note that in the solution just obtained the running
of $G$ is the {\it same} in $A(r)$ and $B(r)$.
The expressions for $A(r)$ and $B(r)$ are consistent with a gradual slow increase in $G$ with distance, in accordance with the formula
\beq
G \; \rightarrow \; G(r) \; = \; 
G \, \left ( 1 \, + \, 
{ a_0 \over 3 \, \pi } \, m^3 \, r^3 \, \ln \, { 1 \over  m^2 \, r^2 }  
\, + \, \dots
\right )
\label{eq:g_small_r3}
\eeq
in the regime $r \gg 2 \, M \, G$,
and therefore of course in agreement with the original result of 
Eqs.~(\ref{eq:grun-k})
or (\ref{eq:grun-k-reg}), namely that the classical laboratory value of $G$ 
is obtained for $ r \ll \xi $.
Note that the correct relativistic small $r$ correction of Eq.~(\ref{eq:g_small_r3})
agrees roughly in magnitude (but not in sign) with the approximate
non-relativistic, Poisson equation result of Eq.~(\ref{eq:phi_small1}).

There are similarities, as well as some rather substantial differences,
with the corresponding QED result of Eq.~(\ref{eq:qed_s}).
In the gravity case, the correction vanishes as $r$ goes to zero: in this
limit one is probing the bare mass, unobstructed by its virtual graviton cloud.
On the other hand, in the QED case, as one approaches the source
one is probing the bare charge, whose magnitude diverges logarithmically for small $r$.

Finally it should be recalled that neither function $A(r)$ or $B(r)$
are directly related to the relativistic potential for particle orbits,
which is given instead by the combination
\beq
V_{eff} (r) \; = \; {1 \over 2 \, A(r)} 
\left [ \, \frac{l^2}{r^2} \, - \, \frac{1}{B(r)} \, + \, 1 \, \right ]
\label{eq:veff}
\eeq
where $l$ is proportional to the orbital angular momentum of the test
particle, as discussed for example in (Hartle 2005).

The running $G$ term acts in a number of ways as a local cosmological
constant term, for which the
$r$ dependence of the vacuum solution for small $r$ is fixed by the nature
of the Schwarzschild solution with a cosmological constant term.
One can therefore wonder what the solutions might look like in $d$
dimensions.
In $d \ge 4$ dimensions the Schwarzschild solution to Einstein
gravity with a cosmological term is (Myers and Perry 1986) 
\beq
A^{-1}(r) = B(r) = 
1 - 2 M G c_d \, r^{3-d} - { 2 \lambda \over (d-2)(d-1) } \, r^2
\eeq
with $c_d = 4 \pi \Gamma({d-1 \over 2}) / (d-2) \pi^{d-1 \over 2}$,
which would suggest, in analogy with the results for $d=4$ given above
that in $d \ge 4 $ dimensions only $\nu=1/(d-1)$ is possible.
This last result would also be in agreement with the exact value
$\nu=0$ found at $d=\infty$ in Sec.~\ref{sec:larged}.

\subsubsection{Cosmological Solutions}
\label{sec:cosm}

A scale dependent Newton's constant will lead to small modifications
of the standard cosmological solutions to the Einstein field
equations.
Here we will provide a brief discussion of what modifications are
expected from the effective field equations on the basis of $G(\Box)$,
as given in Eq.~(\ref{eq:gbox}), which itself originates in
Eqs.~(\ref{eq:grun-k-reg}) and (\ref{eq:grun-k}).

One starts therefore from the quantum effective field equations
of Eq.~(\ref{eq:field1}), 
\beq
R_{\mu\nu} \, - \, \half \, g_{\mu\nu} \, R \, + \, \lambda \, g_{\mu\nu}
\; = \; 8 \pi G  \, \left( 1 + A( \Box ) \right) \, T_{\mu\nu}
\label{field2}
\eeq
with $A(\Box)$ defined in Eq.~(\ref{eq:abox}).
In the Friedmann-Robertson-Walker (FRW) framework these are
applied to the standard homogeneous isotropic metric
\beq
ds^2 \; = \; - dt^2 + a^2(t) \, \left \{ { dr^2 \over 1 - k\,r^2 } 
+ r^2 \, \left( d\theta^2 + \sin^2 \theta \, d\varphi^2 \right)  \right \}
\eeq
It should be noted that there are {\it two} quantum contributions to the
above set of effective field equations. 
The first one arises because of the presence of a non-vanishing 
cosmological constant $\lambda \simeq 1 / \xi^2 $ caused by the
non-perturbative vacuum condensate of Eq.~(\ref{eq:xi_lambda}).
As in the case of standard FRW cosmology, this is expected to be 
the dominant contributions at large times $t$, and gives an exponential
(for $\lambda>0$ or cyclic (for $\lambda < 0$) expansion of the scale factor.

The second contribution arises because of the running of $G$ in the effective field equations,
\beq
G(\Box)  \; = \; G \, \left( 1 + A( \Box ) \right) \; = \; 
G \, \left [ \, 1 + a_0 \left ( \xi^2 \Box \right )^{-{ 1 \over 2 \nu }} \, + \, \dots \, \right ]
\eeq
for $t \ll \xi$, with $\nu \simeq 1/3$ and $a_0>0 $ a calculable 
coefficient of order one [see Eqs.~(\ref{eq:grun-k}) and (\ref{eq:grun-k-reg})].

As a first step in solving the new set of effective field equations, consider
first the {\it trace}
of the field equation in Eq.~(\ref{field2}),
written as 
\beq
\left( 1 \, - \, A( \Box ) \, + \, O ( A( \Box )^2 ) \right) \, R 
\; = \; 8 \pi G  \, T_{\mu}^{\;\;\mu}
\label{eq:naive_rt}
\eeq
where $R$ is the scalar curvature.
Here the operator $A(\Box)$ has been moved over
on the gravitational side, so that it now acts on functions of
the metric only, using the binomial expansion of
$1/( 1 \, + \, A( \Box ) )$.
To proceed further, one needs to compute the effect of $A(\Box)$
on the scalar curvature.
The d'Alembertian operator acting on scalar functions $S(x)$ is given by
\beq
{1 \over \sqrt{g} } \, \partial_\mu \, g^{\mu\nu} \sqrt{g} \, \partial_\nu \, S(x)
\eeq
and for the Robertson-Walker metric, acting on functions of $t$ only,
one obtains a fairly simple result in terms of the scale factor $a(t)$
\beq
- { 1 \over a^3(t) } \, { \partial \over \partial t } \left [ 
a^3(t) { \partial \over \partial t } \right ] \; F(t)
\eeq
As a next step one computes the action of $\Box$ on the scalar curvature $R$, which gives
\beq
-6\,\left[ -2\,k\,\ddot{a}(t) 
- 5\,{\dot{a}}^2(t)\,\ddot{a}(t) + a(t)\,{\ddot{a}}^2(t) 
+ 3\,a(t)\,\dot{a}(t)\,a^{(3)}(t) + a^2(t)\,a^{(4)}(t) \right] / a^3(t)
\eeq
and then $\Box^2$ on $R$ etc.
Since the resulting expressions are of rapidly escalating complexity,
one sets $a(t) = r_0 \, t^{\alpha}$,
in which case one has first for the scalar curvature itself
\beq
R \, = \, 6 \left [
{ k \over r_0^2 \, t^{2\alpha} } +
{ \alpha \left( -1 + 2 \, \alpha \right) \over t^2 } \right ]
\eeq
Acting with $\Box^n$ on the above expression
gives for $k=0$ and arbitrary power $n$ 
\beq
c_n \, 6 \, \alpha \, \left( -1 + 2\,\alpha  \right) t^{-2-2n}
\eeq
with the coefficient $c_n$ given by
\beq
c_n \; = \; 4^n 
{ \Gamma ( n + 1 ) \Gamma ( { 3 \alpha - 1 \over 2  } ) \over
\Gamma ( { 3 \alpha - 1 \over 2 } - n ) }
\eeq
Here use has been made of the relationship
\beq
\left( { d \over d \, z } \right)^{\alpha} \, \left( z \, - \, c \right)^{\beta}
\; = \; { \Gamma ( \beta + 1 ) \over \Gamma ( \beta - \alpha + 1 ) }
\, \left( z \, - \, c \right)^{\beta-\alpha}
\eeq
to analytically continue the above expressions to negative
fractional $n$ (Samko et al 1993; Zavada 1998).
For $n=-1/2\nu$ the correction on the scalar curvature
term $R$ is therefore of the form
\beq
\left [ 1 - a_0 \, c_\nu \, (t/\xi)^{1/\nu} \right ] \, \cdot \, 6 
\, \alpha \, \left( -1 + 2\,\alpha  \right) t^{-2}
\eeq
with 
\beq
c_{\nu} \; = \; 2^{-{1 \over \nu}} 
{ \Gamma ( 1 - {1 \over 2 \nu} ) \Gamma ( { 3 \alpha - 1 \over 2 } ) \over
\Gamma ( { 3 \alpha - 1 \over 2 } + {1 \over 2 \nu } ) }
\label{eq:cr}
\eeq
Putting everything together, one then obtains for the trace
part of the effective field equations
\beq 
\left [
1 - a_0 \, c_{\nu} \, \left ( { t \over \xi } \right )^{1/\nu} 
\, + \, O \left (  ( t / \xi )^{2 / \nu}  \right ) 
\right ] \, 
{ 6 \, \alpha \, \left( 2\,\alpha -1 \right)  \over t^2 } 
\; = \; 8 \pi G \, \rho (t)
\eeq
The new term can now be moved back over to the matter side
in accordance with the structure of the original effective field 
equation of Eq~.(\ref{field2}), and
thus avoids the problem of having to deal with the binomial
expansion of $1/(1 \, + \, A(\Box))$. 
One then has
\beq 
{ 6 \, \alpha \, \left( 2\,\alpha -1 \right) \over t^2 }
\; = \; 8 \pi G \, \left [ 1 + a_0 \, c_{\nu}
\, \left ( { t \over \xi } \right )^{1/\nu}
\, + \, O \left (  ( t / \xi )^{2 / \nu}  \right ) 
\right ] \, \rho (t)
\eeq
which is the Robertson-Walker metric form of Eq.~(\ref{field2}).
If one assumes for the matter density
$\rho(t) \sim \rho_0 \, t^{\beta}$, then matching powers when
the new term starts to take over at larger distances gives the first result
\beq
\beta = -2 - 1/\nu 
\label{eq:beta}
\eeq 
Thus the density decreases {\it faster} in time than the classical value
$(\beta =-2)$ would indicate.
The expansion appears therefore to be accelerating, but before reaching such
a conclusion one needs to determine the time dependence of the scale factor
$a(t)$ (or $\alpha$) as well.

One can alternatively pursue the following exercise in order to check
the overall consistency of the approach.
Consider $\Box^n $ acting on $T_{\mu}^{\;\; \mu} = - \rho(t) $ instead,
as in the trace of the effective field equation Eq.~(\ref{field2})
\beq
R \; = \; - \, 8 \pi G  \, \left( 1 + A( \Box ) \right) \, T_{\mu}^{\;\;\mu}
\eeq
for $\lambda=0$ and $p(t)=0$.
For $\rho(t) = \rho_0 \, t^{\beta}$ and $a(t) = r_0 \, t^{\alpha}$
one finds in this case
\beq
\Box^n \left( - \rho(t) \right) \; \rightarrow \; 4^n (-1)^{n+1}  
{ \Gamma ( {\beta \over 2} + 1 ) 
  \Gamma ( { \beta + 3 \alpha + 1 \over 2 } ) 
\over
  \Gamma ( {\beta \over 2} + 1 - n ) 
  \Gamma ( { \beta + 3 \alpha + 1 \over 2 } - n  )  } \; \rho_0 \, t^{\beta - 2n}
\eeq
which again implies for $n \rightarrow -1/2\nu$ the value
$\beta = -2 - 1/\nu$ as in Eq.~(\ref{eq:beta})
for large(r) times, when
the quantum correction starts to become important (since the left hand side 
of Einstein's equation always goes like $1 / t^2 $, no matter what the value
for $\alpha$ is, at least for k=0).

The next step is to examine the full effective field equations (as opposed
to just their trace part) of Eq.~(\ref{eq:naive_t}) with 
cosmological constant $\lambda=0$,
\beq
R_{\mu\nu} \, - \, \half \, g_{\mu\nu} \, R \, 
\; = \; 8 \pi G  \, \left( 1 + A( \Box ) \right) \, T_{\mu\nu}
\eeq
Here the d'Alembertian operator
\beq
\Box \; = \; g^{\mu\nu} \nabla_\mu \nabla_\nu 
\eeq
acts on a second rank tensor,
\bea
\nabla_{\nu} T_{\alpha\beta} \, = \, \partial_\nu T_{\alpha\beta} 
- \Gamma_{\alpha\nu}^{\lambda} T_{\lambda\beta} 
- \Gamma_{\beta\nu}^{\lambda} T_{\alpha\lambda} \, \equiv \, I_{\nu\alpha\beta}
\nonumber
\eea
\beq 
\nabla_{\mu} \left( \nabla_{\nu} T_{\alpha\beta} \right)
= \, \partial_\mu I_{\nu\alpha\beta} 
- \Gamma_{\nu\mu}^{\lambda} I_{\lambda\alpha\beta} 
- \Gamma_{\alpha\mu}^{\lambda} I_{\nu\lambda\beta} 
- \Gamma_{\beta\mu}^{\lambda} I_{\nu\alpha\lambda} 
\eeq
and would thus seem to require the calculation of 1920 terms,
of which fortunately many vanish by symmetry.
Next one assumes again that $T_{\mu\nu}$ has the perfect fluid form, 
for which one obtains from the action of $\Box$ on $T_{\mu\nu}$
\bea
\left( \Box \, T_{\mu\nu} \right )_{tt} \; & = & \; 
6 \, \left [ \rho (t) \, + \, p(t) \right ]
\, \left ( { \dot{a}(t) \over a(t) } \right )^2
\, - \, 3 \, \dot{\rho}(t) \,  { \dot{a}(t) \over a(t) }
\, - \, \ddot{\rho}(t) 
\nonumber \\
\left( \Box \, T_{\mu\nu} \right )_{rr} \; & = & \; 
{ 1 \over 1 \, - \, k \, r^2 } \left \{
2 \, \left [ \rho (t) \, + \, p(t) \right ] \, \dot{a}(t)^2 
\, - \, 3 \, \dot{p}(t) \, a(t) \, \dot{a}(t) 
\, - \, \ddot{p}(t) \, a (t)^2  \right \}
\nonumber \\
\left( \Box \, T_{\mu\nu} \right )_{\theta\theta} \; & = & \; 
r^2 \, ( 1 \, - \, k \, r^2 ) \, 
\left( \Box \, T_{\mu\nu} \right )_{rr}
\nonumber \\
\left( \Box \, T_{\mu\nu} \right )_{\varphi\varphi} \; & = & \; 
r^2 \, ( 1 \, - \, k \, r^2 ) \, 
\sin^2 \theta \, \left( \Box \, T_{\mu\nu} \right )_{rr}
\label{eq:boxont}
\eea
with the remaining components equal to zero.
Note that a non-vanishing pressure contribution is generated in the effective
field equations, even if one assumes initially a pressureless fluid, $p(t)=0$.
As before, repeated applications of the d'Alembertian $\Box$ to the above expressions leads
to rapidly escalating complexity,
which can only be tamed by introducing some further simplifying assumptions.
In the following we will therefore assume that 
$T_{\mu\nu}$ has the perfect fluid form appropriate
for non-relativistic matter, with a power law behavior for
the density, $\rho(t) = \rho_0 \, t^\beta$, and $p(t)=0$. 
Thus all components of $T_{\mu\nu}$ vanish in the fluid's
rest frame, except the $tt$ one, which is simply $\rho(t)$.
Setting $k=0$ and $a(t) = r_0 \, t^\alpha $ one then finds
\bea
\left( \Box \, T_{\mu\nu} \right )_{tt} \; & = & \; 
\left( 6 \, \alpha^2 - \beta^2 - 3 \, \alpha \, \beta  + \beta \right) \, \rho_0 \,
t^{\beta - 2 }
\nonumber \\
\left( \Box \, T_{\mu\nu} \right )_{rr} \; & = & \; 
2 r_0^2 \, t^{2 \alpha} \alpha^2  \, \rho_0 \, t^{\beta -2 }
\eea
which again shows that the $tt$ and $rr$ components get mixed by the
action of the $\Box$ operator, and that a 
non-vanishing $rr$ component gets generated,
even though it was not originally present.

Higher powers of the d'Alembertian $\Box$ acting on 
$T_{\mu\nu}$ can then be computed as well.
But a comparison with the left hand (gravitational) side of the effective
field equation, which always behaves like $\sim 1 / t^2 $ for $k=0$, shows
that in fact a solution can only be achieved at order $\Box^n$ 
provided the exponent $\beta$ satisfies $\beta = -2 + 2 n $, or
\beq
\beta \; = \; - 2 \, - \, 1 / \nu 
\label{eq:beta1}
\eeq
as was found previously from the trace equation,
Eqs.~(\ref{field2}) and (\ref{eq:beta}).
As a result one obtains a much simpler set of expressions, which
for general $n$ read  
\beq
\left( \Box^n \, T_{\mu\nu} \right )_{tt} \; \rightarrow \; 
c_{tt} ( \alpha, \nu ) \, \rho_0 \, t^{- 2 } 
\eeq
for the $tt$ component, and similarly for the $rr$ component
\beq
\left( \Box^n \, T_{\mu\nu} \right )_{rr} \; \rightarrow \; 
c_{rr} ( \alpha, \nu ) \, r_0^2 \, t^{2 \alpha} \, \rho_0 \, t^{- 2 } 
\eeq
But remarkably one finds for the two coefficients the simple identity
\beq
c_{rr} (\alpha, \nu ) \; = \; \third \, c_{tt} ( \alpha, \nu)
\label{eq:crr}
\eeq
as well as $ c_{\theta\theta} = r^2 \, c_{rr} $ and
$ c_{\varphi\varphi} = r^2 \, \sin^2 \theta \, c_{rr} $.
The identity $ c_{rr} \; = \; \third \, c_{tt} $
implies, from the consistency of the $tt$ and $rr$ effective
field equations at large times,
\beq
\alpha \; = \; { 1 \over 2 }
\label{eq:alpha}
\eeq
One can find a closed form expression
for the coefficients $c_{tt}$ and $c_{rr} = c_{tt}/3$
as functions of $\nu$ which are not particularly
illuminating, except for providing an explicit proof
that they exist.

As a result, in the simplest case, namely for a universe filled with non-relativistic matter ($p$=0), the effective Friedmann equations then have the following appearance 
\bea
{ k \over a^2 (t) } \, + \,
{ \dot{a}^2 (t) \over a^2 (t) }  
& \; = \; & { 8 \pi G(t) \over 3 } \, \rho (t) \, + \, { 1 \over 3 \, \xi^2 }
\nonumber \\
& \; = \; & { 8 \pi G \over 3 } \, \left [ \,
1 \, + \, c_\xi \, ( t / \xi )^{1 / \nu} \, + \, \dots \, \right ]  \, \rho (t)
\, + \, \third \, \lambda 
\label{eq:fried_tt}
\eea
for the $tt$ field equation, and
\bea
{ k \over a^2 (t) } \, + \, { \dot{a}^2 (t) \over a^2 (t) }
\, + \, { 2 \, \ddot{a}(t) \over a(t) } 
& \; = \; & - \, { 8 \pi G \over 3 } \, \left [ \, c_\xi \, ( t / \xi )^{1 / \nu} 
\, + \, \dots \, \right ] \, \rho (t) 
\, + \, \lambda
\label{eq:fried_rr}
\eea
for the $rr$ field equation.
The running of $G$ appropriate for the Robertson-Walker metric,
and appearing explicitly in the first equation, is given by
\beq
G(t) \; = \; G \, \left [ \; 1 \, + \, c_\xi \, 
\left ( { t \over \xi } \right )^{1 / \nu} \, + \, \dots \, \right ]
\label{eq:grun_frw}
\eeq
with $c_\xi$ of the same order as $a_0$ of Eq.~(\ref{eq:grun-k}).
Note that the running of $G(t)$ induces as well an effective pressure term in the second 
($rr$) equation.
\footnote{
We wish to emphasize that we are {\it not} talking here about models with a
time-dependent value of $G$.
Thus, for example, the value of $G \simeq G_c$ at laboratory scales should be taken to
be constant throughout most of the evolution of the universe.} 
One has therefore an effective density given by
\beq
\rho_{eff} (t) \; = \; { G(t) \over G } \, \rho (t)
\label{eq:rho_eff}
\eeq
and an effective pressure
\beq
p_{eff} (t) \; = \; 
{ 1 \over 3 } \, \left ( { G(t) \over G } \, - \, 1 \right ) \, \rho (t) 
\label{eq:p_eff}
\eeq
with $ p_{eff} (t) / \rho_{eff} (t) = \third ( G(t) - G ) / G(t) $.
Strictly speaking, the above results can only be proven if one assumes that
the pressure's time dependence is given by a power law.
In the more general case, the solution of the above equations for various choices of
$\xi$ and $a_0$ has to be done numerically.
Within the FRW framework, the gravitational vacuum polarization term behaves therefore
in some ways (but not all) like a positive pressure term, with 
$p(t) = \omega \, \rho(t)$ and
$\omega=1/3$, which is therefore characteristic of radiation.
One could therefore visualize the gravitational vacuum polarization contribution
as behaving like ordinary radiation, in the form of a dilute virtual
graviton gas: a radiative fluid with an equation of state $p={1 \over 3} \rho$.
But this would overlook the fact that the relationship between
density $\rho (t)$ and scale factor $a(t)$ is quite
different from the classical case.

The running of $G(t)$ in the above equations follows directly
from the basic result of Eq.~(\ref{eq:grun-k}), following 
the more or less unambiguously defined sequence
$G(k^2) \rightarrow G(\Box) \rightarrow G(t)$.
For large times $t \gg \xi$ the form of Eq.~(\ref{eq:grun-k}), and therefore
Eq.~(\ref{eq:grun_frw}), is no longer appropriate, due
to the spurious infrared divergence of Eq.~(\ref{eq:grun-k}) at
small $k^2$.
Indeed from Eq.~(\ref{eq:grun-k-reg}), the infrared regulated version of 
the above expression should read instead
\beq
G(t) \; \simeq \; G \, \left [ \; 1 \, 
+ \, c_\xi \, \left ( { t^2 \over t^2 \, + \, \xi^2 } \right )^{1 \over 2 \nu} \, 
+ \, \dots \; \right ]
\label{eq:grun_frw_m}
\eeq
For very large times $t \gg \xi$ the gravitational
coupling then approaches a constant, finite value 
$G_\infty = ( 1 + a_0 + \dots ) \, G_c $.
The modification of Eq.~(\ref{eq:grun_frw_m}) should apply whenever one considers
times for which $t \ll \xi$ is not valid. 
But since $\xi \sim 1 / \sqrt{\lambda}$ is of the order the size of the visible
universe, the latter regime is largely of academic interest.

It should also be noted that the effective Friedmann equations of Eqs.~(\ref{eq:fried_tt}) 
and (\ref{eq:fried_rr}) also bear a superficial degree of resemblance to what
might be obtained in some scalar-tensor theories of gravity, where the
gravitational Lagrangian is postulated to be some singular function of the scalar curvature (Capoziello et al 2003; Carroll et al 2004; Flanagan 2004).
Indeed in the Friedmann-Robertson-Walker case one has, for the scalar 
curvature in terms of the scale factor,
\beq
R \; = \; 6 \left( k \, + \, {\dot{a}}^2(t) \, + \, a(t) \, \ddot{a}(t) \right) / a^2(t)
\eeq
and for $k=0$ and $a(t) \sim t^{\alpha}$ one has
\beq
R  \; = \; { 6 \, \alpha ( 2 \, \alpha - 1 ) \over t^2 } 
\eeq
which suggests that the quantum correction in Eq.~(\ref{eq:fried_tt}) is,
at this level, nearly indistinguishable from an inverse curvature term of the type
$ ( \xi^2 \, R )^{-1 / 2 \nu }$, 
or $ 1/( 1 \, + \, \xi^2 R )^{1 / 2 \nu }$ if one uses the infrared regulated version.
The former would then correspond the to an effective gravitational action
\beq
I_{eff}  \; \simeq \; { 1 \over 16 \pi G } \int dx \, \sqrt{g} \,
\left ( \, R \, + \, 
{ f \, \xi^{- {1\over\nu} } \over | R |^{{1\over 2 \nu} -1} } 
\, - \, 2 \, \lambda \, \right )
\label{eq:scalar-tens}
\eeq
with $f$ a numerical constant of order one, and $\lambda \simeq 1 / \xi^2 $.
But this superficial resemblance is seen here more as an artifact, due to the 
particularly simple form of the Robertson-Walker metric, with the coincidence
of several curvature invariants not expected to be true in general.
In particular in Eqs.~(\ref{eq:fried_tt}) and (\ref{eq:fried_rr})
it would seem artificial and in fact inconsistent to take 
$\lambda \sim 1 / \xi^2$ to zero
while keeping the $\xi$ in $G(t)$ finite.

\section{CONCLUSIONS and OUTLOOK}
\label{sec:conclusions}

While it is certainly possible that traditional quantum-field theoretic
approaches to quantum gravitation might ultimately fail, it is not clear yet
from the evidence so far that such a conclusion is warranted.
There are certainly aspects of gravity that make it unique among
theories of fundamental interactions, such as the geometric interpretation
and its possible connection with the ultimate nature of space-time. 
These aspects might or might not play a fundamental role in establishing
the theoretical consistency of the quantum theory.
In a more traditional field-theoretic investigation of gravity
the scope might have to be limited as well: the theory might not provide
any new deep insights into fundamental questions (such as why the gravitational
couplings take on particular values), as is the case already in $QED$.
Indeed the following discussion is not incompatible with the belief
that present gravitational theories should eventually be replaced by 
more fundamental ones at distances comparable to the ultraviolet cutoff scale.

The theory of non-renormalizable interactions shows that the usual
approach to quantum gravity, based on straight perturbation theory
in the gravitational coupling applied to four dimensions is essentially flawed,
and leads to fundamentally misleading answers.
The $2+\epsilon$ expansion approach to gravity provides entirely new 
insights into what is most likely the true ground state of the theory,
but faces a tremendous challenge in reaching the physical case of
dimensions four, and furthermore gives few hints as to what the true
nature of the strong coupling ground state arising for $G>G_c$ might be.

Due to its inherent complexity, the lattice approach is generally not 
particularly well suited for analytical investigations; even the simple
task of connecting the lattice weak field limit with the corresponding
continuum result represents a small algebraic tour de force.
Yet the lattice theory appears to provide a concrete and constructive 
proof for the existence of the gravitational functional integral, at least
in the regime where it exists, and perhaps the only reliable framework in
which the issue of the gravitational measure can be properly addressed. 

The existence of a small set of clear, unambiguous analytical results from the $2+\epsilon$ expansions provides the opportunity, as is the case already
in the non-linear sigma model, of using the lattice theory to address a
set of basic issues directly in four dimensions, relying at the same time
on a pre-existing limited theoretical framework of scaling relations and
running coupling scenarios.
The natural underlying assumption is therefore that the four-dimensional
results are qualitatively similar to the analytical results, but
with somewhat different amplitudes and exponents.
The numerical evidence so far suggests that such an identification
is warranted, and is in fact at least up to now completely consistent.

At the same time, the lattice theory provides new essential ingredients,
such as the non-existence (in the Euclidean field theory framework) of 
the weak coupling phase, and the appearance of a renormalization group invariant gravitational correlation length associated with large scale curvature.
Furthermore the specific values of the critical exponents in four
dimensions suggest possible scenarios (such as a non-local effective theory)
by which new non-trivial analytical results in four dimensions might
be obtained. 
In any case, there seems to be a clear prediction that gravitational
couplings will be scale dependent.
One can in fact re-phrase the last sentence even more strongly, to
the effect that it would seem very difficult to accommodate in a
quantum theory of gravity couplings that are not scale dependent.

One might view the non-perturbative lattice results in many respects
as quite unsatisfactory.
Since the lattice theory does not fix the correlation length $\xi$,
the latter remains completely undetermined and has to be fixed
by physical considerations.
Since it relates to curvature, it is natural to take it to be very large
so as to recover agreement with observation.
Consequently there is no clear explanation for the smallness of the cosmological 
constant, which instead has to be tuned to its physical value.
Furthermore, since short-distance cutoff physics essentially
decouples from universal long distance physics, there is no
indication of what specific cutoff mechanism might be operative
at short distances.

There is also the possibility of clear disagreements between theoretical
predictions and observation.
In the strong coupling, physical phase the average curvature for infinitesimal
loops is negative, corresponding to an apparent Euclidean anti-de Sitter phase
at very short distances, comparable to the ultraviolet cutoff.
What happens for large loops is a more difficult and largely open questions,
although there are indications that for large loops the average
curvature is positive instead.
In any case it is not clear yet that the sign of the curvature is a 
truly universal quantity. 
Since the recent distant supernova observations suggest a positive
cosmological constant, there is potential for serious conflict.
The analysis gets clouded further by the fact that in the strong coupling
phase a growing gravitational coupling at large distance 
(corresponding to gravitational anti-screening) leads to cosmic
acceleration and would therefore in part mimic the effects of a
positive cosmological constant.

Could then the weak coupling phase still be physical, in spite of the fact
that it appears pathological in the Euclidean lattice theory?
In principle, as a last measure, such a phase could be reached by 
analytic continuation from the strong coupling phase of the Euclidean theory 
(it is easy to see that for example the $\beta$-function is
expected to be analytic at the fixed point), and the same exponent
$\nu$ would apply to both sides of the transition, as in the non-linear
sigma model.
In this gravitational ``Coulomb'' phase the correlation length 
$\xi$ would be infinite,
the cosmological constant presumably identically zero, and
one would expect gravitational screening on some scale. 
How one would avoid the conclusion that in this phase space-time
is essentially two-dimensional is unclear, at least
to the author.

\begin{acknowledgments}

I would like to thank first of all Ruth Williams for her collaboration on the
many topics covered in this review.
In addition I have benefited over the years from discussions with James Bjorken,
Thibault Damour, Stanley Deser, James Hartle, Giorgio Parisi, Mike Peskin, Tullio Regge, Gabriele Veneziano and Jean Zinn-Justin 
on topics related to the contents of this review, and many other colleagues.
I wish to thank in particular Hermann Nicolai, Robert Schrader and Stefan Theisen for their comments on the manuscript.

This work was supported in part by a research grant from the Max Planck Gesellschaft zur F\"orderung der Wissenschaften.
The author wishes to thank Hermann Nicolai and the 
Max Planck Institut f\"ur Gravitationsphysik (Albert-Einstein-Institut)
in Potsdam for a very warm hospitality.

\end{acknowledgments}






\end{document}